\documentclass{aastex63}

\usepackage{graphicx}
\usepackage{amsmath}

\submitjournal{ApJS}

\shorttitle{Nondiffusive grain-surface and ice chemistry in hot cores}
\shortauthors{Garrod et al.}
\begin{document}

\title{Formation of complex organic molecules in hot molecular cores through nondiffusive grain-surface and ice-mantle chemistry}

\correspondingauthor{Robin T. Garrod}
\email{rgarrod@virginia.edu}

\author[0000-0001-7723-8955]{Robin T. Garrod}
\affiliation{Department of Chemistry, University of Virginia, Charlottesville, VA, 22904, USA}
\affiliation{Department of Astronomy, University of Virginia, Charlottesville, VA, 22904, USA}

\author[0000-0002-4801-436X]{Mihwa Jin}
\affiliation{Department of Chemistry, University of Virginia, Charlottesville, VA, 22904, USA}

\author{Kayla A. Matis}
\affiliation{Department of Chemistry, University of Virginia, Charlottesville, VA, 22904, USA}
\affiliation{Department of Chemical Engineering, University of Virginia, Charlottesville, VA, 22904, USA}

\author[0000-0002-1096-9881]{Dylan Jones}
\affiliation{Department of Chemistry, University of Virginia, Charlottesville, VA, 22904, USA}
\affiliation{Department of Astronomy, University of Virginia, Charlottesville, VA, 22904, USA}

\author[0000-0002-7475-3908]{Eric R. Willis}
\affiliation{Department of Chemistry, University of Virginia, Charlottesville, VA, 22904, USA}

\author[0000-0002-4649-2536]{Eric Herbst}
\affiliation{Department of Chemistry, University of Virginia, Charlottesville, VA, 22904, USA}
\affiliation{Department of Astronomy, University of Virginia, Charlottesville, VA, 22904, USA}

\begin{abstract}

A new, more comprehensive model of gas-grain chemistry in hot molecular cores is presented, in which nondiffusive reaction processes on dust-grain surfaces and in ice mantles are implemented alongside traditional diffusive surface/bulk-ice chemistry. We build on our nondiffusive treatments used for chemistry in cold sources, adopting a standard collapse/warm-up physical model for hot cores. A number of other new chemical model inputs and treatments are also explored in depth, culminating in a final model that demonstrates excellent agreement with gas-phase observational abundances for many molecules, including some (e.g.~methoxymethanol) that could not be reproduced by conventional diffusive mechanisms. Observed ratios of structural isomers methyl formate, glycolaldehyde and acetic acid are well reproduced by the models. The main temperature regimes are identified in which various complex organic molecules (COMs) are formed. Nondiffusive chemistry advances the production of many COMs to much earlier times and lower temperatures than in previous model implementations. Those species may form either as by-products of simple-ice production, or via early photochemistry within the ices while external UV photons can still penetrate. Cosmic ray-induced photochemistry is less important than in past models, although it affects some species strongly over long timescales. Another production regime occurs during the high-temperature desorption of solid water, whereby radicals trapped in the ice are released onto the grain/ice surface, where they rapidly react. Several recently-proposed gas-phase COM-production mechanisms are also introduced, but they rarely dominate. New surface/ice reactions involving CH and CH$_2$ are found to contribute substantially to the formation of certain COMs.

\end{abstract}


\section{Introduction} \label{sec:intro}

Complex organic molecules (COMs), typically defined as carbon-bearing species composed of six or more atoms in total \citep{HvD09}, are now commonly detected toward hot, star-forming cores via their gas-phase rotational emission spectra. The means by which such chemical species are formed continue to be the subject of great interest and debate \citep[e.g.][]{Jorgensen20}. The thermal desorption of dust-grain ice mantles, composed of a selection of simple molecules such as water, CO, CO$_2$ and NH$_3$, as well as the nominally complex molecule methanol (CH$_3$OH), is generally accepted as being integral to the introduction of COMs into the gas phase in hot, star-forming cores. However, the degree to which COMs are produced on the grains themselves, rather than through gas-phase chemistry promoted by the release of simple molecules into the gas, is an area of particular contention.

A leading theory as to the origins of COMs in hot cores is the idea that reactive radicals, produced mainly through cosmic ray-induced photodissociation of the simple ices, could become mobile on the grain surfaces (or on the surface of the ice that forms thereon) as the dust is warmed by the nascent protostar \citep{GH06,GWH08}. Kinetic models of diffusion-driven grain-surface chemistry suggested that the elevated temperatures (i.e.~$\gtrsim$20~K) achieved during the star formation process were essential to COM production. A gradual warm up of the grains allowed the appropriate temperatures to be reached for each species of radical to become mobile, while also allowing a sufficient period of time during which those radicals could be produced through photodissociation. More recent implementations moved from a pure surface chemistry to a so-called three-phase treatment, in which diffusion within the bulk ice (i.e.~in material beneath the ice surface) would allow species in the ice mantles to meet and react \citep{Garrod13a} without requiring them to be present on a surface; bulk-ice species would instead be allowed to diffuse (more slowly), through a three-dimensional random walk within the ice. These thermally-driven diffusive grain chemistry models showed a great deal of success in broadly reproducing observed COM abundances, as well as their relative ratios, although this was dependent on the timescale of the warm-up. 

However, a particular challenge to these models was their inability to reproduce the observed fractional abundance ratios of the important structural isomers methyl formate (HCOOCH$_3$), glycolaldehyde (CH$_2$OHCHO) and acetic acid (CH$_3$COOH); glycolaldehyde was typically overproduced in the ice mantles prior to desorption. Observational comparisons by, e.g., \citet{Taquet16} also suggested an underproduction of methyl formate with respect to methanol in the models; those authors proposed alternative gas-phase mechanisms for methyl formate production, in which the {\em trans}-conformer of protonated methyl formate would be produced through ion-molecule chemistry, to be followed by proton transfer to ammonia that was assumed to lead efficiently to the commonly observed {\em cis}-methyl formate.

Evidence from two other directions has also challenged the idea that a thermally-driven diffusive grain chemistry is necessary for COM production in the interstellar medium. The first is the observation of methyl formate, dimethyl ether (CH$_3$OCH$_3$) and acetaldehyde (CH$_3$CHO) -- three molecules typically associated almost exclusively with hot-core chemistry -- toward pre-stellar cores and other cold environments \citep{Bacmann12,Cernicharo12,Vastel14,JS16}. These molecules did not immediately appear to have sufficiently rapid gas-phase formation mechanisms to explain their presence. In the absence of gas-phase routes, production of those molecules would require efficient grain-surface reactions at very low temperatures ($\sim$10~K), as well as some non-thermal mechanism to desorb the molecules \citep{Cernicharo12,Vastel14}. New gas-phase reactions between radicals, including radiative association processes such as CH$_3$ + CH$_3$O $\rightarrow$ CH$_3$OCH$_3$ + h$\nu$ \citep[][see also Sec. \ref{sec:methods:gas-phase}]{VH13,Balucani15}, have been proposed and have shown some degree of success, although grain-surface routes continue to be investigated \citep{Vasyunin17,Jin20}.

Secondly, laboratory evidence now shows that the hydrogenation of CO on cold surfaces by atomic H leads not only to formaldehyde and methanol production, but to the formation of COMs, through the recombination of radicals intermediate to CO and CH$_3$OH \citep{Fedoseev15,Chuang16}. The molecules so produced may also themselves react with H to alter their state of hydrogenation via hydrogen-atom addition or abstraction. The production of the COM backbone proceeds through a nondiffusive mechanism whereby, on occasion, radicals are produced either adjacent or very nearby to each other, allowing them to react rapidly while minimizing the necessity for higher temperatures to induce diffusion. Further experimental studies suggest that yet more complex and potentially biologically-relevant molecules, including sugars and amino acids, may be produced by similar means \citep{Fedoseev17,Ioppolo20}.

Other laboratory evidence, meanwhile, suggests that bulk diffusion of molecules and/or radicals may not be very efficient \citep{Theule20}, so that even with elevated dust/ice temperatures, diffusion-driven reactions in the bulk ice would be slow. Such apparent bulk diffusion as is observed in experiments may instead be indicative of diffusion within pores or the release of trapped species through cracking of the ice \citep{Ghesquiere18}. Even so, some mechanism that allows radicals produced within the bulk ice to react with each other would appear to be necessary in a theoretical treatment of ice chemistry; it is well known that UV irradiation of organic ices can readily produce COMs such as are observed in the ISM, even at low temperatures \citep[e.g.][]{Oberg09,Schneider19}. The question instead is whether diffusion is required as the mediator of such chemistry in the bulk ices, or whether some other, nondiffusive processes may allow reactions to occur.

Over recent years, treatments for nondiffusive grain-surface reactions have been introduced sporadically into astrochemical models for specific purposes. \citet{GP11} first considered nondiffusive reactions between OH and CO as a solution to the underproduction of CO$_2$ on the grains at low temperatures. In this scenario, an oxygen atom on the grain surface would be found by a diffusive H atom, reacting to produce the radical OH. If the O atom happened to be in contact with a CO molecule at that moment, as determined by a probability based on the overall surface coverage of CO, then OH and CO could react without either of them having to diffuse; thus, the only species required to diffuse to produce CO$_2$ on the grains was atomic H. More recently, \citet{Dulieu19} used a similar approach to model laboratory production of formamide on a surface, based on a reaction between the radical H$_2$NO and formaldehyde (H$_2$CO), in which reaction would proceed when the newly-formed H$_2$NO was in contact with formaldehyde. Those authors found that this nondiffusive mechanism was necessary to explain their experimental results in which atomic H was the only species expected to be mobile on a mixed NO/H$_2$CO surface. The consideration of these nondiffusive processes in rate-based astrochemical models requires the deliberate incorporation of reaction rates that are mathematically distinct from the typical second-order rate equations used for diffusive chemistry.

Although diffusive surface reactions may induce nondiffusive follow-on reactions, as described above for CO$_2$, processes other than diffusion or a preceding reaction may also act to bring reactants together. \citet{Garrod19}, modeling bulk-ice photochemistry in comets, presented a treatment for nondiffusive reactions initiated by photodissociation. The production of molecular photofragments in the presence of a reactive radical allowed a reaction to occur spontaneously, without the need for thermally-activated diffusion.

\citet[][hereafter, JG20]{Jin20} set out a generic formulation to incorporate nondiffusive mechanisms of all kinds into astrochemical kinetics models. Within this framework they tested a selection of new nondiffusive grain-surface/ice-mantle processes in their simulations of cold, pre-stellar core chemistry, with a goal of reproducing observational gas-phase COM abundances. In their models, reactive desorption of surface-produced molecules was responsible for the ejection of COMs into the gas phase; the chemical energy released by surface reactions would allow the product to desorb in a fraction of cases \citep{Garrod07}. 

JG20 included what they called {\em three-body} (3-B) reactions in their surface/ice chemical network, which occur when the product of an initiating reaction happens to be close enough to a potential reaction partner that no further diffusion is required for them to react; this mechanism is essentially the one that has been put forward by Fedoseev and others as a means of forming COMs, based on laboratory evidence, and indeed corresponds to that used by \citet{GP11} to explain the production of CO$_2$ on grains. Such reaction mechanisms are implicitly present in microscopic Monte Carlo kinetic models, and have been found to result in COM formation in simulations of dark-cloud grain chemistry \citep{Chang14}.

In a few selected cases, the JG20 model also allowed the vibrationally-excited product of the initiating reaction to overcome a subsequent reaction barrier (which they termed a {\em three-body excited-formation} reaction, 3-BEF). This was especially important for the production of methyl formate (HCOOCH$_3$), whereby the reaction of H with H$_2$CO would produce excited CH$_3$O that could overcome the barrier to reaction with abundant surface CO, to produce the radical CH$_3$OCO. This radical could then recombine with an H atom to produce methyl formate. This process did not need to be highly efficient; they found an efficiency of 0.1\% (per CH$_3$O formed in the presence of CO) would result in enough methyl formate in the gas phase to reproduce the observational abundance. A statistical treatment of the partition of the energy amongst the available vibrational modes of the excited species appears to provide an acceptable explanation for the variation in efficiency between different combinations of initiating and subsequent reactions.

Following \citet{Garrod19}, the model of JG20 also included nondiffusive reactions induced by the instantaneous formation of UV photodissociation products nearby to potential reaction partners (labeled {\em photodissociation-induced} reactions, PDI). They found that solid-phase abundances of COMs were enhanced within the mantle through this process, even at extremely low ($<$10~K) temperatures. This enhancement did not strongly affect gas-phase abundances, as COMs produced within the mantle would have no easy way to reach the surface layer and desorb, while molecular production rates for this mechanism in the surface layer itself were too weak to influence significantly the gas-phase abundances. They also found that O$_2$ and related species were strongly enhanced in the mantle by this process, and suggested that it could contribute to the large abundances of O$_2$ found in comets.

While the JG20 models focused mainly on the dust-grain {\em surface} chemistry in pre-stellar cores as a means to enrich the cold, gas-phase molecular content, the complete desorption of the ice mantles that occurs during the hot-core stage means that any nondiffusive chemistry in the bulk ice may also be highly influential in the ultimate gas-phase abundances of molecules toward those hot sources. A full treatment of hot-core chemistry must therefore consider all of the new nondiffusive mechanisms that could operate either on the grain/ice surfaces or within the bulk ice mantles, throughout the history of the source.

Here, in order to allow the chemical modeling of hot cores to reflect the most recent experimental and observational evidence for COM production at low temperatures, we present new gas-grain (three-phase) chemical kinetics models in which the nondiffusive mechanisms introduced by JG20 are applied individually or in concert during the collapse and warm-up stages of hot core evolution. The theoretical treatment of the PDI and 3-BEF processes in particular are also further refined beyond those proposed by JG20. The work presented here culminates in a far more comprehensive modeling treatment of the cold and hot chemistry of hot cores than has been attempted before.

While all of the new nondiffusive mechanisms introduced here involve alternative means for reacting species to meet each other (rather than through surface or bulk diffusion), the network of chemical reactions mediated by those mechanisms is broadly similar to that used in previous gas-grain simulations of hot core chemistry. However, several other heretofore untested chemical reactions for certain key species are explored in the present work. A number of pre-existing features of the chemical models are also adjusted and tested. In order to contrast with the effects of the nondiffusive surface processes, we expand our chemical network to include several gas-phase mechanisms for COM production that have been proposed over the past few years. 

Due to the novelty of the model and the breadth of processes it encompasses, we endeavor here to present a correspondingly comprehensive investigation of its various features, and to lay the ground-work for its broader application in future. The paper has been structured to be as modular as possible, to allow the reader to sample different aspects of the modeling selectively. However, particular attention is paid to the results and discussion of the {\tt final} model setup, which represents the culmination of the modeling effort described herein.

The methods employed in 
each of the models are detailed in Sec.~\ref{sec:methods}. The results of the full model grid up to the {\tt final} model are presented in Sec.~\ref{sec:results}; the results of the {\tt final} model, which includes all of the new mechanisms and updates, may be found in Sec.~\ref{sec:results:final}. A description of the temperature-dependent formation of COMs in hot cores using the {\tt final} model is given in Sec.~\ref{results:temps}. A comparison of {\tt final} model results with observational data is provided in  Sec.~\ref{results:obs}. Sec.~\ref{disc} provides a discussion of all of the model findings, with the main Conclusions given in Sec.~\ref{concs}.
Descriptions of certain minor changes to the model methodology, as well as some additional figures and data, are provided in the Appendices.

Readers interested mainly in the results of the finalized chemical models and their comparison with observations are recommended to begin at Sec.~\ref{sec:results:final}.

\section{Methods} \label{sec:methods}

All of the simulations presented here use the three-phase astrochemical model {\em MAGICKAL} \citep{Garrod13a}, to which various additions and alterations are made as described below and in the following subsections. The model uses a system of rate equations, solved using the Gear algorithm, to simulate the coupled gas-phase, grain/ice-surface and bulk-ice mantle chemistry occurring during a two-stage physical treatment of a hot core. The modified-rate method presented by \citet[][``method C'']{Garrod08} is used, which allows stochastic behavior in the surface chemistry to be approximated where required. Grain-surface reaction rates incorporate the back-diffusion (random walk) treatment of \citet{Willis17}. A canonical grain radius of 0.1~$\mu$m is assumed; the number of surface sites per grain is $10^6$.

Grain-surface chemical abundances are coupled with the gas through accretion (adsorption) of atoms and molecules onto the grains, and through both thermal and nonthermal desorption of chemical species back into the gas. The latter includes chemical desorption, in which a newly-formed chemical reaction product may spontaneously desorb from the grain with some small probability; here, the efficiency is determined by the RRK treatment described by \citet{Garrod07}, using a maximum efficiency of 1\%. Photodesorption is also included, but is generally of little importance in these hot core models.

The surface layer (of maximum thickness 1 monolayer) is coupled to the bulk ice phases through the net rate of gain/loss of material to/from the surface (related to the net accretion/desorption rate from/to the gas phase), which either acts to cover up the existing surface, thus rendering it part of the bulk ice, or to expose underlying bulk-ice material, rendering it part of the new surface \citep[see][]{Hasegawa93,GP11}. Bulk diffusion is also considered for some or all species (see Sec.~\ref{sec:methods:bulk-diff}).

The chemical network (including grain-surface parameters) used in all of the models is based on that presented by \citet{Willis20} for isocyanide chemistry, and which includes all the main additions since the butyl cyanide network of \citet{Garrod17}. The present network also includes the methyl isocyanate and N-methyl formamide chemistry of \citet{Belloche17}, and the sulfur chemistry of \citet{Muller16}. Also included are the reactions added by JG20; of particular note is the addition of a hydrogen-abstraction route for dimethyl ether on dust grains. The present network further includes a set of as-yet unpublished reactions for the gas-phase and grain-surface/ice chemistry of ethylene oxide (c-C$_2$H$_4$O) and vinyl alcohol (C$_2$H$_3$OH); see Appendix A for details of these and several other small additions to the chemical network. Initial elemental abundances are those used by \citet{Garrod13a}.

The cosmic-ray ionization rate assumed in all the present models is the standard value used in most past models ($\zeta = 1.3 \times 10^{-17}$ s$^{-1}$), which is applied uniformly under all conditions. Recent simulations \citep{Bonfand19, Barger20, Willis20} have found that the fidelity of model results to observations is typically improved by the use of somewhat elevated and/or extinction-dependent values of $\zeta$ for hot cores such as Sgr B2(N); the standard value is, however, retained here for comparability with earlier models. As well as affecting atoms and molecules in the gas phase, cosmic rays (CR) may also influence the grain-surface and ice-mantle chemistry, through the photodissociation of solid-phase molecules by the secondary UV field induced by CR collisions with gas-phase H$_2$. The photodissociation branching ratios used for the grain-surface chemistry are assumed to be the same as those in the gas phase; the total rates of those processes are initially assumed also to follow the gas-phase values, although this is later adjusted (see Section \ref{sec:results:loPD}).

The two-stage physical treatment used in all models is the same as in our past hot-core models, i.e.~a cold, free-fall collapse from a density $n_\mathrm{H} = 3 \times 10^3$ cm$^{-3}$ to a final value of $2 \times 10^8$ cm$^{-3}$ (nominally appropriate to conditions in Sgr B2N), followed by a static (fixed-density) warm-up from 8 to 400~K. In the first stage, the gas temperature is held at 10~K while the dust temperature falls as a function of the visual extinction following the method of \citet{GP11}, from $\sim$16 to 8~K (or $\sim$14.7 to 8~K), for an initial visual extinction $A_{V,{\mathrm{init}}}=2$ (or 3, see Sec.~\ref{sec:results:extinction}). In the second stage (i.e.~the ``warm-up''), the gas and dust temperatures increase in tandem once the dust temperature exceeds the initial gas temperature of 10~K, on the assumption that the two are well coupled.

For each chemical model setup, a single collapse-stage simulation is run; from this chemical starting point, three separate warm-up stage models are then run, corresponding to a {\em fast}, {\em medium} or {\em slow} warm-up from 8 to 400~K. Following \citet{GH06}, characteristic warm-up timescales of $5 \times 10^{4}$ yr ({\em fast}), $2 \times 10^{5}$ yr ({\em medium}), and $1 \times 10^{6}$ yr ({\em slow}) are used, which technically correspond to the time taken to reach 200~K. Each model continues until a temperature of 400~K is reached.

Starting from the {\tt basic} model, various changes and additions are sequentially applied, as described in the following subsections or as outlined briefly in Sec.~\ref{sec:results} along with their results. Most model results are presented as tabulated final or maximum abundances of selected molecules. Due to the many differences between models, it is not possible to compare every single additional process with every other. However, the work presented here is not intended to {\em choose} between the different models, but rather to document and exhibit their progression to include a much larger and more accurate range of processes than has been used before. 

Major changes and additions made to the {\tt basic} model include: change to the generic sticking coefficient; reduction of solid-phase photodissociation efficiency; change to the treatment of binding energies as influenced by surface H$_2$ content; adoption of ammonia-related proton-transfer reactions in the gas-phase network; restriction of bulk diffusion in ice mantles; Eley-Rideal processes; photodissociation-induced (PDI) surface/bulk reactions; three-body (3-B) surface/bulk reactions; three-body excited formation (3-BEF) surface/bulk reactions; adjustments to atomic H, O, N and generic surface-diffusion barriers; tunneling-mediated bulk diffusion; addition of new gas-phase reactions for COM formation; and addition of new grain-surface/bulk reactions for COM formation. Table \ref{models} indicates which model uses each new method or condition.

\subsection{Basic model} \label{sec:methods:basic}

Aside from a few changes detailed here or in Appendix B, the operation of the {\tt basic} model is well described by \citet{Garrod17}. In this model, chemistry may occur in any of the three chemical phases; gas, grain-surface, and bulk-ice. Grain/ice-surface and bulk-ice reactions in the {\tt basic} model are mediated solely by the diffusion of reactants; on surfaces in particular, this diffusive reaction process is commonly known as the Langmuir-Hinshelwood (L-H) mechanism. Diffusion rates for each species are determined by their diffusion barriers. Surface diffusion barriers are taken uniformly to be $E_{\mathrm{dif}}(i)=0.35 E_{\mathrm{des}}(i)$ for all species, where $E_{\mathrm{des}}(i)$ is the binding energy of species $i$. The {\em basic} model also allows reactions to occur in the bulk, mediated by the bulk-diffusion barriers, which are assumed to be twice the surface-diffusion values, following \citet{Garrod13a}. All of the same reactions occurring in the surface network are allowed to occur in the bulk ice. In cases where reactions have activation energies, their success is dependent on the competition between reaction and diffusion of the species in question, as per \citet{GP11}. If diffusion is slow, the effect of a chemical barrier may be small, even if the barrier is substantial. 

As well as leading to reaction, bulk diffusion also allows atoms or molecules in the bulk ice gradually to reach the surface through a series of diffusion events (hops). This mechanism is especially important in allowing H and H$_2$, produced through photo-dissociation of larger species, to escape the bulk ice. This mechanism uses the random-walk treatment given by \citet{Garrod17}, but is otherwise governed by the same diffusion barriers as the bulk reactions.

\citet{GP11} successfully described the production of solid-phase CO$_2$ on grain surfaces, through the inclusion of a nondiffusive (three-body) treatment for the grain-surface reaction OH + CO $\rightarrow$ CO$_2$ + H. This method is used in the {\tt basic} model, with a comparable mechanism operating in the bulk ice. In subsequent models in which three-body reactions on the grains are treated through a generic approach, the special method is not required.

\subsection{Sticking coefficient} \label{sec:methods:sticking}

Following earlier implementations of gas-grain chemical kinetics using the Ohio State chemical code, {\em MAGICKAL} has until now assumed a blanket sticking coefficient of 0.5 for neutral gas-phase species. Recent evidence \citep{He16} suggests that a value close to unity is more appropriate at the $\sim$10~K temperatures at which most of the atomic and molecular accretion occurs in the models, although a value of around 0.6--0.7 may be better for H$_2$ at these temperatures. In models \texttt{bas\_stk} and beyond, the sticking coefficient is increased to unity for all species.

\subsection{Influence of H$_2$ content on binding energies and diffusion barriers} \label{sec:methods:H2}

\citet{GP11} introduced an adjustment to the binding energies of all surface species, and consequently their diffusion barriers, based on the instantaneous fractional H$_2$ content of the surface layer, $\theta$(H$_2$). Due to the weaker binding to H$_2$, the inclusion of some H$_2$ content would mean faster diffusion and thermal desorption than for a pure amorphous water surface. An assumption was made that the binding (i.e.~desorption) energies of all species in the model on a putative surface composed entirely of H$_2$ would be just 10\% the strength of their binding to amorphous water. To determine rates, the effective binding energy of some surface species would be composed of a fraction $\theta$(H$_2$) taking the 10\% value, and a fraction $1 - \theta$(H$_2$) taking the full binding energy with amorphous water, $E_{\mathrm{des}}$, i.e.
\begin{equation}
E_{\mathrm{des,eff}} = E_{\mathrm{des}} [1 - \theta(\mathrm{H}_{2})] + 0.1 \, E_{\mathrm{des}} \, \theta(\mathrm{H}_{2}) \nonumber
\end{equation}
This effective binding energy would then be implemented in the thermal desorption and diffusion rate calculations. The main effect of this change was a reduction in the surface abundance of H$_2$ itself, avoiding an unphysical build-up of H$_2$ at very low temperatures.

Here, we introduce a more detailed treatment, in which the rates of desorption and diffusion themselves are the direct target of the adjustment. This treatment considers the surface to be composed of a discrete distribution of binding sites of varying strength; the ultimate, averaged rates of diffusion and desorption for any particular chemical species are based on the relative occupation of each type of site.

In this treatment, any surface species is generically assumed to be bound to a total of four binding partners that collectively contribute to the binding energy of the bound species. This arrangement may be pictured as an atom/molecule sitting atop a simple cubic lattice surface, equidistant from its four nearest neighbors. Thus there would be five types of site, composed of 0--4 H$_2$ molecules and 4--0 water molecules (or, in practice, anything other than H$_2$). It is assumed that the surface layer to which the surface species is bound is always fully occupied; either by other atoms/molecules bound to the grain, or by bare-grain atoms that make up any uncovered portions of the grain surface. Assuming that the H$_2$ molecules that contribute to the surface binding are distributed randomly, a binomial distribution is adopted for the fractional populations of each strength of site present on the surface, thus:
\begin{eqnarray}
&& P(4) = \theta^4 \nonumber \\
&& P(3) = 4 \theta^3 (1 - \theta) \nonumber \\
&& P(2) = 6 \theta^2 (1 - \theta)^2 \nonumber \\
&& P(1) = 4 \theta (1 - \theta)^3 \nonumber \\
&& P(0) = (1 - \theta)^4
\end{eqnarray}
\noindent where the number in parentheses indicates the number of H$_2$ molecules, and $\theta = \theta(\mathrm{H}_{2})$. Again using the assumption that binding to an H$_2$ molecule is one tenth the strength of a normal bond, the total binding strength for each type of site, expressed as a fraction of the standard value, is:
\begin{eqnarray}
&& F_{\mathrm{bind}}(4) = 0.1 \nonumber \\
&& F_{\mathrm{bind}}(3) = 0.325 \nonumber \\
&& F_{\mathrm{bind}}(2) = 0.55 \nonumber \\
&& F_{\mathrm{bind}}(1) = 0.775 \nonumber \\
&& F_{\mathrm{bind}}(0) = 1
\end{eqnarray}
These fractions are used to determine the rates of thermal desorption and diffusion specific to each individual surface species from each individual type of site, using standard thermal expressions:
\begin{eqnarray}
&& k_{\mathrm{dif}}(i) = \nu \, \exp \left( -F_{\mathrm{bind}}(i) \, E_{\mathrm{dif}}/T_{\mathrm{dust}} \right) [\mathrm{s}^{-1}] \\
&& k_{\mathrm{des}}(i) = \nu \, \exp \left( -F_{\mathrm{bind}}(i) \, E_{\mathrm{des}}/T_{\mathrm{dust}} \right) [\mathrm{s}^{-1}]
\end{eqnarray}
\noindent where $i$ indicates the number of H$_2$ molecules to which the surface species is bound, $\nu$ is the characteristic vibrational frequency, and $E_{\mathrm{dif}}$ and $E_{\mathrm{des}}$ are the unadjusted diffusion barrier and desorption barrier of that species, respectively.

In order to combine the above species-specific diffusion and desorption rates for each type of surface binding site into an overall, averaged diffusion and desorption rate for that species, it is necessary to calculate the relative amount of time one atom/molecule of that species would spend in each type of site. This occupation lifetime will be determined by the lifetime against diffusion (which will always be faster than thermal desorption), which is simply the inverse of the diffusion rate for the current site type, i.e.~$1/k_{\mathrm{dif}}(i)$. The fraction of sites of type $i$ on the surface as a whole is determined by the binomial distribution described above, so the relative lifetime for each type of site, aggregated over the entire surface, is provided by $P(i)/k_{\mathrm{dif}}(i)$. The fractional occupation, $F_{\mathrm{occ}}(i)$, for a particular type of site is then given by the ratio of each relative lifetime to the sum of such lifetimes over all of the five possible values of $i$, thus:
\begin{eqnarray}
&& F_{\mathrm{occ}}(i) = \frac{ P(i) }{ k_{\mathrm{dif}}(i) } \bigg/ \sum_{j=0}^{4} \frac{ P(j) }{ k_{\mathrm{dif}}(j) }
\end{eqnarray}
The value of $F_{\mathrm{occ}}(i)$ may be considered to represent the probability of finding the species of interest in a site of type $i$ at any given moment. This quantity is not the same as $P(i)$, the surface population of sites of type $i$, precisely because a diffuser will spend less time in a weaker binding site than in a strong one.

With the fractional occupation of each type of site being described as above, the effective diffusion rate for the bound species, to be used in diffusive reaction rates, is then simply a weighted average based on those values, i.e.
\begin{equation}
k_{\mathrm{dif,eff}} = \sum_{i=0}^{4} F_{\mathrm{occ}}(i) \, k_{\mathrm{dif}}(i) = \sum_{i=0}^{4} P(i) \bigg/ \sum_{j=0}^{4} \frac{ P(j) }{ k_{\mathrm{dif}}(j) }
\end{equation}
In a similar fashion, the effective thermal desorption rate for the bound species is given by:
\begin{equation}
k_{\mathrm{des,eff}} = \sum_{i=0}^{4} F_{\mathrm{occ}}(i) \, k_{\mathrm{des}}(i)
\end{equation}
Note that $F_{\mathrm{occ}}$ is based solely on the diffusion rate and should not be computed independently for the desorption rate.

The important feature of this treatment is that the influence of H$_2$ as a component of the surface is treated as a discrete effect, before an average diffusion or desorption rate is calculated over a range of possible binding sites; the old treatment of Garrod \& Pauly used the H$_2$ content to moderate binding in a single, average surface site. This change produces two competing effects: (i) a much faster rate of desorption is accessible via a fraction of surface sites; (ii) the ability of those weak binding sites to increase the overall desorption rate is limited by the shorter period that a diffuser remains in those sites before diffusing out again.

\begin{figure}
{\includegraphics[width=0.5\hsize]{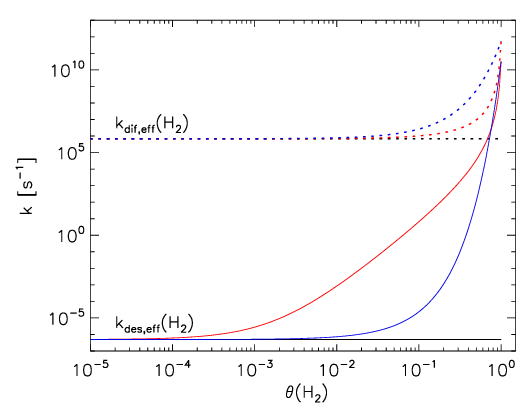}}
\caption{Effective diffusion (dashed) and desorption rates (solid) of H$_2$ on an amorphous H$_2$O surface of varying fractional H$_2$ coverage, $\theta$(H$_2$), at a temperature of 10~K. The black lines indicate the unadjusted rates, blue the method used by \citet{GP11}, and red the new method presented here.}
\label{H2binding}
\end{figure}

Figure \ref{H2binding} shows the desorption and diffusion rates of an H$_2$ molecule using the new method, on a water surface with varying H$_2$ content at a fixed temperature (10~K). Rates calculated using the method of \citet{GP11} are also shown, along with the unadjusted rates for a pure H$_2$O surface. A basic binding energy $E_{\mathrm{des}}$=430~K is assumed for H$_2$ on amorphous water, with $E_{\mathrm{dif}}$=$0.35 E_{\mathrm{des}}$. The variation in the effective diffusion rate with $\theta$(H$_2$) is more subdued using the new method; $k_{\mathrm{dif,eff}}$ is still dominated by the rate of the strongest site until $\theta$(H$_2$) approaches 0.1. However, the rise in the effective desorption rate becomes substantial at a much lower $\theta$(H$_2$) than with the Garrod \& Pauly method, at around $\theta$(H$_2$)=$10^{-3}$, versus close to 0.1 with the old method. The main effect of this change in the chemical models should be further to reduce the surface coverage of H$_2$ at low temperature. Note that, by introducing this new method and thus reducing the surface abundance of H$_2$ itself, the influence of H$_2$ content on the desorption and diffusion rates of {\em other} species is lessened. Also, at temperatures not much greater than 10~K, H$_2$ will be sufficiently sparse on the surface to remove the effect entirely.

The method presented here could be made more complex, although likely with diminishing returns. For example, four equidistant binding partners are assumed here, but this could be extended to five, to include a partner directly beneath the other four (comparable to a BCC lattice arrangement). This could potentially bring the composition of the bulk ice into the calculation, as well as requiring the consideration of binding partners at varying distances from the atom/molecule in question. The influence of such changes could be tested in future. However, the general approach might also be usefully adopted in kinetic simulations of chemistry on surfaces with multiple binding-site strengths irrespective of H$_2$ content; for example, in the consideration of distinct binding energies for a water surface versus a bare grain \citep[cf.][]{Taquet14}, when the grain has only partial coverage with ice.

The effects of the above change on the overall chemical model results are presented in Sec.~\ref{sec:results:H2}.

\begin{deluxetable}{llllllllllllllllll} \label{models}
\tabletypesize{\footnotesize}
\tablecaption{\label{tab-models} Processes/conditions used in each model setup}
\tablewidth{0pt}
\tablehead{\\
\colhead{Model} &  \colhead{\rotatebox{90}{ Diffusion within ice mantle (excl. H \& H$_2$) }}  & \colhead{\rotatebox{90}{$A_{\mathrm{V,init}}=3$ }} &  \colhead{\rotatebox{90}{Sticking coefficient = 1  }} &  \colhead{\rotatebox{90}{ Low photodissociation rates }} &  \colhead{\rotatebox{90}{ New H$_2$-binding treatment }} & \colhead{\rotatebox{90}{ NH$_3$-related proton-transfer reactions }}&  \colhead{\rotatebox{90}{ Eley-Rideal mechanism }} & \colhead{\rotatebox{90}{ Photodissociation-Induced reactions }} & \colhead{\rotatebox{90}{ Recombination of PD products (excl. H \& H$_2$)}} &  \colhead{\rotatebox{90}{ Three-body reactions }} & \colhead{\rotatebox{90}{ Three-body excited-formation reactions }} & \colhead{\rotatebox{90}{ Separate atomic \& molecular diffusion barriers }} & \colhead{\rotatebox{90}{ Adjusted H diffusion \& desorption barriers }} & \colhead{\rotatebox{90}{ Adjusted O \& N desorption barriers }} & \colhead{\rotatebox{90}{ Mantle tunneling diffusion for H and H$_2$ }} & \colhead{\rotatebox{90}{ New gas-phase reactions }} & \colhead{\rotatebox{90}{ New grain-surface/ice reactions }}    }
\startdata
\texttt{basic\_Av2}                                                          & \checkmark     \\ 
\texttt{basic                                                                } & \checkmark  & \checkmark &    \\
\texttt{bas\_stk                                                           } & \checkmark  & \checkmark & \checkmark &    \\
\texttt{bas\_stk\_loPD                                                   } & \checkmark  & \checkmark & \checkmark & \checkmark &    \\
\texttt{bas\_stk\_loPD\_H2                                            } & \checkmark   & \checkmark & \checkmark & \checkmark & \checkmark &   \\
\texttt{bas\_stk\_loPD\_H2\_T16                                    } & \checkmark   & \checkmark & \checkmark & \checkmark & \checkmark & \checkmark &   \\
\texttt{bas\_stk\_loPD\_H2\_T16\_no-bd                         } &                   & \checkmark & \checkmark & \checkmark & \checkmark & \checkmark &   \\
\texttt{bas\_stk\_loPD\_H2\_T16\_no-bd\_ER                   } &                   & \checkmark & \checkmark & \checkmark & \checkmark & \checkmark & \checkmark &   \\
\texttt{bas\_stk\_loPD\_H2\_T16\_no-bd\_PDI                  } &                   & \checkmark & \checkmark & \checkmark & \checkmark & \checkmark && \checkmark &   \\
\texttt{bas\_stk\_loPD\_H2\_T16\_no-bd\_PDI2                } &                   & \checkmark & \checkmark & \checkmark & \checkmark & \checkmark && \checkmark & \checkmark &   \\
\texttt{bas\_stk\_loPD\_H2\_T16\_no-bd\_3B1                 } &                   & \checkmark & \checkmark & \checkmark & \checkmark & \checkmark &&&& \checkmark &   \\
\texttt{bas\_stk\_loPD\_H2\_T16\_no-bd\_3B2                 } &                   & \checkmark & \checkmark & \checkmark & \checkmark & \checkmark &&&& \checkmark &   \\
\texttt{bas\_stk\_loPD\_H2\_T16\_no-bd\_3B3                 } &                   & \checkmark & \checkmark & \checkmark & \checkmark & \checkmark &&&& \checkmark &   \\
\texttt{bas\_stk\_loPD\_H2\_T16\_no-bd\_3B3\_EF           } &                   & \checkmark & \checkmark & \checkmark & \checkmark & \checkmark &&&& \checkmark & \checkmark  \\
\texttt{all                                                                   }  &                   & \checkmark & \checkmark & \checkmark & \checkmark & \checkmark & \checkmark & \checkmark & \checkmark  &  \checkmark  & \checkmark \\
\texttt{all\_Edif                                                            }  &                   & \checkmark & \checkmark & \checkmark & \checkmark & \checkmark & \checkmark & \checkmark & \checkmark  &  \checkmark  & \checkmark  & \checkmark \\
\texttt{all\_Edif\_S17                                                    }  &                   & \checkmark & \checkmark & \checkmark & \checkmark & \checkmark & \checkmark & \checkmark & \checkmark  &  \checkmark  & \checkmark  & \checkmark  & \checkmark \\
\texttt{all\_Edif\_S17\_ON                                             }  &                   & \checkmark & \checkmark & \checkmark & \checkmark & \checkmark & \checkmark & \checkmark & \checkmark  &  \checkmark  & \checkmark  & \checkmark  & \checkmark  & \checkmark \\
\texttt{all\_Edif\_S17\_ON\_mtun                                   }  &                    & \checkmark & \checkmark & \checkmark & \checkmark & \checkmark & \checkmark & \checkmark & \checkmark  &  \checkmark  & \checkmark  & \checkmark  & \checkmark  & \checkmark  & \checkmark \\
\texttt{all\_Edif\_S17\_ON\_mtun\_GP                             }  &                   & \checkmark & \checkmark & \checkmark & \checkmark & \checkmark & \checkmark & \checkmark & \checkmark  &  \checkmark  & \checkmark  & \checkmark  & \checkmark  & \checkmark  & \checkmark  & \checkmark  \\
\texttt{all\_Edif\_S17\_ON\_mtun\_GP\_Gr} ({\tt final})  &                    & \checkmark & \checkmark & \checkmark & \checkmark & \checkmark & \checkmark & \checkmark & \checkmark  &  \checkmark  & \checkmark  & \checkmark  & \checkmark  & \checkmark  & \checkmark  & \checkmark  & \checkmark \\
\enddata
\end{deluxetable}

\subsection{Proton transfer reactions of protonated COMs with gas-phase ammonia} \label{sec:methods:NH3}

In the gas-phase chemistry that occurs subsequent to the desorption of complex organic molecules into the gas phase, the main destruction mechanism of those structures is proton transfer from molecular ions such as H$_3^+$ and H$_3$O$^+$, followed by dissociative recombination of the resultant protonated molecule with electrons. \citet{Geppert06} studied the branching ratios of dissociative recombination of protonated methanol, finding that most branches resulted in the outright destruction of the C--O structure; methanol itself was the product in only a few percent of recombinations. Similar work by those authors confirmed such behavior for larger molecules \citep[e.g.][]{Hamberg10}, indicating that the initiating proton-transfer process would typically lead to the outright destruction of a COM if its protonated form could not be converted back through some process other than electronic recombination.

\citet{Taquet16} introduced alternative gas-phase reactions between protonated COMs and ammonia, NH$_3$, whose proton affinity is relatively large. Taquet et al. argued that, based on existing experimental data, the success rate for simple proton transfer in such cases would be high. In hot cores, following thermal desorption of large amounts of solid-phase ammonia, this would act to increase the gas-phase lifetime of COMs, removing the proton from protonated COMs before dissociative recombination with electrons could occur.

Here we also introduce this mechanism into our network, allowing all COMs with proton affinities lower than that of ammonia to undergo proton transfer. The ion-molecule rates for these reactions are calculated as described by \citet{GWH08}, using the method of \citet{Herbst86}. The same reactions and rates were implemented by \citet{Barger20}, who also provide a Table of selected values. \citet{Taquet16} also considered proton-transfer reactions between a number of complex organic species, including methanol, without the involvement of ammonia. Such reactions are not included in the present work; a more comprehensive network of such processes will be the focus of a future study.

\subsection{Bulk diffusion within ice mantles} \label{sec:methods:bulk-diff}

Earlier gas-grain chemical models of hot cores that included a full chemical treatment, such as \citet{Caselli93}, \citet{GH06} and \citet{GWH08}, were based on a simple two-phase treatment in which the entire grain/ice surface and ice mantle were treated as a single entity, coupled to the gas phase through adsorption and desorption. The model of \cite{Garrod13a} moved to a three-phase approach in which the bulk ice mantle was treated separately, with its own chemistry. This change necessitated a more explicit consideration of chemical reaction mechanisms within the ice; diffusive reactions were assumed to take place as the result of bulk diffusion through a swapping mechanism, justified by combined laboratory and modeling results such as those of \citet{Oberg09b} and \citet{Fayolle11}. 

\citet{Oberg09b} studied the segregation of CO$_2$:H$_2$O and CO:H$_2$O ice mixtures at various temperatures. Segregation in thin ices ($<$100 ML) was partly attributed to surface processes, while there was also evidence of a slower segregation process that was interpreted as resulting from a bulk swapping mechanism. \citet{Fayolle11} studied the desorption of CO and CO$_2$ from water-ice mixtures of varying ratios and thicknesses, as a way to determine the degree of trapping of those molecules during heating. Importantly, they found that the amount of desorption of CO$_2$ around the CO$_2$ desorption temperature of 70~K was independent of the ice thickness, indicating that molecules from a fixed number of upper layers were able to desorb, either as the result of bulk diffusion of CO$_2$ in the upper layers up and out onto the ice surface, or as the result of a large degree of ice porosity that would allow direct surface diffusion and desorption of those molecules. However, the precise kinetics underlying the experiments are complex, and the importance or otherwise of porosity is not well determined. In spite of the evidence of some bulk self-segregation mechanism for CO and CO$_2$ within water ices, it is also unclear whether such mechanisms in general could lead to long-range diffusion in the ice that would promote reaction between relatively sparse radicals.

Many studies have been conducted that consider the production of solid-phase COMs through the UV irradiation of simpler ices \citep[e.g.][]{Allamandola88,Gerakines96,Bennett07}, but in general they have not revealed the importance (or otherwise) of the diffusion of reactants in the bulk ice as a mechanism that brings reactive photo-fragments together. 
However, recent laboratory evidence \citep{Butscher16} suggests that the UV irradiation and subsequent warming of mixed water and organic ices can produce bulk-ice chemical reactions in the 50 -- 70~K temperature range that are driven by the diffusion of radicals that were immobile at lower temperatures. But it is not certain whether such reactions are truly bulk diffusion processes, or whether they are related to re-arrangement or phase changes within the water ice, resulting in the release of trapped radicals either onto the ice surface or into porous structures, where surface diffusion-mediated reactions may occur more easily. It is also possible that 
the re-orientation of the radicals \citep{Martin-Domenech20} or the thermal activation of the reactions themselves could play a role.

\citet{Ghesquiere18} studied reactions between ammonia and CO$_2$ in H$_2$O:NH$_3$:CO$_2$ ice mixtures; while their ices were arguably somewhat thicker (up to around a thousand monolayers) than the expectation for interstellar ices (up to a few hundred monolayers for canonically sized grains), they found that the reactions they observed were more consistent with diffusion along cracks in the ice caused by structural changes related to crystallization (at temperatures $>$120~K) than with bulk diffusion. Indeed, they suggest that such a process may be a possible source of enhanced chemical production in ices during the star-formation process.

Interstellar ices may not be so rich in water that the phase changes seen in the purer mixtures used in laboratory experiments should necessarily be so important; for example, the results provided here, in Section \ref{sec:results}, indicate that water makes up no more than around 50\% of the total ice composition, or around 60 -- 70\% in the first 50 monolayers, once all the major constituents are taken into account. Furthermore, interstellar ices, which are understood to be formed as the result of surface reactions rather than direct deposition, may be much less porous than laboratory ices. The microscopic Monte Carlo kinetics models of \citet{Garrod13b} show that the diffusion of oxygen atoms on a cold grain surface prior to reaction may be sufficient to remove most porosity from water ices formed through surface chemical reactions under dark cloud-like conditions. The release of energy from surface reactions could also reduce porosity further.

Thus, it is still a matter of debate as to whether a substantial degree of bulk diffusion may occur within interstellar ices, or whether there is even the opportunity for internal pore surfaces to provide an alternative means for diffusion within the bulk ice. This brings into question the inclusion of bulk-diffusion driven {\em reactions} in the chemical networks.

From a practical point of view, the inclusion of reactions driven explicitly by a bulk-diffusion mechanism in three-phase models \citep[e.g.][]{Garrod13a} was necessary to facilitate reactions between the radicals produced by photodissociation within the ice. Photodissociation in the bulk would otherwise lead to increasing amounts of unreacted radicals over time. Numerous experiments have of course shown that the UV irradiation of mixed molecular ices leads to photodissociation-induced sub-surface chemical reactions, including the production of COMs \citep[e.g.][]{Oberg09}. Models of these systems must therefore include some mechanism for the products of photodissociation to react with each other, whether or not that explicitly concerns bulk diffusion. Just such a mechanism was included in the models of JG20, who allowed for the immediate reaction of new photo-products with existing radicals in the ice (see Secs.~\ref{sec:methods:non-diff} and \ref{sec:methods:PDI}), without recourse to a separate bulk diffusion mechanism to mediate those reactions. Thus, bulk diffusion-driven reactions can, in the present work, be treated as a distinct mechanism of their own.

For the above reasons, we alter the chemical models (beginning at model {\tt bas\_stk\_loPD\_H2\_T16\_no-bd}) to remove the ability of all but two species in the mantle  (H and H$_2$) to undergo bulk diffusion, regardless of temperature. In models in which this change is made, only bulk-ice reactions that involve one or other of those two species are allowed to occur through the standard diffusive mechanism. Other bulk-ice species may still undergo diffusive reactions, but only as the result of H or H$_2$ diffusion. These are also the only chemical species that are allowed to diffuse out from the bulk-ice phase into the surface layer (Sec.~\ref{sec:methods:basic}). There is experimental evidence, such as that from radiolysis of solid water \citep{Johnson90}, that H$_2$ molecules can diffuse out of the water ice. The ability of H$_2$ and, by inference, atomic H to undergo bulk diffusion may be related to their ability to occupy interstitial positions in the ice structure \citep[see also][]{Chang14}. Section \ref{sec:methods:tunn} discusses the inclusion of tunneling as a means of H and H$_2$ bulk diffusion in the models.

The effects of the switch-off of the generic bulk-diffusion mechanism are presented in Section \ref{sec:results:bulkdiff}.

\subsection{Non-diffusive reactions on dust grains} \label{sec:methods:non-diff}

Following JG20, we selectively and collectively test several new nondiffusive reaction mechanisms occurring on dust-grain/ice surfaces and/or in the bulk-ice mantles. These mechanisms are mostly treated in the same way as by JG20, with some exceptions noted here and below. The nondiffusive mechanisms are: the Eley-Rideal (E-R) process; photodissociation-induced (PDI) reactions; three-body (3-B), or follow-on reactions; and a subset of the latter, three-body excited-formation (3-BEF) reactions. The general formulae used for all of the nondiffusive mechanisms are provided by Eqs.~(6) -- (9) of that paper, the first of which provides the generic form of their reaction rates and is reproduced here:
\begin{equation}
R_{AB} = f_{\mathrm{act}}(AB) \, R_{\mathrm{comp}}(A) \, \frac{N(B)}{N_S} \, + f_{\mathrm{act}}(AB) \, R_{\mathrm{comp}}(B) \, \frac{N(A)}{N_S} \label{eq:rate}
\end{equation}
\noindent where $f_{\mathrm{act}}(AB)$ is an efficiency factor related to any activation energy for the reaction between species $A$ and $B$, $N(i)$ are the populations of those species, $N_S$ is the number of surface binding sites, and $R_{\mathrm{comp}}(i)$ is the ``completion rate'' for the reaction, which depends on the rate of appearance of species $i$ and on the rate at which the subsequent reaction proceeds.
The rate of appearance would correspond to, for example, the accretion rate of species $i$ onto the grain surface for an Eley-Rideal process, or the rate of production of photo-fragment $i$ in the case of a PDI reaction.
Concentrating on the first term of Eq.~(\ref{eq:rate}), the quantity $N(B)/{N_S}$ corresponds to the surface coverage of species B, and represents the probability that some newly arrived/formed species A encounters an atom/molecule of species B with which it may react upon arrival/formation.
If the reaction mechanism in question were, for example, Eley-Rideal, then the first term would represent the rate at which species A arrives from the gas phase and reacts with species B. The second term is the complementary rate at which species B arrives from the gas phase and reacts with species A.

For reactions in the ice mantle, $N_S$ in Eq.~(\ref{eq:rate}) is substituted with $N_M$, the total number of atoms and molecules in the bulk ice mantle (which excludes surface species). The encounter probability $N(B)/{N_M}$ is thus the average fractional population of species B in the bulk ice. The present three-phase model therefore implicitly assumes that the chemical components of the bulk-ice are well mixed.

As mentioned by JG20 (their Sec.~2.2), in cases where barriers to reaction are sufficiently high (and thus reaction rates sufficiently small), alternative processes could interfere with reaction after the reactants have met, such as the arrival of an H atom, or the photodissociation of one or other reactant. The latter, for example, could plausibly occur with a rate around 10$^{-13}$ s$^{-1}$ for a purely CR-induced UV field; this rate would start to become competitive with thermally-activated reactions of energy barrier $\sim$550~K or higher, at $T_{\mathrm{dust}}=10$~K. This competition would thus further reduce the efficiency of the reaction in question. Following the suggestion of JG20, here we include an additional term for each of those processes (dissociation of either species, and the diffusive arrival of an H atom) in the denominator of Eqs.~(5) and (7) of JG20 (note that their Eq. 5 applies both to diffusive and nondiffusive processes). Each additional term is given by the inverse timescale (rate) of that process. In the case of the diffusive arrival of a grain-surface H atom, the timescale is simply the total diffusion, or scanning, rate ($k_{\mathrm{hop}}/N_{S}$) of atomic H (moderated for back-diffusion) multiplied by the surface population of H. (A similar expression may be formulated for competition between bulk processes). It is assumed that, in such a case, the H atom would preferentially react with one of the species in question, ending the possibility of the intended reaction occurring. The production rate of this competing process would already be dealt with through its own individual reaction rate, so no additional rate is required to make it self-consistent. Although in almost every case the arrival of H would indeed interfere with the reaction (certainly those involving an atom or radical reacting with a stable molecule), it is possible that the reaction of H with either species might be inhibited by a barrier, requiring the added competition term to be further tuned; we do not consider this level of detail in the present model. The additional competition terms are included in all applicable surface and mantle reaction calculations.

As in the models of JG20, all reactions that are allowed to occur on grain surfaces through the regular diffusive mechanism are also allowed to occur (on the surface and within the ice mantles) through the various nondiffusive mechanisms (in those models in which the nondiffusive mechanisms are switched on). The reactions themselves may be considered to be the same in either case; the essential difference between mechanisms (excluding 3-BEF) for any one reaction is the means by which the reactants meet.

\subsubsection{Eley-Rideal mechanism (E-R)} \label{sec:methods:E-R}

The Eley-Rideal mechanism involves the adsorption of a gas-phase atom or molecule onto the dust/ice surface such that it is immediately in contact with a reaction partner. This process therefore requires no diffusion to occur, and the driving rates ($R_{\mathrm{comp}}$) for E-R reactions are mainly dependent on the adsorption rate of each species onto the surface. Eq.~\ref{eq:rate}, as formulated, incorporates both possible E-R processes leading to the same reaction, i.e.~the adsorption of species A directly onto species B, and vice versa. As with other nondiffusive processes described below, all reactions in the network are allowed to proceed via the E-R mechanism, including reactions that have activation energy barriers. In the latter case, the completion rate can be substantially lower than the adsorption rate, which typically renders such reactions unimportant. The E-R treatment is described in more detail by JG20.

\subsubsection{Three-body reaction mechanism (3-B)} \label{sec:methods:3-B}

The three-body reaction mechanism, as defined by JG20, is the process by which an initiating reaction produces a new species that immediately meets its own reaction partner nearby and reacts. Although in some cases this may involve all three bodies being contiguous at the moment of the initiating reaction, this is not necessary so long as there is at least some modest degree of nonthermal diffusion of the initial product that can carry it to a reaction partner. Given that almost every reaction in the surface network is substantially exothermic (as compared with the barrier against diffusion), it is likely that most reactions would lead to some degree of translation of the product(s) \citep[see also][]{Fredon17}. In the models presented here, the 3-B mechanism does not explicitly distinguish between an immediate follow-on reaction and one in which there is some small degree of nonthermal diffusion prior to that reaction.

JG20 included the 3-B mechanism with multiple cycles, such that a single 3-B reaction could in turn induce a further follow-on reaction, and so on. Here we employ the same method, but we also test explicitly the influence of the number of reaction cycles considered (1--3).

\subsubsection{Three-body excited-formation mechanism (3-BEF)} \label{sec:methods:3-BEF}

This process constitutes a variation on the basic 3-B mechanism, such that the chemical energy released by the initiating reaction may allow an activation energy barrier to the follow-on reaction to be surmounted instantaneously. Without this process switched on, while the basic 3-B mechanism might bring together the reactants efficiently, the activation energy barrier would not allow them to react rapidly, lessening the impact of the enhanced meeting rate.

Due to the more complicated nature of this process, in which the energy produced by each initiating reaction must be considered, JG20 only incorporated three key examples of the mechanism into their model. Indeed, they suggested that its importance was likely to be small in all cases except those in which a highly exothermic initiating reaction facilitates a barrier-mediated follow-on reaction with a highly-abundant surface species, such as CO. In particular, they found that the production of the methyl formate-precursor radical CH$_3$OCO through this mechanism was sufficient to enhance methyl formate abundance substantially under cold conditions. Testing several efficiency values for that 3-BEF reaction, they determined that only 1 in 1000 meetings of excited CH$_3$O and CO would be required to lead to the formation of CH$_3$OCO in order to reproduce observational gas-phase abundances of methyl formate. They proposed a statistical treatment for the efficiency of 3-BEF reactions based on Rice-Ramsperger-Kassel (RRK) theory.

In the present models we adopt a similar but slightly more detailed treatment that is applied to all initiating/follow-on reaction combinations for which the exothermicity of the initiating reaction is greater than the barrier to the follow-on reaction. If no barrier is present or the barrier is too high to be overcome this way, then the process is dealt with as a simple 3-B process (with a normal thermal or tunneling treatment for the barrier, if present). If a sufficiently low barrier exists, then there is some probability that the 3-BEF mechanism allows the follow-on reaction to proceed immediately, meaning that 3-BEF occurs in only some fraction of cases. Likewise, there is a complementary fraction of cases in which 3-BEF does not proceed, in which (for completeness) the regular (slower) 3-B rate is used instead.

The fraction of 3-B cases in which the 3-BEF mechanism occurs is formulated in the following way for surface reactions (with a slightly different formula for such reactions occurring within the bulk, described further below). Similar to the RRK treatment proposed by \citet{Garrod07} for chemical desorption of surface species, the ability of chemical energy to overcome the activation barrier of the follow-on reaction is dependent on the requisite amount of energy becoming present in the correct vibrational mode, while the total energy in the system is gradually lost to the surface. First, a probability is defined that the requisite energy is instantaneously present in the correct mode:
\begin{equation}\label{RRK}
P_{\mathrm{EF}} = \left[1-\frac{E_\mathrm{A}}{E_\mathrm{reac}}\right]^{s-1}
\end{equation}
\noindent where $s$ is the number of modes, $E_\mathrm{A}$ is the activation energy, and $E_\textrm{reac}$ is the chemical energy released by the reaction. The form of Eq.~(\ref{RRK}) is a standard result in RRK treatments of unimolecular processes \citep[e.g.][]{Forst}.
Crucially, $s$ includes not only the number of vibrational modes within the excited product of the initiating reaction but also three translational modes for the entire molecule, corresponding to the three spatial coordinates. One of these is assumed to be the reaction coordinate. These translational modes may still be considered vibrational, as they constitute the motion of the excited molecule within the potential of its surface binding site.

In order to arrive at the fraction of 3-B cases in which 3-BEF occurs, the above probability must be moderated by competition due to the loss of energy to the surface.
A rate may be defined, $\nu P_{\mathrm{EF}}$, representing the rate at which an energized molecule will obtain sufficient energy in the correct mode for reaction to occur, where $\nu$ is a representative frequency for energy transfer between modes. In the RRK description, $\nu$ may be treated as an average of the characteristic frequencies of all modes. The competing rate of energy loss from the molecule to the surface is defined simply as $\nu$, meaning that energy is lost efficiently to the surface on one vibrational timescale. The probability of the correct mode obtaining sufficient energy before loss to the surface occurs is then equal to the ratio of the rate $\nu P_{\mathrm{EF}}$ to the total rate at which either outcome may occur, i.e.
\begin{equation}
\frac{\nu P_{\mathrm{EF}} }{\nu + \nu P_{\mathrm{EF}} }
 = \frac{P_{\mathrm{EF}} }{1 + P_{\mathrm{EF}} } \nonumber
\end{equation}
\noindent which is approximately equal to $P_{\mathrm{EF}}$ in cases where $P_{\mathrm{EF}} \ll 1$. Based on this probability, we define the fraction of successful 3-BEF cases as:
\begin{equation}
f_{\mathrm{EF}} = \frac{1}{4} \left( \frac{P_{\mathrm{EF}} }{1 + P_{\mathrm{EF}} } \right) \label{eqn-3-BEF}
\end{equation}
\noindent The factor 4 is included to account for the possibility that the excited species diffuses away along one of three other lateral surface diffusion pathways instead of reacting; the choice of four lateral directions is intended to reflect the two directions per spatial coordinate available on the surface, while only one direction along the reaction coordinate will lead to reaction (with the other resulting in surface diffusion). Implicit in this implementation is the assumption that the diffusion barrier is substantially smaller than both the activation energy barrier and the total energy of excitation, meaning that diffusion is always likely, and therefore that even if sufficient energy is present in the reaction coordinate, there is also sufficient energy available in a mode that could result in diffusion on a single vibrational timescale. This condition should be satisfied if the average energy per remaining mode ($s-1$) exceeds the barrier against diffusion, which will typically be true for the highly energetic initiating reactions for which the 3-BEF mechanism would be important (including the one listed below). The probability of outright desorption along a direction perpendicular to the surface is assumed to be negligible compared with the competition provided by diffusion. 

The above formulation produces an efficiency $f_{\mathrm{EF}} = 0.0011$ for the grain-surface 3-BEF process (H + H$_2$CO) + CO $\rightarrow$ CH$_3$O$^*$ + CO $\rightarrow$ CH$_3$OCO, using the energy values adopted by JG20. 

A similar treatment is used in the formulation of 3-BEF processes within the ice mantle (which may be initiated by, for example, a reaction involving atomic H, which is assumed to be mobile within the mantle in all models presented here). However, the factor of 4 is removed in the bulk-ice implementation of Eq. \ref{eqn-3-BEF}, as no reaction products are allowed to diffuse within the mantle in the models in which this mechanism was tested. Note that the 3-BEF process (either on surfaces or in the mantle) is assumed to occur only for 3-B processes initiated by a single-product reaction. Partition of energy between multiple products was deemed too complicated for inclusion in the present implementation.

In contrast to the treatment adopted by JG20, the 3-BEF process is considered for each cycle of the 3-B mechanism, not only the first.

\subsubsection{Photodissociation-induced reaction mechanism (PDI)} \label{sec:methods:PDI}

Following \citet{Garrod19}, JG20 implemented a nondiffusive reaction mechanism related to the instantaneous production of a reactant near to its reaction partner as the result of the photodissociation of some precursor, either on the surface or in the ice mantle. This mechanism in particular provides a direct route for photochemistry to occur in the ice mantles without the requirement for bulk diffusion. The rate at which this instantaneous reaction mechanism occurs depends on the rate at which either reactant is produced via photodissociation (caused either by direct interstellar UV or by cosmic-ray induced photons), and on the abundance of the other reactant in the ice or on the surface. The initiating process of photodissociation will, on most occasions, lead simply to two products, typically radicals and/or atoms, that are not immediately in contact with a reaction partner, and thus do not react through the PDI mechanism. In the fraction of photodissociation events that lead to a follow-on reaction, the presence of a reaction partner in the ice is typically the result of the other dissociation events that were unsuccessful in producing a reaction. The PDI treatment allows all reactions in the network, including those with activation energy barriers, to proceed as follow-on reactions initiated by photodissociation of a precursor to one of the reactants. Neither electronic nor vibrational excitation of photodissociation products is considered.

The above-described implementation, used by JG20, we label simply ``PDI''. Here we also test a refined version of this mechanism, labelled ``PDI2", in which special consideration is given to the PDI process occurring in the ice mantles. If two large, and thus immobile, photodissociation products are formed as the result of a single photodissociation event, they have the opportunity to react with contiguous reaction partners. However, if neither finds a reaction partner, which is the most common outcome, then it must be assumed that they recombine with each other in the ice mantle, as they do not have the ability to diffuse away (at least without substantial structural rearrangement of the surrounding ice). In the PDI2 implementation, therefore, if either one or other of the two (immobile) photodissociation products does not react with something else, then the products are allowed immediately to recombine. This recombination thus reduces the effective photodissociation rates in the ice, after reactions have been considered. The initiating rates of bulk-ice molecule photodissociation, prior to this correction, correspond to dissociation alone, regardless of subsequent outcomes. Since those solid-phase photodissociation rates are based on gas-phase rates, in which spontaneous recombination would not occur, this treatment should be entirely self-consistent.

An exception is made to the above-described recombination requirement only in the case where one of the products of photodissociation is either H or H$_2$, which are allowed to diffuse within the mantle, allowing them each to escape from their counterparts even if no follow-on reaction occurs. Classical mechanical considerations would suggest that the lighter product, H or H$_2$, would carry away the majority of any surplus energy produced by the photodissociation, making it more likely that the light species could escape via diffusion. Molecular dynamics calculations were carried out by \citet{Andersson08}, who studied the outcome of photodissociation of water molecules in the upper monolayers of water ice. In their deepest layer (6 ML), they found that roughly 40\% of H$_2$O dissociations led to recombination, a similar number resulted in the trapping of the OH and H separately somewhere in the ice matrix, with the remainder leading to desorption of H and trapping of OH; the total fraction of H escaping without recombination was thus around 60\%. Our PDI2 implementation assumes that 100\% of H from H$_2$O photodissociation (for example) escapes from the OH (while we do not consider direct desorption from the surface or upper ice layers as the result of photo-dissociation). A test model was run (not shown) in which only 60\% of all H-producing photodissociations of any species in the bulk ice would lead to separation of the products. However, the differences from the basic PDI2 results were minor. Thus, considering the substantial additional cost of calculation required to treat all H-producing photodissociation processes individually (which involves correlating each dissociation mechanism with every possible follow-on reaction that might result), the regular PDI2 mechanism was deemed sufficient for present purposes. 

Note that the PDI2 method assumes that there is no immediate recombination of surface species (as distinct from mantle species) following photodissociation, as the products are free to diffuse on the surface. Thus both photo-products will remain available for surface reactions if they do not immediately undergo a follow-on reaction with another nearby reaction partner. The PDI and PDI2 methods are therefore identical in their treatment of surface chemistry.

\subsection{Generic diffusion barriers}  \label{sec:methods:barrs}

Although desorption (i.e.~binding) energies may be measured in the laboratory, the barriers against diffusion on a surface are more difficult to ascertain directly. Past models of gas-grain chemistry have typically assumed the surface diffusion barrier for each chemical species to take a fixed ratio with respect to its desorption energy. \citet{GH06} and \citet{GWH08} assumed a factor $E_{\mathrm{dif}}/E_{\mathrm{des}}$=0.5, while \citet{GP11} determined that values less than 0.4 provided the best results for models involving CO diffusion; \citet{Garrod13a} used 0.35 for all surface species. Calculations by \citet{Kars14} determined values for CO and CO$_2$ of 0.31 and 0.39, respectively. Experimental work has yielded results that highlight a distinction between atomic and molecular diffusion barrier ratios. \citet{Minissale16} suggest using a ratio for atoms in particular of 0.55 while retaining a lower value for molecules.

JG20 adopted a high ratio for atoms, of 0.6, but tested the generic 0.35 value also; they found that the slower diffusion of H atoms on the grains produced with the higher value was beneficial to the action of nondiffusive reaction mechanisms, which rely on the fractional coverage of radicals.

In the present models, we initially adopt a uniform $E_{\mathrm{dif}}/E_{\mathrm{des}}$=0.35, as in our most recent hot core models. We then test a ratio 0.55 for all surface atoms, as per \citet{Minissale16}. Following \citet{Garrod13a}, the ratio for {\em bulk} diffusion is assumed to be twice the value of the surface ratio, whatever that may be. For atoms, this would provide a bulk diffusion barrier 10\% greater than the surface binding energy.

\subsubsection{Atomic binding energies}  \label{sec:methods:bind}

In past models, atomic binding energies have been poorly defined; a value of 800~K has frequently been adopted for both O and N on water ice. Although other atoms, such as carbon, also have poorly defined values, the binding energies of O and N have recently been investigated in the laboratory. \citet{He15} obtained an experimental value of 1660~K ($\pm 60$~K) for the former. \citet{Minissale16} determined binding energies for O and N atoms on amorphous water ice; their value for O was consistent with that of \citet{He15}. Here we apply that value in selected models, in tandem with a value of 650~K for N, guided by the results of \citet{Minissale16} under the assumption of a diffusion barrier set to 0.55 of the desorption energy.

The desorption energy and diffusion barrier of hydrogen have also been the subject of debate over the years. In past models, we have assumed a desorption energy of 450~K on amorphous water ice (ASW), with a diffusion barrier ratio of 0.35. Recently, \citet{Sen17} carried out calculations for H on water surfaces including ASW. They found an average classical barrier height of 21~meV ($\sim$243~K) for diffusion of H on ASW, with an average binding energy of 57~meV ($\sim$661~K). We employ these values in selected models -- again, with a bulk diffusion barrier for H of twice its surface value.

\subsubsection{Bulk diffusion of H and H$_2$ via tunneling}  \label{sec:methods:tunn}

The influence of tunneling on hydrogen atom diffusion on surfaces is likely to be small, as suggested by \citet{Sen17} and as assumed in this and past models. However, the higher barriers expected for diffusion of H (and H$_2$) in the bulk ice could mean that tunneling plays a more substantial role in the mantle. To test this possible effect, we activate a lower limit to the rate of a single diffusion event within the mantle, based on a rectangular-barrier tunneling treatment. The width of the barrier is set to 3.2~\AA, the approximate width of a water molecule, to represent the distance between interstitial sites within amorphous water ice. If temperatures are low enough that the tunneling rate for H or H$_2$ is faster than its thermal rate, then the tunneling rate instead is used in the calculation of both the diffusive chemical rate and the rate at which that species can escape from the ice mantle and onto the ice surface.

\subsection{New gas-phase reactions}  \label{sec:methods:gas-phase}

The subsections above are largely focused on newly-added grain-surface and bulk-ice processes. For comparison with alternative mechanisms proposed by other authors, a selection of gas-phase processes is added to the network used in certain models. Firstly, a radiative association reaction for the production of dimethyl ether is included:
\begin{eqnarray}
\mathrm{CH}_3 + \mathrm{CH}_3\mathrm{O} \rightarrow \mathrm{CH}_3\mathrm{OCH}_3 + \mathrm{h}\nu \nonumber
\end{eqnarray}
\noindent for which a rate of $3 \times 10^{-11}$ cm$^{3}$~s$^{-1}$ is assumed. This rate pre-supposes that radiative association proceeds in this case with an efficiency of around 10\% per collision at 10~K, with an inverse temperature dependence; more detailed calculations for this process are under way \citep{Tennis21}.

Related to this mechanism, we include the rates used by \citet{Ruaud15} for the reaction between OH and methanol, which are based on values obtained by \citet{Shannon13} and \citet{Atkinson04}:
\begin{eqnarray}
\mathrm{OH} + \mathrm{CH}_3\mathrm{OH} & \rightarrow & \mathrm{CH}_3\mathrm{O} + \mathrm{H}_2\mathrm{O} \nonumber \\
                     & \rightarrow & \mathrm{CH}_2\mathrm{OH} + \mathrm{H}_2\mathrm{O} \nonumber
\end{eqnarray}
which provides one of the precursors for dimethyl ether production via radiative association. We also adopt the rate value determined by \citet{Skouteris19} for proton transfer between protonated dimethyl ether and ammonia (see Table \ref{gp-rxns}).

\citet{Balucani15} suggested a gas-phase mechanism to allow conversion of a radical related to dimethyl ether into methyl formate, {\em viz.}
\begin{eqnarray}
\mathrm{CH}_3\mathrm{O}\mathrm{CH}_2 + \mathrm{O} & \rightarrow & \mathrm{HCOOCH}_3 + \mathrm{H} \nonumber
\end{eqnarray}
\noindent with the radical being produced via hydrogen abstraction from dimethyl ether by a Cl or F atom. Due to the sparseness of our network for those two elements, we choose not to include such abstraction processes. The CH$_3$OCH$_2$ radical nevertheless exists in our network; it may be produced on grains as the result of several possible H-abstraction processes, or in the gas phase, either through H-abstraction from dimethyl ether by OH \citep{Shannon14} or through a proton-transfer reaction between ionized dimethyl ether and ammonia (see Sec.~\ref{sec:methods:NH3}). To test any influence this radical may have in producing methyl formate, the above reaction is added to the network with a rate $2 \times 10^{-10}$ cm$^{3}$~s$^{-1}$ \citep{Balucani15}.

A molecule of particular interest in hot core chemistry is formamide (NH$_2$CHO). \citet{Barone15} suggested that the gas-phase reaction between the radical NH$_2$ and formaldehyde could produce this species:
\begin{eqnarray}
\mathrm{NH}_2 + \mathrm{H}_2\mathrm{CO} \rightarrow \mathrm{NH}_2\mathrm{CHO} + \mathrm{H} \nonumber
\end{eqnarray}

\citet{Skouteris17} conducted further calculations, deriving rates that approach the collisional rate at low temperatures. We adopt their modified Arrhenius rate fit. \citet{Ayouz19} calculated rates for the electronic recombination of protonated formamide, which we also employ here, using our existing branching ratios.

The new reactions and rates described here, which are adopted in selected models, are detailed in Table \ref{gp-rxns}.

\begin{deluxetable}{llllc} \label{gp-rxns}
\tabletypesize{\footnotesize}
\tablecaption{Gas-phase reactions related to COMs, added to the network in selected model runs.}
\tablewidth{0pt}
\tablecolumns{5}
\tablehead{
\colhead{Reaction} & \colhead{} & \colhead{} & \colhead{Rate} & \colhead{Reference} \\
            \colhead{} & \colhead{} & \colhead{} & \colhead{(cm$^{3}$~s$^{-1}$)}
}
\startdata
CH$_3$  +  CH$_3$O                   & $\rightarrow$ &      CH$_3$OCH$_3$                            &   $3.00 \times 10^{-11}$ & a \\
CH$_3$OCH$_2$ + O                   & $\rightarrow$ &      HCOOCH$_3$ + H                           &   $2.00 \times 10^{-10}$ & b \\
OH  +  CH$_3$OH                        & $\rightarrow$ &      CH$_3$O  +  H$_2$O                     &   $6.00 \times 10^{-13} \ (T/300)^{-1.2}$ & c \\
OH  +  CH$_3$OH                        & $\rightarrow$ &      CH$_2$OH  +  H$_2$O                   &   $3.10 \times 10^{-12} \ \exp(-360/T)$ & d \\
CH$_3$OCH$_4^+$  +  NH$_3$ & $\rightarrow$ &      CH$_3$OCH$_3$  +  NH$_4^+$  &   $9.67 \times 10^{-10}$ & e \\
NH$_2$ +  H$_2$CO                     & $\rightarrow$ &      NH$_2$CHO  +  H                          &   $7.79 \times 10^{-15} \ (T/300)^{-2.56} \ \exp(-4.88/T)$ & f \\
NH$_2$CHOH$^+$ +  e$^-$         & $\rightarrow$ &      NH               +  H$_2$O  +  H         &   $3.75 \times 10^{-7} \ (T/300)^{-0.5}$ & g \\
NH$_2$CHOH$^+$ +  e$^-$         & $\rightarrow$ &      NH$_2$        +  HCO  +  H              &   $3.75 \times 10^{-7} \ (T/300)^{-0.5}$ & g \\
NH$_2$CHOH$^+$ +  e$^-$         & $\rightarrow$ &      NH$_2$        +  H$_2$CO               &   $4.17 \times 10^{-8} \ (T/300)^{-0.5}$ & g \\
NH$_2$CHOH$^+$ +  e$^-$         & $\rightarrow$ &      NH$_2$CHO  +  H                          &   $4.17 \times 10^{-8} \ (T/300)^{-0.5}$ & g \\
\enddata
\tablecomments{Temperatures are in units of K. References: $^a$estimate, see text ;$^b$\citet{Balucani15}; $^c$\citet{Shannon13}, \citet{Ruaud15};
$^d$\citet{Atkinson04}, \citet{Ruaud15};
$^e$\citet{Skouteris19}; 
$^f$\citet{Skouteris17}; 
$^g$\citet{Ayouz19}.}
\end{deluxetable}

\subsection{New grain-surface/ice-mantle reactions}  \label{sec:methods:grain}

In the final set of updates to the model and network, which are applied only in the {\tt final} model, we introduce a large number of grain-surface/mantle reactions that have not been considered before in our hot-core chemical networks, and which may occur through both diffusive and nondiffusive mechanisms. Parameters and treatments relating to a selection of other surface/mantle reactions are also adjusted. The new and/or adjusted reactions are shown in Tables \ref{gs-rxns} -- \ref{misc-rxns}, with details of their implementation provided below.

\subsubsection{Glycolaldehyde-related chemistry and directionality of surface reactions}

The formation and destruction of glycolaldehyde (CH$_2$(OH)CHO) in hot cores has presented a problem for previous gas-grain chemical models, due to its overproduction relative to its structural isomer, methyl formate (HCOOCH$_3$). The introduction of grain-surface radical-radical chemistry into hot-core models \citep{GWH08} allowed both species to be produced primarily through the addition of the HCO radical on grain surfaces to one of the photo-fragments of methanol, CH$_2$OH or CH$_3$O, during the warm-up stage of hot-core evolution. However, while observations indicate that gas-phase methyl formate is around an order of magnitude more abundant than glycolaldehyde, past gas-grain models have exhibited the inverse relationship. This strong disagreement has been taken to suggest that the production of glycolaldehyde through radical addition could be inefficient \citep{Garrod13a}, or that gas-phase mechanisms may play a larger role in producing methyl formate \citep[e.g.][]{Taquet16}. However, another area of the surface chemistry that has been inadequately explored is the link between glyoxal, glycolaldehyde and ethylene glycol.

Laboratory studies by \cite{Fedoseev15} and \cite{Chuang16} have highlighted the possibility for glyoxal, HCOCHO, to be formed at low temperatures through H-atom addition to solid-phase CO, which produces reactive HCO radicals that may combine in place. These same studies showed also that glyoxal itself may be hydrogenated to glycolaldehyde, and thence to ethylene glycol, (CH$_2$OH)$_2$, by continued H-atom addition. This cold mechanism would allow glycolaldehyde (and ethylene glycol) to be formed at a much earlier, colder stage of the star-formation process than if a thermally-triggered, photodissociation-induced radical-addition mechanism were the sole formation process. However, a complete treatment of the chemical relationship between the three species must include not only the forward H-addition reactions from glyoxal all the way to ethylene glycol, but the H-abstraction routes that reverse this process. The introduction of these hydrogen addition and abstraction reactions into the network provides possible loss and gain routes for glycolaldehyde on the grains that could affect its ultimate gas-phase ratio with methyl formate.

Recently, \cite{Simons20} published a microscopic Monte Carlo kinetics modeling study concentrating on the glyoxal--ethylene-glycol system. Their network added several intermediate radicals and included new reactions involving hydrogenation and H-abstraction, some mediated by activation energies. Here, we similarly incorporate such reactions into our network, although some of the specific activation-energy barriers used here follow our existing network, which already included some of the reaction steps. The full network of grain-surface/mantle reactions for this system is shown in Table \ref{gs-rxns}. Some of the intermediate radicals relating to the new reactions were not previously present, and thus required incorporation into the full gas-grain network to allow their possible destruction in the gas phase \citep[see ][for a description of the method]{GWH08}.

Crucially, the models of \cite{Simons20} take particular account of the directionality of reactions, including those involving activation energy barriers. The direction in which (in this case) a hydrogen atom approaches the target molecule will affect which functional group is attacked, and thus the product of the reaction and the activation energy barrier \citep{AB18}. This directional selectivity is also relevant to atom-radical reactions \citep{Lamberts18}. The implementation of such directional behavior is arguably more straightforward for microscopic Monte Carlo kinetics models such as that of Simons et al., but rate equation-based kinetics models such as {\em MAGICKAL} can also be adapted to incorporate directional behavior.

The method used in the {\tt final} model to take account of the relative orientations of reacting grain-surface (or bulk-ice) atoms/molecules is described below. This method is most relevant to reactions involving large molecules with more than one functional group; methanol (CH$_3$OH) may be taken as an example. Grain-surface hydrogen can react with surface methanol, abstracting another hydrogen atom to produce H$_2$ and either of the radicals CH$_2$OH or CH$_3$O, corresponding to the loss of H from either the methyl or the hydroxyl group, respectively. Each of these mechanisms assumes a different activation energy barrier. 

In the existing treatment used in the {\tt basic} model and most others presented in this paper, the two reaction mechanisms directly compete with each other in all cases in which H and CH$_3$OH meet on the surface. The two reactions also compete directly with diffusion of either species away from the other, which, if successful, terminates the opportunity to react. The fraction of successful meetings resulting in the production of CH$_3$O + H$_2$ would be described in this case by:
\begin{eqnarray}\label{comp}
\frac{ \nu \, \kappa(\mathrm{CH}_{3}\mathrm{O} + \mathrm{H}_{2}) }{  \nu \, \kappa(\mathrm{CH}_{3}\mathrm{O} + \mathrm{H}_{2}) + \nu \, \kappa(\mathrm{CH}_{2}\mathrm{OH} + \mathrm{H}_{2}) + k_{\mathrm{hop}}(\mathrm{H}) + k_{\mathrm{hop}}(\mathrm{CH}_{3}\mathrm{OH}) }
\end{eqnarray}

\noindent where $\nu$ is the characteristic reaction attempt frequency, $\kappa(\mathrm{CH}_{3}\mathrm{O} + \mathrm{H}_{2})$ is the reaction probability per attempt (given by a Boltzmann factor or tunneling expression) for the reaction producing CH$_3$O and H$_2$, $k_{\mathrm{hop}}$(H) is the single-site hopping rate of atomic H, and so on.

The new treatment instead identifies two cases of competition in which the reaction may occur. Firstly, in some fraction of meetings ({\em case i}), only one or other of the two reactions is allowed, corresponding to the attack of the hydrogen atom on one end or other of the molecule, while the other is inaccessible. In this way, each of the two reactions may proceed independently of the other, through their own {\em case i} reactions that are accessed by approaches from different directions. Once the H and CH$_3$OH have met on the surface with the correct orientation for the CH$_3$O-producing reaction alone to be accessible (for example), its competition is described by the expression:
\begin{eqnarray}\label{no-comp}
\frac{ \nu \, \kappa(\mathrm{CH}_{3}\mathrm{O} + \mathrm{H}_{2}) }{  \nu \, \kappa(\mathrm{CH}_{3}\mathrm{O} + \mathrm{H}_{2}) + k_{\mathrm{hop}}(\mathrm{H}) + k_{\mathrm{hop}}(\mathrm{CH}_{3}\mathrm{OH}) }
\end{eqnarray}

Secondly, in some fraction of meetings ({\em case ii}), the two reactions are allowed to compete with each other directly, as before (Eq. \ref{comp}); this corresponds to the H atom arriving from some intermediate direction in which both reaction pathways are accessible.

To quantify the balance of the different competition scenarios, each possible barrier-mediated reaction is assigned an additional two parameters. Firstly, the fraction of directions, $F_{\mathrm{dir}}$, from which a particular reaction (of specified products and barrier) is accessible is assigned a value ranging from 0--1. This parameter may be considered to represent the fraction of all possible relative orientiations of the two reactants in which that reaction is allowed to occur ({\em cases i} and {\em ii}). From a technical point of view, this fraction is applied to the reaction rate as a direct multiplier to the meeting rate of the two species, whether through a diffusive meeting or by other means. Secondly, of that fraction of meetings in which that reaction is allowed ($F_{\mathrm{dir}}$), a fraction $F_{\mathrm{comp}}$ is set in competition with all other possible processes (i.e.~{\em case ii}). For the remaining $1-F_{\mathrm{comp}}$ meetings in which that reaction is allowed ({\em case i}), the reaction competes only with diffusion (however, see Sec.~\ref{sec:methods:non-diff} for a description of additional non-reactive processes other than diffusion that are also considered in the full competition treatment).

These $F$ parameters can also be trivially applied to a reaction (with or without a barrier) in which no other alternative reaction process is considered. In that case, $F_{\mathrm{comp}}$ is redundant, while $F_{\mathrm{dir}}$ simply indicates the fraction of orientations from which that one reaction pathway is accessible.

In cases where two or more barrier-mediated reactions are considered possible between the same reactants, both $F_{\mathrm{dir}}$ and $F_{\mathrm{comp}}$ for any particular reaction in that set may in principle take any value from 0--1. Furthermore, the sum of $F_{\mathrm{dir}}$ values for all the different available barrier-mediated reactions between the same pair of reactants may collectively exceed unity, as more than one reaction pathway may be accessible from the same direction. The only constraint on the choice of parameters for the entire set of reactions of the same two reactants is that:
\begin{equation}
\underset{\mathrm{all} \, j}{\max} [F_{\mathrm{dir}}(j) \, F_{\mathrm{comp}}(j)] + \sum_{\mathrm{all} \, j} F_{\mathrm{dir}}(j) \, (1 - F_{\mathrm{comp}}(j)) \leq 1
\end{equation}
\noindent which ensures that the fractional orientation space over which reactions may occur between any one pair of reactants, summed over all $j$ reactions, will never exceed unity. If the above expression is less than unity then there is, by definition, some fraction of orientation space in which no reactions are accessible (i.e.~the reactants meet but no reaction can occur). 

To employ the old, usual competition method between alternative reactions, values of $F_{\mathrm{dir}}=1$ and $F_{\mathrm{comp}}=1$ would be chosen for each reaction, meaning that all reactions are accessible from all directions and that their rates all compete directly with each other. Note that for the fraction of cases, $F_{\mathrm{comp}}$, in which the reaction in question competes with any other available reactions, it does so with all other possible reactions simultaneously. This simplifies the calculations somewhat, in cases where there are multiple possible reaction pathways available. 

The effect of the new treatment is that, in cases where the relative orientations of the reactants are correct, only one barrier-mediated reaction is allowed; this benefits reaction pathways that would otherwise compete poorly against alternative, lower-barrier reactions between the same reactants. Meanwhile, in that fraction of orientiations in which both/all reactions are accessible, the fastest would win out.

Table \ref{gs-rxns} shows the $F_{\mathrm{dir}}$ and $F_{\mathrm{comp}}$ values adopted for each of the new or adjusted reactions, where relevant. For the H + CH$_3$OH $\rightarrow$ H$_2$ + CH$_2$OH/CH$_3$O reactions, $F_{\mathrm{dir}}=0.5$ and $F_{\mathrm{comp}}=0.5$ are adopted for both branches, which reproduces the approach taken by \cite{Simons20}. For the reactions between an H-atom and glycolaldehyde, we allow all the different barrier-mediated reactions to take the same orientation parameters, while we convert the reaction rates of Simons et al. to activation energy barriers such that the rectangular-barrier tunneling treatment adopted here produces the same ultimate rates. For consistency, similar orientation parameters are used for a selection of comparable reactions in our network.

Table \ref{gs-rxns} also includes values for the branching ratios, BR, assumed for certain pairs of barrierless reactions; these correspond to different possible reaction outcomes unrelated to the initial orientation of the reactants. Note that the branching ratio, as distinct from $F_{\mathrm{dir}}$, should be applied at the end of the rate calculation. This small distinction in treatment is relevant when considering modified rates for diffusive reactions, or when calculating the rates of nondiffusive reaction processes.


\startlongtable
\begin{deluxetable}{lllrlll} \label{gs-rxns}
\tabletypesize{\footnotesize}
\tablecaption{Grain-surface reactions related to COMs, adjusted or added to the network in selected model runs.}
\tablewidth{0pt}
\tablehead{
\colhead{Reaction} & \colhead{} & \colhead{} & \colhead{$E_A$ (K)} & \colhead{$F_{\mathrm{comp}}$} & \colhead{$F_{\mathrm{dir}}$} & \colhead{BR}
}
\startdata
H  +    CH$_3$OH                  & $\rightarrow$ &  H$_2$  +  $\overset{\centerdot}{\rm C}$H$_2$OH        &  4380  &    0.5$^a$ &  0.5$^a$ & --   \\
H  +    CH$_3$OH                  & $\rightarrow$ &  H$_2$  +  CH$_3$$\overset{\centerdot}{\rm O}$           &  6640  &    0.5$^a$ &  0.5$^a$ & --   \\
H  +    HCOOCH$_3$              & $\rightarrow$ &   H$_2$              +  CH$_3$O$\overset{\centerdot}{\rm C}$O   &  3970  &     0.5  &  0.5  & --   \\
H  +    HCOOCH$_3$              & $\rightarrow$ &   H$_2$              +  HCOO$\overset{\centerdot}{\rm C}$H$_2$ &  4470   &     0.5  &  0.5  & --   \\
H  +    HCOCHO                     & $\rightarrow$ &   H$_2$  +  $\overset{\centerdot}{\rm C}$OCHO              &    600$^a$  &     1  &  0.333$^a$  & --   \\
H  +    HCOCHO                     & $\rightarrow$ &   CH$_2$$\overset{\centerdot}{\rm O}$CHO                            &  1100$^b$  &     1  &  0.333$^a$  & --   \\
H  +    HCOCHO                     & $\rightarrow$ &   $\overset{\centerdot}{\rm C}$HOHCHO                                 &  2520$^b$  &     1  &  0.333$^a$  & --   \\
H  +    CH$_2$(OH)CHO         & $\rightarrow$ &   H$_2$              +  CH$_2$($\overset{\centerdot}{\rm O}$)CHO &    992$^a$  &     0.75  &  0.5  & --   \\
H  +    CH$_2$(OH)CHO         & $\rightarrow$ &   H$_2$              +  $\overset{\centerdot}{\rm C}$H(OH)CHO      &  1530$^a$  &     0.75  &  0.5  & --   \\
H  +    CH$_2$(OH)CHO         & $\rightarrow$ &   H$_2$              +  CH$_2$(OH)$\overset{\centerdot}{\rm C}$O &    374$^a$  &     0.75  &  0.5  & --   \\
H  +    CH$_2$(OH)CHO         & $\rightarrow$ &   CH$_2$(OH)CH$_2$$\overset{\centerdot}{\rm O}$                   &  1100$^a$  &     0.75  &  0.5  & --   \\
H  +    CH$_2$(OH)CHO         & $\rightarrow$ &   CH$_2$(OH)$\overset{\centerdot}{\rm C}$HOH                        &  2520$^a$  &     0.75  &  0.5  & --   \\
H  +    (CH$_2$OH)$_2$        & $\rightarrow$ &   H$_2$  +  CH$_2$(OH)CH$_2$$\overset{\centerdot}{\rm O}$    &  3400$^c$  &       --       &    --     & --   \\
H  +    (CH$_2$OH)$_2$        & $\rightarrow$ &   H$_2$  +  CH$_2$(OH)$\overset{\centerdot}{\rm C}$HOH         &  3400$^c$  &       --       &    --     & --   \\
$\overset{\centerdot}{\rm C}$H$_3$  +  HCOOCH$_3$       & $\rightarrow$ &   CH$_4$             +  CH$_3$O$\overset{\centerdot}{\rm C}$O   &  5040  &     0.5  &  0.5  & --   \\
$\overset{\centerdot}{\rm C}$H$_3$  +  HCOOCH$_3$       & $\rightarrow$ &   CH$_4$             +  HCOO$\overset{\centerdot}{\rm C}$H$_2$ &  5800  &     0.5  &  0.5  & --   \\
CH$_3$$\overset{\centerdot}{\rm O}$ + HCOOCH$_3$       & $\rightarrow$ &   CH$_3$OH        +  CH$_3$O$\overset{\centerdot}{\rm C}$O    &  3240$^c$  &     0.5  &  0.5  & --   \\
CH$_3$$\overset{\centerdot}{\rm O}$ + HCOOCH$_3$       & $\rightarrow$ &   CH$_3$OH        +  HCOO$\overset{\centerdot}{\rm C}$H$_2$  &  3200$^c$  &     0.5  &  0.5  & --   \\
CH$_3$$\overset{\centerdot}{\rm O}$ +  CH$_2$(OH)CHO & $\rightarrow$ &   CH$_3$OH        +  CH$_2$(OH)$\overset{\centerdot}{\rm C}$O  &  1580$^c$  &     0.5  &  0.5  & --   \\
CH$_3$$\overset{\centerdot}{\rm O}$ +  CH$_2$(OH)CHO & $\rightarrow$ &   CH$_3$OH        +  CH$_2$($\overset{\centerdot}{\rm O}$)CHO  &  3470$^c$  &     0.5  &  0.5  & --   \\
CH$_3$$\overset{\centerdot}{\rm O}$ +  CH$_2$(OH)CHO & $\rightarrow$ &   CH$_3$OH        +  $\overset{\centerdot}{\rm C}$H(OH)CHO       &  3470$^c$  &     0.5  &  0.5  & --   \\
CH$_3$$\overset{\centerdot}{\rm O}$ + (CH$_2$OH)$_2$ & $\rightarrow$ &   CH$_3$OH + CH$_2$(OH)CH$_2$$\overset{\centerdot}{\rm O}$  &  3470$^c$  &       --       &    --     & --   \\
CH$_3$$\overset{\centerdot}{\rm O}$ + (CH$_2$OH)$_2$ & $\rightarrow$ &   CH$_3$OH + CH$_2$(OH)$\overset{\centerdot}{\rm C}$HOH       &  3470$^c$  &       --       &    --     & --   \\
$\overset{\centerdot}{\rm N}$H$_2$ +   HCOOCH$_3$       & $\rightarrow$ &   NH$_3$            +  CH$_3$O$\overset{\centerdot}{\rm C}$O    &  3130$^c$  &     0.5  &  0.5  & --   \\
$\overset{\centerdot}{\rm N}$H$_2$ +   HCOOCH$_3$       & $\rightarrow$ &   NH$_3$            +  HCOO$\overset{\centerdot}{\rm C}$H$_2$  &  5020$^c$  &     0.5  &  0.5  & --   \\
$\overset{\centerdot}{\rm N}$H$_2$ +   CH$_2$(OH)CHO  & $\rightarrow$ &   NH$_3$            +  CH$_2$(OH)$\overset{\centerdot}{\rm C}$O  &  1360$^c$  &     0.5  &  0.5  & --   \\
$\overset{\centerdot}{\rm N}$H$_2$ +   CH$_2$(OH)CHO  & $\rightarrow$ &   NH$_3$            +  CH$_2$($\overset{\centerdot}{\rm O}$)CHO  &  3500$^c$  &     0.5  &  0.5  & --   \\
$\overset{\centerdot}{\rm N}$H$_2$ +   CH$_2$(OH)CHO  & $\rightarrow$ &   NH$_3$            +  $\overset{\centerdot}{\rm C}$H(OH)CHO       &  3500$^c$  &     0.5  &  0.5  & --   \\
$\overset{\centerdot}{\rm N}$H$_2$ +   (CH$_2$OH)$_2$  & $\rightarrow$ &   NH$_3$ + CH$_2$(OH)CH$_2$$\overset{\centerdot}{\rm O}$     &  3500$^c$  &       --       &    --     & --   \\
$\overset{\centerdot}{\rm N}$H$_2$ +   (CH$_2$OH)$_2$  & $\rightarrow$ &   NH$_3$ + CH$_2$(OH)$\overset{\centerdot}{\rm C}$HOH          &  3500$^c$  &       --       &    --     & --   \\
$\overset{\centerdot}{\rm O}$H  +  CH$_3$OH                  & $\rightarrow$ &   H$_2$O           +  $\overset{\centerdot}{\rm C}$H$_2$OH       &   359  &     0.5  &  0.5  & --   \\
$\overset{\centerdot}{\rm O}$H  +  CH$_3$OH                  & $\rightarrow$ &   H$_2$O           +  CH$_3$$\overset{\centerdot}{\rm O}$         &   852  &     0.5  &  0.5  & --   \\
$\overset{\centerdot}{\rm O}$H  +   HCOOCH$_3$             & $\rightarrow$ &   H$_2$O           +  CH$_3$O$\overset{\centerdot}{\rm C}$O     &   597$^c$  &     0.5  &  0.5  & --   \\
$\overset{\centerdot}{\rm O}$H  +   HCOOCH$_3$             & $\rightarrow$ &   H$_2$O           +  HCOO$\overset{\centerdot}{\rm C}$H$_2$   &   597$^c$  &     0.5  &  0.5  & --   \\
$\overset{\centerdot}{\rm O}$H  +   CH$_2$(OH)CHO        & $\rightarrow$ &   H$_2$O           +  CH$_2$(OH)$\overset{\centerdot}{\rm C}$O   &   597$^c$  &     0.5  &  0.5  & --   \\
$\overset{\centerdot}{\rm O}$H  +   CH$_2$(OH)CHO        & $\rightarrow$ &   H$_2$O           +  CH$_2$($\overset{\centerdot}{\rm O}$)CHO   &  2530$^c$  &    0.5  &  0.5  & --   \\
$\overset{\centerdot}{\rm O}$H  +   CH$_2$(OH)CHO        & $\rightarrow$ &   H$_2$O           +  $\overset{\centerdot}{\rm C}$H(OH)CHO        &  2530$^c$  &    0.5  &  0.5  & --   \\
$\overset{\centerdot}{\rm O}$H  +    (CH$_2$OH)$_2$       & $\rightarrow$ &   H$_2$O   +  CH$_2$(OH)CH$_2$$\overset{\centerdot}{\rm O}$  &  2530$^c$  &       --       &    --     & --   \\
$\overset{\centerdot}{\rm O}$H  +    (CH$_2$OH)$_2$       & $\rightarrow$ &   H$_2$O   +  CH$_2$(OH)$\overset{\centerdot}{\rm C}$HOH       &  2530$^c$  &       --       &    --     & --   \\
H    +   $\overset{\centerdot}{\rm C}$OCHO                             & $\rightarrow$ &  HCOCHO                                     &       0  & --             &  --    & 0.5 \\
H    +   $\overset{\centerdot}{\rm C}$OCHO                             & $\rightarrow$ &  H$_2$CO          +              CO      &       0  & --             &  --    & 0.5 \\
$\overset{\centerdot}{\rm C}$H$_2$OH  +  $\overset{\centerdot}{\rm C}$H(OH)CHO             & $\rightarrow$ &  CH$_2$(OH)CHO  +     H$_2$CO      &       0  & --             &  --    & -- \\
$\overset{\centerdot}{\rm C}$H$_2$OH  +  CH$_2$($\overset{\centerdot}{\rm O}$)CHO        & $\rightarrow$ &  CH$_2$(OH)CHO  +     H$_2$CO      &       0  & --             &  --    & -- \\
$\overset{\centerdot}{\rm C}$H$_2$OH + CH$_2$(OH)CH$_2$$\overset{\centerdot}{\rm O}$ & $\rightarrow$ &  (CH$_2$OH)$_2$  +  H$_2$CO      &       0  & --             &  --    & -- \\
$\overset{\centerdot}{\rm C}$H$_2$OH  +  CH$_2$(OH)$\overset{\centerdot}{\rm C}$HOH    & $\rightarrow$ &  (CH$_2$OH)$_2$  +  H$_2$CO      &       0  & --             &  --    & -- \\
CH$_3$$\overset{\centerdot}{\rm O}$    +  $\overset{\centerdot}{\rm C}$H(OH)CHO             & $\rightarrow$ &  CH$_2$(OH)CHO  +      H$_2$CO     &       0  & --             &  --    & -- \\
CH$_3$$\overset{\centerdot}{\rm O}$  +  CH$_2$(OH)CH$_2$$\overset{\centerdot}{\rm O}$ & $\rightarrow$ &  (CH$_2$OH)$_2$  +   H$_2$CO     &       0  & --             &  --    & -- \\
CH$_3$$\overset{\centerdot}{\rm O}$    +  CH$_2$($\overset{\centerdot}{\rm O}$)CHO        & $\rightarrow$ &  CH$_2$(OH)CHO  +   H$_2$CO     &       0  & --             &  --    & -- \\
CH$_3$$\overset{\centerdot}{\rm O}$    +  CH$_2$(OH)$\overset{\centerdot}{\rm C}$HOH    & $\rightarrow$ &  (CH$_2$OH)$_2$  +  H$_2$CO      &       0  & --             &  --    & -- \\
$\overset{\centerdot}{\rm C}$OOH         +  $\overset{\centerdot}{\rm C}$H(OH)CHO             & $\rightarrow$ &  CH$_2$(OH)CHO  +        CO$_2$     &       0  & --             &  --    & -- \\
$\overset{\centerdot}{\rm C}$OOH         +  CH$_2$($\overset{\centerdot}{\rm O}$)CHO        & $\rightarrow$ &  CH$_2$(OH)CHO  +        CO$_2$     &       0  & --             &  --    & -- \\$\overset{\centerdot}{\rm C}$OOH       +  CH$_2$(OH)CH$_2$$\overset{\centerdot}{\rm O}$ & $\rightarrow$ &  (CH$_2$OH)$_2$   +    CO$_2$     &       0  & --             &  --    & -- \\
$\overset{\centerdot}{\rm C}$OOH         +  CH$_2$(OH)$\overset{\centerdot}{\rm C}$HOH    & $\rightarrow$ &  (CH$_2$OH)$_2$   +    CO$_2$     &       0  & --             &  --    & -- \\
H            +  CH$_2$(OH)CH$_2$$\overset{\centerdot}{\rm O}$  & $\rightarrow$ &  (CH$_2$OH)$_2$                         &       0  & --             &  --    & 0.5 \\
H            +  CH$_2$(OH)CH$_2$$\overset{\centerdot}{\rm O}$  & $\rightarrow$ &  H$_2$  +  CH$_2$(OH)CHO        &       0  & --             &  --    & 0.5 \\
H               +  CH$_2$(OH)$\overset{\centerdot}{\rm C}$HOH    & $\rightarrow$ &  (CH$_2$OH)$_2$                         &       0  & --             &  --    & 0.5 \\
H               +  CH$_2$(OH)$\overset{\centerdot}{\rm C}$HOH    & $\rightarrow$ &  H$_2$  +  CH$_2$(OH)CHO        &       0  & --             &  --    & 0.5 \\
H$\overset{\centerdot}{\rm C}$O           +  $\overset{\centerdot}{\rm C}$H(OH)CHO             & $\rightarrow$ &  CH$_2$(OH)CHO  +               CO      &       0  & --             &  --    & -- \\
H$\overset{\centerdot}{\rm C}$O        +  CH$_2$(OH)CH$_2$$\overset{\centerdot}{\rm O}$  & $\rightarrow$ &  (CH$_2$OH)$_2$   +           CO      &       0  & --             &  --    & -- \\
H$\overset{\centerdot}{\rm C}$O           +  CH$_2$($\overset{\centerdot}{\rm O}$)CHO        & $\rightarrow$ &  CH$_2$(OH)CHO  +               CO      &       0  & --             &  --    & -- \\
H$\overset{\centerdot}{\rm C}$O           +  CH$_2$(OH)$\overset{\centerdot}{\rm C}$HOH    & $\rightarrow$ &  (CH$_2$OH)$_2$   +           CO      &       0  & --             &  --    & -- \\
\enddata
\tablecomments{For clarity, radical sites are indicated with a dot. Activation energy barriers are taken from the network of \citet{Garrod13a} unless otherwise marked. For purposes of quantum tunneling, barriers are assumed to take the default width of 1 \AA. References: $^a$ \citet{Simons20}; $^b$ based on equivalent process for CH$_2$(OH)CHO; $^c$ estimate, based on the Evans-Polanyi relation \citep[e.g.][]{Dean99}, where applicable.}
\end{deluxetable}

\subsubsection{Methylene (CH$_2$) reactions}\label{sec:CH2-rxns}

JG20 found that even the inclusion of nondiffusive grain-surface chemical processes was not sufficient to reproduce the measured gas-phase abundance of dimethyl ether (CH$_3$OCH$_3$) in pre-stellar cores. One of the solutions that they suggested to remedy this was the reaction of methanol with ground-state (i.e.~triplet) methylene, CH$_2$; this reactive diradical would, in this scenario, abstract a hydrogen atom from one or other end of the methanol (CH$_3$OH) molecule. The resulting pair of radicals, CH$_3$ and CH$_3$O/CH$_2$OH, would then immediately recombine to produce either dimethyl ether or ethanol (C$_2$H$_5$OH).

Here we introduce these reactions into the network (see Table \ref{CH2-rxns}), to test their influence during both the cold collapse stage and the warm-up stage of hot-core chemistry. For simplicity, we assume that there is no direct competition between the CH$_3$O- or CH$_2$OH-producing pathways in the abstraction of H from methanol by methylene, aside from the orientation-related efficiencies ($F_{\mathrm{dir}}=0.5$). The barriers for either process are taken from the gas-phase values determined by \citet{Tsang87}, with a standard H-tunneling treatment used to determine the surface/mantle rates; the latter therefore involve a substantial degree of uncertainty. For either set of products, a posterior branching ratio of 0.5 is used to provide the fraction of cases in which the radicals immediately recombine to produce a stable molecule.

To ensure that the production rates of dimethyl ether and ethanol are not unfairly biased by a lack of other methylene reactions, similar reaction processes have also been added to the network, with a concentration on reaction partners that are abundant on grains and/or are expected to have a small barrier to reaction with methylene.

As noted by \citet{Sim20}, experimental evidence suggests that the grain-surface reaction C + H$_2$ $\rightarrow$ CH$_2$ has only a minimal or zero activation energy barrier \citep{Kras16, Henning19}, in contrast to the long-standing value of 2500~K used in this and earlier models. In addition to the other changes to methylene chemistry, a zero barrier is adopted for this reaction. One effect of this change is to reduce the availability of CH on grain surfaces, by providing an abundant alternative reaction partner for C atoms other than atomic H (also see Sec.~\ref{sec:CH-rxns}).

\startlongtable
\begin{deluxetable}{lllrrr} \label{CH2-rxns}
\tabletypesize{\footnotesize}
\tablecaption{Grain-surface reactions involving methylene (CH$_2$) added to the network or adjusted in selected model runs.}
\tablewidth{0pt}
\tablehead{
\colhead{Reaction} & \colhead{} & \colhead{} & \colhead{$E_A$ (K)} & \colhead{$F_{\mathrm{dir}}$} & \colhead{BR}
}
\startdata
C            +  H$_2$           &  $\rightarrow$  &  {\"C}H$_2$                                                                                                                  &  0$^a$  & -- & --\\ 
{\"C}H$_2$  +  CH$_3$OH     &  $\rightarrow$  &  $\overset{\centerdot}{\rm C}$H$_3$  +  $\overset{\centerdot}{\rm C}$H$_2$OH           &  3610$^b$  & 0.5 & 0.5  \\ 
{\"C}H$_2$  +  CH$_3$OH     &  $\rightarrow$  &  C$_2$H$_5$OH                                                                                                        &  3610$^b$  & 0.5 & 0.5  \\ 
{\"C}H$_2$  +  CH$_3$OH     &  $\rightarrow$  &  CH$_3$$\overset{\centerdot}{\rm O}$ +  $\overset{\centerdot}{\rm C}$H$_3$              &  3490$^b$  & 0.5 & 0.5  \\ 
{\"C}H$_2$  +  CH$_3$OH     &  $\rightarrow$  &  CH$_3$OCH$_3$                                                                                                       &  3490$^b$  & 0.5 & 0.5  \\ 
{\"C}H$_2$  +  CH$_4$         &  $\rightarrow$  &  $\overset{\centerdot}{\rm C}$H$_3$ +   $\overset{\centerdot}{\rm C}$H$_3$                &  5050$^c$  & 1.0    & 0.5 \\ 
{\"C}H$_2$  +  CH$_4$         &  $\rightarrow$  &  C$_2$H$_6$                                                                                                             &  5050$^c$  & 1.0    & 0.5 \\ 
{\"C}H$_2$  + CO       &  $\rightarrow$  &  CH$_2$CO                                                                                                                          &  2870$^d$  & 1.0    & 1.0 \\ 
{\"C}H$_2$  +  H$_2$CO       &  $\rightarrow$  &  $\overset{\centerdot}{\rm C}$H$_3$ +   H$\overset{\centerdot}{\rm C}$O                    &    817$^e$  & 1.0    & 0.5 \\ 
{\"C}H$_2$  +  H$_2$CO       &  $\rightarrow$  &  CH$_3$CHO                                                                                                              &   817$^e$  & 1.0    & 0.5 \\ 
{\"C}H$_2$  +  C$_2$H$_6$  &  $\rightarrow$  &  $\overset{\centerdot}{\rm C}$H$_3$  +  $\overset{\centerdot}{\rm C}$H$_2$CH$_3$     &  3980$^e$  & 1.0    &  0.5 \\ 
{\"C}H$_2$  +  C$_2$H$_6$  &  $\rightarrow$  &  C$_3$H$_8$                                                                                                             &  3980$^e$  & 1.0    &  0.5 \\ 
{\"C}H$_2$  +  C$_3$H$_8$  &  $\rightarrow$  &  $\overset{\centerdot}{\rm C}$H$_3$ +   C$_3$H$_7$                                                   &  3600$^f$  & 0.5 & 0.5  \\ 
{\"C}H$_2$  +  C$_3$H$_8$  &  $\rightarrow$  &  C$_4$H$_{10}$                                                                                                         &  3600$^f$  & 0.5 & 0.5  \\ 
{\"C}H$_2$  +  C$_3$H$_8$  &  $\rightarrow$  &  $\overset{\centerdot}{\rm C}$H$_3$  +   CH$_3$$\overset{\centerdot}{\rm C}$HCH$_3$  &  3760$^f$  & 0.5 & 0.5  \\ 
{\"C}H$_2$  +  C$_3$H$_8$  &  $\rightarrow$  &  i-C$_4$H$_{10}$                                                                                                        &  3760$^f$  & 0.5 & 0.5  \\ 
{\"C}H$_2$  +  CH$_3$CHO   &  $\rightarrow$  &  $\overset{\centerdot}{\rm C}$H$_3$  +  CH$_3$$\overset{\centerdot}{\rm C}$O             &  1770$^f$  & 1.0    &  0.5 \\ 
{\"C}H$_2$  +  CH$_3$CHO   &  $\rightarrow$  &  CH$_3$COCH$_3$                                                                                                       &  1770$^f$  & 1.0    &  0.5 \\ 
\enddata
\tablecomments{For clarity, radical sites are indicated with a dot. Methylene is assumed to be in its ground state, and thus a diradical. For purposes of quantum tunneling, barriers are assumed to take the default width of 1 \AA. $F_{\mathrm{comp}}$ is set to zero for all the reactions listed. References: 
$^a$\citet{Sim20};
$^b$\citet{Tsang87};
$^c$\citet{Bohland85};
$^d$\citet{Jin20} estimate, based on CH$_3$ + CO reaction;
$^e$\citet{Wang06};
$^f$\citet{Tsang88}.
}
\end{deluxetable}

\begin{deluxetable}{lllrl} \label{CH-rxns}
\tabletypesize{\footnotesize}
\tablecaption{Grain-surface reactions involving methylidyne (CH) added to the network, or with adjusted barriers, used in selected model runs.}
\tablewidth{0pt}
\tablehead{
\colhead{Reaction} & \colhead{} & \colhead{} & \colhead{$E_A$ (K)} & \colhead{BR}
}
\startdata
$\overset{\centerdot}{\rm C}$H  +   H$_2$               & $\rightarrow$ &   CH$_3$                              &  0$^a$  &  --  \\ 
$\overset{\centerdot}{\rm C}$H  +   C$_2$H$_2$      & $\rightarrow$ &   C$_3$H$_3$                        &  0$^b$  &  --  \\ 
$\overset{\centerdot}{\rm C}$H  +   C$_2$H$_4$      & $\rightarrow$ &   C$_3$H$_5$                        &  0$^c$  &  --  \\ 
$\overset{\centerdot}{\rm C}$H  +   C$_2$H$_5$      & $\rightarrow$ &   C$_3$H$_6$                        &  0  &  --  \\
$\overset{\centerdot}{\rm C}$H  +   C$_2$H$_6$      & $\rightarrow$ &   $\overset{\centerdot}{\rm C}$H$_2$CH$_2$CH$_3$   &  0  &  0.5  \\
$\overset{\centerdot}{\rm C}$H  +   C$_2$H$_6$      & $\rightarrow$ &   CH$_3$$\overset{\centerdot}{\rm C}$HCH$_3$          &  0  &  0.5  \\
$\overset{\centerdot}{\rm C}$H  +   C$_3$H$_2$      & $\rightarrow$ &   C$_4$H$_3$                        &  0  &  --  \\
$\overset{\centerdot}{\rm C}$H  +   C$_3$H$_4$      & $\rightarrow$ &   C$_4$H$_5$                        &  0  &  --  \\
$\overset{\centerdot}{\rm C}$H  +   C$_3$H$_6$      & $\rightarrow$ &   C$_4$H$_7$                        &  0  &  --  \\
$\overset{\centerdot}{\rm C}$H  +   C$_3$H$_8$      & $\rightarrow$ &   C$_4$H$_9$                        &  0  &  0.5  \\
$\overset{\centerdot}{\rm C}$H  +   C$_3$H$_8$      & $\rightarrow$ &   i-C$_4$H$_9$                      &  0  &  0.5  \\
$\overset{\centerdot}{\rm C}$H  +   CH$_4$             & $\rightarrow$ &   C$_2$H$_5$                        &  0$^d$  &  --  \\ 
$\overset{\centerdot}{\rm C}$H  +   CO                    & $\rightarrow$ &   H$\overset{\centerdot}{\rm C}$CO                           &  0$^e$  &  --  \\ 
$\overset{\centerdot}{\rm C}$H  +   H$\overset{\centerdot}{\rm C}$O                  & $\rightarrow$ &   {\"C}H$_2$  +  CO                     &  0  &  --  \\
$\overset{\centerdot}{\rm C}$H  +   H$_2$CO           & $\rightarrow$ &   CH$_2$CHO                         &  0$^f$  &  --  \\ 
$\overset{\centerdot}{\rm C}$H  +   $\overset{\centerdot}{\rm C}$H$_2$OH         & $\rightarrow$ &   C$_2$H$_3$OH                    &  0  &  --  \\
$\overset{\centerdot}{\rm C}$H  +   CH$_3$$\overset{\centerdot}{\rm O}$           & $\rightarrow$ &   {\"C}H$_2$  +  H$_2$CO            &  0  &  --  \\
$\overset{\centerdot}{\rm C}$H  +   CH$_3$OH         & $\rightarrow$ &   C$_2$H$_4$OH                    &  0$^g$  &  --  \\ 
$\overset{\centerdot}{\rm C}$H  +   CO$_2$             & $\rightarrow$ &   H$\overset{\centerdot}{\rm C}$O  +  CO                          &  345$^h$  &  --  \\ 
$\overset{\centerdot}{\rm C}$H  +   H$_2$O             & $\rightarrow$ &   $\overset{\centerdot}{\rm C}$H$_2$OH                           &  0$^i$  &  --  \\ 
$\overset{\centerdot}{\rm C}$H  +   H$_2$O$_2$      & $\rightarrow$ &   {\"C}H$_2$  +  O$_2$H              &  0  &  --  \\
$\overset{\centerdot}{\rm C}$H  +   NH$_3$             & $\rightarrow$ &   CH$_2$NH  +  H                    &  0$^j$  &  --  \\ 
$\overset{\centerdot}{\rm C}$H  +   NH$_2$OH         & $\rightarrow$ &   {\"C}H$_2$  +  H$\overset{\centerdot}{\rm N}$OH                 &  0  &  --  \\
$\overset{\centerdot}{\rm C}$H  +   $\overset{\centerdot}{\rm C}$H$_2$NH$_2$  & $\rightarrow$ &   {\"C}H$_2$  +  CH$_2$NH           &  0  &  --  \\
$\overset{\centerdot}{\rm C}$H  +   CH$_3$NH$_2$  & $\rightarrow$ &   {\"C}H$_2$  +  $\overset{\centerdot}{\rm C}$H$_2$NH$_2$    &  0$^k$  &  --  \\ 
$\overset{\centerdot}{\rm C}$H  +   O$_3$               & $\rightarrow$ &   H$\overset{\centerdot}{\rm C}$O  +  O$_2$                     &  0  &  --  \\
\enddata
\tablecomments{For clarity, and where necessary, radical sites are indicated with a dot. References: 
$^a$\citet{Fulle97}; 
$^b$\citet{Thiesemann97}; 
$^c$\citet{Thiesemann01}; 
$^d$\citet{Blitz97};
$^e$\citet{Baulch92};
$^f$\citet{Zarbanick88};
$^g$\citet{Johnson00};
$^h$\citet{Berman82};
$^i$\citet{Hickson2013};
$^j$\citet{Blitz12};
$^k$\citet{Zarbanick89}
}
\end{deluxetable}

\subsubsection{Methylidyne (CH) reactions}\label{sec:CH-rxns}

The radical CH is an important precursor on the path of C-atom conversion to methane on grain surfaces, and this has typically been the main focus of its inclusion in interstellar grain-surface chemical networks. However, experiments show also that CH may participate in gas-phase reactions with stable species, with little to no apparent activation energy barrier \citep[e.g.][]{Hickson2013}. \citet{Blitz12}, in a combined experimental and theoretical study, showed that CH may react with ammonia (NH$_3$), producing NH$_2$CH + H in $\sim$96\% of cases. Crucially, \citet{Ioppolo20} showed that such a reaction should be important in the production of radicals serving as precursors to glycine (NH$_2$CH$_2$COOH), suggesting also that glycine could be formed in substantial quantities in cold interstellar clouds/cores. Following \citet{Ioppolo20}, we adopt this reaction as a barrierless surface/mantle process for all models; however, in the {\tt final} model we assume products NH$_2$CH + H (per Blitz et al.), as opposed to the single product NH$_2$CH$_2$ used in the Ioppolo et al. models.

As with CH$_2$, to ensure that the CH reaction with ammonia is not unduly biased by the sparsity of the network for methylidyne, a number of new surface/mantle CH reactions are added, with a focus on abundant surface reaction partners and/or those expected to have minimal barriers against reaction with CH. Unless the reaction products are otherwise determined (as per the ammonia reaction), the new CH reactions are assigned appropriate single products where such already exist in the network; the reaction is otherwise assumed to proceed as an abstraction process. Table \ref{CH-rxns} gives references for reactions in which there is no apparent experimental/theoretical barrier. Unreferenced reactions in the table nevertheless do not show experimental evidence of any substantial barrier and are therefore assumed to behave similarly.

The adoption of the reaction CH + H$_2$ $\rightarrow$ CH$_3$ with a zero barrier is based on the experimental high-pressure limit for this process \citep{Fulle97,Brownsword97}, which should be most appropriate for reactions on a grain (in the absence of explicit measurements or calculations for surfaces). This reaction provides a means to form CH$_3$ on the grains without the involvement of the precursor CH$_2$.

\begin{deluxetable}{lllrr} \label{misc-rxns}
\tabletypesize{\footnotesize}
\tablecaption{Miscellaneous grain-surface reactions added to the network or adjusted in selected model runs.}
\tablewidth{0pt}
\tablehead{
\colhead{Reaction} & \colhead{} & \colhead{} & \colhead{$E_A$ (K)} & BR
}
\startdata
$\overset{\centerdot}{\rm N}$H$_2$  +  H$_2$CO    & $\rightarrow$ &  NH$_2$CHO    +  H     &  1900  & --  \\
O  +  CH$_3$O$\overset{\centerdot}{\rm C}$H$_2$  & $\rightarrow$ &  HCOOCH$_3$  +  H     &       0  & --  \\
$\overset{\centerdot}{\rm O}$H  +  CO                                                              & $\rightarrow$ &  CO$_2$  +  H             &     80  &  0.99 \\
$\overset{\centerdot}{\rm O}$H  +  CO                                                              & $\rightarrow$ &  COOH                  &     80  &  0.01 \\
\enddata
\tablecomments{For clarity, radical sites are indicated with a dot.}
\end{deluxetable}

\subsubsection{Other miscellaneous reactions}\label{methods:GR}

The reaction NH$_2$ + H$_2$CO $\rightarrow$ NH$_2$CHO + H, which is added to the gas-phase network (see Sec.~\ref{sec:methods:gas-phase}), may also plausibly occur on the grains, although inclusion in the surface network requires an explicit activation energy value. \citet{Vazart16}, with a view to gas-phase formation, studied this reaction system computationally, finding that at low energies the reaction should proceed effectively, with a modest barrier to be overcome following capture. To calculate the rate at which this reaction proceeds on grain surfaces following the meeting of the two reactants, we assume that the system begins in the weak van der Waals complex shown in their Figure 9. This leaves a barrier of around 15.8 kJ/mol, or 1900~K, to be overcome. In our model, this barrier is treated thermally, and is in competition with the diffusion of both reactants. This treatment remains crude and would be improved by energy-level calculations on a water cluster, for example. This and the other miscellaneous reactions added to the {\tt final} model are shown in Table \ref{misc-rxns}.

For completeness, here the O + CH$_3$OCH$_2$ $\rightarrow$ HCOOCH$_3$ + H reaction that was added to the gas-phase chemistry is also included in the surface/mantle network, with no activation-energy barrier.

In all of the models presented here, the grain-surface/ice reaction of CO + OH may produce either COOH or CO$_2$ + H as products. This reaction likely involves a small entrance barrier, assumed here to be 80~K \citep{Ruffle01}, producing COOH$^*$; this excited intermediate may either relax to COOH through vibrational interactions with the surface, or it may eject a hydrogen atom via tunneling, to produce CO$_2$. The latter is, in general, the main mechanism of surface CO$_2$ production in the models \citep{GP11}. The reaction of atomic H with stabilized COOH is assumed to produce either formic acid or CO$_2$ + H$_2$, in a ratio of 50:50 (which is held the same in all models). This process therefore provides an additional mechanism for CO$_2$ production that still derives from the reaction of CO and OH.

In most of the models presented in this paper, a ratio of 50:50 between products COOH and CO$_2$ + H is assumed for the CO + OH reaction, following \citet{Ioppolo20}. In the {\tt final} model, the branching ratio is tuned to 99:1 in favor of CO$_2$ production, primarily as a means to fix the overproduction of HCOOH on the grains (see Sec.~\ref{sec:results:final}). Experimental evidence indeed suggests that CO$_2$ can be efficiently formed through the CO + OH $\rightarrow$ CO$_2$ + H pathway; \citet{Oba10} conducted experiments in which non-energetic OH radicals and CO molecules were co-deposited onto an Al surface. Although some COOH was detected, the lack of H present in the experimental setup indicates that reaction of COOH with mobile H was not the formation mechanism for CO$_2$. Experiments by \citet{Watanabe02} and \cite{Watanabe07} indicate that excited OH produced by water photodissociation can also react efficiently with CO to form CO$_2$ in CO/H$_2$O ice mixtures. Molecular dynamics simulations by \citet{Arasa13} sought to reproduce those experiments and to determine the relative production rates of CO$_2$ versus COOH within water ice, but found that COOH$^*$ was around 100 times more likely to stabilize via energy exchange with the surrounding ice before decay to CO$_2$ could occur. This result did not appear to agree with the experiments, although their rate of CO consumption was consistent. Further reactions of COOH with other species could not be ruled out as an explanation for the efficiency of CO$_2$ production in the experiments.

In the present models, most OH that goes on to react with CO is produced as the result of the grain-surface reaction H + O $\rightarrow$ OH, which is exothermic by around 4.4~eV. In cases where the OH + CO reaction is induced by the 3-B or 3-BEF mechanism, the OH is thus likely to be highly excited, which would allow a more rapid ejection of H from the COOH$^*$ complex. The precise 99:1 value chosen here is nevertheless entirely empirical.



\section{Results} \label{sec:results}

Table \ref{tab-models} indicates the different features of each model presented here. Models ranging from {\tt basic\_Av2} to {\tt bas\_stk\_loPD\_H2\_T16\_no-bd} include only the diffusive surface/mantle chemistry; each model adds one new feature to the model above. The next six models in the table are used to test only one individual nondiffusive surface/mantle process at a time, ending with {\tt bas\_stk\_loPD\_H2\_T16\_no-bd\_3B3}. Model {\tt bas\_stk\_loPD\_H2\_T16\_no-bd\_3B3\_EF} adopts both the 3-BEF and standard 3-B mechanism (assuming three cycles). Model {\tt all} is the first in the table to use all of the nondiffusive mechanisms combined, along with all of the other prior changes to the models. Further changes/additions are progressively made to model {\tt all} until the {\tt final} version of the model, also labeled {\tt all\_Edif\_S17\_ON\_mtun\_GP\_Gr}, is reached. 

In Tables \ref{tab-st1-ice1} -- \ref{tab-gas6}, selected chemical abundance results are presented from each model, or from the literature for the same species in comparable observational sources. Tables \ref{tab-st1-ice1} -- \ref{tab-st1-gas} in particular show data from/related to the stage-1 (cold collapse) models; Table \ref{tab-st1-ice1} shows modeled abundances with respect to water ice of simple ice species, at the end of the collapse stage; Table \ref{tab-obs-ice1} shows observational data for the same species, where available; Table \ref{tab-st1-ice2} shows the modeled abundances of solid-phase oxygen-bearing complex organics; Table \ref{tab-st1-gas} shows the modeled {\em peak} fractional abundances with respect to total hydrogen of the same complex organic molecules in the gas phase, some of which have been detected in cold cores; Table \ref{tab-st1-gas} includes observational values for those species, where available, for the prestellar core L1544.

Tables \ref{tab-gas1} -- \ref{tab-gas3} show modeled peak gas-phase abundances with respect to total hydrogen for oxygen-bearing COMs in the stage-2 models, corresponding to the different warm-up timescales; each of the three timescales shares the same stage-1 pre-cursor model. Tables \ref{tab-gas4} -- \ref{tab-gas6} show stage-2 results for a selection of nitrogen-bearing species.

Figures \ref{basic_medium} -- \ref{final_slow} plot a selection of time-dependent molecular abundances for various models in stages 1 and 2.

Later figures and tables correspond solely to the {\tt final} model results, and are described in more detail from Sec.~\ref{sec:results:final} onward. Table \ref{tab-output} in particular shows the peak gas-phase abundances and corresponding system temperatures for a wide range of molecules, produced by each of the three warm-up timescales used in the {\tt final} model.

\subsection{Initial visual extinction during collapse} \label{sec:results:extinction}

As in past models, the gas density during the collapse stage is governed by an isothermal, one-dimensional freefall collapse treatment based on that used by \cite{Rawlings92}. The visual extinction follows the relationship $A_{V} \propto n_{H}^{2/3}$, which varies over time. The initial visual extinction, $A_{V,\mathrm{init}}$, and the initial density of 3000 cm$^{-3}$ are chosen to approximate a low density medium from which a core is formed. In past models, $A_{V,\mathrm{init}}$ has been taken as either 2 or 3 magnitudes. Here, two versions of the basic model are presented, the first using $A_{V,\mathrm{init}}$=2 ({\tt basic\_Av2}), which was used in the most recent published models, and the other with a value of 3 ({\tt basic}). 

The change has a significant effect on the solid-phase abundance of CO during the collapse stage. Table \ref{tab-st1-ice1} shows a final CO abundance with respect to water of around 51\% for the {\tt basic\_Av2} model. The review paper by \citet{Boogert15}, from which some data is reproduced in Table \ref{tab-obs-ice1}, indicates a median observed value for MYSOs of 7\%, with a maximum value of 26\%. The model abundance is therefore substantially higher than the measured value in the high-mass case. Boogert et al. find a median value of 18\% for LYSOs, but with observed values as high as 85\% and an upper quartile value of 35\%, indicating rather better agreement with the model value. As noted by \citet{Garrod17}, the volatility of CO may be a cause of the discrepancy in the high-mass case if temperatures are somewhat elevated \citep[see also][]{Oberg11}, which in the models they are not. The predominance of extremely low dust temperatures in the low-mass case could also explain an extreme degree of CO freeze-out onto the dust grains in some observed sources, greater even than that produced by the present models. 

In order primarily to reduce the final abundance of solid-phase CO to something more in line with observations of both low- and high-mass sources, the {\tt basic} and subsequent models adopt a value $A_{V,\mathrm{init}}$=3. The CO/H$_2$O fraction for model {\tt basic} is around 40\%. This reduction is caused partly by the smaller fraction of gas-phase carbon that is in ionic form, due to the lower photodissociation rate; this somewhat reduces the rate at which CO is produced in the gas phase, which may then be accreted onto the dust grains. Also, the somewhat lower initial dust temperature that corresponds to the higher extinction (i.e.~$\sim$14.7~K versus $\sim$16~K) allows grain-surface chemistry, including the production of water and methane (CH$_4$) from O and C, to proceed more effectively at early times in the collapse stage by allowing longer surface lifetimes (against thermal desorption) for atomic H. A corresponding increase in the CH$_4$/H$_2$O ratio is seen in model {\tt basic}, while the fractional abundance of water itself (with respect to total H) is also around 9\% greater, leading to a reduction in all molecular fractions with respect to H$_2$O. 

As with CO, the higher extinction produces a fall in CO$_2$ and CH$_3$OH ratios with respect to water. Although CO$_2$ is on the low side as a result, at 10.6\% versus observational minimum values of 11 and 12\% for MYSOs and LYSOs, respectively, the modeled CH$_3$OH/H$_2$O ratio of 6.3\% is now closer to the observational mean of 5\% in both cases. The final solid-phase abundances of O$_2$, H$_2$O$_2$ and HCN are also seen to be reduced with the increase in visual extinction.

The solid- and gas-phase abundances of COMs are generally small during the collapse stage of both the {\tt basic\_Av2} and {\tt basic} models, although the absolute solid-phase abundance of ethanol (C$_2$H$_5$OH) is seen to reach a value comparable with gas-phase abundances in hot cores, i.e.~around 10$^{-8} n_{\mathrm{H}}$ (Table \ref{tab-st1-ice2}), in the higher visual extinction case. This is caused by increased production of the pre-cursor radical C$_2$H$_5$O on grain surfaces, through the addition of O to C$_2$H$_5$. Tables \ref{tab-st1-ice2} and \ref{tab-st1-gas} also indicate that, while the solid-phase abundance of methanol (CH$_3$OH) is a little lower in the {\tt basic} model, its peak gas-phase abundance during the cold collapse stage, which is rather high, is reduced by around a factor of 2. This methanol is formed on the grain surfaces and desorbs through chemical desorption. The gas-phase abundances of COMs in the stage-2 warm-up models show minor changes between the {\tt basic\_Av2} and {\tt basic} models. The fractional abundances of COMs larger than methanol produced by these models generally fall short of observational values, when compared with literature data for the representative prestellar core L1544.



\begin{deluxetable}{lllllllllll}
\tabletypesize{\footnotesize}
\tablecaption{\label{tab-st1-ice1} Final solid-phase abundances with respect to H$_2$O of simple or common ice species during stage 1 (collapse).}
\tablewidth{0pt}
\tablehead{  \colhead{Model} &  \colhead{CO} &  \colhead{CO$_2$} &  \colhead{CH$_4$} &  \colhead{H$_2$CO} &  \colhead{CH$_3$OH} &  \colhead{HCOOH} &  \colhead{NH$_3$} &  \colhead{HCN} &  \colhead{O$_2$} &  \colhead{H$_2$O$_2$} }
\startdata
\texttt{basic\_Av2}                                     & 5.12(-1) & 1.25(-1) & 1.65(-2) & 3.56(-2) & 9.00(-2) & 1.03(-2) & 1.97(-1) & 8.66(-4) & 7.34(-5) & 5.32(-4) \\ 
\texttt{basic                                        }  & 4.04(-1) & 1.06(-1) & 6.83(-2) & 3.22(-2) & 6.30(-2) & 1.01(-2) & 1.85(-1) & 4.16(-3) & 2.17(-5) & 2.39(-4) \\
\texttt{bas\_stk                                     }  & 3.05(-1) & 1.07(-1) & 1.06(-1) & 3.29(-2) & 4.38(-2) & 9.61(-3) & 1.94(-1) & 6.72(-3) & 2.08(-5) & 2.81(-4) \\
\texttt{bas\_stk\_loPD                               }  & 3.14(-1) & 1.23(-1) & 1.05(-1) & 2.72(-2) & 5.89(-2) & 1.08(-2) & 1.97(-1) & 6.42(-3) & 5.23(-5) & 6.15(-4) \\
\texttt{bas\_stk\_loPD\_H2                           }  & 3.22(-1) & 1.50(-1) & 9.74(-2) & 2.43(-2) & 6.27(-2) & 1.54(-2) & 1.97(-1) & 6.31(-3) & 8.34(-5) & 7.14(-4) \\
\texttt{bas\_stk\_loPD\_H2\_T16                      }  & 3.22(-1) & 1.50(-1) & 9.74(-2) & 2.43(-2) & 6.27(-2) & 1.54(-2) & 1.97(-1) & 6.31(-3) & 8.34(-5) & 7.14(-4) \\
\texttt{bas\_stk\_loPD\_H2\_T16\_no-bd               }  & 3.44(-1) & 1.64(-1) & 1.03(-1) & 1.42(-2) & 5.51(-2) & 1.87(-2) & 1.91(-1) & 6.24(-3) & 1.40(-6) & 1.49(-6) \\
\texttt{bas\_stk\_loPD\_H2\_T16\_no-bd\_ER           }  & 3.44(-1) & 1.64(-1) & 1.03(-1) & 1.42(-2) & 5.51(-2) & 1.87(-2) & 1.91(-1) & 6.24(-3) & 1.40(-6) & 1.49(-6) \\
\texttt{bas\_stk\_loPD\_H2\_T16\_no-bd\_PDI          }  & 3.22(-1) & 1.76(-1) & 1.02(-1) & 1.53(-2) & 5.00(-2) & 1.85(-2) & 1.98(-1) & 5.65(-3) & 4.12(-5) & 5.59(-4) \\
\texttt{bas\_stk\_loPD\_H2\_T16\_no-bd\_PDI2         }  & 3.07(-1) & 1.93(-1) & 5.33(-2) & 1.50(-2) & 4.90(-2) & 1.85(-2) & 1.49(-1) & 5.97(-3) & 1.40(-6) & 3.88(-6) \\
\texttt{bas\_stk\_loPD\_H2\_T16\_no-bd\_3B1          }  & 3.44(-1) & 1.76(-1) & 9.90(-2) & 1.30(-2) & 4.77(-2) & 1.66(-2) & 1.90(-1) & 6.49(-3) & 2.06(-5) & 4.98(-4) \\
\texttt{bas\_stk\_loPD\_H2\_T16\_no-bd\_3B2          }  & 3.44(-1) & 1.76(-1) & 9.74(-2) & 1.31(-2) & 4.73(-2) & 1.64(-2) & 1.91(-1) & 6.40(-3) & 1.94(-5) & 4.05(-4) \\
\texttt{bas\_stk\_loPD\_H2\_T16\_no-bd\_3B3          }  & 3.44(-1) & 1.76(-1) & 9.73(-2) & 1.31(-2) & 4.72(-2) & 1.64(-2) & 1.91(-1) & 6.40(-3) & 1.92(-5) & 3.97(-4) \\
\texttt{bas\_stk\_loPD\_H2\_T16\_no-bd\_3B3\_EF      }  & 3.39(-1) & 1.91(-1) & 9.81(-2) & 1.27(-2) & 4.56(-2) & 1.79(-2) & 1.93(-1) & 6.47(-3) & 2.48(-5) & 3.89(-4) \\
\texttt{all                                          }  & 3.06(-1) & 2.23(-1) & 5.25(-2) & 1.33(-2) & 4.26(-2) & 1.79(-2) & 1.48(-1) & 5.83(-3) & 2.07(-5) & 1.65(-5) \\
\texttt{all\_Edif                                    }  & 2.86(-1) & 1.75(-1) & 6.81(-2) & 2.26(-2) & 3.53(-2) & 7.68(-3) & 1.37(-1) & 2.01(-3) & 3.51(-3) & 3.00(-3) \\
\texttt{all\_Edif\_S17                               }  & 3.21(-1) & 1.92(-1) & 6.80(-2) & 1.82(-2) & 3.34(-2) & 9.96(-3) & 1.41(-1) & 1.86(-3) & 3.30(-3) & 3.00(-3) \\
\texttt{all\_Edif\_S17\_ON                           }  & 3.33(-1) & 1.97(-1) & 7.00(-2) & 1.85(-2) & 3.44(-2) & 1.04(-2) & 1.05(-1) & 1.96(-3) & 3.72(-3) & 2.98(-3) \\
\texttt{all\_Edif\_S17\_ON\_mtun                     }  & 3.40(-1) & 1.99(-1) & 7.15(-2) & 1.87(-2) & 4.03(-2) & 1.17(-2) & 1.04(-1) & 2.38(-3) & 4.43(-3) & 2.93(-3) \\
\texttt{all\_Edif\_S17\_ON\_mtun\_GP                 }  & 3.40(-1) & 1.99(-1) & 7.15(-2) & 1.87(-2) & 4.03(-2) & 1.17(-2) & 1.04(-1) & 2.38(-3) & 4.43(-3) & 2.93(-3) \\
\texttt{all\_Edif\_S17\_ON\_mtun\_GP\_Gr} ({\tt final}) & 3.34(-1) & 2.07(-1) & 5.58(-2) & 1.83(-2) & 5.89(-2) & 8.12(-4) & 1.73(-1) & 3.06(-4) & 3.34(-3) & 2.47(-3) \\
\enddata
\tablecomments{$A(B)=A \times 10^B$.}
\end{deluxetable}

\begin{deluxetable}{lccccccc}
\tabletypesize{\footnotesize}
\tablecaption{\label{tab-obs-ice1} Literature observational solid-phase abundances as a percentage of H$_2$O ice for molecules shown in Table \ref{tab-st1-ice1} (where available), toward high mass and low-mass YSOs; data are obtained from the review of \citet{Boogert15}.}
\tablewidth{0pt}
\tablehead{  \colhead{Source} &    \colhead{CO} &  \colhead{CO$_2$} &  \colhead{CH$_4$} &  \colhead{H$_2$CO} &  \colhead{CH$_3$OH} &  \colhead{HCOOH} &  \colhead{NH$_3$} }
\startdata
MYSOs                         &  $7^{15}_{4}$   &   $19^{25}_{12}$  &   $-$             &   $-$              &   $9^{23}_{5}$      &   $4^{5}_{3}$    &   $-$         \\
                              &  $3 - 26$       &   $11-27$         &   $1-3$           &   $\sim$$2-7$      &   $(<$$3)-31$       &   $(<$$0.5)-6$   &   $\sim$$7$   \\
LYSOs                         &  $21^{35}_{12}$ &   $28^{37}_{23}$  &   $4.5^{6}_{3}$   &   $-$              &   $6^{12}_{5}$      &   $-$            &   $6^{8}_{4}$ \\
                              &  $(<$$3) - 85$  &   $12-50$         &   $1-11$          &   $\sim$$6$        &   $(<$$1)-25$       &   $(<$$0.5)-4$   &   $3-10$      \\
\enddata
\tablecomments{Median, lower and upper quartile values are shown in the upper line for each class of object, with the observed range of values shown below. The reader should refer to \citet{Boogert15} for further details.}
\end{deluxetable}

\begin{deluxetable}{lllllllll}
\tabletypesize{\footnotesize}
\tablecaption{\label{tab-st1-ice2} Final solid-phase fractional abundances with respect to total hydrogen of selected COMs during stage 1 (collapse).}
\tablewidth{0pt}
\tablehead{  \colhead{Model}  &  \colhead{CH$_3$OH} &  \colhead{HCOOH} &  \colhead{CH$_3$CHO} &  \colhead{HCOOCH$_3$} &  \colhead{CH$_2$(OH)CHO} &  \colhead{CH$_3$COOH} &  \colhead{CH$_3$OCH$_3$} &  \colhead{C$_2$H$_5$OH} }
\startdata
\texttt{basic\_Av2}                                                         & 1.42(-5) & 1.63(-6) & 6.86(-11) & 6.36(-15) & 2.90(-17) & 1.17(-12) & 2.92(-11) & 1.46(-9) \\
\texttt{basic                                                           }      & 1.09(-5) & 1.73(-6) & 2.58(-10) & 8.13(-15) & 1.85(-17) & 3.28(-13) & 4.23(-11) & 1.07(-8) \\ 
\texttt{bas\_stk                                                      }      & 7.93(-6) & 1.74(-6) & 2.14(-10) & 7.61(-15) & 3.14(-17) & 5.17(-13) & 5.55(-11) & 1.71(-8) \\
\texttt{bas\_stk\_loPD                                             }      & 1.07(-5) & 1.97(-6) & 3.14(-10) & 8.85(-15) & 2.27(-17) & 1.19(-11) & 7.65(-10) & 2.70(-8) \\
\texttt{bas\_stk\_loPD\_H2                                      }       & 1.09(-5) & 2.68(-6) & 2.72(-10) & 1.17(-14) & 1.92(-17) & 1.13(-11) & 8.58(-10) & 2.19(-8) \\
\texttt{bas\_stk\_loPD\_H2\_T16                              }       & 1.09(-5) & 2.68(-6) & 2.72(-10) & 1.55(-14) & 1.92(-17) & 1.13(-11) & 8.59(-10) & 2.19(-8) \\
\texttt{bas\_stk\_loPD\_H2\_T16\_no-bd                  }        & 9.29(-6) & 3.16(-6) & 5.30(-12) & 1.59(-14) & 3.21(-20) & 5.15(-14) & 2.85(-12) & 7.77(-9) \\
\texttt{bas\_stk\_loPD\_H2\_T16\_no-bd\_ER           }         & 9.29(-6) & 3.16(-6) & 5.31(-12) & 1.59(-14) & 1.55(-18) & 5.37(-14) & 2.85(-12) & 7.78(-9) \\
\texttt{bas\_stk\_loPD\_H2\_T16\_no-bd\_PDI          }        & 8.55(-6) & 3.16(-6) & 5.61(-9) & 2.90(-10) & 1.42(-9) & 2.61(-9) & 2.72(-9) & 5.39(-8) \\
\texttt{bas\_stk\_loPD\_H2\_T16\_no-bd\_PDI2        }         & 8.37(-6) & 3.17(-6) & 7.79(-9) & 2.34(-10) & 1.16(-9) & 3.36(-9) & 3.85(-9) & 4.18(-7) \\
\texttt{bas\_stk\_loPD\_H2\_T16\_no-bd\_3B1          }        & 8.06(-6) & 2.80(-6) & 2.94(-8) & 2.26(-9) & 9.97(-9) & 1.15(-9) & 2.75(-9) & 4.23(-8) \\
\texttt{bas\_stk\_loPD\_H2\_T16\_no-bd\_3B2          }        & 8.01(-6) & 2.77(-6) & 3.10(-8) & 2.67(-9) & 1.21(-8) & 1.61(-9) & 3.26(-9) & 5.01(-8) \\
\texttt{bas\_stk\_loPD\_H2\_T16\_no-bd\_3B3          }        & 8.01(-6) & 2.77(-6) & 3.09(-8) & 2.68(-9) & 1.22(-8) & 1.68(-9) & 3.34(-9) & 5.08(-8) \\
\texttt{bas\_stk\_loPD\_H2\_T16\_no-bd\_3B3\_EF    }        & 7.61(-6) & 2.98(-6) & 4.04(-8) & 1.49(-8) & 1.18(-8) & 3.82(-9) & 4.13(-9) & 5.10(-8) \\
\texttt{all                                                                }      & 7.09(-6) & 2.98(-6) & 1.21(-8) & 1.58(-8) & 1.81(-8) & 7.85(-9) & 4.43(-9) & 4.36(-7) \\
\texttt{all\_Edif                                                        }      & 5.81(-6) & 1.26(-6) & 2.16(-8) & 1.47(-7) & 1.20(-7) & 2.36(-8) & 4.42(-8) & 3.84(-7) \\
\texttt{all\_Edif\_S17                                                }      & 5.45(-6) & 1.63(-6) & 2.27(-8) & 8.21(-8) & 6.68(-8) & 2.54(-8) & 4.69(-8) & 3.26(-7) \\
\texttt{all\_Edif\_S17\_ON                                         }      & 5.47(-6) & 1.66(-6) & 2.29(-8) & 8.24(-8) & 6.78(-8) & 2.56(-8) & 4.69(-8) & 3.26(-7) \\
\texttt{all\_Edif\_S17\_ON\_mtun                               }      & 6.45(-6) & 1.86(-6) & 7.48(-9) & 7.51(-8) & 5.45(-8) & 1.34(-8) & 2.66(-8) & 2.93(-7) \\
\texttt{all\_Edif\_S17\_ON\_mtun\_GP                        }      & 6.45(-6) & 1.86(-6) & 7.48(-9) & 7.51(-8) & 5.45(-8) & 1.33(-8) & 2.65(-8) & 2.92(-7) \\
\texttt{all\_Edif\_S17\_ON\_mtun\_GP\_Gr} ({\tt final})        & 9.58(-6) & 1.32(-7) & 1.99(-8) & 1.12(-7) & 8.33(-8) & 5.67(-9) & 8.07(-8) & 6.87(-7) \\
\enddata
\tablecomments{$A(B)=A \times 10^B$.}
\end{deluxetable}

\begin{deluxetable}{lllllllll}
\tabletypesize{\footnotesize}
\tablecaption{\label{tab-st1-gas} Peak gas-phase fractional abundances with respect to total hydrogen of selected COMs during stage 1 (collapse). Literature values for selected species are also shown for the prestellar core L1544.}
\tablewidth{0pt}
\tablehead{  \colhead{Model} &  \colhead{CH$_3$OH} &  \colhead{HCOOH} &  \colhead{CH$_3$CHO} &  \colhead{HCOOCH$_3$} &  \colhead{CH$_2$(OH)CHO} &  \colhead{CH$_3$COOH} &  \colhead{CH$_3$OCH$_3$} &  \colhead{C$_2$H$_5$OH} }
\startdata
\texttt{basic\_Av2}                                                         & 1.22(-7) & 2.75(-9) & 7.78(-12) & 1.74(-14) & 5.97(-18) & 8.06(-14) & 1.17(-11) & 3.19(-12) \\
\texttt{basic                                                           }      & 7.54(-8) & 2.56(-9) & 5.06(-11) & 5.60(-14) & 2.26(-18) & 1.31(-14) & 4.03(-12) & 3.98(-11) \\
\texttt{bas\_stk                                                      }      & 6.49(-8) & 1.87(-9) & 4.85(-11) & 6.82(-14) & 3.13(-18) & 1.24(-14) & 2.59(-12) & 4.73(-11) \\
\texttt{bas\_stk\_loPD                                             }       & 6.57(-8) & 1.84(-9) & 5.13(-11) & 6.85(-14) & 3.02(-18) & 8.75(-15) & 2.71(-12) & 3.42(-11) \\
\texttt{bas\_stk\_loPD\_H2                                      }       & 6.56(-8) & 2.46(-9) & 5.45(-11) & 6.88(-14) & 3.02(-18) & 8.75(-15) & 2.71(-12) & 3.42(-11) \\
\texttt{bas\_stk\_loPD\_H2\_T16                              }       & 6.56(-8) & 2.46(-9) & 5.47(-11) & 8.45(-14) & 3.02(-18) & 8.75(-15) & 2.76(-12) & 3.42(-11) \\
\texttt{bas\_stk\_loPD\_H2\_T16\_no-bd                  }        & 6.35(-8) & 2.57(-9) & 5.47(-11) & 1.05(-13) & 2.72(-18) & 8.90(-15) & 2.62(-12) & 3.43(-11) \\
\texttt{bas\_stk\_loPD\_H2\_T16\_no-bd\_ER           }         & 6.35(-8) & 2.57(-9) & 5.47(-11) & 1.05(-13) & 2.73(-18) & 8.90(-15) & 2.62(-12) & 3.43(-11) \\
\texttt{bas\_stk\_loPD\_H2\_T16\_no-bd\_PDI          }         & 6.35(-8) & 2.57(-9) & 5.47(-11) & 1.05(-13) & 5.52(-17) & 8.89(-15) & 2.62(-12) & 3.40(-11) \\
\texttt{bas\_stk\_loPD\_H2\_T16\_no-bd\_PDI2        }         & 6.36(-8) & 2.57(-9) & 5.47(-11) & 1.05(-13) & 5.52(-17) & 8.88(-15) & 2.62(-12) & 3.38(-11) \\
\texttt{bas\_stk\_loPD\_H2\_T16\_no-bd\_3B1         }         & 6.02(-8) & 2.69(-9) & 5.81(-11) & 2.60(-12) & 2.75(-12) & 8.81(-15) & 2.51(-12) & 3.22(-11) \\
\texttt{bas\_stk\_loPD\_H2\_T16\_no-bd\_3B2          }        & 6.01(-8) & 2.70(-9) & 5.82(-11) & 2.86(-12) & 3.16(-12) & 8.81(-15) & 2.51(-12) & 3.22(-11) \\
\texttt{bas\_stk\_loPD\_H2\_T16\_no-bd\_3B3          }        & 6.01(-8) & 2.70(-9) & 5.82(-11) & 2.87(-12) & 3.18(-12) & 8.81(-15) & 2.51(-12) & 3.22(-11) \\
\texttt{bas\_stk\_loPD\_H2\_T16\_no-bd\_3B3\_EF    }        & 5.82(-8) & 2.95(-9) & 2.92(-9) & 2.52(-9) & 2.91(-12) & 8.81(-15) & 8.00(-12) & 3.22(-11) \\
\texttt{all                                                                }      & 5.83(-8) & 2.95(-9) & 2.92(-9) & 2.52(-9) & 2.92(-12) & 8.79(-15) & 8.01(-12) & 3.19(-11) \\
\texttt{all\_Edif                                                        }      & 6.00(-8) & 1.79(-9) & 2.83(-9) & 1.94(-9) & 3.02(-10) & 2.66(-12) & 3.48(-10) & 1.83(-11) \\
\texttt{all\_Edif\_S17                                                }      & 3.34(-8) & 2.04(-9) & 2.76(-9) & 1.14(-9) & 1.40(-10) & 2.30(-12) & 1.57(-10) & 8.87(-12) \\
\texttt{all\_Edif\_S17\_ON                                         }      & 3.20(-8) & 2.03(-9) & 2.79(-9) & 1.11(-9) & 1.38(-10) & 2.52(-12) & 1.46(-10) & 9.02(-12) \\
\texttt{all\_Edif\_S17\_ON\_mtun                               }      & 3.20(-8) & 2.03(-9) & 2.79(-9) & 1.11(-9) & 1.38(-10) & 2.52(-12) & 1.46(-10) & 9.02(-12) \\
\texttt{all\_Edif\_S17\_ON\_mtun\_GP                        }      & 3.20(-8) & 2.03(-9) & 2.79(-9) & 1.11(-9) & 1.38(-10) & 2.52(-12) & 1.45(-10) & 9.02(-12) \\
\texttt{all\_Edif\_S17\_ON\_mtun\_GP\_Gr} ({\tt final})        & 3.25(-8) & 1.89(-10) & 3.40(-9) & 1.15(-9) & 2.68(-10) & 2.35(-12) & 1.74(-8) & 3.54(-10) \\
\hline
Observations: L1544 & \\
\citet{Vastel14} & 6(-9) & 1(-10) & 1(-10) & $\leq$2(-10) & -- & -- & $\leq$2(-10) & -- \\
\citet{JS16}$^a$     &       & --     & 2.1(-10) & 1.5(-10) & -- & -- & 5.1(-11) & -- \\
\enddata
\tablecomments{$A(B)=A \times 10^B$. $^a$Values correspond to the methanol peak of L1544.}
\end{deluxetable}

\begin{deluxetable}{lllllllll}
\tabletypesize{\footnotesize}
\tablecaption{\label{tab-gas1} Peak gas-phase fractional abundances with respect to total hydrogen of selected COMs during stage 2 with a {\em fast} warm-up.}
\tablewidth{0pt}
\tablehead{  \colhead{Model} &  \colhead{CH$_3$OH} &  \colhead{HCOOH} &  \colhead{CH$_3$CHO} &  \colhead{HCOOCH$_3$} &  \colhead{CH$_2$(OH)CHO} &  \colhead{CH$_3$COOH} &  \colhead{CH$_3$OCH$_3$} &  \colhead{C$_2$H$_5$OH}
}
\startdata
\texttt{basic\_Av2}                                                          & 1.67(-5) & 1.59(-6) & 5.46(-8) & 2.62(-7) & 7.80(-7) & 4.18(-9) & 8.24(-8) & 5.38(-8) \\
\texttt{basic                                                             }     & 1.37(-5) & 1.66(-6) & 1.07(-7) & 3.46(-7) & 8.18(-7) & 5.94(-9) & 7.74(-8) & 9.48(-8) \\
\texttt{bas\_stk                                                        }     & 1.12(-5) & 1.71(-6) & 1.98(-7) & 4.71(-7) & 9.22(-7) & 9.73(-9) & 7.52(-8) & 1.29(-7) \\
\texttt{bas\_stk\_loPD                                               }     & 1.32(-5) & 1.95(-6) & 6.65(-8) & 2.81(-7) & 6.94(-7) & 3.86(-9) & 5.98(-8) & 9.11(-8) \\
\texttt{bas\_stk\_loPD\_H2                                        }      & 1.27(-5) & 2.72(-6) & 4.56(-7) & 1.48(-7) & 3.81(-7) & 8.15(-8) & 1.17(-7) & 3.02(-7) \\
\texttt{bas\_stk\_loPD\_H2\_T16                                }      & 1.28(-5) & 2.71(-6) & 4.56(-7) & 1.51(-7) & 3.89(-7) & 8.29(-8) & 1.04(-7) & 3.04(-7) \\
\texttt{bas\_stk\_loPD\_H2\_T16\_no-bd                    }       & 8.31(-6) & 2.58(-6) & 6.73(-10) & 2.09(-11) & 4.48(-15) & 1.60(-11) & 1.22(-8) & 9.58(-9) \\
\texttt{bas\_stk\_loPD\_H2\_T16\_no-bd\_ER              }      & 8.35(-6) & 2.60(-6) & 7.01(-10) & 2.11(-11) & 5.23(-15) & 2.11(-11) & 1.23(-8) & 9.62(-9) \\
\texttt{bas\_stk\_loPD\_H2\_T16\_no-bd\_PDI             }      & 9.30(-6) & 3.08(-6) & 1.62(-8) & 4.92(-10) & 1.04(-11) & 2.43(-9) & 2.01(-8) & 7.38(-8) \\
\texttt{bas\_stk\_loPD\_H2\_T16\_no-bd\_PDI2           }      & 1.02(-5) & 3.12(-6) & 1.19(-8) & 4.12(-10) & 8.13(-11) & 4.30(-9) & 2.93(-8) & 4.39(-7) \\
\texttt{bas\_stk\_loPD\_H2\_T16\_no-bd\_3B1            }      & 7.94(-6) & 2.61(-6) & 3.47(-8) & 2.34(-9) & 2.37(-11) & 1.45(-10) & 1.50(-8) & 5.88(-8) \\
\texttt{bas\_stk\_loPD\_H2\_T16\_no-bd\_3B2            }      & 7.98(-6) & 2.62(-6) & 3.73(-8) & 2.77(-9) & 3.05(-11) & 2.87(-10) & 1.57(-8) & 6.93(-8) \\
\texttt{bas\_stk\_loPD\_H2\_T16\_no-bd\_3B3            }      & 7.99(-6) & 2.62(-6) & 3.73(-8) & 2.78(-9) & 3.09(-11) & 3.07(-10) & 1.58(-8) & 7.02(-8) \\
\texttt{bas\_stk\_loPD\_H2\_T16\_no-bd\_3B3\_EF      }      & 7.62(-6) & 2.82(-6) & 5.00(-8) & 1.51(-8) & 3.07(-11) & 6.57(-10) & 1.58(-8) & 7.03(-8) \\
\texttt{all                                                                  }     & 8.51(-6) & 2.93(-6) & 1.68(-8) & 1.65(-8) & 1.14(-9) & 9.06(-9) & 2.38(-8) & 4.56(-7) \\
\texttt{all\_Edif                                                          }     & 8.74(-6) & 1.19(-6) & 3.28(-8) & 1.79(-7) & 8.55(-8) & 2.67(-8) & 7.38(-8) & 3.94(-7) \\
\texttt{all\_Edif\_S17                                                  }     & 8.34(-6) & 1.52(-6) & 3.43(-8) & 1.02(-7) & 2.77(-8) & 2.82(-8) & 7.23(-8) & 3.40(-7) \\
\texttt{all\_Edif\_S17\_ON                                           }     & 8.42(-6) & 1.55(-6) & 3.34(-8) & 1.02(-7) & 2.64(-8) & 2.92(-8) & 8.18(-8) & 3.39(-7) \\
\texttt{all\_Edif\_S17\_ON\_mtun                                 }     & 8.42(-6) & 1.85(-6) & 1.11(-8) & 8.50(-8) & 2.01(-8) & 1.48(-8) & 5.96(-8) & 2.91(-7) \\
\texttt{all\_Edif\_S17\_ON\_mtun\_GP                          }     & 8.38(-6) & 1.85(-6) & 1.10(-8) & 1.05(-7) & 2.01(-8) & 1.48(-8) & 5.85(-8) & 2.91(-7) \\
\texttt{all\_Edif\_S17\_ON\_mtun\_GP\_Gr} ({\tt final})         & 1.08(-5) & 1.31(-7) & 3.92(-8) & 1.88(-7) & 9.43(-8) & 5.81(-9) & 1.12(-7) & 6.58(-7) \\
\enddata
\tablecomments{$A(B)=A \times 10^B$.}
\end{deluxetable}

\begin{deluxetable}{lllllllll}
\tabletypesize{\footnotesize}
\tablecaption{\label{tab-gas2} Peak gas-phase fractional abundances with respect to total hydrogen of selected COMs during stage 2 with a {\em medium} warm-up.}
\tablewidth{0pt}
\tablehead{  \colhead{Model} &  \colhead{CH$_3$OH} &  \colhead{HCOOH} &  \colhead{CH$_3$CHO} &  \colhead{HCOOCH$_3$} &  \colhead{CH$_2$(OH)CHO} &  \colhead{CH$_3$COOH} &  \colhead{CH$_3$OCH$_3$} &  \colhead{C$_2$H$_5$OH}
}
\startdata
\texttt{basic\_Av2}                                                          & 1.46(-5) & 1.14(-6) & 1.18(-7) & 7.93(-8) & 4.46(-7) & 1.88(-8) & 1.44(-7) & 1.43(-7) \\
\texttt{basic                                                           }       & 1.22(-5) & 1.12(-6) & 2.62(-7) & 2.00(-7) & 4.70(-7) & 2.61(-8) & 1.26(-7) & 2.26(-7) \\
\texttt{bas\_stk                                                      }       & 1.01(-5) & 1.09(-6) & 4.69(-7) & 2.96(-7) & 5.37(-7) & 3.41(-8) & 1.22(-7) & 2.78(-7) \\
\texttt{bas\_stk\_loPD                                             }       & 1.25(-5) & 1.60(-6) & 1.78(-7) & 2.18(-7) & 4.96(-7) & 1.56(-8) & 1.19(-7) & 1.87(-7) \\
\texttt{bas\_stk\_loPD\_H2                                      }        & 1.19(-5) & 2.23(-6) & 4.46(-7) & 1.22(-7) & 2.75(-7) & 1.03(-7) & 1.26(-7) & 3.57(-7) \\
\texttt{bas\_stk\_loPD\_H2\_T16                              }        & 1.23(-5) & 2.24(-6) & 4.46(-7) & 1.39(-7) & 2.98(-7) & 1.08(-7) & 1.12(-7) & 3.65(-7) \\
\texttt{bas\_stk\_loPD\_H2\_T16\_no-bd                  }         & 8.35(-6) & 2.57(-6) & 3.51(-10) & 6.32(-11) & 3.36(-14) & 1.83(-10) & 3.92(-8) & 1.09(-8) \\
\texttt{bas\_stk\_loPD\_H2\_T16\_no-bd\_ER            }         & 8.43(-6) & 2.60(-6) & 3.63(-10) & 6.42(-11) & 4.40(-14) & 2.33(-10) & 3.98(-8) & 1.13(-8) \\
\texttt{bas\_stk\_loPD\_H2\_T16\_no-bd\_PDI           }         & 9.21(-6) & 2.93(-6) & 1.98(-8) & 9.23(-10) & 1.57(-10) & 2.86(-9) & 5.67(-8) & 8.54(-8) \\
\texttt{bas\_stk\_loPD\_H2\_T16\_no-bd\_PDI2         }         & 1.01(-5) & 3.03(-6) & 2.04(-8) & 8.53(-10) & 9.55(-11) & 5.73(-9) & 8.08(-8) & 4.43(-7) \\
\texttt{bas\_stk\_loPD\_H2\_T16\_no-bd\_3B1          }         & 7.85(-6) & 2.48(-6) & 3.88(-8) & 2.59(-9) & 1.50(-10) & 6.18(-10) & 4.04(-8) & 6.33(-8) \\
\texttt{bas\_stk\_loPD\_H2\_T16\_no-bd\_3B2          }         & 7.92(-6) & 2.47(-6) & 4.18(-8) & 3.07(-9) & 1.93(-10) & 1.07(-9) & 4.19(-8) & 7.37(-8) \\
\texttt{bas\_stk\_loPD\_H2\_T16\_no-bd\_3B3          }         & 7.92(-6) & 2.47(-6) & 4.18(-8) & 3.09(-9) & 1.96(-10) & 1.12(-9) & 4.20(-8) & 7.47(-8) \\
\texttt{bas\_stk\_loPD\_H2\_T16\_no-bd\_3B3\_EF    }         & 7.57(-6) & 2.67(-6) & 5.84(-8) & 1.53(-8) & 2.04(-10) & 2.02(-9) & 4.03(-8) & 7.47(-8) \\
\texttt{all                                                                }       & 8.66(-6) & 2.84(-6) & 2.82(-8) & 1.83(-8) & 2.37(-9) & 1.08(-8) & 6.44(-8) & 4.60(-7) \\
\texttt{all\_Edif                                                        }       & 8.83(-6) & 1.16(-6) & 4.36(-8) & 1.82(-7) & 1.08(-7) & 3.12(-8) & 1.25(-7) & 3.99(-7) \\
\texttt{all\_Edif\_S17                                                }       & 8.20(-6) & 1.48(-6) & 4.44(-8) & 1.22(-7) & 6.55(-8) & 3.61(-8) & 1.13(-7) & 3.51(-7) \\
\texttt{all\_Edif\_S17\_ON                                         }       & 8.34(-6) & 1.51(-6) & 4.40(-8) & 1.20(-7) & 7.99(-8) & 3.66(-8) & 1.38(-7) & 3.58(-7) \\
\texttt{all\_Edif\_S17\_ON\_mtun                               }       & 8.25(-6) & 1.83(-6) & 1.56(-8) & 1.08(-7) & 5.31(-8) & 1.99(-8) & 1.16(-7) & 2.99(-7) \\
\texttt{all\_Edif\_S17\_ON\_mtun\_GP                        }       & 8.19(-6) & 1.83(-6) & 1.56(-8) & 1.21(-7) & 5.30(-8) & 1.98(-8) & 1.18(-7) & 2.98(-7) \\
\texttt{all\_Edif\_S17\_ON\_mtun\_GP\_Gr} ({\tt final})         & 1.05(-5) & 1.29(-7) & 7.62(-8) & 1.97(-7) & 1.14(-7) & 6.17(-9) & 1.77(-7) & 6.79(-7) \\
\enddata
\tablecomments{$A(B)=A \times 10^B$.}
\end{deluxetable}

\begin{deluxetable}{lllllllll}
\tabletypesize{\footnotesize}
\tablecaption{\label{tab-gas3} Peak gas-phase fractional abundances with respect to total hydrogen of selected COMs during stage 2 with a {\em slow} warm-up.}
\tablewidth{0pt}
\tablehead{  \colhead{Model} &  \colhead{CH$_3$OH} &  \colhead{HCOOH} &  \colhead{CH$_3$CHO} &  \colhead{HCOOCH$_3$} &  \colhead{CH$_2$(OH)CHO} &  \colhead{CH$_3$COOH} &  \colhead{CH$_3$OCH$_3$} &  \colhead{C$_2$H$_5$OH}
}
\startdata
\texttt{basic\_Av2}                                                          & 6.99(-6) & 4.38(-8) & 6.09(-8) & 1.36(-9) & 7.05(-8) & 4.13(-8) & 8.61(-8) & 2.50(-7) \\
\texttt{basic                                                             }     & 5.99(-6) & 3.51(-8) & 3.32(-8) & 1.72(-9) & 5.16(-8) & 5.64(-8) & 7.12(-8) & 3.66(-7) \\
\texttt{bas\_stk                                                        }     & 5.65(-6) & 2.84(-8) & 8.41(-8) & 3.11(-9) & 4.98(-8) & 7.24(-8) & 6.28(-8) & 4.57(-7) \\
\texttt{bas\_stk\_loPD                                               }     & 9.36(-6) & 6.31(-7) & 6.30(-8) & 1.40(-8) & 1.36(-7) & 7.10(-8) & 1.03(-7) & 4.68(-7) \\
\texttt{bas\_stk\_loPD\_H2                                        }      & 8.92(-6) & 8.76(-7) & 1.22(-7) & 8.36(-9) & 1.00(-7) & 1.69(-7) & 9.79(-8) & 5.65(-7) \\
\texttt{bas\_stk\_loPD\_H2\_T16                                }      & 1.11(-5) & 9.27(-7) & 1.62(-7) & 6.38(-8) & 1.56(-7) & 1.98(-7) & 1.07(-7) & 6.00(-7) \\
\texttt{bas\_stk\_loPD\_H2\_T16\_no-bd                    }       & 7.14(-6) & 1.99(-6) & 2.19(-8) & 8.61(-11) & 7.00(-13) & 2.01(-9) & 5.87(-8) & 1.28(-8)\\
\texttt{bas\_stk\_loPD\_H2\_T16\_no-bd\_ER              }       & 7.33(-6) & 2.01(-6) & 2.25(-8) & 8.86(-11) & 7.75(-13) & 2.44(-9) & 6.17(-8) & 1.32(-8) \\
\texttt{bas\_stk\_loPD\_H2\_T16\_no-bd\_PDI             }       & 8.53(-6) & 2.52(-6) & 1.21(-7) & 1.13(-8) & 1.63(-9) & 1.83(-8) & 9.56(-8) & 1.34(-7) \\
\texttt{bas\_stk\_loPD\_H2\_T16\_no-bd\_PDI2           }       & 8.31(-6) & 2.78(-6) & 6.25(-8) & 8.52(-9) & 2.77(-10) & 3.09(-8) & 1.09(-7) & 4.54(-7) \\
\texttt{bas\_stk\_loPD\_H2\_T16\_no-bd\_3B1            }       & 7.26(-6) & 1.88(-6) & 9.21(-8) & 4.15(-9) & 1.95(-9) & 2.67(-9) & 6.76(-8) & 7.73(-8) \\
\texttt{bas\_stk\_loPD\_H2\_T16\_no-bd\_3B2            }       & 7.41(-6) & 1.88(-6) & 9.96(-8) & 5.16(-9) & 2.53(-9) & 5.39(-9) & 7.10(-8) & 8.98(-8) \\
\texttt{bas\_stk\_loPD\_H2\_T16\_no-bd\_3B3            }       & 7.43(-6) & 1.88(-6) & 9.99(-8) & 5.22(-9) & 2.58(-9) & 5.69(-9) & 7.13(-8) & 9.09(-8) \\
\texttt{bas\_stk\_loPD\_H2\_T16\_no-bd\_3B3\_EF      }       & 7.12(-6) & 2.03(-6) & 1.46(-7) & 1.96(-8) & 2.69(-9) & 8.06(-9) & 7.00(-8) & 9.07(-8) \\
\texttt{all                                                                  }      & 6.79(-6) & 2.53(-6) & 1.25(-7) & 3.32(-8) & 5.99(-9) & 3.57(-8) & 8.38(-8) & 4.66(-7) \\
\texttt{all\_Edif                                                          }      & 7.22(-6) & 1.06(-6) & 1.41(-7) & 2.35(-7) & 1.74(-7) & 6.73(-8) & 1.90(-7) & 4.03(-7) \\
\texttt{all\_Edif\_S17                                                  }      & 6.47(-6) & 1.34(-6) & 1.46(-7) & 1.49(-7) & 1.16(-7) & 6.79(-8) & 1.58(-7) & 3.65(-7) \\
\texttt{all\_Edif\_S17\_ON                                           }      & 6.52(-6) & 1.37(-6) & 1.45(-7) & 1.46(-7) & 1.29(-7) & 6.93(-8) & 1.84(-7) & 3.70(-7) \\
\texttt{all\_Edif\_S17\_ON\_mtun                                 }      & 6.55(-6) & 1.65(-6) & 8.28(-8) & 1.29(-7) & 6.53(-8) & 4.33(-8) & 1.63(-7) & 2.98(-7) \\
\texttt{all\_Edif\_S17\_ON\_mtun\_GP                          }      & 6.47(-6) & 1.65(-6) & 8.28(-8) & 1.39(-7) & 6.53(-8) & 4.32(-8) & 1.61(-7) & 2.98(-7) \\
\texttt{all\_Edif\_S17\_ON\_mtun\_GP\_Gr} ({\tt final})          & 8.31(-6) & 1.07(-7) & 2.51(-7) & 2.51(-7) & 6.42(-8) & 9.17(-9) & 2.18(-7) & 6.61(-7) \\
\enddata
\tablecomments{$A(B)=A \times 10^B$.}
\end{deluxetable}

\begin{deluxetable}{lllllllll}
\tabletypesize{\footnotesize}
\tablecaption{\label{tab-gas4} Peak gas-phase fractional abundances with respect to total hydrogen of selected N-bearing species during stage 2 with a {\em fast} warm-up.}
\tablewidth{0pt}
\tablehead{  \colhead{Model} &  \colhead{HCN} &  \colhead{CH$_3$CN} &  \colhead{C$_2$H$_3$CN} &  \colhead{C$_2$H$_5$CN} &  \colhead{NH$_2$CN} &  \colhead{HNCO} &  \colhead{NH$_2$CHO} &  \colhead{CH$_3$NH$_2$}
}
\startdata
\texttt{basic\_Av2}                                                         & 3.03(-7) & 1.90(-9) & 1.44(-9) & 1.46(-8) & 5.57(-11) & 3.37(-7) & 1.33(-8) & 1.08(-7) \\
\texttt{basic                                                             }    & 8.43(-7) & 5.99(-9) & 2.45(-9) & 1.72(-8) & 4.00(-11) & 3.97(-7) & 1.22(-8) & 2.83(-7) \\
\texttt{bas\_stk                                                        }    & 1.30(-6) & 7.87(-9) & 3.01(-9) & 2.17(-8) & 3.37(-11) & 5.34(-7) & 9.04(-9) & 4.17(-7) \\
\texttt{bas\_stk\_loPD                                               }    & 1.24(-6) & 1.21(-8) & 2.52(-9) & 1.57(-8) & 4.04(-12) & 3.05(-7) & 8.42(-10) & 2.20(-6) \\
\texttt{bas\_stk\_loPD\_H2                                        }     & 1.23(-6) & 1.46(-8) & 2.62(-9) & 1.38(-8) & 4.61(-12) & 3.53(-7) & 1.67(-9) & 2.32(-6) \\
\texttt{bas\_stk\_loPD\_H2\_T16                                }     & 1.20(-6) & 2.17(-8) & 1.81(-9) & 1.39(-8) & 4.63(-12) & 3.53(-7) & 1.69(-9) & 2.32(-6) \\
\texttt{bas\_stk\_loPD\_H2\_T16\_no-bd                    }      & 1.70(-6) & 7.44(-9) & 1.77(-9) & 1.16(-9) & 9.34(-13) & 1.18(-12) & 1.18(-13) & 1.83(-6) \\
\texttt{bas\_stk\_loPD\_H2\_T16\_no-bd\_ER              }      & 1.70(-6) & 7.46(-9) & 1.78(-9) & 1.28(-9) & 8.58(-13) & 1.02(-12) & 1.28(-13) & 1.84(-6) \\
\texttt{bas\_stk\_loPD\_H2\_T16\_no-bd\_PDI             }      & 1.14(-6) & 9.73(-9) & 1.98(-9) & 3.41(-9) & 3.09(-9) & 4.35(-8) & 1.21(-8) & 2.07(-6) \\
\texttt{bas\_stk\_loPD\_H2\_T16\_no-bd\_PDI2           }      & 1.14(-6) & 2.38(-8) & 7.60(-10) & 1.19(-8) & 1.05(-8) & 2.78(-8) & 1.51(-8) & 1.18(-5) \\
\texttt{bas\_stk\_loPD\_H2\_T16\_no-bd\_3B1            }      & 1.42(-6) & 6.85(-9) & 2.42(-9) & 3.45(-9) & 6.74(-10) & 1.06(-8) & 1.12(-8) & 1.87(-6) \\
\texttt{bas\_stk\_loPD\_H2\_T16\_no-bd\_3B2            }      & 1.37(-6) & 6.96(-9) & 2.61(-9) & 3.71(-9) & 8.04(-10) & 9.98(-9) & 1.62(-8) & 1.90(-6) \\
\texttt{bas\_stk\_loPD\_H2\_T16\_no-bd\_3B3            }      & 1.37(-6) & 6.97(-9) & 2.62(-9) & 3.73(-9) & 8.07(-10) & 9.89(-9) & 1.65(-8) & 1.90(-6) \\
\texttt{bas\_stk\_loPD\_H2\_T16\_no-bd\_3B3\_EF      }      & 1.36(-6) & 6.91(-9) & 2.69(-9) & 4.22(-9) & 7.85(-10) & 9.90(-9) & 1.73(-8) & 1.89(-6) \\
\texttt{all                                                                  }     & 1.08(-6) & 2.29(-8) & 7.13(-10) & 1.37(-8) & 9.40(-9) & 2.73(-8) & 1.84(-8) & 1.16(-5) \\
\texttt{all\_Edif                                                          }     & 8.59(-7) & 1.64(-8) & 1.40(-9) & 1.11(-8) & 5.50(-9) & 1.44(-7) & 6.73(-8) & 9.67(-6) \\
\texttt{all\_Edif\_S17                                                  }     & 7.48(-7) & 1.25(-8) & 1.11(-9) & 8.87(-9) & 3.92(-9) & 1.29(-7) & 4.85(-8) & 9.77(-6) \\
\texttt{all\_Edif\_S17\_ON                                           }     & 5.68(-7) & 1.21(-8) & 6.00(-9) & 7.91(-8) & 3.98(-9) & 8.32(-8) & 3.32(-8) & 9.74(-6) \\
\texttt{all\_Edif\_S17\_ON\_mtun                                 }     & 5.77(-7) & 1.08(-8) & 6.04(-9) & 8.25(-8) & 3.26(-9) & 7.61(-8) & 1.82(-8) & 1.13(-5) \\
\texttt{all\_Edif\_S17\_ON\_mtun\_GP                          }     & 5.77(-7) & 1.08(-8) & 6.04(-9) & 8.25(-8) & 3.26(-9) & 7.61(-8) & 1.96(-8) & 1.13(-5) \\
\texttt{all\_Edif\_S17\_ON\_mtun\_GP\_Gr} ({\tt final})         & 1.29(-7) & 1.70(-9) & 6.11(-9) & 1.02(-7) & 4.07(-10) & 8.72(-8) & 1.88(-8) & 3.73(-7) \\
\enddata
\tablecomments{$A(B)=A \times 10^B$.}
\end{deluxetable}

\begin{deluxetable}{lllllllll}
\tabletypesize{\footnotesize}
\tablecaption{\label{tab-gas5} Peak gas-phase fractional abundances with respect to total hydrogen of selected N-bearing species during stage 2 with a {\em medium} warm-up.}
\tablewidth{0pt}
\tablehead{  \colhead{Model} &  \colhead{HCN} &  \colhead{CH$_3$CN} &  \colhead{C$_2$H$_3$CN} &  \colhead{C$_2$H$_5$CN} &  \colhead{NH$_2$CN} &  \colhead{HNCO} &  \colhead{NH$_2$CHO} &  \colhead{CH$_3$NH$_2$}
}
\startdata
\texttt{basic\_Av2}                                                          & 8.39(-7) & 3.03(-9) & 5.27(-9) & 3.13(-8) & 2.10(-10) & 1.96(-8) & 7.45(-11) & 2.55(-8) \\
\texttt{basic                                                           }       & 1.46(-6) & 9.70(-9) & 9.90(-9) & 4.10(-8) & 1.87(-10) & 2.32(-8) & 8.79(-11) & 1.74(-7) \\
\texttt{bas\_stk                                                      }       & 1.89(-6) & 1.51(-8) & 1.09(-8) & 4.79(-8) & 1.70(-10) & 3.40(-8) & 1.28(-10) & 3.01(-7) \\
\texttt{bas\_stk\_loPD                                             }       & 1.66(-6) & 2.98(-8) & 1.41(-8) & 7.57(-8) & 3.97(-11) & 2.46(-8) & 2.13(-10) & 1.68(-6) \\
\texttt{bas\_stk\_loPD\_H2                                      }        & 1.62(-6) & 3.19(-8) & 1.53(-8) & 8.19(-8) & 4.41(-11) & 4.41(-8) & 2.75(-10) & 1.77(-6) \\
\texttt{bas\_stk\_loPD\_H2\_T16                              }        & 1.59(-6) & 4.28(-8) & 7.64(-9) & 8.48(-8) & 4.50(-11) & 4.41(-8) & 3.37(-10) & 1.78(-6) \\
\texttt{bas\_stk\_loPD\_H2\_T16\_no-bd                  }         & 2.23(-6) & 2.39(-8) & 1.65(-9) & 3.21(-9) & 1.24(-12) & 3.65(-12) & 9.84(-13) & 1.63(-6) \\
\texttt{bas\_stk\_loPD\_H2\_T16\_no-bd\_ER            }         & 2.23(-6) & 2.41(-8) & 1.67(-9) & 3.41(-9) & 1.20(-12) & 3.47(-12) & 1.17(-12) & 1.66(-6) \\
\texttt{bas\_stk\_loPD\_H2\_T16\_no-bd\_PDI           }         & 1.74(-6) & 2.21(-8) & 2.61(-9) & 5.14(-9) & 2.87(-9) & 7.84(-8) & 2.03(-8) & 1.78(-6) \\
\texttt{bas\_stk\_loPD\_H2\_T16\_no-bd\_PDI2         }         & 1.88(-6) & 3.90(-8) & 2.45(-9) & 1.47(-8) & 1.04(-8) & 5.43(-8) & 2.69(-8) & 1.12(-5) \\
\texttt{bas\_stk\_loPD\_H2\_T16\_no-bd\_3B1          }         & 1.96(-6) & 2.10(-8) & 2.62(-9) & 5.14(-9) & 6.13(-10) & 1.53(-8) & 2.08(-8) & 1.65(-6) \\
\texttt{bas\_stk\_loPD\_H2\_T16\_no-bd\_3B2          }         & 1.90(-6) & 2.07(-8) & 2.86(-9) & 5.25(-9) & 7.28(-10) & 1.40(-8) & 2.45(-8) & 1.66(-6) \\
\texttt{bas\_stk\_loPD\_H2\_T16\_no-bd\_3B3          }         & 1.90(-6) & 2.06(-8) & 2.87(-9) & 5.26(-9) & 7.31(-10) & 1.39(-8) & 2.46(-8) & 1.66(-6) \\
\texttt{bas\_stk\_loPD\_H2\_T16\_no-bd\_3B3\_EF    }         & 1.90(-6) & 2.02(-8) & 2.91(-9) & 5.73(-9) & 7.12(-10) & 1.37(-8) & 2.45(-8) & 1.66(-6) \\
\texttt{all                                                                }        & 1.80(-6) & 3.59(-8) & 2.36(-9) & 1.63(-8) & 9.33(-9) & 5.31(-8) & 3.29(-8) & 1.12(-5) \\
\texttt{all\_Edif                                                        }        & 1.94(-6) & 3.57(-8) & 2.06(-9) & 1.16(-8) & 5.24(-9) & 1.69(-7) & 8.28(-8) & 8.32(-6) \\
\texttt{all\_Edif\_S17                                                }        & 1.73(-6) & 2.95(-8) & 1.63(-9) & 9.17(-9) & 3.78(-9) & 1.59(-7) & 7.39(-8) & 8.76(-6) \\
\texttt{all\_Edif\_S17\_ON                                         }        & 1.52(-6) & 2.99(-8) & 1.05(-8) & 7.84(-8) & 3.86(-9) & 1.02(-7) & 5.53(-8) & 8.80(-6) \\
\texttt{all\_Edif\_S17\_ON\_mtun                               }        & 1.48(-6) & 2.70(-8) & 1.07(-8) & 8.24(-8) & 3.20(-9) & 9.33(-8) & 2.46(-8) & 1.02(-5) \\
\texttt{all\_Edif\_S17\_ON\_mtun\_GP                        }        & 1.48(-6) & 2.69(-8) & 1.07(-8) & 8.24(-8) & 3.20(-9) & 9.33(-8) & 2.66(-8) & 1.02(-5) \\
\texttt{all\_Edif\_S17\_ON\_mtun\_GP\_Gr} ({\tt final})         & 7.08(-7) & 8.20(-9) & 1.04(-8) & 1.10(-7) & 4.04(-10) & 1.15(-7) & 2.55(-8) & 3.39(-7) \\
\enddata
\tablecomments{$A(B)=A \times 10^B$.}
\end{deluxetable}

\begin{deluxetable}{lllllllll}
\tabletypesize{\footnotesize}
\tablecaption{\label{tab-gas6} Peak gas-phase fractional abundances with respect to total hydrogen of selected N-bearing species during stage 2 with a {\em slow} warm-up.}
\tablewidth{0pt}
\tablehead{  \colhead{Model} &  \colhead{HCN} &  \colhead{CH$_3$CN} &  \colhead{C$_2$H$_3$CN} &  \colhead{C$_2$H$_5$CN} &  \colhead{NH$_2$CN} &  \colhead{HNCO} &  \colhead{NH$_2$CHO} &  \colhead{CH$_3$NH$_2$}
}
\startdata
\texttt{basic\_Av2}                                                          & 4.60(-6) & 3.18(-9) & 7.21(-9) & 4.13(-8) & 4.47(-11) & 1.36(-9) & 1.18(-10) & 3.88(-9) \\
\texttt{basic                                                             }     & 7.89(-6) & 6.19(-9) & 1.36(-8) & 7.34(-8) & 5.01(-11) & 1.66(-9) & 1.70(-10) & 4.04(-9) \\
\texttt{bas\_stk                                                        }     & 1.11(-5) & 1.06(-8) & 2.31(-8) & 1.28(-7) & 4.56(-11) & 2.10(-9) & 2.53(-10) & 6.81(-9) \\
\texttt{bas\_stk\_loPD                                               }      & 1.26(-5) & 2.35(-8) & 4.14(-8) & 2.85(-7) & 7.36(-11) & 5.07(-9) & 9.06(-10) & 8.01(-8) \\
\texttt{bas\_stk\_loPD\_H2                                        }      & 1.18(-5) & 2.55(-8) & 3.80(-8) & 2.63(-7) & 1.50(-10) & 6.22(-9) & 1.23(-9) & 7.68(-8) \\
\texttt{bas\_stk\_loPD\_H2\_T16                                }      & 6.93(-6) & 5.59(-8) & 4.76(-8) & 3.33(-7) & 1.75(-10) & 3.06(-9) & 2.46(-9) & 7.49(-8) \\
\texttt{bas\_stk\_loPD\_H2\_T16\_no-bd                    }       & 1.32(-5) & 8.08(-8) & 2.06(-9) & 1.08(-8) & 7.52(-17) & 1.48(-11) & 1.23(-11) & 8.16(-7) \\
\texttt{bas\_stk\_loPD\_H2\_T16\_no-bd\_ER              }       & 1.32(-5) & 8.19(-8) & 2.06(-9) & 1.11(-8) & 9.05(-17) & 1.37(-11) & 1.32(-11) & 8.30(-7) \\
\texttt{bas\_stk\_loPD\_H2\_T16\_no-bd\_PDI             }       & 1.18(-5) & 8.04(-8) & 4.02(-9) & 1.32(-8) & 3.11(-9) & 2.11(-7) & 4.00(-8) & 7.85(-7) \\
\texttt{bas\_stk\_loPD\_H2\_T16\_no-bd\_PDI2           }       & 1.07(-5) & 8.72(-8) & 7.76(-9) & 1.87(-8) & 1.11(-8) & 1.67(-7) & 5.64(-8) & 9.05(-6) \\
\texttt{bas\_stk\_loPD\_H2\_T16\_no-bd\_3B1            }       & 1.26(-5) & 7.95(-8) & 2.71(-9) & 1.24(-8) & 4.50(-10) & 2.88(-8) & 1.07(-8) & 8.05(-7) \\
\texttt{bas\_stk\_loPD\_H2\_T16\_no-bd\_3B2            }       & 1.24(-5) & 7.89(-8) & 2.93(-9) & 1.27(-8) & 5.45(-10) & 2.62(-8) & 1.40(-8) & 7.99(-7) \\
\texttt{bas\_stk\_loPD\_H2\_T16\_no-bd\_3B3            }       & 1.24(-5) & 7.89(-8) & 2.94(-9) & 1.27(-8) & 5.47(-10) & 2.61(-8) & 1.41(-8) & 7.99(-7) \\
\texttt{bas\_stk\_loPD\_H2\_T16\_no-bd\_3B3\_EF      }       & 1.24(-5) & 7.87(-8) & 2.78(-9) & 1.26(-8) & 5.38(-10) & 2.57(-8) & 1.45(-8) & 7.92(-7) \\
\texttt{all                                                                  }      & 9.92(-6) & 7.67(-8) & 7.19(-9) & 1.97(-8) & 1.01(-8) & 1.62(-7) & 6.56(-8) & 8.79(-6) \\
\texttt{all\_Edif                                                          }      & 1.12(-5) & 9.80(-8) & 4.55(-9) & 1.50(-8) & 4.71(-9) & 2.29(-7) & 1.44(-7) & 4.49(-6) \\
\texttt{all\_Edif\_S17                                                  }      & 1.04(-5) & 8.07(-8) & 4.24(-9) & 1.48(-8) & 3.73(-9) & 2.38(-7) & 1.45(-7) & 5.19(-6) \\
\texttt{all\_Edif\_S17\_ON                                           }      & 9.95(-6) & 9.27(-8) & 1.69(-8) & 8.04(-8) & 3.80(-9) & 1.60(-7) & 1.17(-7) & 5.35(-6) \\
\texttt{all\_Edif\_S17\_ON\_mtun                                 }      & 1.02(-5) & 8.84(-8) & 1.77(-8) & 8.69(-8) & 3.39(-9) & 1.41(-7) & 2.16(-8) & 6.29(-6) \\
\texttt{all\_Edif\_S17\_ON\_mtun\_GP                          }      & 1.02(-5) & 8.70(-8) & 1.77(-8) & 8.69(-8) & 3.39(-9) & 1.41(-7) & 2.28(-8) & 6.29(-6) \\
\texttt{all\_Edif\_S17\_ON\_mtun\_GP\_Gr} ({\tt final})         & 6.86(-6) & 7.04(-8) & 1.98(-8) & 1.50(-7) & 5.05(-10) & 2.03(-7) & 5.32(-8) & 2.57(-7) \\
\enddata
\tablecomments{$A(B)=A \times 10^B$.}
\end{deluxetable}


\subsection{Hot-core chemistry in the {\tt basic} model}

Some of the main features of the behavior of the {\tt basic} model are described here, to give context to the subsequent changes and additions. The {\tt basic} model is very simliar to past published models, and does {\em not} include any of the new nondiffusive grain-surface/ice-mantle reaction mechanisms.

Figure \ref{basic_medium} shows the time-dependent evolution of a selection of molecules during stage 2 of the {\tt basic} model, using the {\em medium} warm-up timescale. Panel (a) shows the behavior of some simple ice components; the more volatile CO and CH$_4$ desorb strongly between around 25--30~K. Note that no trapping of these species occurs, as they are free to diffuse within the bulk ice, allowing them to reach the ice surface when thermal bulk diffusion comes sufficiently rapid. Water and ammonia, along with methanol (panel b) desorb completely by around 120~K. Solid-phase formaldehyde is lost from the grains between 40--50~K, and is gradually destroyed in the gas phase until its abundance is replenished as the result of the destruction of other species at higher temperatures. HNCO, also shown in panel (b), is formed in abundance in the solid phase, but as it approaches its desorption temperature and diffuses out of the bulk ice and onto the surface, it becomes susceptible to H-abstraction reactions with surface radicals, reducing its abundance. When it desorbs at $\sim$60~K, what remains of it is then destroyed in the gas phase or through re-accretion onto grain surfaces where it may still react with surface radicals. Its gas-phase abundance picks up again at higher temperatures as the result of the destruction of larger species. Acetaldehyde (CH$_3$CHO) is formed in similar quantities as HNCO on the grains, through the additional of CH$_3$ and HCO radicals between around 20--40~K. However, as it begins to desorb from the grains, its reactions with surface atoms and radicals are slower than those of HNCO, due to larger activation energy barriers, meaning that it is mainly destroyed by gas-phase processes. It therefore shows a much less rapid decline following desorption. The acetaldehyde present in the gas phase at high temperatures is mostly that which survives from earlier times. Formic acid in this model is produced at low temperatures on the grains, through the reaction of OH with CO; the product, COOH, is hydrogenated to produce formic acid. There is also some contribution from the addition of HCO and OH radicals under warm conditions ($\sim$20 -- 40~K), but it is small compared to the cold OH + CO route (however, compare with model {\tt final}, Sec.~\ref{sec:results:final}).

Panel (c) shows the evolution of oxygen-bearing species such as methyl formate (MF, HCOOCH$_3$) and dimethyl ether (DME, CH$_3$OCH$_3$); these two are formed on the grains at temperatures around 20--40~K, primarily through the addition of either the HCO or CH$_3$ radical to CH$_3$O, the latter of which derives directly from the photodissociation of methanol. Much of this production of MF and DME occurs within the ice mantle, mediated by bulk diffusion. 
Both MF and DME desorb from the grains at somewhat lower temperatures than water ice; in the {\em slow} warm-up timescale model (not shown) this produces a substantial decline in their abundances through gas-phase ion-molecule destruction until water, methanol, and various other species desorb from the grains at higher temperatures, reducing the availability of destructive ions. Dimethyl ether then experiences a resurgence through its effective gas-phase formation mechanism involving methanol and protonated methanol. The abundance of gas-phase glycolaldehyde (GA, CH$_2$(OH)CHO) is substantially higher than that of MF, as seen in past models that include GA. This dominance is caused by the greater production of the CH$_2$OH radical on the grains, versus CH$_3$O, via photodissociation of solid-phase methanol, by a factor 5:1.

Panel (d) of Fig. \ref{basic_medium} shows the abundances of selected nitrogen-bearing species in the {\tt basic} model. Formamide (NH$_2$CHO) is mainly formed in the ice mantles, through the addition of NH$_2$ and HCO radicals. However, it is strongly destroyed while still present in the ice, prior to complete desorption. H-atom abstraction by various radicals produces NH$_2$CO, which may further react with other radicals to form larger COMs. Methyl cyanide (CH$_3$CN) and ethyl cyanide (C$_2$H$_5$CN) are both formed on the grains; quantities on the order of 10$^{-9} n_{\mathrm{H}}$ are formed during the collapse stage, through hydrogenation of C$_2$N on the grains and through the addition of CH$_2$ to the CH$_2$CN radical, whose product is further hydrogenated to C$_2$H$_5$CN. At higher temperatures ($\sim$40 -- 80~K) during stage 2, methyl and ethyl cyanide are further enhanced through surface chemistry related to the release of HCN into the gas phase and the subsequent re-accretion of gas-phase HCN-destruction products \citep[for more detail, see][]{Garrod17}. On the grains, vinyl cyanide (C$_2$H$_3$CN) is quickly hydrogenated to ethyl cyanide. Once ethyl cyanide is relased into the gas phase at high temperatures, its destruction leads to the production of vinyl cyanide, as the result of electronic dissociative recombination of protonated ethyl cyanide.

\subsection{Sticking coefficient} \label{sec:results:sticking}

Model \texttt{bas\_stk} assumes a sticking coefficient of unity for all species, versus the previous value of 0.5 (model {\tt basic}). This change affects the solid-phase abundance of CO as a fraction of water content at the end of the collapse stage (see Table \ref{tab-st1-ice1}), which falls from $\sim$40\% to $\sim$31\%, putting it in even better agreement with observational values. The solid-phase methanol ratio with respect to water also falls somewhat, while the methane ratio is increased. Similar to the explanation relating to higher visual extinction (Sec.~\ref{sec:results:extinction}), the more rapid adsorption of C and O atoms in particular allows water and methane ice to be formed in greater abundance, prior to the build-up of the gas-phase CO that is then deposited onto the grains. The CO$_2$ ratio with respect to water is largely unaffected by the higher sticking coefficient; this molecule, which is formed from CO deposited onto the grains, is produced more strongly at early times, when the water and methane abundances are still growing, so it suffers less from the somewhat lower availability of gas-phase CO at late times that is caused by the more rapid loss of C and O from the gas phase in this model. The abundance of HCN is somewhat increased using the higher sticking coefficient value.

The abundances of solid-phase and gas-phase COMs are only marginally affected during the collapse stages (Tables \ref{tab-st1-ice2} \& \ref{tab-st1-gas}), for those species abundant enough to be important. Solid-phase ethanol (C$_2$H$_5$OH) abundance increases by around 70\% in the collapse stage, through the reaction of atomic O on the grains with the radical C$_2$H$_5$, to form an ethanol precursor. 

During the stage-2 warm-up, most COMs show only a modest difference in peak gas-phase abundance between models {\tt basic} and {\tt bas\_stk}, with some rising and some falling. Acetaldehyde (CH$_3$CHO) shows the strongest change, increasing by a factor 2.5 in the {\em slow} warm-up timescale model (Tables \ref{tab-gas1}--\ref{tab-gas3}). Several of the nitrile species show increases approaching a factor of 2 (Tables \ref{tab-gas4}--\ref{tab-gas6}).

\subsection{Photodissociation rates of solid-phase species} \label{sec:results:loPD}

\citet{Kalvans18} ran chemical models to investigate the effects of varying the efficiency of the photodissociation of solid-phase molecules. Gas-grain chemical models typically assume, in the absence of other information, that the rates and products of the photodissociation of solid-phase species are the same as those for the gas phase. The rates for the latter have typically been calculated based on known cross sections, integrated over the appropriate UV field (i.e.~ISRF or the secondary CR-induced field). The models of Kalv{\={a}}ns showed that observational molecular abundances were most accurately reproduced using efficiencies less than unity; a generic value 0.3 was recommended. In our models, beginning with model {\tt bas\_stk\_loPD}, we adopt this efficiency.

Comparing the results of models {\tt bas\_stk} and {\tt bas\_stk\_loPD} shown in Tables \ref{tab-st1-ice1}--\ref{tab-gas6}, the effect on most simple solid-phase species during the collapse stage is relatively small, with ratios with respect to water varying by up to a few tens of percent in most cases. O$_2$ and H$_2$O$_2$ show a stronger enhancement, by a little over a factor of 2. The final abundance of water itself is essentially unchanged, increasing by $\sim$0.3\%.

Solid-phase abundances of certain COMs in the collapse stage are substantially increased; acetic acid (AA, CH$_3$COOH) and dimethyl ether abundances are raised by factors $\sim$23 and $\sim$14, respectively. Early in the cold collapse stage of models {\tt bas\_stk} and {\tt bas\_stk\_loPD}, these species may be formed through the meeting of CH$_2$ with radicals COOH and CH$_3$O in the bulk ice; this produces new precursor radicals CH$_2$COOH and CH$_3$OCH$_2$ that can be hydrogenated to AA and DME. The reduction in the methane photodissociation rate within the bulk ice in model {\tt bas\_stk\_loPD} leads to less CH$_2$ production, but also -- crucially -- less production of H and OH, which can otherwise react with the CH$_2$. The result is an outsized enhancement in the abundances of COMs whose production relies strongly (in these particular models) on bulk reactions involving CH$_2$. As the collapse stage progresses and temperatures fall, CH$_2$ is no longer mobile in the ices and production of AA and DME ceases. The lower photodissociation rates in model {\tt bas\_stk\_loPD} benefit the survival of those species over long time periods.

Peak gas-phase abundances of COMs during the collapse phase are little affected by the change, as photodissociation has little effect either on the degree of nonthermal desorption of molecules from the grain surfaces nor on the overall rates of surface COM production.

In the {\em fast} and {\em medium} warm-up models, the reduction of solid-phase photodissociation rates tends to reduce the production of COMs, which in these models are mainly formed as the result of the recombination of photo-fragments in the ices at elevated temperatures. However, in the {\em slow} warm-up model, the destruction of COMs through photodissociation is a more important influence on their final abundances, such that (in general) COM abundances are greater in model {\tt bas\_stk\_loPD} than in model {\tt bas\_stk}. Variations in COM abundances for all of the stage-2 warm-up timescales are generally on the order of a factor of a few, but can be larger, and can go in either direction. Of particular note, the final abundance of methylamine (CH$_3$NH$_2$) in the {\em fast} and {\em medium} warm-up models is more than 5 times higher using the reduced photodissociation rates, while it is around 12 times higher in the {\em slow} warm-up case, for reasons related to those described above.

\subsection{H$_2$ treatment} \label{sec:results:H2}

Model {\tt bas\_stk\_loPD\_H2} adopts the new treatment for the modification of surface binding energies and diffusion barriers due to surface H$_2$ content.

Modest changes in abundance for the main ice constituents (Table \ref{tab-st1-ice1}) are found as the result of a substantial reduction in H$_2$ coverage on the ice surface in model {\tt bas\_stk\_loPD\_H2}, toward the end of the stage 1 run. The fraction of H$_2$ with respect to water on the surface, which achieves its maximum at the end of the models, reaches only 2\% with the new treatment, versus 76\% using model {\tt bas\_stk\_loPD}. There is a much smaller fraction of H$_2$ in the ice mantle itself in either model, but its abundance is two orders of magnitude lower at the end of the model run with the new H$_2$ treatment, i.e.~$2 \times 10^{-13}$ versus $2 \times 10^{-11}$ with respect to total hydrogen. This corresponds to less than one H$_2$ molecule in the ice mantle, versus around 16 without the new treatment. However, the abundance of H$_2$ inside the ice mantle reaches its maximum before the end time is reached; in both models, the peak abundance of H$_2$ in the mantle is achieved at time $9.2 \times 10^{5}$ yr, i.e.~22,000 yr before the end time. In model {\tt bas\_stk\_loPD}, the peak value is $\sim$19,700 H$_2$ molecules (i.e.~$\sim$10$^{-4}$ of the total ice mantle). In model {\tt bas\_stk\_loPD\_H2}, the peak is only $\sim$1 H$_2$ molecule, due to the smaller fraction of H$_2$ in the surface layer as the ice is growing.

As well as being incorporated into the ice as it forms, H$_2$ within the mantle is also produced in situ, and may diffuse out to the surface or react with other species within the mantle. (The drop in H$_2$ abundance in the mantle toward the end of the models, as mentioned above, is caused by this gradual diffusion of those molecules out of the bulk and onto the surface, where they may escape to the gas phase). The smaller H$_2$ abundance within the ice when using the new treatment reduces the rate at which CH$_2$, a photodissociation product of methane, can react repetitively with H$_2$ to form CH$_3$ and thence CH$_4$. As a result, a larger quantity of CH$_2$ and CH$_3$ is retained throughout the stage 1 model run, which then can react (via thermal diffusion) during the warm-up stage with the radical COOH to form acetic acid, or with HCO to form acetaldehyde, beginning around 15~K. These and other species that are formed in the mantle through the addition of CH$_3$ to other radicals, such as ethanol (C$_2$H$_5$OH), are also somewhat enhanced in their gas-phase abundances during the warm-up stage. Note that these effects are not seen in models in which the bulk diffusion of species other than H and H$_2$ is switched off (Sec.~\ref{sec:results:bulkdiff}). No major effects (i.e.~order of magnitude changes) are seen in the abundances of other molecules as a result of the new H$_2$ treatment.

\subsection{Proton transfer reactions} \label{results:methods:NH3}

Model {\tt bas\_stk\_loPD\_H2\_T16} includes gas-phase proton-transfer reactions between protonated COMs and ammonia, following the suggestion of \citet{Taquet16}. The effects of this change are seen to be important only under conditions in which gas-phase COMs are abundant. Differences between models {\tt bas\_stk\_loPD\_H2} and {\tt bas\_stk\_loPD\_H2\_T16} are therefore negligible during the collapse phase.

Considering the stage-2 results shown in Tables \ref{tab-gas1}--\ref{tab-gas3}, for the two fastest warm-up stages the difference in peak abundances caused by the additional reactions is very small. In the {\em slow} warm-up models, the peak abundances of methanol and several other species are somewhat enhanced, while methyl formate undergoes a much stronger enhancement, by approximately 8 times. In Tables \ref{tab-gas4}--\ref{tab-gas6}, the nitrogen-bearing species show some modest changes up or down; the peak vinyl cyanide (C$_2$H$_3$CN) abundance falls by as much as a factor of two in the two faster warm-up models using the new reactions, while it increases a little in the {\em slow} warm-up model.

\begin{figure*}[t]
\center
         {\includegraphics[width=0.48\hsize]{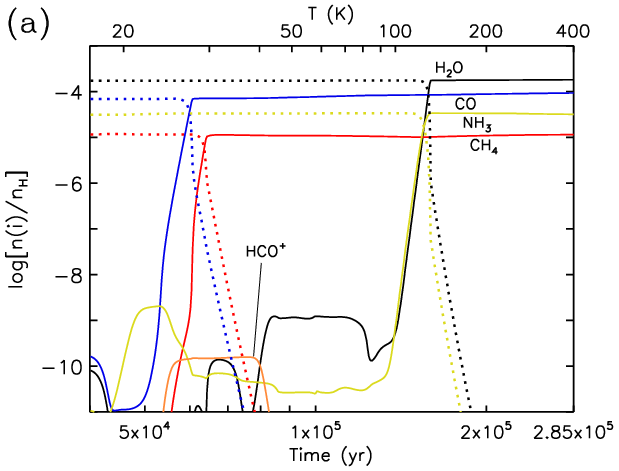}}
         {\includegraphics[width=0.48\hsize]{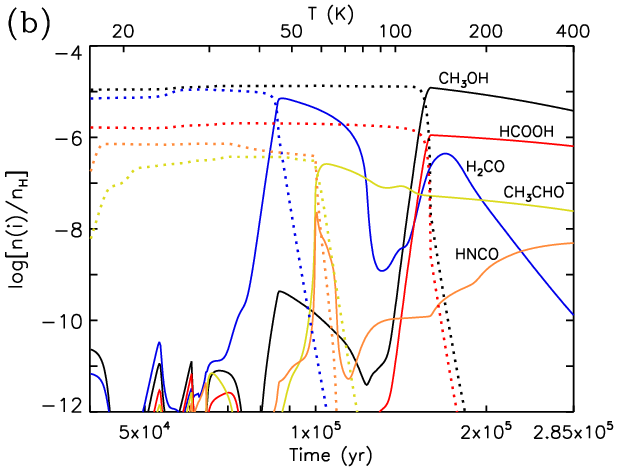}}
         {\includegraphics[width=0.48\hsize]{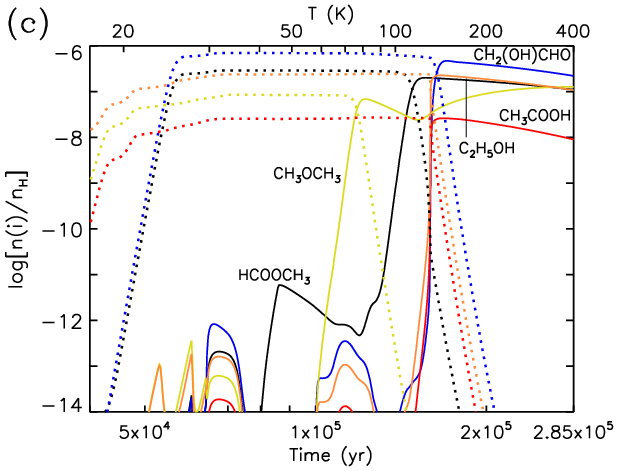}}
         {\includegraphics[width=0.48\hsize]{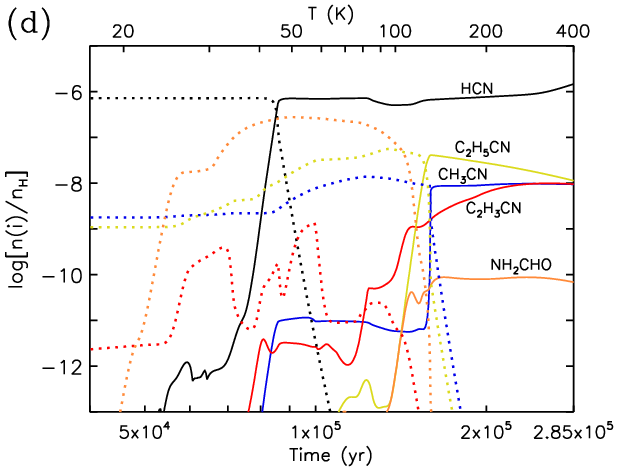}}
\caption{Selected chemical abundances with respect to total hydrogen. Results correspond to model {\tt basic} stage 2, using the {\em medium} warm-up timescale. Solid lines indicate gas-phase abundances. Dotted lines of the same color indicate the same species on the dust grains.}
\label{basic_medium}
\end{figure*}

\begin{figure*}[t]
\center
         {\includegraphics[width=0.48\hsize]{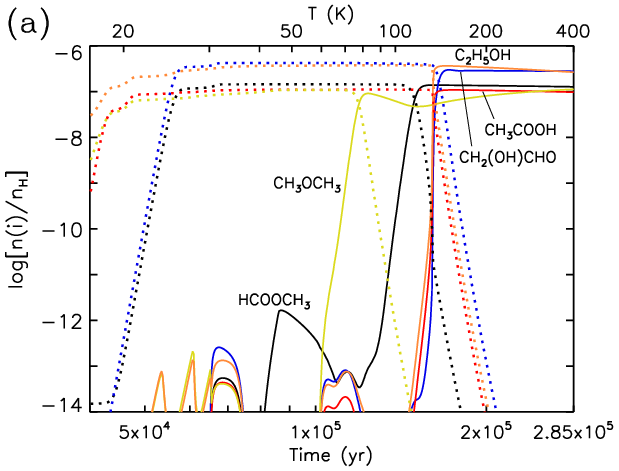}}
         {\includegraphics[width=0.48\hsize]{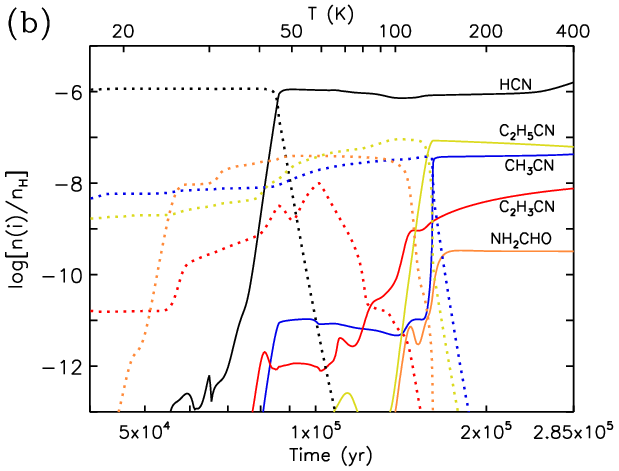}}
\caption{Description as per Fig.~\ref{basic_medium}. Results correspond to model {\tt bas\_stk\_loPD\_H2\_T16} stage 2, using the {\em medium} warm-up timescale. In comparison with the {\tt basic} model, this setup includes a uniform sticking coefficient of 1, reduced photodissociation efficiency for solid-phase species, a new treatment for the effect of H$_2$ coverage on surface binding energies, and includes new gas-phase proton-transfer reactions of large ions with ammonia.}
\label{T16_medium}
\end{figure*}

The main effect of the additional reactions in fact concerns the survival timescales of COMs in the gas phase, and not their peak abundance values. Fig.~\ref{T16_medium} shows the stage-2 gas-phase abundances of selected COMs in model {\tt bas\_stk\_loPD\_H2\_T16} using the {\em medium} warm-up timescale. These results may be crudely compared with those of Fig.~\ref{basic_medium}, which corresponds to the {\tt basic} model. The COM abundance profiles are generally flatter with respect to time with the new reactions included, and indeed lead to the maintenance of COM abundances very close to their peak values as the model progresses. The main destruction mechanism for most of these COMs consists of protonation by molecular ions, followed by dissociative electronic recombination. Reaction of the protonated COMs instead with ammonia returns the COMs unscathed.

In the {\em slow} warm-up model (not shown in Fig.~\ref{T16_medium}), the release of methyl formate from the grains ahead of water and other abundant solid-phase species exposes it immediately to destruction by molecular ions. Without the ammonia reactions, this model has sufficient time for methyl formate to be substantially destroyed in the gas phase while it is still being gradually desorbed from the grains, leading to a lower peak abundance value.

Subsequent models include the additional ammonia-related reactions.

\subsection{Switch-off of bulk diffusion within the ice mantles} \label{sec:results:bulkdiff}

Model {\tt bas\_stk\_loPD\_H2\_T16\_no-bd} corresponds to the case with no diffusion in the bulk ice (with the exception of H and H$_2$, which are exempt from this change). Comparing with the preceding model, {\tt bas\_stk\_loPD\_H2\_T16}, most of the simple ices in Table \ref{tab-st1-ice1} show only modest changes in their final fractional abundances. Formaldehyde and methanol abundances are a little lower, but in keeping with observations. The abundances of O$_2$ and H$_2$O$_2$, however, are markedly different, falling by factors of 60 and 480, respectively, without bulk diffusion. When bulk diffusion is switched on, most O$_2$ in the mantle is produced as the result of diffusion-driven atomic oxygen reactions: O + OH $\rightarrow$ O$_2$H and O + O$_2$H $\rightarrow$ O$_2$ + OH, with photodissociation of O$_2$H also providing an alternative to the second step. The reaction of a bulk-ice H atom with O$_2$H produces H$_2$O$_2$. The oxygen atoms are produced by photodissociation of CO$_2$ molecules in the ice, while photodissociation of water provides the OH radical. Thus, while in practice being driven by photodissociation, the mediating reactions leading to O$_2$ and H$_2$O$_2$ are diffusive. When bulk diffusion by all but H and H$_2$ is switched off (with no nondiffusive processes included), production of O$_2$ and related species drops precipitously.

Although the production of COMs in the ices during stage 1 is already quite weak, Table \ref{tab-st1-ice2} shows that even those fall drastically in most cases. Formic acid (HCOOH) is little affected and in fact rises somewhat when bulk diffusion is switched off; it is mainly formed on the grain surfaces in tandem with CO$_2$, through the reaction of OH with CO \citep[which occurs through the nondiffusive method provided by][see Sec~\ref{sec:methods:basic}]{GP11}. Ethanol also remains moderately abundant in the ices following the change, as it has surface production mechanisms involving CH$_2$, which can be mobile on the grains early in stage 1, when the dust temperatures are a little higher than the minimum. The influence of bulk diffusion on gas-phase COM abundances in stage 1 is minimal.

\begin{figure*}[t]
\center
         {\includegraphics[width=0.48\hsize]{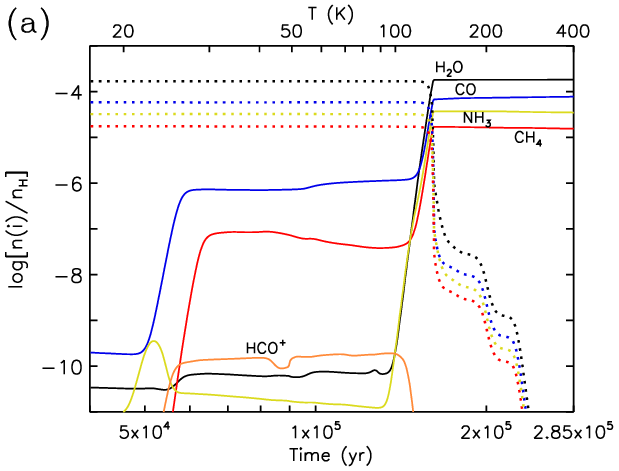}}
         {\includegraphics[width=0.48\hsize]{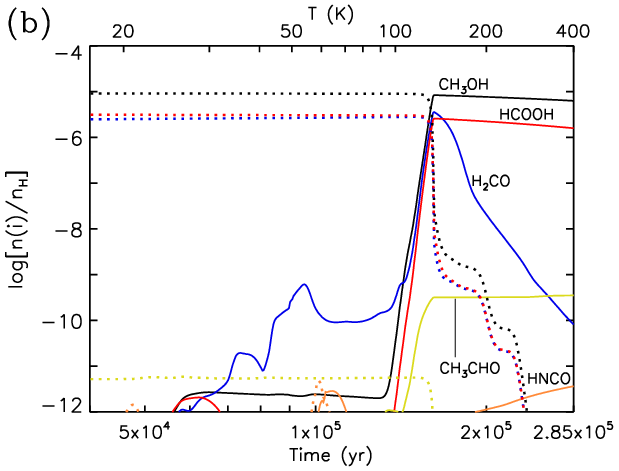}}
         {\includegraphics[width=0.48\hsize]{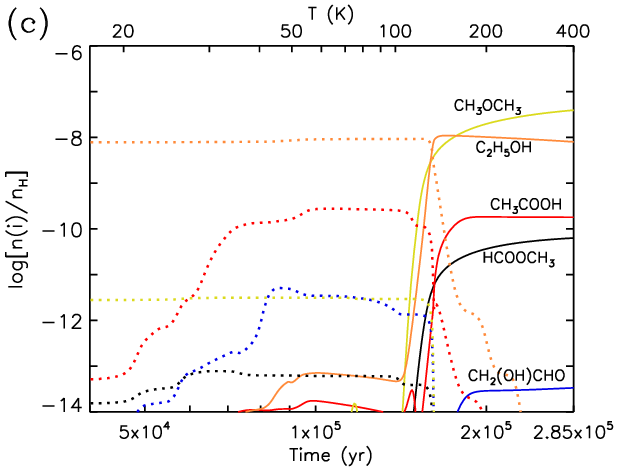}}
         {\includegraphics[width=0.48\hsize]{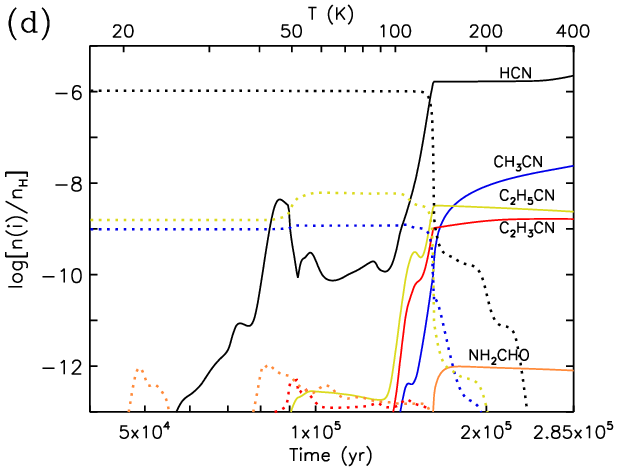}}
\caption{Description as per Fig.~\ref{basic_medium}. Results correspond to model {\tt bas\_stk\_loPD\_H2\_T16\_no-bd} stage 2, using the {\em medium} warm-up timescale. In comparison with the {\tt bas\_stk\_loPD\_H2\_T16} model (shown in Fig.~\ref{T16_medium}), this setup has bulk diffusion switched off for all species except H and H$_2$.}
\label{no-md_medium}
\end{figure*}

The stage-2 peak gas-phase abundances of COMs shown in Tables \ref{tab-gas1} -- \ref{tab-gas6} are strongly dependent on the inclusion of bulk diffusion, many falling by more than an order of magnitude (in some cases, several) when this process is switched off. Fig.~\ref{no-md_medium} shows the evolution of stage-2 abundances for the {\tt bas\_stk\_loPD\_H2\_T16\_no-bd} model. In panels (c) and (d) in particular, the strong production of COMs in the ices in the $\sim$20 -- 100~K regime is now extremely weak. Acetic acid and ethyl cyanide retain some degree of production in the ices at elevated temperatures, but not enough to replicate the abundances produced in the earlier models. Ethanol, which does achieve high gas-phase abundances, is seen to inherit most of its substantial abundance from stage 1. Dimethyl ether also achieves large abundances, but in this case as the result of gas-phase production relating to methanol desorption from the grains.

Fig.~\ref{no-md_medium} also demonstrates the different desorption behavior of simple species when bulk diffusion is switched off; all substantial desorption of ice-mantle material now occurs around the period when water desorbs at $\sim$120~K. The volatiles CO and CH$_4$ are still lost from the grains at the same low temperatures as before, but in relatively small quantities; this loss constitutes on the order of 1\% of their total solid-phase abundances. That is, material is lost only from the upper few monolayers of the ice during the strongest periods of low-temperature desorption of volatiles. The remaining material is trapped within the water and other strongly-bound species, as it is unable to reach the surface layer through bulk diffusion. Molecules with rather higher binding energies than water, e.g.~glycolaldehyde, remain on the grains to yet higher temperatures. Small fractions of the more volatile species (as well as water itself) remain on the grains even at temperatures above the water desorption temperature, as they are trapped by the small quantities of species with extremely high binding energies. By around 300~K, the desorption of all remaining ice-mantle material is essentially complete.

\begin{figure*}[t]
\center
         {\includegraphics[width=0.48\hsize]{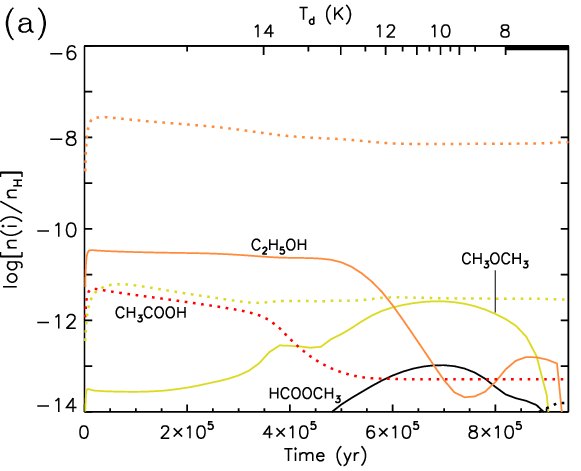}}
         {\includegraphics[width=0.48\hsize]{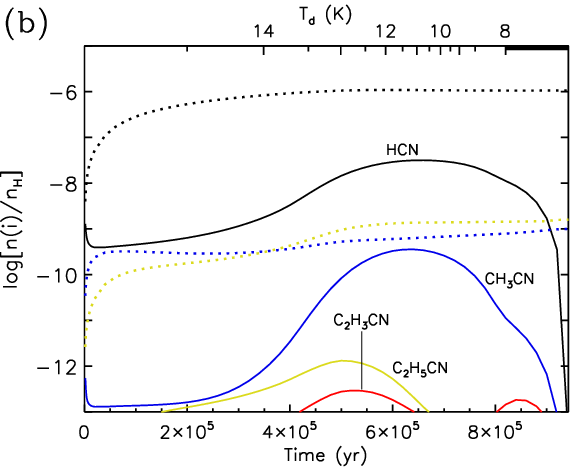}}
\caption{Selected chemical abundances with respect to total hydrogen, for model {\tt bas\_stk\_loPD\_H2\_T16\_no-bd} stage 1 (collapse), in which bulk diffusion is switched off for all species except H and H$_2$. Solid lines indicate gas-phase abundances. Dotted lines of the same color indicate the same species on the dust grains. The peak dust temperature is $\sim$14.7~K. The minimum dust temperature is 8~K.}
\label{no-md_collapse}
\end{figure*}

Fig.~\ref{no-md_collapse} shows the gas- and solid-phase abundances of selected species, including various COMs, during the stage-1 cold collapse. As in the other figures, solid lines indicate gas-phase species, while dotted lines of the same color indicate the same species in the solid phase. In several cases, the gas-phase species' abundances are too small to show up in the plots (in which case the solid-phase species are directly labeled). The production of ethanol on the grains is seen to occur at very early times, which would correspond to deposition into the very deepest ice layers. Some purely grain-surface production of methyl formate and glycolaldehyde occurs at late times (low temperatures) in stage 1, but it is not sufficient to reach observational gas-phase abundances of those species.

It is clear, based on the models examined so far (which include only diffusive mechanisms for reactions on the grains), that the production of COMs through purely surface diffusive processes is insufficient to account for their observed abundances in hot cores.


\subsection{Eley-Rideal (E-R) mechanism} \label{sec:results:ER}

Model {\tt bas\_stk\_loPD\_H2\_T16\_no-bd\_ER} uses the E-R mechanism as described in Sec.~\ref{sec:methods:E-R}, and is compared with model {\tt bas\_stk\_loPD\_H2\_T16\_no-bd}. There are some changes in COM abundances during stage 2 with the Eley-Rideal mechanism switched on, by as much as 20\% in selected cases. However, for the most part, the effects of the E-R mechanism are negligible. The abundances of simple, solid-phase species are unchanged by the inclusion of E-R.

\subsection{Photodissociation-induced reactions} \label{sec:results:PDI}

\subsubsection{Basic method (PDI)}\label{basicPDI}

Model {\tt bas\_stk\_loPD\_H2\_T16\_no-bd\_PDI} uses the photodissociation-induced reaction mechanism labeled PDI, as described in Sec.~\ref{sec:methods:PDI}, and is again compared with model {\tt bas\_stk\_loPD\_H2\_T16\_no-bd}. The results shown in Table \ref{tab-st1-ice1} indicate that most of the simple ice species are only weakly affected by the inclusion of immediate reactions for photodissociation products in the ice. The only substantial changes discernible are the return of O$_2$ and H$_2$O$_2$ to sizeable abundances. Both are of the same order of magnitude produced in the last model to include bulk diffusion, model {\tt bas\_stk\_loPD\_H2\_T16}, although a little lower. Most of this solid-phase O$_2$ is now formed by the reaction O + O $\rightarrow$ O$_2$, which occurs as a PDI process. H$_2$O$_2$ now also forms via the PDI reaction OH + OH $\rightarrow$ H$_2$O$_2$.

Table \ref{tab-st1-ice2} shows solid-phase COM abundances at the end of the collapse stage. Here, it can be seen that all COMs enjoy a notable increase in their final abundances in the bulk ice with the PDI mechanism switched on. While dimethyl ether and ethanol increase by only a factor of 2--3, all other COMs in the table rise by orders of magnitude; indeed they now exceed not only the values obtained with bulk diffusion switched off, but also those with it switched on. However, most of those abundances are still on the order of $10^{-9}$ with respect to total H, leaving them a little shy of observed gas-phase abundances in hot cores. The peak gas-phase abundances of these species during the collapse stage, shown in Table \ref{tab-st1-gas}, are largely unaffected, as the PDI mechanism is most important in the bulk ice rather than on the ice surface itself, from which molecules could escape directly into the gas.

The stage-2 abundances for O-bearing COMs shown in Tables \ref{tab-gas1} -- \ref{tab-gas3} show a similar trend as for the solid-phase COMs at the end of stage 1, although the gas-phase peak values in this case do not reach the high values produced in models that include bulk diffusion. However, with increasing stage-2 warm-up timescale, the peak gas-phase abundances increase further, rising most strongly for the {\em slow} warm-up timescale. The additional time provided by stage 2 allows further production of COMs through photodissociation-induced reactions. While some of the peak COM abundances produced by the {\em medium} warm-up timescale model may arguably agree with observed values, methyl formate only reaches substantial abundances ($10^{-8} n_{\mathrm{H}}$) in the {\em slow} model. The production of methyl formate in this model occurs through the reaction HCO + CH$_3$O $\rightarrow$ HCOOCH$_3$, although the HCO radical is primarily formed through the photodissociation of the abundant HCOOH molecule rather than some process involving H$_2$CO or CO. 

It is noteworthy also that the abundance of glycolaldehyde in the gas phase in all stage-2 models is much lower than that achieved in the solid phase at the end of stage 1. This is caused by the substantially higher binding energy of glycolaldehyde than water; when water itself desorbs, glycolaldehyde gathers increasingly on what is left of the ice surface, but the dust temperature is not initially high enough to cause it to desorb. However, it is susceptible to H-abstraction reactions, both with H adsorbed from the gas phase and with surface radicals, producing the CH$_2$(OH)CO radical, which further reacts with mobile radicals to produce larger molecules. By the time the dust temperature is high enough for glycolaldehyde to desorb rapidly, much has already been converted to other species. It is notable in this model that, although the abundance of acetic acid is relatively high, the ratio of methyl formate to glycolaldehyde is the correct order of magnitude with respect to observational values, for all warm-up timescales.

The N-bearing species shown in Tables \ref{tab-gas4} -- \ref{tab-gas6} show increases in gas-phase abundance, when compared with model {\tt bas\_stk\_loPD\_H2\_T16\_no-bd}. In the case of NH$_2$CN, HNCO and NH$_2$CHO, the increases are very large, and exceed even the values produced by model {\tt bas\_stk\_loPD\_H2\_T16}, typically by at least an order of magnitude. These species, therefore, appear to be strongly affected by the inclusion of the new PDI mechanism. This is true for all stage-2 warm-up timescales.


\subsubsection{Advanced method (PDI2)}

As described in Sec.~\ref{sec:methods:PDI}, the PDI2 mechanism is the same as PDI except that, within the ice mantles, if no reaction partner is found by one or other of the newly-formed photo-products then those products recombine immediately to re-form the original molecule (assuming that neither photo-product is hydrogen). The main overall effect of the PDI2 treatment, versus PDI, is that the production of unreacted radicals in the ice is reduced, because such species can only survive if their partner product reacts with something else nearby. This distinction has a major effect on the abundances of O$_2$ and H$_2$O$_2$ during stage 1, whose final solid-phase abundances return approximately to the values achieved in model {\tt bas\_stk\_loPD\_H2\_T16\_no-bd} (no PDI mechanism and no bulk diffusion). The abundance of H$_2$O$_2$ is still slightly enhanced, and O$_2$ and H$_2$O$_2$ are still formed primarily via the PDI2 mechanism. The final collapse-stage abundances of NH$_3$ and CH$_4$ in the ice also fall somewhat using PDI2, to $\sim$15\% and $\sim$5\%, respectively -- down from roughly 20\% and 10\%. The fractional abundance of CO falls a little, with a commensurate increase in CO$_2$ abundance; this is caused by the decreased efficiency of CO$_2$ photodissociation using the PDI2 treatment. The peak gas-phase abundances of COMs during stage 1 are again essentially unchanged by PDI2.

\begin{figure}[t]
\center
         {\includegraphics[width=0.48\hsize]{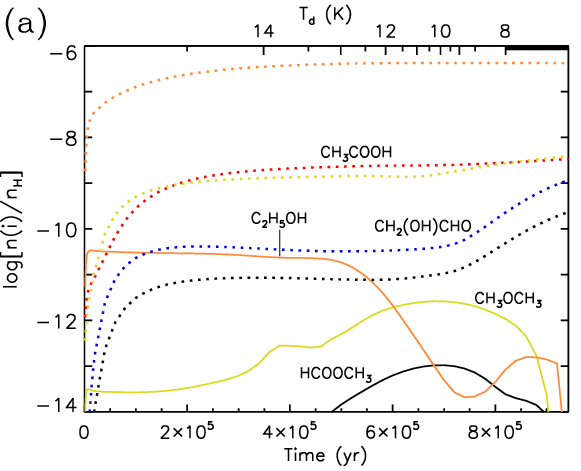}}
         {\includegraphics[width=0.48\hsize]{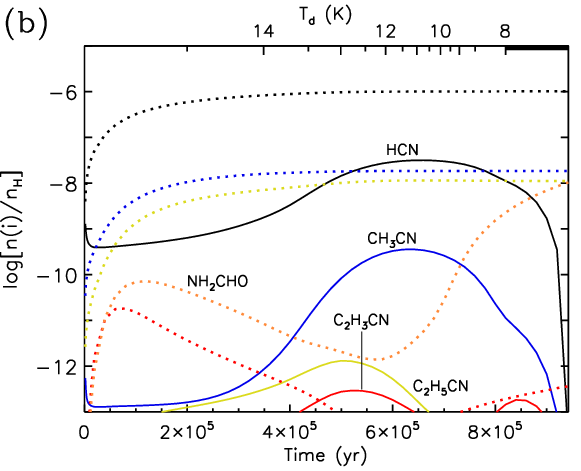}}
\caption{Description as per Fig.~\ref{no-md_collapse}. Results correspond to model {\tt bas\_stk\_loPD\_H2\_T16\_no-bd\_PDI2} stage 1 (collapse). In comparison with the {\tt bas\_stk\_loPD\_H2\_T16\_no-bd} model (shown in Fig.~\ref{no-md_collapse}), this setup includes the PDI2 mechanism in which unreacted photoproduct pairs within the bulk ice are allowed spontaneously to recombine.}
\label{PDI2_collapse}
\end{figure}

The peak gas-phase abundance of methyl formate during stage 2 is slightly lower with the PDI2 treatment in effect, although the same general trend is seen as for basic PDI, in which longer timescales result in larger values. Acetic acid abundances also follow their PDI trend, although with somewhat higher absolute abundances. The behavior of glycolaldehyde is different; its peak abundance is higher in the {\em fast} warm-up model, but lower in the other two, so that it maintains more stable peak abundances between the different warm-up timescales. Dimethyl ether abundances are moderately increased by the PDI2 treatment. However, ethanol increases by factors of around 3--6, leading to stable maximum gas-phase abundances close to $5 \times 10^{-7} n_{\mathrm{H}}$ for all warm-up timescales. Almost all of this ethanol is inherited from the cold collapse stage (see Fig.~\ref{PDI2_collapse}), formed through diffusive surface reactions of atomic O with the C$_2$H$_5$ radical, followed by H addition. The reason that ethanol shows increased abundances is that its photodissociation in the bulk ice is much less effective under the PDI2 treatment. In model {\tt bas\_stk\_loPD\_H2\_T16\_no-bd}, ethanol can be photodissociated but has no mechanism to recombine. In model {\tt bas\_stk\_loPD\_H2\_T16\_no-bd\_PDI}, it is dissociated but has a mechanism allowing it to be re-formed from the appropriate radicals in the ices. In model {\tt bas\_stk\_loPD\_H2\_T16\_no-bd\_PDI2}, its photodissociation is far less efficient, because its photo-fragments will frequently recombine in the ice immediately following dissociation; in cases where photo-dissociation produces an H atom and a large radical, which results in the escape of the H atom and no immediate recombination, another H atom soon arrives (through bulk diffusion) to re-form ethanol.

Peak gas-phase abundances of most of the N-bearing species in stage 2 are altered (up or down) by factors as large as 2--3. The most significant effect is experienced by methylamine (CH$_3$NH$_2$), which increases by around a factor of 10, reaching fractional abundances on the order of $10^{-5}$ for all warm-up timescales. These values correspond to a substantial fraction of the carbon and nitrogen budgets, and are inherited from stage 1, hence the decreases in methanol and ammonia in the bulk ice. The reason for greater apparent production of methylamine is, as with ethanol, a decrease in the photodissociation efficiency for pathways leading to two immobile fragments (i.e.~CH$_3$ + NH$_2$).

It is apparent that the consideration of recombination of bulk-ice photo-products in the models is important to the regulation of the efficiency of COM production and destruction by UV photons.

\subsection{Three-body Reaction Mechanisms} \label{sec:results:3B}

Models {\tt bas\_stk\_loPD\_H2\_T16\_no-bd\_3B1/2/3} involve the inclusion of the basic three-body reaction mechanism, allowing between one and three cycles of follow-on reactions to take place. In the {\tt 3B1} case, after all diffusive processes have been evaluated, the production rates of the various grain surface and mantle species produced are used to construct the rates of nondiffusive follow-on (three-body) reactions caused by the appearance of some of those products in proximity to possible reaction partners. In the {\tt 3B2} case, the production rates of those follow-on reactions are used to construct further follow-on reaction rates -- and again, in the {\tt 3B3} case.

Model {\tt bas\_stk\_loPD\_H2\_T16\_no-bd\_3B3\_EF} includes the basic three-body mechanism as well as the excited formation mechanism that can take effect in cases where the follow-on reactions have activation energy barriers that can be overcome by the energy of formation of the reaction product that initiates that follow-on reaction.

\subsubsection{Basic three-body mechanism (3-B)}

Comparing with model {\tt bas\_stk\_loPD\_H2\_T16\_no-bd}, the basic three-body mechanism has only a minor effect on most simple ice species during stage-1, except for O$_2$ and H$_2$O$_2$; these again rise to values of comparable magnitude to those achieved when full bulk diffusion is switched on. The first round of three-body reactions is by far the most important; indeed O$_2$ and H$_2$O$_2$ fall a little when the second and third round of reactions are included.

During stage 1 there are substantial increases in solid-phase abundance of COMs containing functional groups derived from CO, such as glycolaldehyde, methyl formate, acetaldehyde, dimethyl ether and ethanol. Abundances tend to increase moderately with inclusion of multiple rounds of three-body reactions. The increases in COM abundances produced during stage 1 tend to carry through to the gas-phase abundances of stage 2 in a similar way as described for the PDI mechansims. Methyl formate enhancement by the 3-B process is stronger in both stages than the PDI mechanisms, with the peak gas-phase MF abundance in the {\em slow} warm-up model reaching around $5 \times 10^{-9} n_{\mathrm{H}}$. The effect on N-bearing species is not too dissimilar from the PDI case, although NH$_2$CN in particular is not strongly enhanced by the 3-B mechanism. This is because when the three-body mechanism is the only nondiffusive mechanism available in the bulk ice, it relies solely on diffusive production (at least in the first round) for the formation of precursor species. None of the precursors of NH$_2$CN, in particular NH$_2$ and CN, have strong diffusive production mechanisms in the ice mantles, which rely on H diffusion.

\subsubsection{Three-body excited formation mechanism (3-BEF)}

\begin{figure}[t]
\center
         {\includegraphics[width=0.48\hsize]{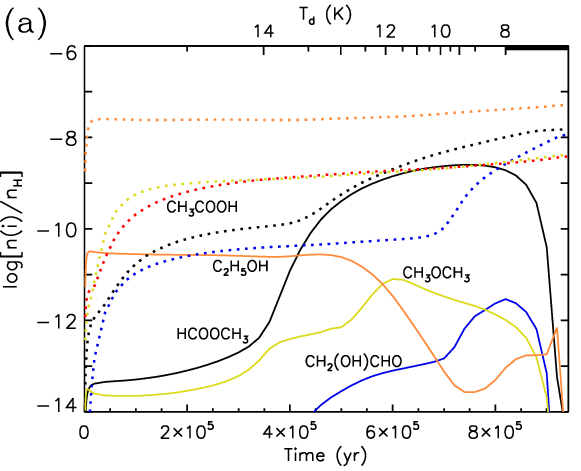}}
         {\includegraphics[width=0.48\hsize]{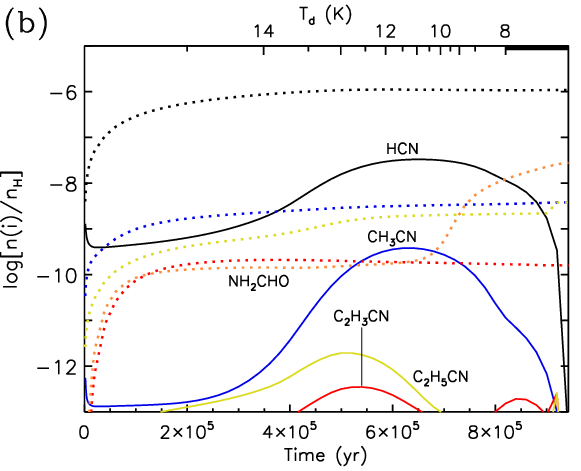}}
\caption{Description as per Fig.~\ref{no-md_collapse}. Results correspond to model {\tt bas\_stk\_loPD\_H2\_T16\_no-bd\_3B3\_EF} stage 1 (collapse). In comparison with the {\tt bas\_stk\_loPD\_H2\_T16\_no-bd} model (shown in Fig.~\ref{no-md_collapse}), this setup includes three cycles of the three body reaction mechanism, with the excited-formation three-body process also switched on.}
\label{3BR3-EF_collapse}
\end{figure}

Simple solid-phase molecular abundances are only modestly affected, but certain key COMs are strongly enhanced by the 3-BEF process, such as acetaldehyde and methyl formate. The latter now achieves an abundance during stage 1 that is comparable to observed gas-phase abundances in hot cores, i.e.~around $10^{-8} n_{\mathrm{H}}$. Peak gas-phase abundances of those species during stage 1 are also enhanced, to values around $10^{-9} n_{\mathrm{H}}$, although gas-phase dimethyl ether is not similarly enhanced, increasing by only a factor of a few (see Fig.~\ref{3BR3-EF_collapse}). The solid-phase abundances of MF and acetaldehyde achieved during stage 1 produce substantial gas-phase abundances in the stage-2 models, and indeed reach values comparable to observed fractional abundances; the modeled peak gas-phase abundances actually exceed the observational values somewhat. DME also achieves a gas-phase value that falls less than an order of magnitude short of the observations.

It is notable that, of all the nondiffusive mechanisms individually tested, only the 3-BEF mechanism is strong enough to reproduce observable fractional abundances of methyl formate in hot cores. It is also the only process that can produce any substantial gas-phase abundance of MF under the cold conditions of stage 1.


\subsection{Combined effects of all nondiffusive processes} \label{sec:results:combi}

Model {\tt all} incorporates all of the model changes up to model {\tt bas\_stk\_loPD\_H2\_T16\_no-bd\_3B3\_EF}, including all of the nondiffusive processes that were previously tested individually. The photodissociation-induced reaction process uses the PDI2 treatment. The three-body reaction rates are calculated last of all in each evaluation of the chemical rates so that, in the first 3-B cycle, the three-body reaction rates are based on appearance rates generated by all other diffusive and nondiffusive reaction processes. The appearance rates used in the second and third cycles derive entirely from the preceding cycles of three-body reactions. Again, the 3-BEF process is carried out in tandem with the basic 3-B process.

Comparing with model {\tt bas\_stk\_loPD\_H2\_T16\_no-bd}, the solid-phase abundances of CO, CH$_4$ and NH$_3$ with respect to water appear to behave the same way as in the {\tt PDI2} model, in which the CH$_3$NH$_2$ abundance is also enhanced. The abundance of CO$_2$ also rises to $\sim$22\% with respect to water ice. The O$_2$ and H$_2$O$_2$ solid-phase abundances are both higher than in the model with no nondiffusive processes, but only O$_2$ approaches the maximum value achieved for the individual 3-B models; H$_2$O$_2$ is somewhat lower than when the 3-B/3-BEF processes alone are operative, due to the reduction in unreacted photo-fragments caused by the PDI2 treatment (note that the photodissociation process itself is present in all models, including 3-B/3-BEF). In general, the solid-phase COMs during stage 1 achieve values roughly equal to the maximum values attained in the individual nondiffusive-mechanism models. Methyl formate, glycolaldehyde, and acetic acid are a little higher even than in the 3-BEF case. Gas-phase COM abundances in stage 1 are fairly representative of the 3-BEF values.

In stage 2 of the {\tt all} models, the peak gas-phase abundance of glycolaldehyde is much higher than in the individual nondiffusive models. This is the result of two effects; as explained above, the abundance of glycolaldehyde declines on the grain surfaces shortly prior to its desorption, caused by the attack of mobile surface radicals. In the {\tt all} model, fewer radicals of any kind are retained in the ice, due to the PDI2 treatment. Meanwhile, the total production of glycolaldehyde in the ice is greater, due to the 3-B mechanism, which makes it more difficult for the surface radicals to destroy such a large fraction of it. Acetic acid benefits from the same general behavior, albeit to a smaller degree, although in this case it is the photodissociation-induced reaction mechanism itself that causes the original enhancement in CH$_3$COOH abundance in the ice. Again, for several other COMs, including both O- and N-bearing species, the peak gas-phase abundances achieved are a reflection of the individual nondiffusive production mechanisms that showed the greatest influence individually.


\subsection{Atomic diffusion barrier adjustments}

Model {\tt all\_Edif} uses different generic $E_{\mathrm{dif}}$:$E_{\mathrm{des}}$ ratios for grain-surface diffusion of atoms versus molecules; the {\tt all} model (and all preceding versions) used a generic ratio of 0.35 for both. The new model assumes a ratio 0.55 for all atoms, while retaining the 0.35 value for molecules (see Sec.~\ref{sec:methods:barrs}). Note that the change in diffusion barrier applies to atomic H, the main diffuser at low temperatures. Its bulk diffusion barrier is also increased by the same factor.

The effects of this change on the stage-1 ice abundances are noticeable but modest for most simple species; formic acid abundance falls by a factor of around two, while the main ice components experience smaller changes. Due to the higher diffusion barrier for atomic hydrogen of 247.5~K (versus 157.5~K in the {\tt all} model), the surface diffusion of atomic H becomes as many as 5 orders of magnitude slower. However, in response, the surface abundances of the reaction partners of H increase substantially, as the lifetime against reaction is that much longer. The net result for simple hydrides such as H$_2$O, CH$_4$ and NH$_3$ is that their overall production rates do not change very much.

However, the abundances of both O$_2$ and H$_2$O$_2$ are enhanced by two orders of magnitude in model {\tt all\_Edif}, each rising to around 0.3\% with respect to solid water. This is also related to the slower diffusion of surface atomic H, which gives O$_2$ a longer surface lifetime against destruction. The O$_2$ on the surface originates in the gas phase, and is formed through the reaction O + OH $\rightarrow$ O$_2$ + H. O$_2$ reaches as high as $\sim$2\% coverage of the ice surface at late times in the collapse stage, and this high fraction is preserved in the ice mantle as the ice continues to grow. 

Slower H diffusion also causes a higher surface abundance of atomic O, which can rise to a few percent coverage, leading to the transfer of atomic O into the bulk ice. Although this oxygen is not able to diffuse within the bulk, the further production of O and OH through photodissociation of CO$_2$ and water, respectively, leads to the spontaneous appearance of those species in the presence of atomic O, allowing O$_2$ or O$_2$H also to form via the PDI mechanism within the ices. This in turn leads to the production of O$_3$ and H$_2$O$_2$, through PDI reactions with O and H, respectively. The collective O$_2$, O$_2$H, H$_2$O$_2$ and O$_3$ content of the ice reaches 2.3\% of the bulk water abundance by the end of stage 1 for model {\tt all\_Edif}. As noted by JG20, the PDI production of O$_2$ and related species, when high atomic diffusion barriers are assumed, seems to agree with the high fractional O$_2$ content required in interstellar ices if they are to be the direct precursors of O$_2$-rich cometary ice.

As noted also by JG20, the adoption of a high atomic diffusion barrier increases the surface lifetimes of radicals related to CO, H$_2$CO and CH$_3$OH, increasing the rates at which nondiffusive surface reaction mechanisms convert them into oxygen-bearing COMs. The abundances of methyl formate and glycolaldehyde both reach peak solid-phase values on the order of $10^{-7}$ with respect to total hydrogen, which is ample to explain gas-phase abundances observed in hot cores. Acetic acid rises to abundances on the order of $10^{-8}$. We note that in this model, the solid-phase abundances of all three of these structural isomers are now in the correct sequence of increasing abundance, with acetic acid at the lowest and methyl formate the greatest. Dimethyl ether abundance rises to $10^{-8}$, although its structural isomer ethanol falls a little.

The gas-phase abundances of COMs achieved during the cold collapse also begin to appear closer to the values observed in pre-stellar cores. Chemical desorption is responsible for the transfer of a fraction of the surface-formed COMs into the gas phase. Dimethyl ether remains around an order of magnitude lower than methyl formate and acetaldehyde, but achieves a very acceptable match to the observational values.

The correct order of methyl formate (MF), glycolaldehyde (GA) and acetic acid (AA) seen in the solid phase is preserved in their stage-2 gas-phase abundances -- again, albeit with their ratios less stark than is typical in the observational measurements. Based on the individual peak gas-phase values of each, the MF:GA ratio is no greater than 2 for all warm-up timescales, while the GA:AA ratio is no greater than 3.5. Other COM abundances seem appropriate to hot-core sources but not substantially different from the values obtained in the {\tt all} models. Of the nitrogen-bearing species, HNCO and NH$_2$CHO are both somewhat enhanced in the {\tt all\_Edif} models, as is vinyl cyanide.

\subsection{Adjustment of the H diffusion barrier and binding energy}

Model {\tt all\_Edif\_S17} adopts the diffusion barrier and binding energy of atomic H on amorphous solid water suggested by \citet{Sen17}, overriding the values used for H in model {\tt all\_Edif} (although the values for other atoms are retained). In fact, although the new atomic H binding energy, at 661~K, is substantially greater than the 450~K used in the preceding models, the diffusion barrier, at 243~K, is almost identical to that used in model {\tt all\_Edif}, which took a value 0.55 $\times$ 450~K  = 247.5~K. 

Models {\tt all\_Edif} and {\tt all\_Edif\_S17} therefore share rather similar behavior, although the abundances of solid phase CO and CO$_2$ are a little higher in the latter case, while formaldehyde is a little lower. Other solid-phase abundances in stage 1 also show small variations; methyl formate and glycolaldehyde are a little lower in model {\tt all\_Edif\_S17}. Variations between the two models are attributable to the longer lifetime against desorption of surface atomic hydrogen.

The abundances of COMs during stage 2 show minor variations; the peak gas-phase abundance of methyl formate and glycolaldehyde are somewhat lower here, following the solid-phase values. The N-bearing species display only minor differences.

We note that the major changes resulting from model {\tt all\_Edif}, described above, are equally major in model {\tt all\_Edif\_S17} when comparing with the earlier model {\tt all}. The arbitrary order in which we have applied the changes to the models should not be mistaken for a lack of importance of the findings of \citet{Sen17}.

\subsection{Adjustment to atomic O and N binding energies and diffusion barriers}

In model {\tt all\_Edif\_S17\_ON}, the desorption barrier of O is raised to 1660~K, while that of N is lowered to 650~K, as described in Sec.~\ref{sec:methods:bind}, under the assumption of a diffusion barrier set to a fraction 0.55 of the desorption energy. Comparing with the preceding model, {\tt all\_Edif\_S17}, the effect on all solid-phase and gas-phase oxygen-bearing species shown in Tables \ref{tab-st1-ice1}-\ref{tab-gas3} is very minor. The effect on some of the nitrogen-bearing species is rather stronger in some cases; solid-phase ammonia abundance is reduced in stage 1, from $\sim$14 to $\sim$11\% of water. Most of this reduction is accounted for by the diffusion-mediated reaction between N and O atoms on the grain/ice surfaces, producing nitric oxide, NO. The faster N diffusion, combined with the slower diffusion rate of H present in models {\tt all\_Edif} and {\tt all\_Edif\_S17}, makes this reaction more competitive. The grain/ice-surface NO radicals thus produced may further react with atomic N, again through the L-H (i.e.~diffusive) mechanism, forming nitrous oxide, N$_2$O. The N$_2$O produced on the surface is then incorporated into the bulk ice, taking a final abundance of $\sim$0.8\% with respect to solid water. Note that, while photodissociation mechanisms for nitrous oxide are included, the grain-surface chemical network for this species is otherwise quite sparse, thus it could plausibly lack important destruction mechanisms. However, the gas-phase destruction reactions of N$_2$O with partners such as H, O and OH involve substantial activation-energy barriers, so destruction on the grains by similar mechanisms could be limited in effect.

Also particularly affected by the change in O and N surface parameters are the peak stage-2 gas-phase abundances of vinyl and ethyl cyanide, C$_2$H$_3$CN and C$_2$H$_5$CN, both increasing by around an order of magnitude. Indeed, the abundances of both of these species are enhanced also during stage 1. The increased diffusivity of surface N allows it to react with C$_3$ accreted from the gas phase, producing C$_3$N. The latter may then react with atomic H to produce HC$_3$N. Reactions with atomic H still dominate the surface chemistry of C$_3$, although N may also react with the product C$_3$H, again forming HC$_3$N. This species may be further hydrogenated to vinyl and ethyl cyanide, with the latter being the main beneficiary of the overall process. The effect on other nitriles, either in stage 1  or 2, appears to be very small.

\subsection{Tunneling-mediated bulk diffusion of H and H$_2$}

As described in Sec.~\ref{sec:methods:tunn}, in model {\tt all\_Edif\_S17\_ON\_mtun} the bulk diffusion rates of H and H$_2$ are allowed to be no slower than some minimum value that represents tunneling between interstitial sites within the ice structure. The tunneling rates per individual site-to-site diffusion event are calculated to be $1.44$~s$^{-1}$ and $0.0334$~s$^{-1}$, for H and H$_2$ respectively. Atomic hydrogen reaches its limit at temperatures below $\sim$17~K, so bulk-ice tunneling of H is effective throughout stage 1. H$_2$ reaches its own limit at $\sim$9.4~K. These limiting rates are applied both to the rates of reactions mediated by bulk diffusion of H or H$_2$ and to the rates at which either species may diffuse from the bulk into the surface layer.

The effect of the change on the abundant, simple ice components during stage 1 is fairly minor; methanol and O$_2$, for example, exhibit small rises. The effect on solid-phase COMs during stage 1 is a little stronger, especially in the case of acetaldehyde, which falls by a factor $\sim$3, and of acetic acid and dimethyl ether, whose final solid-phase abundances are roughly halved. Methyl formate, glycolaldehyde and ethanol fall more modestly.

However, the effect on the abundance of atomic H itself is drastic. In model {\tt all\_Edif\_S17\_ON}, H comprises a maximum fraction of around 0.05 with respect to water; in model {\tt all\_Edif\_S17\_ON\_mtun} this value falls to $\sim$$1.5 \times 10^{-13}$. The fraction of H$_2$, meanwhile, remains small in both models (less than one H$_2$ molecule per grain), falling by a factor of $\sim$100 in the tunneling model. In either model, it should be noted that the abundance of H (in particular) is not indicative of its overall success or failure to react, only its {\em lifetime} against reaction. It is produced almost entirely by photodissociation of abundant species such as H$_2$O, and any individual hydrogen atom will typically react almost as soon as it is produced. Almost none of the atomic hydrogen produced in the bulk ice manages to escape without reacting, while the fractional loss rate of H$_2$ to the surface is around 2\%, with the rest being lost to reactions (which can in turn produce atomic H).

Barrier-mediated reactions occurring through nondiffusive mechanisms, for example the photodissociation-induced reaction H + CO $\rightarrow$ HCO (where the H may result from water photodissociation), are made slightly less efficient by the ability of the H atom to diffuse away prior to reaction. One result of this is a less effective production of CH$_3$CHO through the PDI reaction of CH$_3$ and HCO. Dimethyl ether and acetic acid, which also rely somewhat on nondiffusive reactions among radicals within the bulk ice, are affected for similar reasons.

The effects on complex organic molecules in stage 2 are qualitatively similar to those of stage 1, although dimethyl ether in particular is less affected, due to its effective gas-phase production mechanism. The chemistry of the grain/ice surfaces are essentially unchanged by the introduction of bulk tunneling.

\subsection{New gas-phase reactions} \label{sec:results:gas-phase}

Model {\tt all\_Edif\_S17\_ON\_mtun\_GP} includes the additional gas-phase reactions described in Sec.~\ref{sec:methods:gas-phase}, which correspond to the production of several important COMs, {\em viz.} methyl formate, dimethyl ether and formamide. The new reactions have no discernible effect on simple ice abundances, and the difference in solid-phase COM abundances during stage 1 is also negligible. Perhaps more surprising is that no significant difference in the gas-phase abundances of the species in question is apparent during stage 1. The main production mechanism for gas-phase DME is recombination of atomic H with the radical CH$_3$OCH$_2$ on grain surfaces, followed by immediate chemical desorption. The gas-phase radiative association reaction between radicals CH$_3$ and CH$_3$O never exceeds 1\% of overall production of DME during stage 1, except at the end of the run, when its abundance is very low (10$^{-14} n_{\mathrm{H}}$). A similar degree of importance is found for the gas-phase reaction of O with CH$_3$OCH$_2$ to produce methyl formate, and the main production route for gas-phase CH$_3$OCH$_2$ is generally found to be the reaction of CH$_2$ with CH$_3$O on grain surfaces, followed by chemical desorption. NH$_2$CHO reaches a stage-1 peak gas-phase abundance of less than 10$^{-11} n_{\mathrm{H}}$, to which the NH$_2$ + H$_2$CO reaction does not contribute substantially.

During stage 2, the production of DME shows a slightly larger change, but still on the $\sim$1\% level, and in the downward direction for both the {\em fast} and {\em slow} warm-up models. The maximum absolute gain is around $2 \times 10^{-9} n_{\mathrm{H}}$. The increase in peak methyl formate abundance is much more substantial; in the {\em fast} model, the gain is around $2 \times 10^{-8} n_{\mathrm{H}}$ over the {\tt all\_Edif\_S17\_ON\_mtun} model, although this corresponds to only 20\% of the total, which is otherwise produced on the grains. The longer warm-up timescales show slightly smaller absolute and relative increases. The additional production of MF comes through the O + CH$_3$OCH$_2$ reaction, during the hot stage when ice mantles have desorbed; however, the CH$_3$OCH$_2$ itself is formed almost entirely through the reaction of ionized dimethyl ether, CH$_3$OCH$_3^+$, with ammonia -- a reaction that is only postulated but has not, to our knowledge, been studied experimentally. This reaction, along with various others, was added to the network as described in Sec.~\ref{sec:methods:NH3}, but involves proton transfer not from a protonated stable molecule but from an ionized stable molecule (which in the models is formed via CR-induced UV photoionization of DME). Several other such processes have been studied in the laboratory and indeed have long been included in the chemical networks, e.g.~CH$_4^+$ + NH$_3$ $\rightarrow$ CH$_3$ + NH$_4^+$, for which a competitive charge-transfer reaction route also exists \citep{Smith77}. In the case of ionized DME, the efficiency of the process as adopted in our chemical network could plausibly be lower by a factor of a few, which would make MF production through the gas-phase mechanism less important.

The increase in peak gas-phase NH$_2$CHO abundance in stage 2 is around 1 -- $2 \times 10^{-9} n_{\mathrm{H}}$, corresponding to roughly 10\% of the total, with the remainder coming directly from the grains. The reaction NH$_2$ + H$_2$CO $\rightarrow$ NH$_2$CHO + H is indeed responsible for the greater gas-phase production. This mechanism is aided by the greater availability of H$_2$CO in the gas phase caused by the removal from the model of bulk diffusion for species other than H and H$_2$, resulting in the trapping of H$_2$CO in the ice until high temperatures are achieved, rather than desorbing from the grains at around 40~K. The NH$_2$ radical is formed through the CR-induced UV photodissociation of ammonia and CH$_3$NH$_2$ molecules released from the grains at high temperatures. Although the production of formamide through the gas-phase mechanism is overpowered by production/release from the ices, in the absence of the latter, the gas-phase process alone could produce substantial quantities of NH$_2$CHO under the conditions modeled here (in which both reactants are simultaneously produced in significant amounts).

\section{Results of the {\tt final} model} \label{sec:results:final}

Model {\tt all\_Edif\_S17\_ON\_mtun\_GP\_Gr}, which is also described more succinctly as the ``{\tt final}'' model setup, includes additional grain-surface/ice reactions and treatments as per Sec.~\ref{sec:methods:grain}, and constitutes the culmination of all of the updates to the models applied, individually or collectively, in this paper. 
The additional features of the grain-surface/ice chemistry applied uniquely in the {\tt final} model may be summarized as (i) the completion of the glycolaldehyde network; (ii) the addition of methylene and methylidyne reactions; (iii) the addition of grain-surface/ice reactions for NH$_2$CHO and HCOOCH$_3$, corresponding to the same reactions in the gas phase; and (iv) the adjustment of the product branching ratio of the CO + OH reaction.

\subsection{Simple molecules}\label{sec:results:simple}

Table \ref{tab-st1-ice1}, as before, shows the final stage-1 abundances of simple solid-phase species on the grains, as a fraction of solid water. The largest obvious change compared with the preceding model ({\tt all\_Edif\_S17\_ON\_mtun\_GP}) is the abundance of formic acid (HCOOH) in the ice, which is now below 0.1\%; this is caused by the change to the branching ratio of the CO + OH reaction, as described in Sec.~\ref{methods:GR}, which is adjusted to favor CO$_2$ production (versus COOH), with a ratio 99:1 (previously 50:50). Although the solid-phase HCOOH fractional abundance of 1--2\% (with respect to water) in the preceding models would not apparently violate existing observational upper limits \citep{Boogert15}, the preservation of such a large quantity of solid-phase formic acid through to the hot stage produces excessive gas-phase abundances as compared with observations. The branching ratio adjustment brings the peak gas-phase values to within a factor of two of observational values (see Sec.~\ref{results:obs} and Table \ref{tab:obs}). However, it has only a modest influence on the production rate of surface CO$_2$, and the {\tt final} model shows no substantial difference in solid-phase CO$_2$ abundance versus the preceding model.

Ammonia takes a slightly higher abundance in the ice than in model {\tt all\_Edif\_S17\_ON\_mtun\_GP} (17\% versus 10\%), which is nevertheless in line with some earlier models. The increase is the result of the preferential production of CH$_2$ on the grain surfaces, through the barrierless reaction of atomic C with H$_2$, rather than forming CH through the reaction of C and H. The removal of the barrier to reaction with the much more abundant H$_2$ allows CH production to be skipped over. This strongly reduces the availablility of CH, and therefore minimizes the barrierless reaction of CH with NH$_3$ \citep[][and also Table \ref{CH-rxns}]{Blitz12}, which is otherwise a major destruction mechanism for ammonia (as well as contributing to glycine production).

\begin{figure}
\center
         {\includegraphics[width=0.5\hsize]{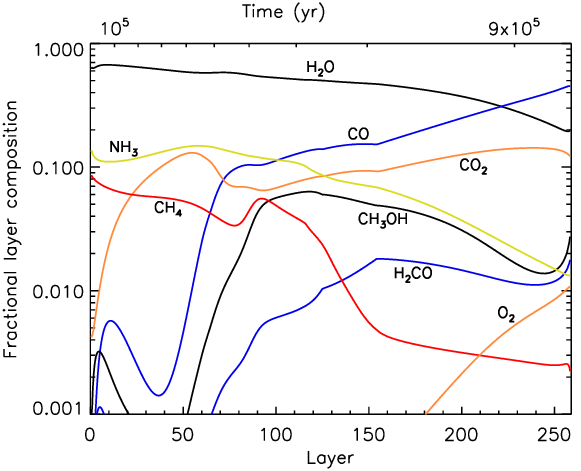}}
\caption{Layer-by-layer ice composition produced during stage 1 (collapse) of the {\tt final} model setup. All abundances are given as a fraction of the total ice composition (by number), and are local to each layer (rather than cumulative over the total ice thickness formed at a given time). Abundances shown for any particular layer or time in the model (top axis) are those recorded at the moment that that material is incorporated into the bulk ice. Material in the bulk continues to be processed after its incorporation. 
}
\label{final_layers}
\end{figure}

Fig.~\ref{final_layers} shows the composition of the ice as a function of layer depth. Water is seen to be the main component of each ice layer, until late times when CO takes over, reaching nearly 50\% of the total ice in the upper layers (right-hand side of Fig.~\ref{final_layers}). Much of the growth in CO fraction with respect to water in the bulk ice occurs at this late stage. As in past models \citep[see][]{GP11}, CO$_2$ dominates over CO at early times, when the dust temperatures are higher and CO and OH may react on grain/ice surfaces through the Langmuir-Hinshelwood (diffusive) mechanism. When the temperature falls to around 12~K, CO is no longer sufficiently mobile to compete for OH purely through diffusive means, and the 3-B mechanism takes over, at which point the local CO fraction begins to dominate over CO$_2$. Also of note is the growth of O$_2$ in the ice at late times, rising to around 1\% of the total; related species (not shown) such as H$_2$O$_2$, O$_2$H and O$_3$ become similarly abundant, collectively reaching fractions with respect to total water of more than 1\% as averaged over the entire ice mantle. 

Of interest also is the abundance of radicals and atoms within the ice mantles (not shown in Fig.~\ref{final_layers}) during the collapse stage. The OH radical, which is capable of easily abstracting hydrogen from various stable species, reaches around 0.02\% of the ice composition by number, at the end of the stage-1 run. NH$_2$, which is also responsible for H-abstraction from various species, reaches a value around half that of OH. The methanol-related radicals CH$_3$O and CH$_2$OH together reach close to 0.8\% by number. Oxygen atoms attain a final value of 0.4\% of the ice composition. Atomic carbon, due to its ability to react rapidly with H$_2$, which in this model is mobile within the bulk ice, maintains a minimal abundance, on the order of one atom in total throughout the ice mantle.


\begin{figure*}
\center
         {\includegraphics[width=0.48\hsize]{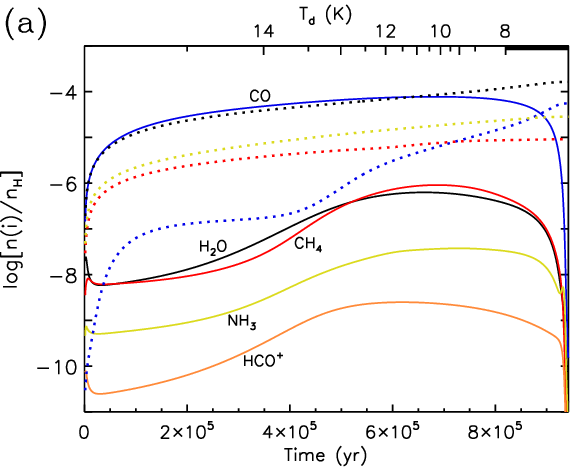}}
         {\includegraphics[width=0.48\hsize]{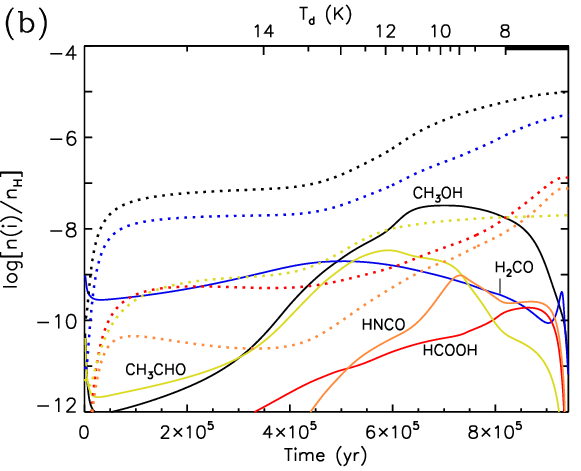}}
         {\includegraphics[width=0.48\hsize]{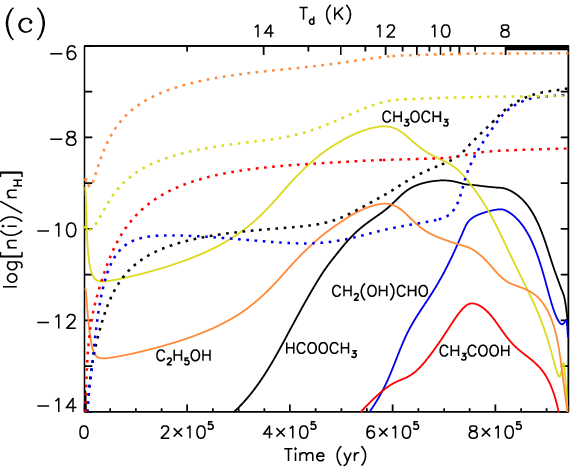}}
         {\includegraphics[width=0.48\hsize]{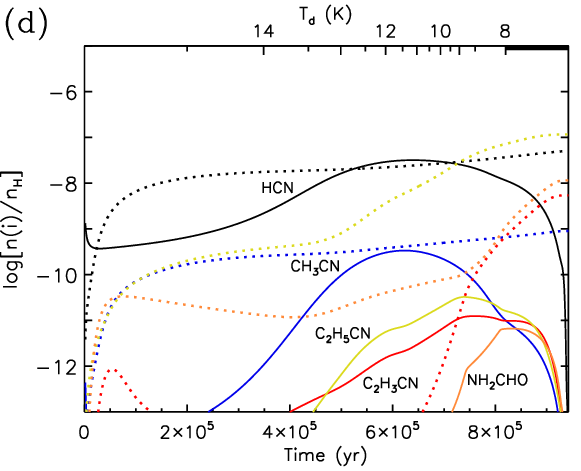}}
\caption{Description as per Fig.~\ref{no-md_collapse}. Results correspond to stage 1 (collapse) of the model {\tt final}.}
\label{final_collapse}
\end{figure*}

\begin{figure*}
\center
         {\includegraphics[width=0.48\hsize]{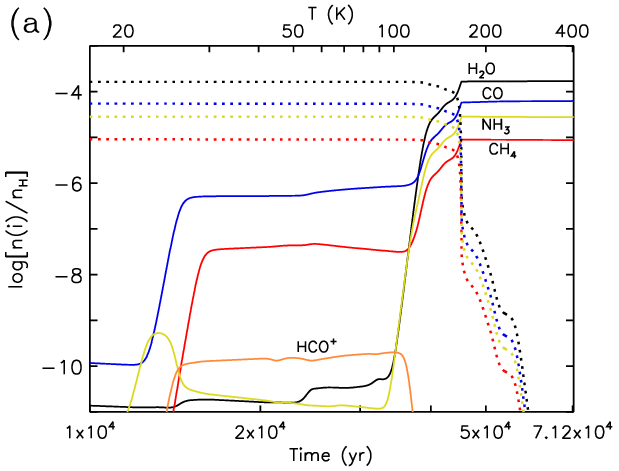}}
         {\includegraphics[width=0.48\hsize]{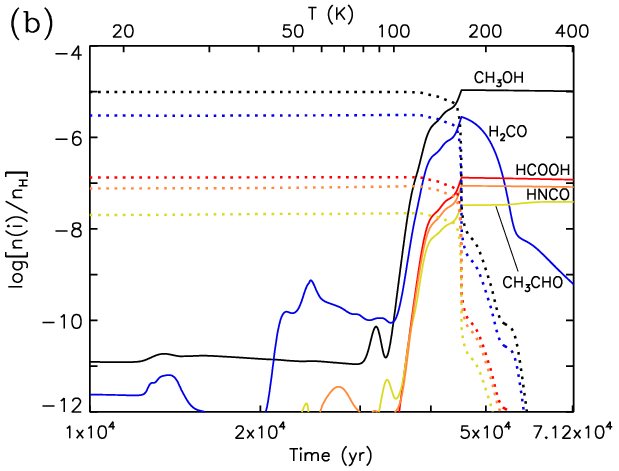}}
         {\includegraphics[width=0.48\hsize]{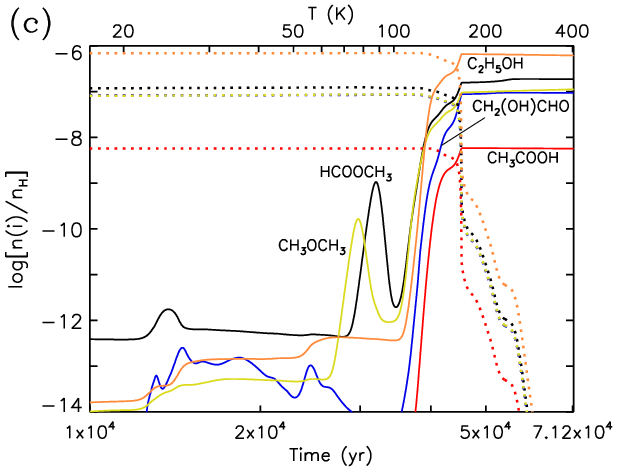}}
         {\includegraphics[width=0.48\hsize]{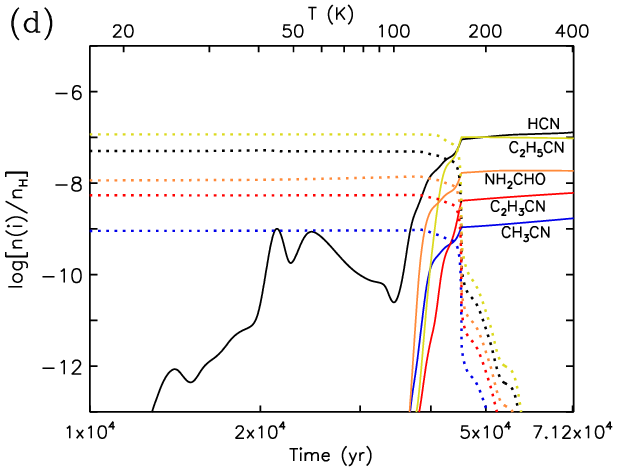}}
\caption{Description as per Fig.~\ref{basic_medium}. Results correspond to model {\tt final} stage 2, using the {\em fast} warm-up timescale.}
\label{final_fast}
\end{figure*}

\begin{figure*}
\center
         {\includegraphics[width=0.48\hsize]{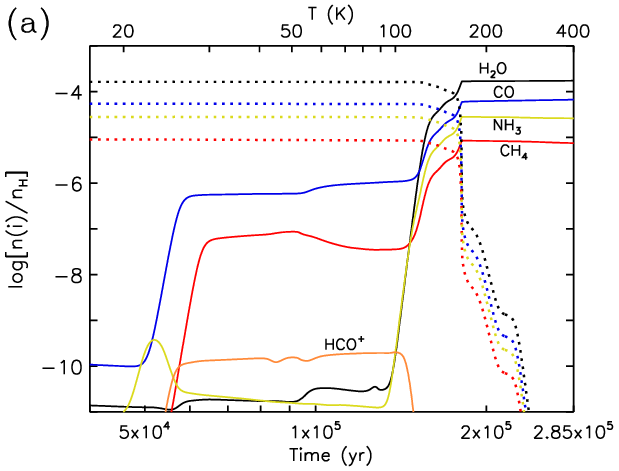}}
         {\includegraphics[width=0.48\hsize]{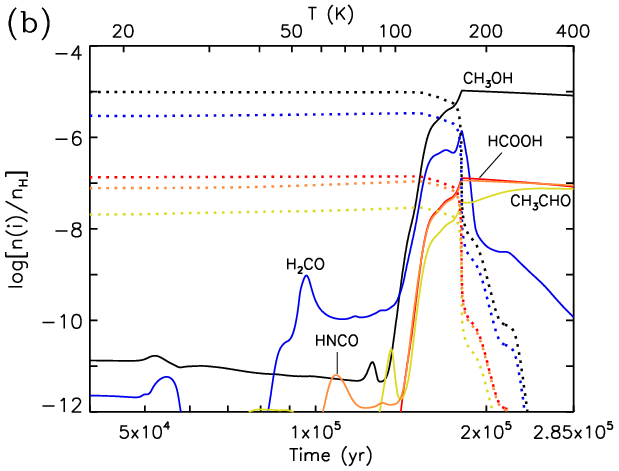}}
         {\includegraphics[width=0.48\hsize]{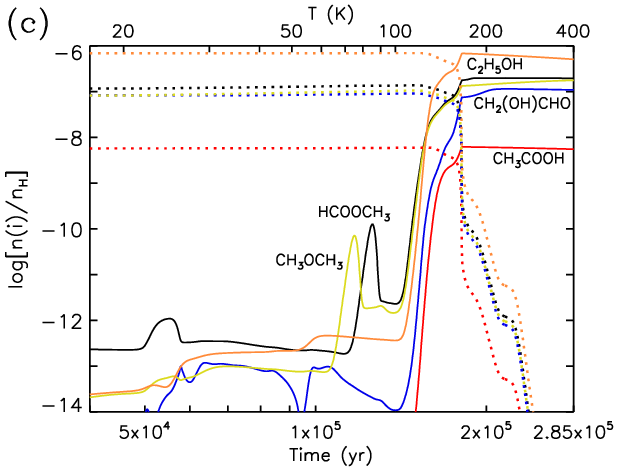}}
         {\includegraphics[width=0.48\hsize]{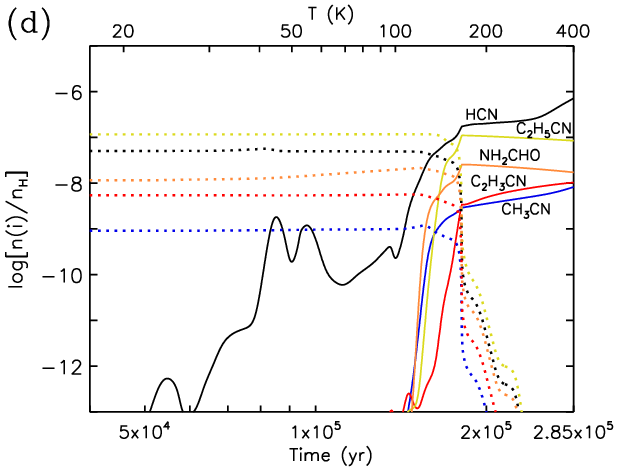}}
\caption{Description as per Fig.~\ref{basic_medium}. Results correspond to model {\tt final} stage 2, using the {\em medium} warm-up timescale.}
\label{final_medium}
\end{figure*}

\begin{figure*}
\center
         {\includegraphics[width=0.48\hsize]{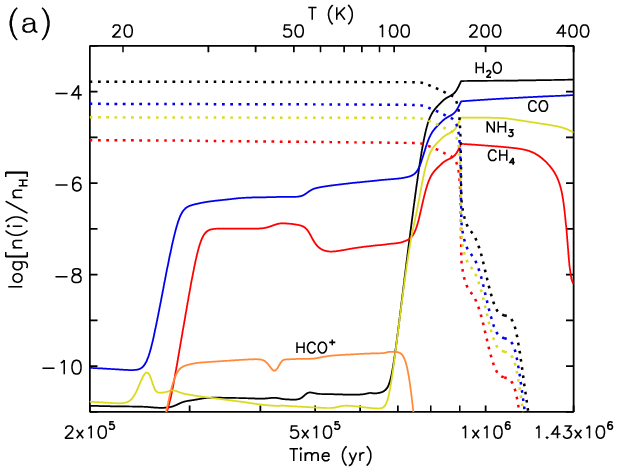}}
         {\includegraphics[width=0.48\hsize]{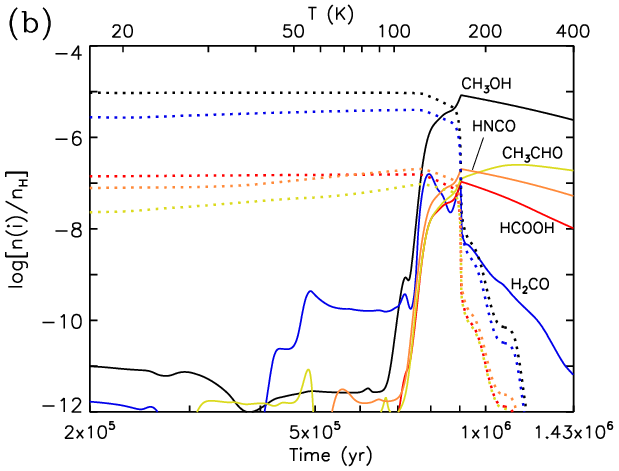}}
         {\includegraphics[width=0.48\hsize]{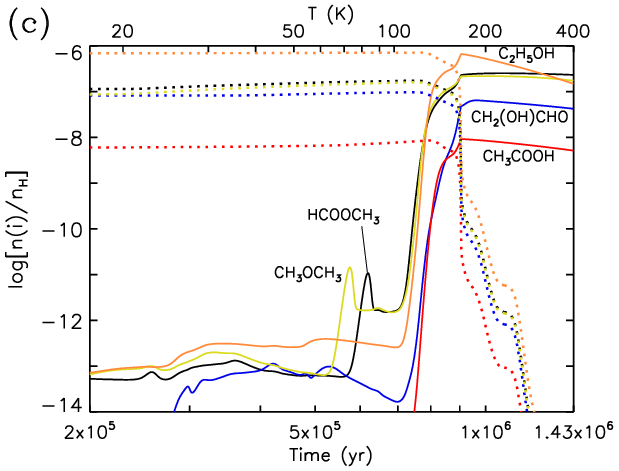}}
         {\includegraphics[width=0.48\hsize]{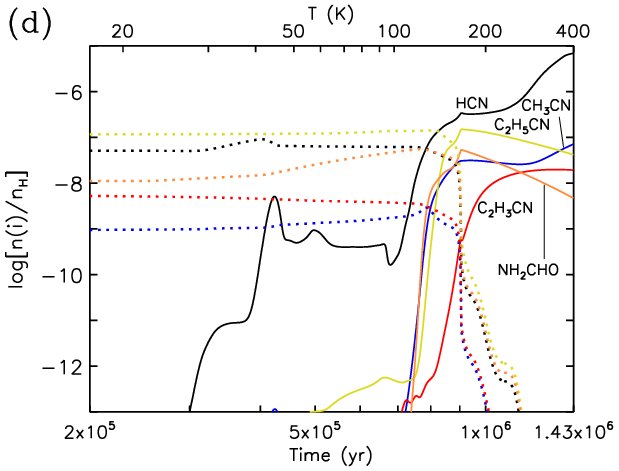}}
\caption{Description as per Fig.~\ref{basic_medium}. Results correspond to model {\tt final} stage 2, using the {\em slow} warm-up timescale.}
\label{final_slow}
\end{figure*}


\subsection{Complex molecules during the cold collapse}

Fig.~\ref{final_collapse} shows the behavior of a selection of simple and complex molecules during stage-1 (collapse) of the {\tt final} model, with gas-phase species indicated by solid lines, and ice species indicated by dotted lines; all species shown are plotted with respect to total hydrogen in the model. While the abundances of most solid-phase species continue to grow over time, the gas-phase abundances of many fall off toward the end time as the gas density increases and freeze-out becomes more extreme.

Through the course of the stage-1 evolution, COMs such as acetaldehyde (CH$_3$CHO), dimethyl ether (CH$_3$OCH$_3$) and methyl formate (HCOOCH$_3$) reach substantial gas-phase abundances, achieving their peak values at times prior to the end of stage 1; those peak values are compatible with the detected abundances toward prestellar sources \citep[e.g.][]{JS16}. It is not the purpose of the present paper to pursue detailed comparison with such sources, which was done at length by JG20 (who showed that modeling the column densities of the molecules is more meaningful than a crude comparison of their fractional abundances with respect to total H). However, we note that the models of JG20 had trouble producing sufficient dimethyl ether in the gas phase. Here, DME abundances exceed even those of methyl formate, as the direct result of the inclusion of methylene (CH$_2$) reactions with surface methanol. As methylene is assumed to be in its ground (triplet) state, a barrier is present for this reaction. H-abstraction from either the methyl or hydroxyl group is followed (in 50\% of cases) by immediate recombination of the resulting radical with the newly-formed CH$_3$ to produce either dimethyl ether or ethanol (see Sec.~\ref{sec:CH2-rxns}). This mechanism, under the influence also of the other changes to methylene and methylidyne chemistry in the network, enhances solid-phase DME and ethanol abundances by more than a factor of 2. Chemical desorption allows both to enter the gas phase from the grain/ice surface, with DME taking a higher rate of desorption due to its weaker surface binding \citep[3675~K vs. 6260~K;][]{Garrod13a}. The RRK treatment for chemical desorption proposed by \citet{Garrod07}, which is used in this model, provides a desorption efficiency of 0.066\% for the H + C$_2$H$_4$OH$ \rightarrow$ C$_2$H$_5$OH recombination reaction, versus 0.20\% for H + CH$_3$OCH$_2$ $\rightarrow$ CH$_3$OCH$_3$. As noted by JG20, chemical desorption is further amplified by repetitive abstraction and recombination of atomic H with surface COMs. Although further tuning of the DME-related chemistry may be necessary, the reaction of methylene with methanol on grains appears ample to reproduce gas-phase DME abundances under cold conditions.

Table \ref{tab-st1-ice2} shows that the solid-phase abundances of several other COM species are raised by around a factor of 2 in the collapse stage of the {\tt final} model (versus the preceding model). Acetaldehyde is enhanced by the reaction of methylene with formaldehyde (H$_2$CO) shown in Table \ref{CH2-rxns}. Formic acid falls by a little more than an order of magnitude versus the preceding model, as explained above (Sec.~\ref{sec:results:simple}). The abundance of acetic acid (CH$_3$COOH) goes down with it, because its main surface formation mechanism was previously the addition of CH$_3$ (or CH$_2$) to the COOH radical. Acetic acid formation in the {\tt final} model is dominated instead by addition of the OH and CH$_3$CO radicals; the latter is formed by H-abstraction from acetaldehyde.

Much of the growth in glycolaldehyde and methyl formate on the grains is seen to occur in the late/cold stages of collapse (Fig.~\ref{final_collapse}, panel c); the additional material survives mostly intact up until the ice mantles desorb at higher temperatures. Acetic acid is in fact more abundant on the grains than either of these species at the earliest times; they instead begin to dominate on the grains when temperatures fall to $\sim$10~K or lower, around the time when CO begins to grow substantially in its relative abundance in the ice (Fig. \ref{final_layers}). A number of N-bearing species also become abundant in the ice toward the end of the collapse stage, such as ethyl cyanide (C$_2$H$_5$CN), vinyl cyanide (C$_2$H$_3$CN) and formamide (NH$_2$CHO). The latter in particular shows a peak gas-phase abundance that is around twice as large as in model {\tt all\_Edif\_S17\_ON\_mtun\_GP}, which did not include the NH$_2$ + H$_2$CO in the grain-surface/ice network. The dominant production mechanisms for important COMs are discussed in more detail in Sec.~\ref{results:temps}.

\subsection{Complex molecules during the warm-up}\label{COM-warm-up}

Tables \ref{tab-gas1}--\ref{tab-gas6} again show the peak gas-phase abundances of selected oxygen- and nitrogen-bearing COMs during each of the {\em fast}, {\em medium} and {\em slow} warm-up stages. Following past papers \citep[e.g.][]{Garrod13a}, the peak gas-phase abundances produced by the {\tt final} model for a large selection of COMs and other molecules that may be detected in hot cores are shown in Table \ref{tab:abuns}; the corresponding system temperature is shown alongside each peak fractional abundance value.

Figs. \ref{final_fast}--\ref{final_slow} show the time-dependent behavior of the abundances of selected molecules during the warm-up stages. Again, as this model includes bulk diffusion only for H and H$_2$, the majority of solid phase material is trapped by the mostly-water ice until temperatures exceed around 120~K. However, the loss of the ice mantles is only completed at temperatures of around 160~K or more; the range of temperatures over which significant desorption occurs is therefore extended beyond that of the earlier models. This effect applies even to water, which is the most abundant constituent of the ices under cold conditions. It is due to the presence in the ice of substantial amounts of large molecules, such as hydroxylamine (NH$_2$OH) and hydrocarbons including C$_3$H$_8$ and C$_4$H$_{10}$, whose binding energies are yet higher than that of water. Sufficient quantities of these molecules are present (collectively up to around 10 monolayers) that even water may be trapped at temperatures beyond its nominal sublimation temperature. The increase in production of these molecules is due to the inclusion of nondiffusive production mechanisms, acting at various stages prior to desorption. As a result of the trapping, some species do not reach their gas-phase peak abundances until around the 160~K mark or higher, although much of the grain-surface material may have already desorbed prior to the achievement of that ultimate peak value.

Around 1\% of solid-phase CO and CH$_4$ are released at low temperatures; when these molecules desorb from the surface layer, they are replaced by material from the ice mantle beneath, some of which is also CO, CH$_4$ or other volatile molecules. The volatiles continue to desorb and be replaced by a mixture of volatile and non-volatile material from the bulk, until the surface layer of the ice is composed only of molecules that cannot desorb until higher temperatures are reached. The remaining CO and CH$_4$ are thus trapped in the bulk at this stage.

Past models have shown a substantial release of formaldehyde (H$_2$CO) at temperatures around 40~K; this effect too is reduced, although it is still present due to losses from the upper layers. Most formaldehyde is released with the other ices at higher temperatures, although it is rapidly destroyed in the gas phase via ion-molecule reactions.

Most COMs also present their main abundance peaks at high temperatures, when water, methanol, and other strongly-bound ice components thermally desorb. Methyl formate and dimethyl ether, due to their lower binding energies (lower than water) show modest early peaks as the result of the desorption of molecules originating in the upper ice layers. This effect is most pronounced in the {\em fast} warm-up model.

Much of the COM material that is released into the gas phase from the grains appears to be present already at the beginning of the warm-up stage; the dotted lines in the figures, which show the solid-phase abundances, appear quite flat until the ice mantles desorb. This is in strong contrast to past models \citep[e.g.][]{GH06, GWH08, Garrod13a} in which the production of COMs on grains peaked sharply at temperatures above $\sim$20~K, due to the increased mobility of radicals on the surfaces or within the bulk ices, 
with different radicals becoming mobile at different temperatures. The removal of bulk diffusion as a feature of the model, along with the addition of new nondiffusive mechanisms, breaks this strong temperature dependence. Reactions between radicals in the bulk ice and on the ice surface continue to drive the production of COMs. However, this radical chemistry is now distributed through several much broader temperature regimes: (i) at low temperatures and early times, mainly through three-body surface reactions; (ii) over a wide range of intermediate temperatures, via photodissociation-induced reactions in the ice mantles; and (iii) at high temperatures, as the thermal desorption of solid-phase molecules into the gas phase reveals radicals that were locked into the ice and are now available for rapid diffusive reactions on the grain/ice surfaces. The latter process, in general, has only a small effect compared especially to (i).

The process of gradual photodissociation-induced COM production, labeled (ii) above, is more evident in the longer-timescale models. In Fig.~\ref{final_slow}, panels (b) and (d), the solid-phase abundances especially of CH$_3$CHO, HNCO and NH$_2$CHO are seen to rise over long periods. Sec.~\ref{results:temps} discusses these mechanisms in more detail.

The {\em slow} warm-up model also exhibits a clearer distinction between the peak gas-phase abundance of methyl formate and glycolaldehyde, with the former rising in abundance between its pre- and post-desorption values, and the latter falling, exacerbating the pre-existing solid-phase disparity. As discussed in Sec.~\ref{sec:results:PDI}, glycolaldehyde abundance suffers as it has a higher binding energy than water, so that it remains on the grain surfaces to be attacked by the emerging, formerly-trapped radicals. Methyl formate has a lower binding energy than water, so that as soon as it emerges on the ice surface as material above it is desorbed, it too may immediately leave the surface, reducing its exposure to destructive radicals.

It is notable that in the {\tt final} models, the relative ratios of the peak abundances of structural isomers methyl formate, glycolaldehyde and acetic acid are in the correct order, and are close to the appropriate proportion. Past models based on grain-surface production of these molecules have failed to reproduce the observational ratios. Aside from the behavior of glycolaldehyde and methyl formate described above, the {\tt final} models also show success in achieving lower abundances of acetic acid, caused by the weaker production of formic acid in the ices at the earliest stages of the models. Sec.~\ref{results:obs} provides a more detailed comparison with observations.

As in past models, the production of certain N-bearing species in particular has a stronger time/temperature dependence during/after the desorption of the ice mantles. HCN, CH$_3$CN and C$_2$H$_3$CN each grow as model temperatures increase. For HCN, although there is an intermediate-temperature plateau value around $10^{-10}$--$10^{-9} n_{\mathrm{H}}$, and another rise following mantle desorption, gas-phase mechanisms become strongly active around 300~K, so that HCN abundance peaks at the end of the model ($\sim$400~K). Important mechanisms are discussed in Sec.~\ref{results:temps}.


\startlongtable
\begin{deluxetable}{lclcrclcrclcr}\label{tab:abuns}
\tabletypesize{\scriptsize}
\tablecaption{\label{tab-output} Peak gas-phase fractional abundances and corresponding temperatures of a selection of molecules for each warm-up timescale, using the {\tt final} model setup.}
\tablewidth{0pt}
\tablehead{ &\\
 && \multicolumn{3}{c}{Fast} && \multicolumn{3}{c}{Medium} && \multicolumn{3}{c}{Slow} \\
\cline{3-5} \cline{7-9} \cline{11-13}
\colhead{Molecule} && \colhead{$n[i]/n_{\mathrm{H}}$} && \colhead{T (K)} && \colhead{$n[i]/n_{\mathrm{H}}$} && \colhead{T (K)} && \colhead{$n[i]/n_{\mathrm{H}}$} && \colhead{T (K)}
}
\startdata
H$_2$O && 1.7(-4) && 396 && 1.7(-4) && 395 && 1.8(-4) && 400  \\
CO && 6.2(-5) && 400 && 6.7(-5) && 398 && 8.4(-5) && 400  \\
CO$_2$ && 3.4(-5) && 169 && 3.4(-5) && 167 && 3.5(-5) && 166  \\
CH$_4$ && 8.9(-6) && 169 && 8.5(-6) && 167 && 7.2(-6) && 166  \\
HCO$^+$ && 2.0(-10) && 98 && 2.0(-10) && 100 && 2.1(-10) && 97  \\
H$_2$CO && 2.8(-6) && 166 && 1.4(-6) && 165 && 1.6(-7) && 129  \\
CH$_3$OH && 1.1(-5) && 166 && 1.0(-5) && 166 && 8.3(-6) && 166  \\
NH$_3$ && 2.8(-5) && 169 && 2.8(-5) && 167 && 2.7(-5) && 166  \\
H$_2$O$_2$ && 4.1(-7) && 166 && 3.7(-7) && 165 && 7.2(-8) && 165  \\
OH && 3.4(-10) && 165 && 1.1(-10) && 55 && 1.0(-10) && 110  \\
O$_2$ && 5.6(-7) && 166 && 4.9(-7) && 165 && 9.1(-8) && 165  \\
O$_3$ && 6.9(-7) && 171 && 4.3(-7) && 167 && 4.7(-8) && 166  \\
NO && 9.7(-7) && 167 && 8.1(-7) && 165 && 3.7(-7) && 143  \\
HNO && 2.3(-9) && 165 && 2.0(-10) && 165 && 2.8(-11) && 56  \\
HNCO && 8.7(-8) && 167 && 1.1(-7) && 166 && 2.0(-7) && 166  \\
OCN && 7.1(-10) && 165 && 3.2(-10) && 127 && 1.0(-9) && 337  \\
SO && 4.2(-8) && 295 && 3.2(-8) && 234 && 1.2(-8) && 166  \\
SO$_2$ && 1.5(-8) && 397 && 1.7(-8) && 206 && 1.2(-8) && 171  \\
HCS && 1.5(-10) && 220 && 1.2(-10) && 173 && 7.2(-11) && 135  \\
H$_2$CS && 1.8(-8) && 397 && 3.8(-8) && 398 && 7.7(-8) && 377  \\
CH$_3$SH && 3.1(-9) && 166 && 3.1(-9) && 166 && 3.7(-9) && 166  \\
C$_2$H$_5$SH && 1.8(-10) && 166 && 2.0(-10) && 166 && 2.5(-10) && 166  \\
C$_2$H && 2.9(-11) && 91 && 2.0(-10) && 93 && 1.9(-10) && 89  \\
C$_2$H$_2$ && 2.5(-8) && 400 && 2.4(-8) && 188 && 3.4(-8) && 367  \\
C$_2$H$_4$ && 6.5(-8) && 400 && 5.0(-8) && 193 && 4.8(-8) && 166  \\
C$_2$H$_6$ && 4.4(-7) && 168 && 4.8(-7) && 166 && 5.7(-7) && 166  \\
C$_3$H$_6$ && 1.1(-7) && 400 && 2.5(-7) && 398 && 4.9(-7) && 368  \\
C$_3$H$_8$ && 1.9(-6) && 169 && 1.9(-6) && 167 && 1.9(-6) && 167  \\
CN && 8.6(-11) && 43 && 1.7(-10) && 43 && 3.9(-10) && 42  \\
HCN && 1.3(-7) && 400 && 7.1(-7) && 398 && 6.9(-6) && 400  \\
HNC && 1.9(-9) && 127 && 4.1(-9) && 118 && 3.7(-9) && 117  \\
H$_2$CN && 6.4(-12) && 29 && 3.4(-12) && 29 && 7.4(-12) && 28  \\
CH$_3$CN && 1.7(-9) && 400 && 8.2(-9) && 398 && 7.0(-8) && 400  \\
CH$_3$NC && 3.0(-12) && 140 && 1.9(-12) && 134 && 7.5(-13) && 134  \\
HC$_3$N && 2.0(-8) && 400 && 1.7(-7) && 398 && 6.7(-7) && 400  \\
C$_2$H$_3$CN && 6.1(-9) && 400 && 1.0(-8) && 398 && 2.0(-8) && 343  \\
C$_2$H$_5$CN && 1.0(-7) && 168 && 1.1(-7) && 166 && 1.5(-7) && 166  \\
n-C$_3$H$_7$CN && 2.7(-10) && 190 && 9.1(-10) && 197 && 2.6(-9) && 214  \\
i-C$_3$H$_7$CN && 2.1(-10) && 195 && 3.9(-10) && 205 && 5.4(-10) && 200  \\
n-C$_4$H$_9$CN && 4.3(-9) && 198 && 5.5(-9) && 208 && 2.8(-8) && 210  \\
i-C$_4$H$_9$CN && 4.4(-10) && 211 && 3.5(-9) && 217 && 3.1(-8) && 209  \\
s-C$_4$H$_9$CN && 2.5(-8) && 196 && 2.1(-8) && 191 && 4.4(-9) && 185  \\
t-C$_4$H$_9$CN && 7.8(-11) && 199 && 7.9(-11) && 193 && 9.4(-11) && 186  \\
CH$_2$CO && 8.1(-9) && 348 && 2.9(-8) && 235 && 1.2(-7) && 166  \\
CH$_3$CHO && 3.9(-8) && 379 && 7.6(-8) && 305 && 2.5(-7) && 257  \\
c-C$_2$H$_4$O && 2.9(-9) && 195 && 3.7(-9) && 170 && 7.3(-9) && 168  \\
C$_2$H$_3$OH && 2.8(-10) && 396 && 4.4(-10) && 351 && 1.3(-9) && 166  \\
C$_2$H$_5$CHO && 3.6(-9) && 169 && 3.6(-9) && 166 && 4.2(-9) && 166  \\
CH$_2$NH && 4.4(-8) && 400 && 5.3(-8) && 246 && 5.1(-8) && 157  \\
CH$_3$NH$_2$ && 3.7(-7) && 166 && 3.4(-7) && 165 && 2.6(-7) && 166  \\
NH$_2$CHO && 1.9(-8) && 255 && 2.5(-8) && 168 && 5.3(-8) && 166  \\
NH$_2$CN && 4.1(-10) && 166 && 4.0(-10) && 166 && 5.0(-10) && 166  \\
NH$_2$OH && 3.9(-6) && 174 && 3.6(-6) && 168 && 2.9(-6) && 166  \\
CH$_3$NCO && 1.1(-11) && 166 && 1.2(-11) && 166 && 3.6(-11) && 166  \\
CH$_3$NHCHO && 2.6(-10) && 175 && 2.9(-10) && 170 && 2.5(-10) && 166  \\
C$_2$H$_5$OH && 6.6(-7) && 169 && 6.8(-7) && 167 && 6.6(-7) && 166  \\
CH$_3$OCH$_3$ && 1.1(-7) && 400 && 1.8(-7) && 398 && 2.2(-7) && 200  \\
HCOOCH$_3$ && 1.9(-7) && 261 && 2.0(-7) && 216 && 2.5(-7) && 225  \\
HCOOC$_2$H$_5$ && 2.0(-10) && 169 && 2.4(-10) && 167 && 3.9(-10) && 166  \\
HCOOH && 1.3(-7) && 166 && 1.3(-7) && 166 && 1.1(-7) && 166  \\
CH$_3$COOH && 5.8(-9) && 172 && 6.2(-9) && 169 && 9.2(-9) && 166  \\
C$_2$H$_5$COOH && 4.0(-10) && 178 && 4.9(-10) && 173 && 4.3(-10) && 168  \\
CH$_3$COCHO && 4.1(-10) && 171 && 5.2(-10) && 168 && 1.0(-9) && 166  \\
(CH$_3$)$_2$CO && 6.4(-9) && 172 && 7.0(-9) && 169 && 1.4(-8) && 166  \\
CH$_3$C(O)NH$_2$ && 3.6(-9) && 166 && 3.5(-9) && 165 && 4.2(-9) && 166  \\
CH$_3$OCOCH$_3$ && 1.1(-9) && 169 && 1.2(-9) && 166 && 1.6(-9) && 166  \\
CH$_2$(OH)COCH$_3$ && 1.0(-9) && 181 && 1.0(-9) && 177 && 1.1(-9) && 169  \\
HOCOOH && 8.0(-9) && 195 && 8.1(-9) && 191 && 8.4(-9) && 184  \\
HCOCOCHO && 8.4(-9) && 173 && 8.8(-9) && 170 && 1.0(-8) && 166  \\
OHCCHO && 1.1(-7) && 369 && 1.1(-7) && 181 && 1.3(-7) && 167  \\
OHCCOOH && 2.2(-8) && 168 && 2.2(-8) && 168 && 2.3(-8) && 166  \\
NH$_2$COCHO && 1.7(-8) && 168 && 1.7(-8) && 166 && 1.6(-8) && 166  \\
NH$_2$COOH && 3.1(-8) && 213 && 3.1(-8) && 207 && 3.0(-8) && 200  \\
NH$_2$C(O)NH$_2$ && 8.7(-9) && 209 && 8.2(-9) && 203 && 9.0(-9) && 196  \\
NH$_2$CH$_2$CHO && 3.8(-10) && 167 && 4.0(-10) && 165 && 4.7(-10) && 166  \\
NH$_2$CH$_2$CN && 9.4(-11) && 167 && 1.0(-10) && 166 && 1.2(-10) && 166  \\
NH$_2$CH$_2$COOH && 3.1(-10) && 236 && 3.5(-10) && 228 && 2.2(-10) && 221  \\
NH$_2$C$_2$H$_5$ && 1.5(-8) && 187 && 2.1(-8) && 180 && 1.8(-8) && 173  \\
(NH$_2$)$_2$ && 9.0(-11) && 165 && 1.8(-11) && 165 && 3.2(-12) && 165  \\
CH$_3$ONH$_2$ && 2.4(-7) && 172 && 2.5(-7) && 173 && 2.6(-7) && 166  \\
CH$_2$(OH)NH$_2$ && 2.2(-7) && 212 && 2.1(-7) && 207 && 1.8(-7) && 200  \\
CH$_3$OCOCHO && 1.4(-8) && 169 && 1.4(-8) && 168 && 1.3(-8) && 166  \\
CH$_3$OCOOH && 2.7(-8) && 185 && 2.7(-8) && 180 && 2.7(-8) && 174  \\
CH$_3$OCONH$_2$ && 2.4(-8) && 178 && 2.3(-8) && 172 && 1.9(-8) && 166  \\
(CH$_3$O)$_2$CO && 7.2(-9) && 173 && 6.9(-9) && 169 && 6.1(-9) && 166  \\
CH$_3$OCOCH$_2$OH && 4.5(-9) && 208 && 1.9(-9) && 196 && 1.1(-10) && 187  \\
CH$_3$OOH && 1.6(-7) && 170 && 1.6(-7) && 168 && 1.6(-7) && 166  \\
(CH$_3$O)$_2$ && 2.3(-7) && 166 && 2.1(-7) && 166 && 1.5(-7) && 166  \\
CH$_3$OCH$_2$OH && 1.4(-7) && 188 && 1.0(-7) && 180 && 6.4(-8) && 174  \\
(CH$_2$OH)$_2$ && 6.0(-8) && 244 && 1.1(-8) && 217 && 2.2(-12) && 166  \\
CH$_2$(OH)CHO && 9.4(-8) && 241 && 1.1(-7) && 233 && 6.4(-8) && 187  \\
CH$_2$(OH)COCHO && 4.0(-9) && 225 && 1.9(-9) && 230 && 1.6(-10) && 196  \\
CH$_2$(OH)COOH && 1.2(-8) && 244 && 7.1(-9) && 237 && 1.5(-9) && 216  \\
CH$_2$(OH)CONH$_2$ && 2.6(-8) && 237 && 2.4(-8) && 229 && 1.5(-8) && 221  \\
(CH$_2$OH)$_2$CO && 9.4(-13) && 210 && 2.5(-13) && 143 && 1.1(-13) && 10  \\
CH$_2$(OH)$_2$ && 4.9(-8) && 196 && 3.4(-8) && 189 && 1.7(-8) && 177  \\
\enddata
\tablecomments{$A(B)=A \times 10^B$.}
\end{deluxetable}


The peak abundance of methylamine (CH$_3$NH$_2$) during stage 2 of the {\tt final} models is again lower than the corresponding value in other models beginning at model {\tt all}, although it is a little higher than the {\tt basic} models, especially for the {\em slow} warm-up case (in which the {\tt basic} model is particularly deficient in CH$_3$NH$_2$). The decline is partly due to the adjustment of the barrierless reaction CH + NH$_3$ to give the correct products CH$_2$NH + H, as per \citet{Blitz12}. However, the assignment of a zero barrier to the reaction C + H$_2$ $\rightarrow$ CH$_2$ in the {\tt final} model allows much of the production of CH to be stepped over in favor of CH$_2$, thus lowering CH$_3$NH$_2$ production also. In the {\tt final} model, methylamine is mostly formed on the grains through the PDI mechanism, at very early times in the collapse stage (see also Sec.~\ref{amides}).

\section{Temperature regimes for the formation of COMs}\label{results:temps}

\begin{figure*}
\includegraphics[width=0.999\hsize]{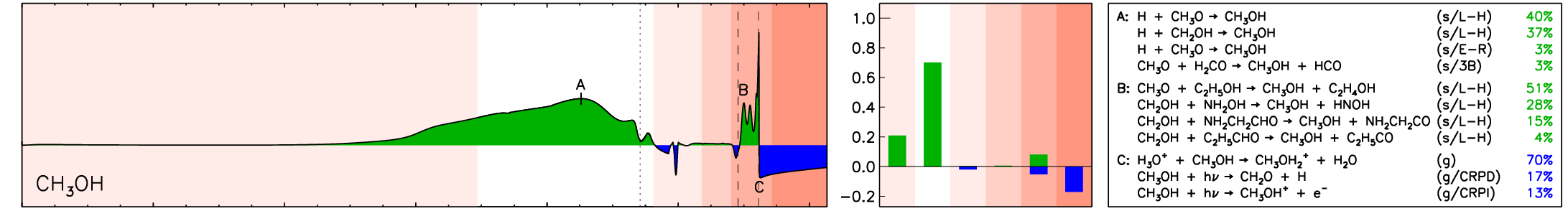}

\includegraphics[width=0.999\hsize]{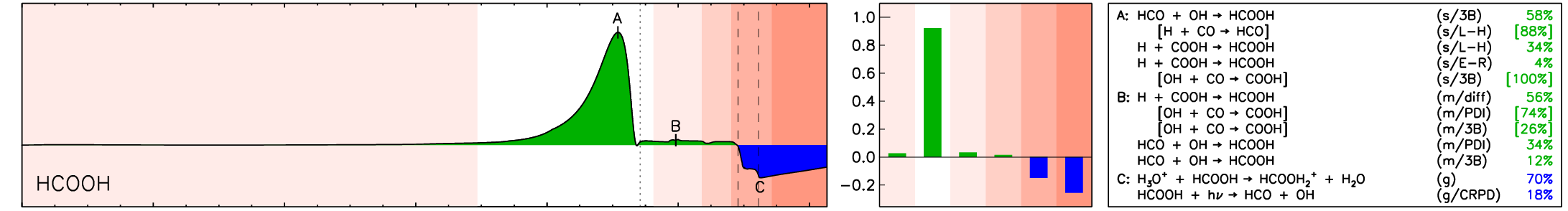}

\includegraphics[width=0.999\hsize]{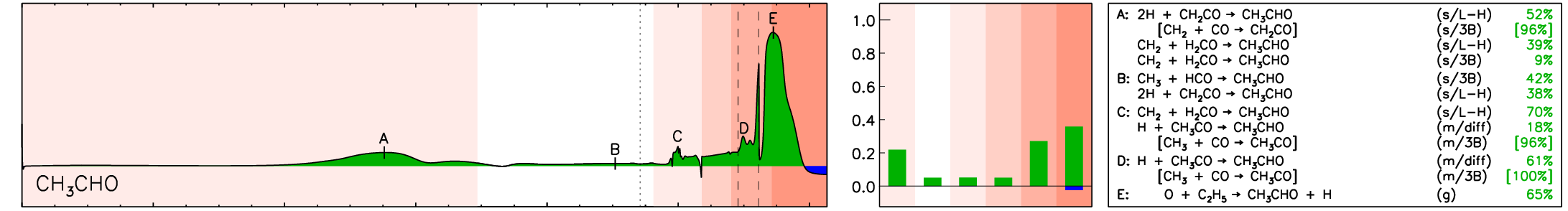}

\includegraphics[width=0.999\hsize]{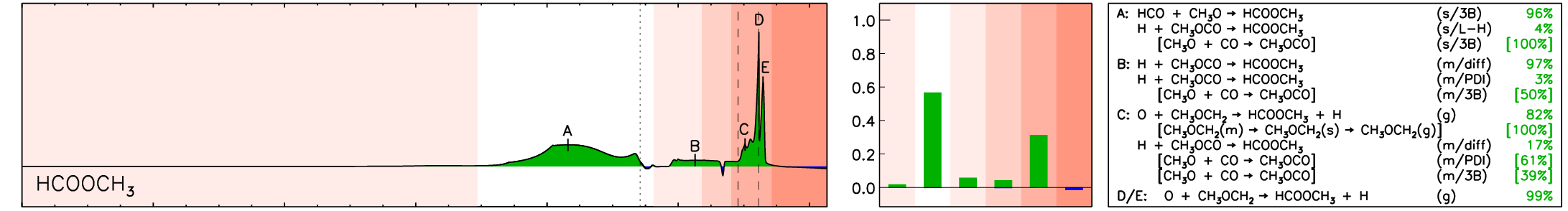}

\includegraphics[width=0.999\hsize]{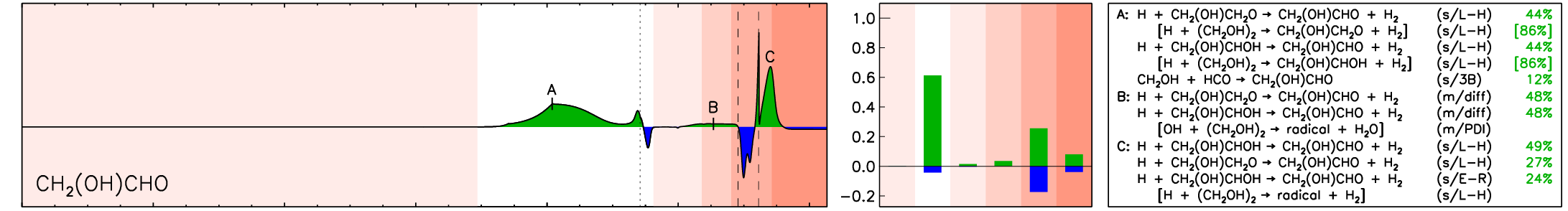}

\includegraphics[width=0.999\hsize]{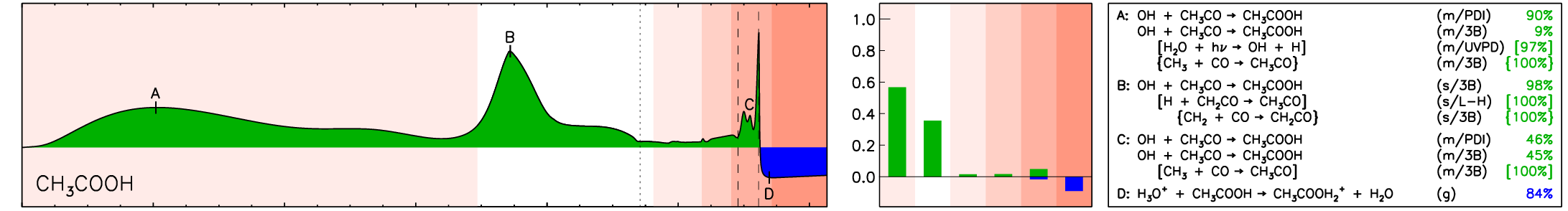}

\includegraphics[width=0.999\hsize]{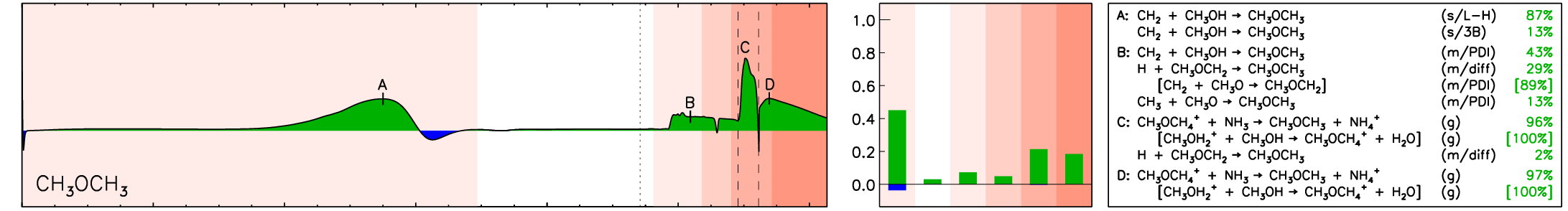}

\includegraphics[width=0.999\hsize]{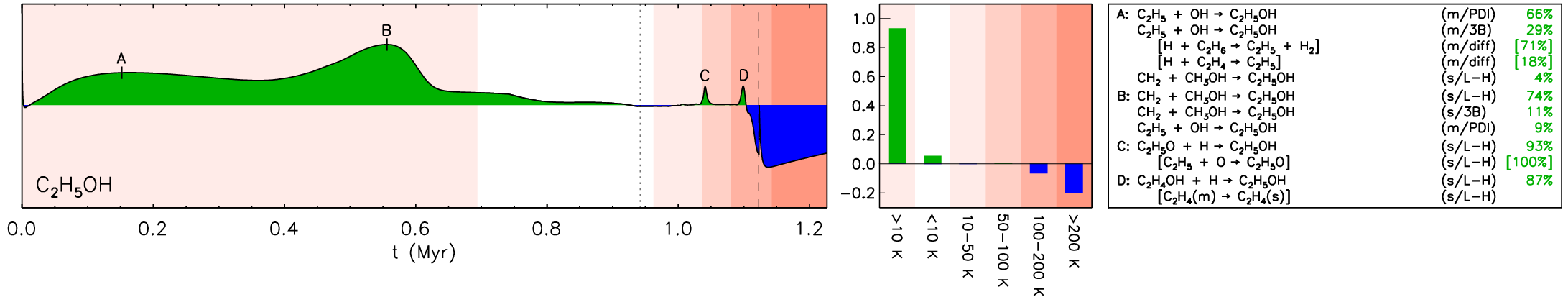}
\caption{{\bf {\em Left:}} Net rate of change (arbitrary units) in the abundances of selected oxygen-bearing molecules, summed over all chemical phases, during stages 1 and 2. Data correspond to the {\tt final} model, using the {\em medium} warm-up timescale in stage 2; the vertical dotted line indicates the start of the warm-up phase. Net gain is shown in green, net loss in blue. Vertical dashed lines indicate the onset and end-point of water desorption. Background colors indicate the dust temperature regime; from left to right these are: $>$10~K, $<$10~K, 10--50~K, 50--100~K, 100--200~K, 200--400~K. The initial dust temperature is $\sim$14.7~K. {\bf {\em Middle:}} Net rates of change integrated over each temperature range. Positive (formation) and negative (destruction) rates are integrated independently; both are normalized to the total integrated formation rate. {\bf {\em Right:}} The main formation (green) and destruction mechanisms (blue) at points indicated in the left panels (see text).}
\label{roc1}
\end{figure*}

Although useful for showing the general abundance behavior of molecules in the models, Figs.~\ref{final_collapse}--\ref{final_slow} do not show so clearly the times/temperatures at which certain molecules are primarily formed. It is instructive instead to plot the net formation rate for each molecule over time, summed over all physical phases (i.e.~gas, surface and mantle), as shown in the left-hand panels of Figs.~\ref{roc1} and \ref{roc2} for the {\em medium} timescale {\tt final} model. By summing the net formation/destruction rates over {\em all} phases, the transfer of material from one to another is removed as a consideration, so that the most important periods of production can be discerned. The panels indicate the net formation rates for various molecules, consecutively through stages 1 and 2, in arbitrary units normalized to the highest rate achieved; comparison of the absolute values obtained for different molecules is therefore not meaningful. The rates are plotted linearly against time, so that the area under the curve may be used to indicate the total production over a chosen period. Net formation and destruction are shown in green and blue, respectively. A vertical dotted line indicates the threshold between stages 1 and 2, while approximate temperature regimes are indicated by the background shading. Vertical dashed lines indicate the approximate beginning and end points of water-ice desorption from the grains, which occur at dust temperatures of $\sim$114~K and $\sim$164~K, respectively, in these models.

The bar plots in the middle panels of Figs.~\ref{roc1} and \ref{roc2} show the integrated net formation rates over each of the temperature regimes marked in the left panels; note that these regimes are not equal in their temperature span nor in the time spent in each -- their temperature ranges are chosen merely for convenience in describing the behavior of the chemistry. The bars are normalized to the total integrated positive formation rate, i.e.~the green bars correspond only to the integrated green portions of the left panels, and they sum to 100\% of those green portions over the combined stages 1 and 2. In cases where positive (green) and negative (blue) rates are achieved in the same temperature regime (in the left panels), both an upward green bar and a downward blue bar are shown. Table \ref{tab:rates}, in Appendix \ref{app-figs}, provides the integrated positive formation-rate data (i.e.~green) for a larger selection of molecules in each temperature regime.

The right-hand panels of Figs.~\ref{roc1} and \ref{roc2} show the main formation or destruction mechanisms at certain key moments marked by the letters in the left-hand panels. (If indicator lines are not otherwise marked in the plots, the letters are placed directly above or below the feature of interest). Note that the letter-labels used for each molecule do not correspond to the same moments for other species. For each reaction or process indicated, the approximate percentage contribution to the total formation (green) or destruction (blue) is also shown. Percentages are calculated to reflect the net contributions once any chemical closed-loop behavior is removed. Indented entries, marked in square or curly brackets, show the mechanisms that contribute to the production of a particular reactant that is used up in the main reactions; percentages shown for these contributing processes correspond to the production rate of a specific main reactant. A process with no percentage shown may be understood to be the sole/dominant contributor (at that moment). Each reaction shown in the right-hand panels is given a code, indicating whether it occurs in the gas phase (g), on the surface (s), or in the ice mantle (m), with more detailed descriptions provided where applicable, including Langmuir-Hinshelwood (L-H, i.e.~diffusion-driven surface reactions), Eley-Rideal (E-R), three-body reactions (3B), photodissociation-induced reactions (PDI, which in these models use the PDI2 treatment), reactions involving bulk diffusion of H or H$_2$ (diff), or photodissociation caused by external photons (UVPD) or the CR-induced UV field (CRPD).

Similar figures associated with the {\em fast} and {\em slow} timescale {\tt final} models may be found in Appendix \ref{app-figs} (Figs.~\ref{roc3}-\ref{roc6}).  In Sec.~\ref{res:timescales} we discuss variations from the {\em medium} warm-up timescale model, produced by using different model timescales.

\subsection{Methanol}

Methanol (CH$_3$OH) is seen, as expected, to be formed on cold grain surfaces mainly via the Langmuir-Hinshelwood (diffusive) mechanism (s/L-H), although other processes contribute a small amount to certain steps, including Eley-Rideal. Most of this formation occurs at temperatures below 10~K, although it begins at slightly higher temperatures (earlier times). Around 90\% of methanol is produced during these early, cold stages.

Much later in the hot-core evolution, at much higher temperatures, the remaining $\sim$10\% of methanol production occurs. Although there are several peaks within this late-time spike, all of them occur in the period when the water ice is desorbing strongly from the grains, as marked by the dashed lines. (The individual peaks are mainly associated with the loss of other species with binding energies similar to that of water, contributing further to the overall loss of the mantles). As ice material is rapidly lost via thermal desorption from the grain/ice surface, CH$_3$O and CH$_2$OH radicals are uncovered that were produced or locked in the ice at earlier times. As they are freed onto the surface, the dust temperatures are already high enough that those radicals are highly mobile and are able to desorb thermally; atomic H on the grains is very scarce for the same reason. Therefore, instead of reacting directly with H, the mobile methanol-related radicals attack stable molecules on the (ice) surface, abstracting hydrogen from them to form methanol that ultimately desorbs into the gas phase. The dust temperatures of 100~K or so are sufficient for some of these barrier-mediated H-abstraction reactions to proceed thermally. Once water has fully desorbed, most of the trapped radicals are also gone and the net production of methanol ends. The destruction of methanol in the gas phase now takes over the net rate, occurring mostly through reactions with H$_3$O$^+$ ions, which are formed from the desorbed water. Note that even as methanol was being formed on the grains, it was also being destroyed in the gas phase, reducing the net production rate apparent in the figure.

\subsection{Formic acid}

Formic acid (HCOOH), in the {\tt final} model implementation, is formed only partially as the result of the three-body surface reaction between CO and OH, followed by hydrogenation of the COOH radical. As seen at point ``A'', the three-body reaction between OH and HCO radicals is dominant; the latter is formed via diffusive H addition to CO. Grain/ice surface three-body processes tend to become more important at very low temperatures, as the diffusion rate of atomic H becomes lower, producing a commensurate rise in the surface coverage of reactive radicals including HCO, which makes their reaction through the 3-B process more likely. Interestingly, almost all of the formic acid produced in the model is formed under these very cold conditions.

Contrary to the models of \citet{GH06} and \citet{GWH08}, formic acid is not formed strongly at intermediate temperatures on the grains. In the {\em medium} timescale warm-up model, only around 4\% of the net-positive formation occurs in the 10--100~K regime. The majority of this production, as seen at point ``B'', occurs within the bulk ice through photodissociation-induced reactions. This is the result of CR-induced UV photodissociation of water in close proximity to a CO molecule, producing an OH radical that may immediately react with CO to form COOH (which then reacts with a mobile H atom). Neither the OH or CO is assumed to be mobile in the bulk ice, so there is little competition to their reaction when they are produced in close proximity. Formic acid does not show any strong production on the grain surfaces during the water-desorption period, although the irregular shape of the negative (blue) net production rate at that point indicates that some surface production from trapped radicals is occurring. Following desorption from the grains, formic acid is destroyed mainly through gas-phase ion-molecule reactions that are followed by dissociative electronic recombination.

\subsection{Acetaldehyde and ketene}

Unlike methanol and formic acid, acetaldehyde (CH$_3$CHO) is quite evenly produced throughout the different temperature regimes. Around 21\% of its net-positive production occurs at very early times, through grain-surface reactions at dust temperatures greater than 10~K, while another 5\% occurs later in the collapse stage, in the lowest temperature regime. Under those cold conditions, much of the acetaldehyde is formed through the hydrogenation of ketene, CH$_2$CO, which is formed through the newly-added surface reaction of methylene (CH$_2$) with CO via the 3-B mechanism. Due to its rapid hydrogenation by mobile atomic H on the dust/ice surface, ketene (not shown in the figures, but present in Table \ref{tab:rates}) does not achieve substantial net formation during the cold stages. However, its abundance in the ice finally grows beginning at a temperature around 55~K through the photodissociation-induced reaction of CH$_2$ with CO in the bulk, which continues steadily until the completion of water desorption \citep[see also the discussion of][on the topic of CH$_2$CO formation]{Jin20}.

The cold formation of acetaldehyde also has a large contribution from the grain-surface reaction of CH$_2$ with formaldehyde (driven by methylene diffusion) and through the barrierless addition of the radicals CH$_3$ and HCO (via 3-B reactions). Once the build-up of the ice mantles is essentially complete, production of acetaldehyde continues on the grains at later times and higher temperatures, largely through 3-B reactions in the bulk ice between CH$_3$ and CO followed by hydrogenation by diffusive H. Although the CH$_3$ + CO reaction is itself a 3-B process, the production of the CH$_3$ that initiates it occurs mainly through H-abstraction from methane by OH; the latter is a PDI process driven by the photodissociation of water. Thus, the formation of acetaldehyde in the bulk ice is, ultimately, driven by photodissociation.

At these elevated temperatures, a limited contribution is also made by diffusive surface reactions of CH$_2$ with H$_2$CO; this mechanism undergoes a brief spike at point ``C'' when substantial amounts of methane desorb from the grains at around 25~K. Ion-molecule destruction of the gas-phase CH$_4$ produces a moderate, short-lived build-up of CH$_2$ in the gas, which can re-adsorb briefly onto the grains to participate in surface reactions.

When large-scale water desorption occurs at high temperatures, acetaldehyde trapped in the ice is released. During this period, some production via the gas-phase reaction O + C$_2$H$_5$ $\rightarrow$ CH$_3$CHO + H also occurs. This becomes the major production mechanism once the ice mantles have desorbed fully into the gas.

The broad range of mechanisms producing acetaldehyde in this model stand in contrast to the results of past, diffusion-only chemical models in which the radical-addition reaction CH$_3$ + HCO dominated the formation. Although this reaction is indeed important late in the cold collapse stage (occuring on grain/ice surfaces through the three-body mechanism), methylene (CH$_2$) reactions with either CO or H$_2$CO make a more consistent contribution.

\subsection{Methyl formate, glycolaldehyde and acetic acid}

\begin{figure*}
\includegraphics[width=0.999\hsize]{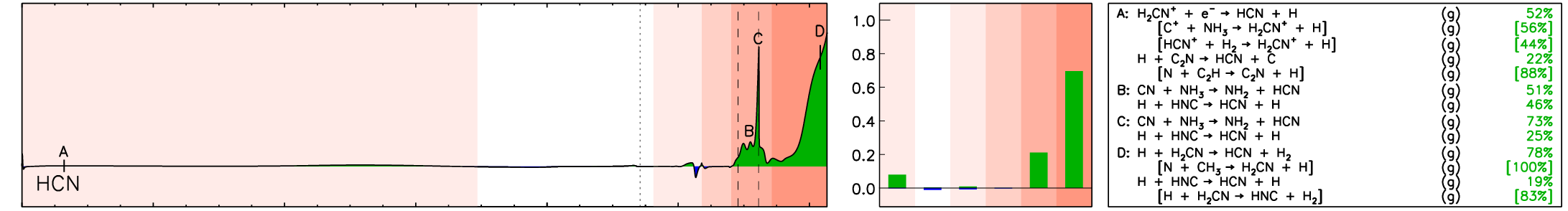}

\includegraphics[width=0.999\hsize]{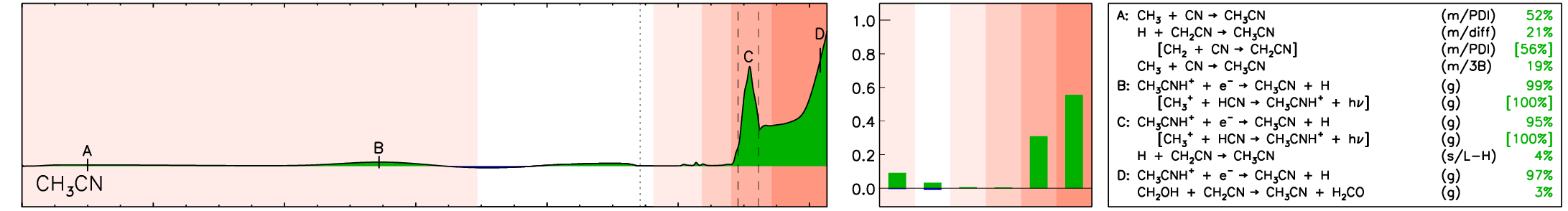}

\includegraphics[width=0.999\hsize]{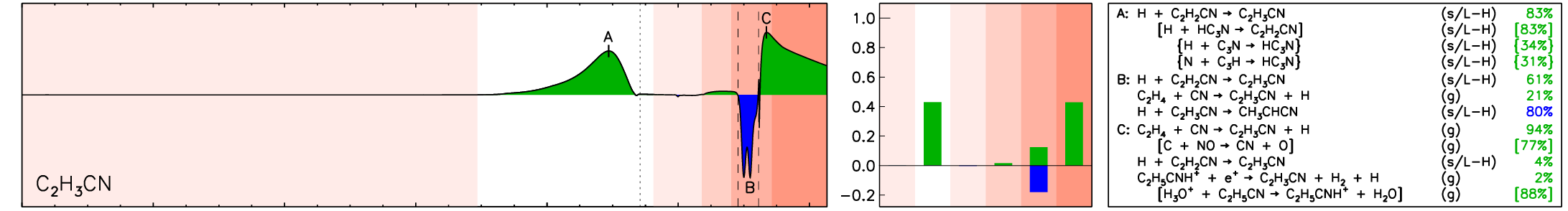}

\includegraphics[width=0.999\hsize]{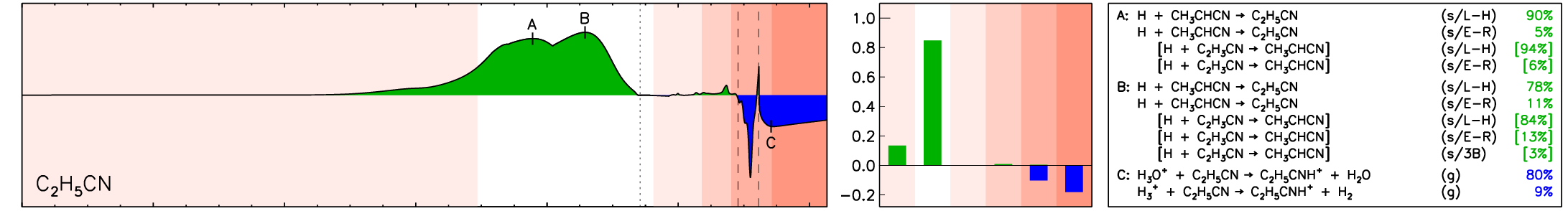}

\includegraphics[width=0.999\hsize]{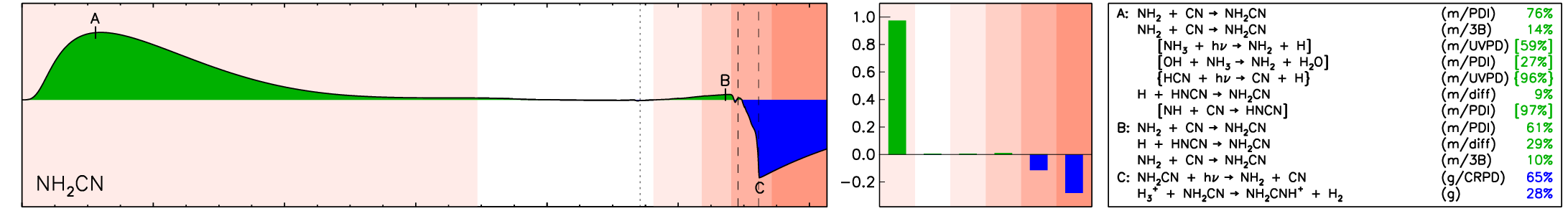}

\includegraphics[width=0.999\hsize]{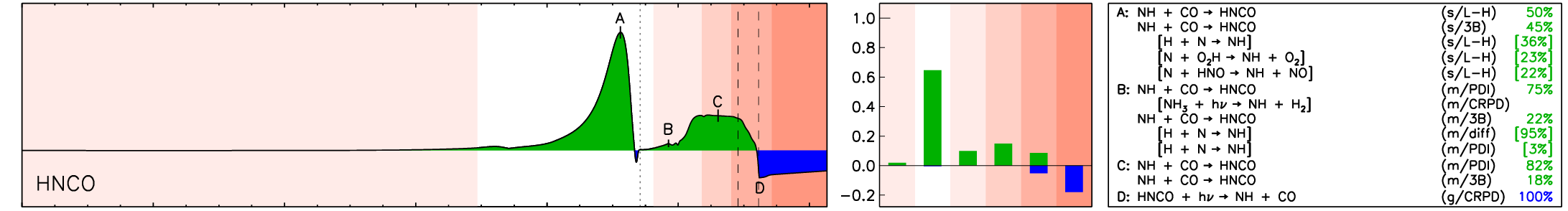}

\includegraphics[width=0.999\hsize]{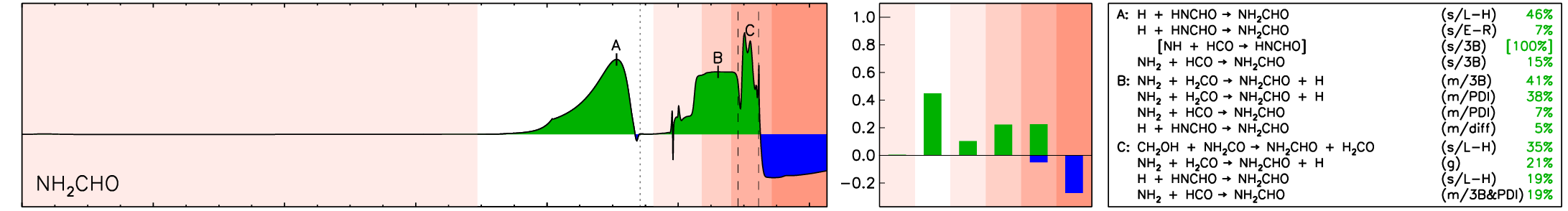}

\includegraphics[width=0.999\hsize]{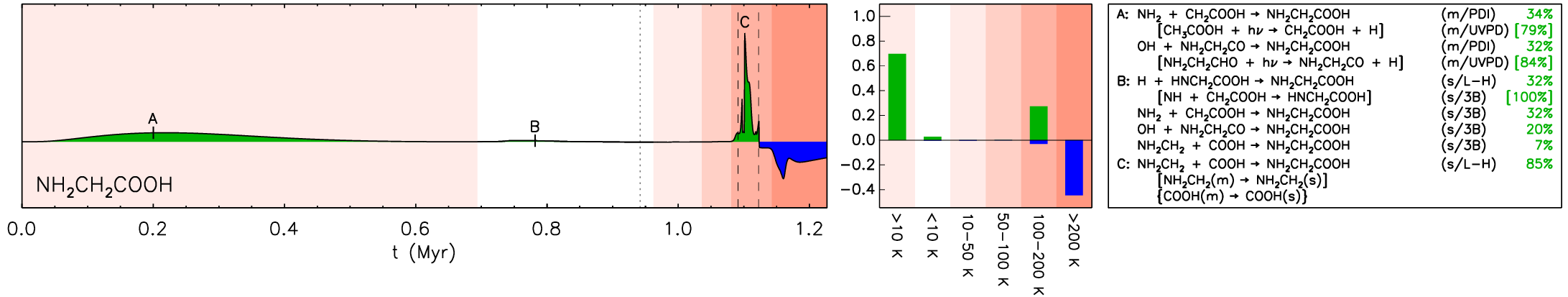}
\caption{Net rate of change in the abundances of selected nitrogen-bearing molecules, summed over all chemical phases; description as per Fig.~\ref{roc1}. Data correspond to the {\tt final} model, using the {\em medium} warm-up timescale in stage 2.}
\label{roc2}
\end{figure*}

The structural isomers methyl formate (HCOOCH$_3$), glycolaldehyde (CH$_2$(OH)CHO), and acetic acid (CH$_3$COOH) are all seen to have cold formation routes that contribute substantially to their total production. Methyl formate and glycolaldehyde in particular share similar production profiles, especially in the earlier/colder periods, and both are formed predominantly at temperatures below 10~K. (Note again that the formation rates shown in Figs.~\ref{roc1} and \ref{roc2} are plotted on arbitrary scales that vary between different molecules). At point ``A'', methyl formate is produced on the grain surfaces mainly through 3-B reactions of the radicals HCO and CH$_3$O. In fact, as time progresses further, the three-body excited-formation (3-BEF) reaction CH$_3$O + CO $\rightarrow$ CH$_3$OCO (introduced by JG20), which is followed by hydrogenation, takes over as the main production route of methyl formate. This changeover in the dominant mechanism is caused not so much by the changing dust temperature, which stabilizes at 8~K throughout most of the MF production, but by the increasing gas density that causes much faster deposition of gas-phase material onto the grains, altering the populations and production rates of radicals such as CH$_3$O and HCO, while the surface coverage of CO steadily grows (see Fig.~\ref{final_layers}).

The production of glycolaldehyde at low temperatures appears more complicated, due to the possibility of a number of forward and backward reactions that allow interconversion between glyoxal (HCOCHO), glycolaldehyde (GA), and ethylene glycol (EG, (CH$_2$OH)$_2$), some of which are shown in the right-hand panel of Fig.~\ref{roc1}. As a result, glycolaldehyde production has contributions from the conversion of the other two; the percentages shown in the right-hand panel reflect the net production of glycolaldehyde once H-addition/abstraction loops are removed. In fact, interconversion between glyoxal and glycolaldehyde is found to be a net loss for glycolaldehyde, although it gains through the net destruction of ethylene glycol, as well as having some direct formation of its own. The underlying production mechanisms for all three species (GA, EG and glyoxal) rely on 3-B surface radical-addition reactions of HCO or CH$_2$OH radicals, either with themselves or with each other. Over the course of the cold stages of the model, all three such reactions contribute to the formation of the same underlying carbon/oxygen backbone to a similar degree. These nondiffusive processes are driven by the initiating diffusive surface reactions H + CO $\rightarrow$ HCO and H + CH$_3$OH $\rightarrow$ CH$_2$OH + H$_2$, the latter of which has an activation-energy barrier. Note that, because the hydrogenation of formaldehyde on the grain surfaces strongly favors the production of CH$_3$O, rather than CH$_2$OH, there is no substantial 3-BEF formation of the glycolaldehyde pre-cursor CH$_2$(OH)CO (through the reaction CH$_2$OH + CO $\rightarrow$ CH$_2$(OH)CO), while the corresponding mechanism for methyl formate production is very strong. This, especially, leads to the somewhat lower absolute abundance of glycolaldehyde produced on the grains at low temperatures, versus that of methyl formate.

At somewhat higher temperatures, modest production of methyl formate (point ``B'') is achieved as the result of three-body reactions, this time within the bulk ice. These are driven again by the reaction of H with formaldehyde, producing CH$_3$O that can react immediately with CO. Although the 3-BEF mechanism is helpful to this formation process, the immobility of the CH$_3$O radicals that are formed in proximity to CO greatly improves the chances of a reaction through the regular 3-B process. At similar times, a smaller amount of glycolaldehyde is produced through repetitive H-abstraction from ethylene glycol, by either H or OH produced by cosmic ray-induced photodissociation of water. EG also receives a gradual boost through the addition of bulk-ice CH$_2$OH radicals with each other, via the PDI reaction mechanism acting on methanol.

As noted earlier (Sec.~\ref{basicPDI}), the surface chemistry occurring when water is strongly desorbing from the grains favors methyl formate production and survival, and disfavors glycolaldehyde. (Similar behavior can also be seen clearly in the plots for the {\em slow} warm-up model, in Fig.~\ref{roc5}). The methyl formate has a low binding energy compared with water, allowing MF molecules to desorb rapidly from the grain surface as soon as they are uncovered during the gradual loss of water. Glycolaldehyde has a higher binding energy than either MF or water; as a result, the mobile radicals that are released from the bulk ice onto the surface as water desorbs, along with atomic H from the gas phase, can abstract hydrogen from the aldehyde end of the glycolaldehyde molecule. However, further reactions of the resulting CH$_2$(OH)CO with other surface radicals leads to the production of yet more complex species, such as the sugar dihydroxyacetone (DHA), (CH$_2$OH)$_2$CO. The latter is destroyed similarly to GA, but the network does not allow further chemical complexity to be traced beyond DHA, and so this molecule is simply broken down. Its modeled abundance is therefore unlikely to be very reliable without an extension to the network.

During the water-desorption stage, while there is some grain-surface production of methyl formate (occurring largely in the mantle through the 3-B mechanism), the newly-added gas-phase mechanism O + CH$_3$OCH$_2$ $\rightarrow$ HCOOCH$_3$ + H \citep{Balucani15} becomes its main production route. Around 30\% of total net-production of methyl formate throughout the course of the model occurs through this gas-phase process in the 100--200~K temperature regime. However, while the water-ice is still strongly desorbing, the CH$_3$OCH$_2$ radical is not primarily formed in the gas phase; rather, this radical is released from the bulk ice onto the surface and rapidly desorbs into the gas phase. Once the water has mostly left the grains, the CH$_3$OCH$_2$ is sourced from the reaction of OH with gas-phase dimethyl ether \citep{Shannon14}.

After water has fully desorbed, glycolaldehyde remains for a while longer on the grain surfaces due to its higher binding energy. While ethylene glycol is also still present on the grains, due to its own high binding energy, H-abstraction from EG by H atoms continues to produce some glycolaldehyde that is ultimately desorbed into the gas.

Acetic acid (AA) shows very different chemical behavior from that of its structural isomers throughout the model runs. Although there is a small contribution during the water-desorption process, early-time production dominates. In all temperature regimes it is the addition of OH to the radical CH$_3$CO, either in the bulk ice or on the surface, that produces acetic acid. Production at the earliest times occurs in the bulk ice, starting with the 3-BEF reaction of CH$_3$ with CO in the bulk, although H + CH$_2$CO $\rightarrow$ CH$_3$CO also makes a contribution. Reaction with the OH radical then proceeds via the PDI process, driven by UV photodissociation of water; note that this photodissociation is caused explicitly by external UV photons. The strong, early AA production tracks somewhat with the build-up of methane in the bulk ice (shown in Fig.~\ref{final_layers}), because the latter is the ultimate source of the CH$_3$ and CH$_2$ required for its formation via photodissociation-induced reactions. It gradually drops off as the visual extinction increases during the collapse process and external dissociating photons are blocked. This early peak in acetic acid production is therefore strongly associated with the photo-processing of simple ices by external UV photons.

Acetic acid undergoes a second low-temperature production spike (point ``B''), this time as the result of surface reactions (which are unrelated to external UV processing). This spike appears as CO surface coverage on the grains rises with falling temperatures, when its conversion to CO$_2$ becomes less efficient. C atoms that are accreted onto the grains rapidly react with H$_2$ to form CH$_2$, which in turn reacts with surface CO via the 3-B mechanism, forming ketene. Reaction of ketene with diffusive surface H produces the CH$_3$CO radical, which reacts with OH. The spike in grain-surface AA production at point ``B'' drops off again with falling gas-phase atomic carbon abundance.

Acetic acid production continues at a modest rate up until the end of the collapse stage. At later times, a small degree of production is achieved, especially during water desorption, although the reactions still occur mainly within the bulk ice.

\subsection{Dimethyl ether and ethanol}

Dimethyl ether (DME) has long been known to have an efficient gas-phase production route; the reaction of methanol with protonated methanol turns out to be responsible for roughly half of net DME production in the model examined here. However, the inclusion of triplet methylene (CH$_2$) reactions with methanol in the grain-surface network provides an important cold process for DME production on grains. Methanol is sufficiently abundant, and methylene sufficiently mobile, that this reaction can occur effectively through the Langmuir-Hinshelwood mechanism, supplemented by 3-B reactions. Dimethyl ether production occurs within the bulk ice at later times, occurring mostly through photochemistry. However, pure bulk-ice photochemistry is generally far weaker than the cold, grain/ice-surface mechanism; this may be contrasted with past models \citep[e.g.][]{GWH08,Garrod13a} in which production on the grains at intermediate temperatures was much stronger. Longer model timescales, however, allow photochemistry to contribute more strongly. It is likely that higher cosmic-ray ionization rates also would produce a stronger photochemistry, by providing a stronger secondary UV field.

When the water-ice mantles begin to desorb strongly from the grains, accompanied by methanol, the production of DME from methanol in the gas phase takes over as the main contributor. This continues for as long as substantial quantities of methanol are still present in the gas phase. The final step in the ion-molecule production of DME is proton transfer to ammonia \citep[see also][]{Taquet16}.

Ethanol is formed almost entirely at the earliest stages of the models; rather like acetic acid, it is produced strongly in the mantles, as the result of photochemistry driven by external UV photons that penetrate the 3 magnitudes of visual extinction. C$_2$H$_5$ radicals in the bulk ice, formed by diffusive reactions of atomic H with ethane (C$_2$H$_6$) and ethylene (C$_2$H$_4$) that are formed on the surfaces and retained within the bulk ice, react with OH produced by photodissociation of water in the mantles.

A little later, while this bulk-ice process continues to an extent, the new reactions of methylene with methanol on the grain/ice surface take over as the main ethanol formation process.

The two early/cold processes combined dominate all ethanol production, in models of all timescales. Strong production of ethanol may therefore be understood as the result of early-time chemistry that produces hydrocarbons effectively, combined with low-extinction conditions that allow the hydrocarbon/water-ice mixture to undergo fast photochemistry. Note that the efficiency of neither the methylene reactions with methanol, nor of H-atom abstraction or addition to hydrocarbons in the bulk ice, are very well defined in the models.

\subsection{Hydrogen cyanide and methyl cyanide}

Hydrogen cyanide (HCN) and methyl cyanide (CH$_3$CN) both exhibit rather similar behavior in their net production rate plots (Fig.~\ref{roc2}); they each have early-stage production totalling on the order of 10\%, but are then dominated by production in the very late/hot stages of the models. HCN is produced at early times through a selection of gas-phase mechanisms, some of which is then adsorbed onto the grains and incorporated into the ice mantles. However, methyl cyanide is initially formed in the ice mantles, through photochemistry induced by the initially low visual extinction during the collapse stage (at point ``A''). Non-diffusive reactions of CN in the mantle are mostly made possible by the photodissociation of HCN to produce the CN radical, inducing a PDI process. A little later on (at point ``B''), gas-phase reactions involving hydrogen cyanide become the dominant processes for CH$_3$CN production: the radiative association of CH$_3^+$ with HCN is followed by electronic dissociative recombination of protonated methyl cyanide. Note that ammonia is not sufficiently abundant in the gas phase at this stage to be an important agent of proton transfer. 

Although outweighed in overall influence by later, hot chemistry, the early production mechanisms detailed above are in fact very effective in producing large quantities of each molecule, especially on the grains, as seen in Fig.~\ref{final_collapse}; the production of each is quite steady throughout the collapse stage.

The release of water from the dust grains at high temperatures allows other species to escape, and during this period HCN undergoes renewed net production in the gas phase. The release of methane and related radicals from the ice mantles onto the grain surfaces and then into the gas phase ultimately acts to increase the abundance of atomic carbon in the gas phase during the desorption stage. This carbon reacts with NO that is also released from the grains, to produce HCN. This process is accompanied by a spike in HCN conversion to methyl cyanide. HCN and CH$_3$CN achieve around 21\% and 30\%, respectively, of their total production during this period.

In the very high temperature regime, at above $\sim$300~K, H atoms can overcome the activation energy barriers to react efficiently with NH$_3$, NH$_2$ and NH, abstracting hydrogen gradually to release atomic N. This N reacts with CH$_3$ in the gas to produce HCN, HNC and H$_2$CN, the latter of which may be converted to HCN also.

Through the close connection of CH$_3$CN with HCN, both in the gas and on the grains, these species are expected to trace very similar regions and conditions, at least during the hot stage following ice mantle desorption.

\subsection{Vinyl cyanide and ethyl cyanide}

As discussed by \citet{Garrod17}, vinyl cyanide (C$_2$H$_3$CN) and ethyl cyanide (C$_2$H$_5$CN) have a close chemical link; however, in the present model they are formed on the grains mainly in different temperature regimes than before, and the gas-phase relationship is weakened.

Much of the chemistry resulting in net production of both species occurs on grain surfaces through standard diffusive H-addition reactions -- in particular, the repetitive hydrogenation of C$_3$N molecules that originate in the gas phase. Thus, all ethyl cyanide production at early times passes through vinyl cyanide. Production of ethyl cyanide on grain surfaces ramps up especially when hydrocarbon production in the gas phase is rapid, at which point the gas density is beginning to grow substantially, which increases molecular deposition of species such as C$_3$N and C$_3$H onto the grains. While gas-phase atomic-H abundance also remains very high, full conversion to ethyl cyanide is efficient. At slightly later times, as the availability of gas-phase H begins to fall and the grain surface has more radicals with which H may react, H on the grains becomes more scarce, allowing more grain-surface vinyl cyanide to survive through to its incorporation into the bulk ices.

Production of either vinyl or ethyl cyanide at intermediate temperatures (i.e.~during the third and fourth temperature regimes) is minimal, in contrast to the results of \citet{Garrod17}; those authors found that combined surface and gas-phase chemistry beginning at temperatures around 40~K was highly effective in increasing ethyl cyanide abundances, following the thermal desorption of HCN. In the present models, the far more modest release of HCN into the gas phase at that temperature, caused by the switch-off of bulk diffusion that would otherwise allow HCN in the ice mantles to reach the surface layer and desorb, means that there is less free CN produced in the gas; the lack of CN reduces the subsequent production of other molecules on the grains when the CN radicals re-adsorb (briefly) to the grain/ice surfaces. In fact, at these temperatures, a degree of production of vinyl and ethyl cyanide continues in the bulk ices via repetitive, diffusive H addition to species trapped in the mantles. However, because such large quantities of both molecules are produced at early times, these later bulk-ice processes do not contribute significantly to the totals.

During water desorption, vinyl and ethyl cyanide are brought to the surface, and there some of the vinyl cyanide is converted to ethyl cyanide prior to its own desorption. After this, ethyl cyanide undergoes net destruction in the gas phase, as there are no efficient routes to its formation. Vinyl cyanide continues to grow in abundance at this stage; however, while \citet{Garrod17} noted that much of its high-temperature/late-time gas-phase production was a direct result of ethyl cyanide destruction, in the present models this is not the case. The inclusion of proton-transfer reactions between molecular ions (including protonated ethyl cyanide) and ammonia far reduces the influence of electronic dissociative recombination under post-desorption conditions. This makes gas-phase ion-molecule destruction mechanisms generally less important, and reduces the influence of the dissociative recombination of protonated ethyl cyanide, one of whose products is assumed to be vinyl cyanide in our network. Instead, much of the growth in vinyl cyanide abundance at late times is the result of gas-phase reactions of CN with ethylene (C$_2$H$_4$). Due to the far weaker gas-phase conversion of ethyl cyanide to vinyl cyanide, the ratio of the latter to the former does not reach such a high value in the present models as compared with those of \citet{Garrod17}.

\subsection{Isocyanic acid and formamide}\label{isocy}

Isocyanic acid (HNCO) and formamide (NH$_2$CHO) are the subject of great interest at present, and appear to have a strong correlation based on observational data \citep[e.g.][]{Jorgensen20}. In the present models they also show a strong correlation in the periods at which they are most strongly formed, and indeed in their net-production profiles. Both show an initial peak in production toward the end of the collapse stage, when dust temperatures are very low. Their formation mechanisms have similarities; both are formed on grain surfaces through the addition of NH to either CO or HCO; in the latter case, the resulting radical, HNCHO, further reacts with mobile H to produce NH$_2$CHO. The surface 3-B reaction between NH and HCO is driven by the production of each of these radicals in proximity to the other. The direct addition of surface radicals NH$_2$ and HCO via the 3-B mechanism also contributes somewhat to the production of formamide at this stage. A substantial fraction of the total isocyanic acid and formamide produced in these models is formed at this early stage. It is noteworthy that, as prescribed by the chemical network used here, there is no mechanism for the NH$_2$CO radical to be converted into formamide through direct H-atom addition (see Sec.~\ref{app2}), although there are indeed alternative reactions between this and other radicals that can produce formamide.

Later, especially in the 50--100~K temperature regime, PDI and 3-B mechanisms within the bulk ice are effective in driving production of both species. For isocyanic acid, this is again through the reaction of NH with CO. For formamide, the reaction between NH$_2$ and H$_2$CO -- adopted in the grain-surface/ice network for full consistency with the gas-phase chemistry -- is dominant. Both NH and NH$_2$ are photoproducts of ammonia, while CO and formaldehyde abundances are related through the production of the latter from the former at earlier times. Thus, while the chemistry of HNCO and NH$_2$CHO are not directly connected in the network, they follow parallel paths based on the same parent species in the ice. The bulk-ice production of both species, which ultimately is driven by photochemistry, is responsible for roughly 20--30\% of total production. However, the longer timescale model run allows more substantial production via bulk-ice photochemistry; see Sec.~\ref{res:timescales}.

During the water desorption stage, HNCO production on the grains ramps down monotonically as the total amount of bulk ice falls. HNCO has a desorption energy that is substantially less than that of water, so it desorbs rapidly from the surface once uncovered by the loss of H$_2$O. Ion-molecule destruction is quite slow, due to preferential proton transfer to ammonia. The most important net gas-phase destruction mechanism is cosmic ray-induced UV photodissociation.

Formamide abundance, however, grows more strongly as the water-ice mantle is lost, through several disparate mechanisms. Trapped NH$_2$CO radicals are brought to the surface where they may be more easily hydrogenated to formamide; this occurs through reactions with radicals. NH$_2$CHO also continues to be formed in the bulk ice via direct addition of the NH$_2$ and HCO radicals through both the PDI and 3-B mechanisms. Gas-phase production of formamide also ramps up at this point, as formaldehyde, ammonia and the radical NH$_2$ are released from the grains and into the gas phase.

\subsection{Cyanamide}\label{cyanamide}

The net production profile of cyanamide (NH$_2$CN) indicates that production of this molecule occurs almost entirely in the very earliest stages of the model, when visual extinctions and gas densities are low. External UV photons induce a photochemistry in the young ice mantle. HCN and NH$_3$ photodissociation are the main drivers of cyanamide production, forming CN and NH$_2$ radicals that recombine (via PDI), although H-abstraction from ammonia by OH (produced by water photodissociation) also makes a contribution. The other photoproduct of ammonia in this model, the radical NH, can also react with CN; this is followed by hydrogen addition to complete the complement of bulk-ice cyanamide formation pathways.

Unlike ethanol and acetic acid, which also show strong production at very early times due to ice-mantle photochemistry, cyanamide has no significant later production mechanisms. This means that all of the production of NH$_2$CN should be strongly dependent on the initial visual extinction behavior of the nascent hot core, and the degree of ice build-up achieved at those early times. We compare the model behavior with observational data for NH$_2$CN in Sec.~\ref{results:obs}.

\subsection{Glycine}

Glycine (NH$_2$CH$_2$COOH), the simplest amino acid, is included in the model using the network introduced by \citet{Garrod13a}; the present model allows production to be driven by nondiffusive surface and ice-mantle chemistry, as well as L-H surface reactions, although diffusive reactions between radicals in the bulk ice is switched off, as per Sec.~\ref{sec:results:bulkdiff}.

The production of glycine occurs at two main points in this model; very early, through external-UV photochemistry, and during the water desorption stage, through surface recombination of radicals trapped in the bulk ice. A small contribution is made at temperatures below 10~K, through surface reactions. The first contribution is strongly related to the early photochemical production of acetic acid in the bulk ice, which may itself be photodissociated to produce CH$_2$COOH, with which the NH$_2$ radical can react. Another contribution at this time comes via glycinal (NH$_2$CH$_2$CHO), which is formed through the addition of HCO and CH$_2$NH$_2$ radicals; glycinal is photodissociated to produce NH$_2$CH$_2$CO, which reacts with OH.

Around 2\% of total glycine production occurs on the grain/ice surfaces, at slightly later times in the collapse stage, when temperatures are very low. Similar reactions, here driven by the three-body mechanism, cause this production. We note that the reaction CH$_2$NH$_2$ + COOH $\rightarrow$ NH$_2$CH$_2$COOH, cited by \citet{Ioppolo20} as an important production route for glycine on the grains, contributes only a small amount -- around 7\% of the production occurring under the lowest temperature conditions. This point is explored further in the Discussion section.

The later formation mechanism, which is more important in the {\em medium} and {\em slow} warm-up timescale models than in the {\em fast} model, occurs when water desorbs, uncovering radicals trapped in the ice that are mobile as soon as they reach the surface. In this case, it is indeed the CH$_2$NH$_2$ + COOH $\rightarrow$ NH$_2$CH$_2$COOH reaction that produces most of the glycine at this stage.

\subsection{Effects of warm-up timescale}\label{res:timescales}

The descriptions above concern mainly the {\em medium} warm-up timescale models of the {\tt final} model setup. The collapse-stage chemistry is shared between all three warm-up models, thus the absolute contribution to COM production from the early, cold stage is identical in each case. Note that the production rates shown in Figs.~\ref{roc1}--\ref{roc2} and \ref{roc3}--\ref{roc6}, for each molecule in each model, are normalized individually, thus cannot be compared in absolute terms. Table \ref{tab:abuns} indicates the absolute peak fractional abundance values achieved.

The most marked difference between the different timescale models is the greater or lesser influence of bulk-ice chemistry, which in most cases is directly or indirectly associated with photodissociation. Longer periods of irradiation allow more radicals or more H atoms to be produced, leading to more nondiffusive reactions. In the longer timescale models especially, the additional time spent by molecules in the gas phase under hot conditions leads to a more complete destruction of the more complex species by the end of the model, even if there is a greater degree of net production attained early on.

A strong timescale dependence is notable for ethylene glycol (not shown in figures), which achieves far lower peak abundances in the {\em slow} model. A long, slow desorption process for water allows EG to be destroyed on the grains by radical attack. The higher desorption temperature for ethylene glycol (versus water) also makes it susceptible to destruction for longer.

Also noteworthy is the stronger production of N-bearing and cyanide molecules in the longer-timescale model runs. This effect is particularly apparent for CH$_3$CN, for example, but it applies to other, related species. Figs.~\ref{final_fast}--\ref{final_slow} demonstrate this clearly for methyl cyanide; its abundance on the grains is very similar between the models, up until the desorption of the ice mantles. At this point in the {\em slow} model, its gas-phase abundance grows far beyond that preserved in the solid phase. Its growth, following mantle desorption, tracks that of HCN in each model. Ethyl and vinyl cyanide also show a similar dependence on timescale, although in the case of ethyl cyanide this is not caused by gas-phase processing.

The models presented here assume a canonical value for the cosmic-ray ionization rate. The adoption of a larger value could substantially increase the influence of bulk-ice processes, making them more effective over shorter time periods \citep[see also][]{Barger20}.

\begin{deluxetable}{lccccc}\label{tab:obs}
\tabletypesize{\footnotesize}
\tablecaption{Comparison of model results with observational data collated in the review by \citet{Jorgensen20}, obtained from the PILS \citep{Jorgensen16} and EMoCA \citep{Belloche16} surveys. Observational column density ratios with respect to methanol are provided for the low-mass source IRAS 16293-2422B and the high-mass source Sgr B2(N2), using methanol column densities of $1 \times 10^{19}$ cm$^{-2}$ \citep{Jorgensen18} and $4 \times 10^{19}$ cm$^{-2}$ \citep{Belloche16}, respectively.
Model ratios correspond to the peak gas-phase abundance of each molecule divided by the peak gas-phase methanol abundance, for each of the {\tt final} model warm-up timescales (see Table \ref{tab:abuns}). Chemical species not included in the models are omitted.}
\tablewidth{0pt}
\tablehead{  \colhead{Molecule/CH$_3$OH} &  \colhead{IRAS 16293B} &  \colhead{Sgr B2(N2)} &  \colhead{{\em Fast} model} &  \colhead{{\em Medium} model} &  \colhead{{\em Slow} model} \\
  \colhead{} &  \colhead{(PILS)} &  \colhead{(EMoCA)} &  \colhead{} &  \colhead{} &  \colhead{} }
\startdata
H$_2$CO & 1.9(-1) & -- & 2.6(-1) & 1.3(-1) & 1.9(-2) \\
HCOOH & 5.6(-3) & -- & 1.2(-2) & 1.2(-2) & 1.3(-2) \\
CH$_2$CO & 4.8(-3) & 8.3(-3) & 7.5(-4) & 2.8(-3) & 1.5(-2) \\
C$_2$H$_5$OH & 2.3(-2) & 5.0(-2) & 6.1(-2) & 6.5(-2) & 8.0(-2) \\
CH$_3$OCH$_3$ & 2.4(-2) & 5.5(-2) & 1.0(-2) & 1.7(-2) & 2.6(-2) \\
HCOOCH$_3$ & 2.6(-2) & 3.0(-2) & 1.7(-2) & 1.9(-2) & 3.0(-2) \\
CH$_2$(OH)CHO & 3.4(-3) & 3.3(-3) & 8.7(-3) & 1.1(-2) & 7.7(-3) \\
CH$_3$COOH & 2.8(-4) & $<$5.0(-4) & 5.4(-4) & 5.9(-4) & 1.1(-3) \\
CH$_3$CHO & 1.2(-2) & 1.1(-2) & 3.6(-3) & 7.3(-3) & 3.0(-2) \\
c-C$_2$H$_4$O & 5.4(-4) & -- & 2.7(-4) & 3.5(-4) & 8.8(-4) \\
CH$_3$OCH$_2$OH & 1.4(-2) & $<$1.7(-2) & 1.3(-2) & 9.8(-3) & 7.7(-3) \\
(CH$_2$OH)$_2$ & 9.9(-3) & -- & 5.5(-3) & 1.0(-3) & 2.7(-7) \\
(CH$_3$)$_2$CO & 1.7(-3) & 1.0(-2) & 5.9(-4) & 6.6(-4) & 1.7(-3) \\
C$_2$H$_5$CHO & 2.2(-4) & -- & 3.3(-4) & 3.4(-4) & 5.0(-4) \\
HNCO & 3.7(-3) & 5.0(-2) & 8.1(-3) & 1.1(-2) & 2.4(-2) \\
NH$_2$CHO & 1.0(-3) & 8.8(-2) & 1.7(-3) & 2.4(-3) & 6.4(-3) \\
CH$_3$NCO & 4.0(-4) & 5.5(-3) & 1.1(-6) & 1.1(-6) & 4.4(-6) \\
CH$_3$CN & 4.0(-3) & 5.5(-2) & 1.6(-4) & 7.8(-4) & 8.5(-3) \\
CH$_3$NC & 2.0(-5) & 2.5(-4) & 2.8(-7) & 1.8(-7) & 9.1(-8) \\
HC$_3$N & 1.8(-5) & 8.8(-3) & 1.8(-3) & 1.6(-2) & 8.1(-2) \\
C$_2$H$_3$CN & 7.4(-5) & 1.1(-2) & 5.7(-4) & 9.9(-4) & 2.4(-3) \\
C$_2$H$_5$CN & 3.6(-4) & 1.6(-1) & 9.4(-3) & 1.1(-2) & 1.8(-2) \\
CH$_3$C(O)NH$_2$ & 2.5(-4) & 3.5(-3) & 3.4(-4) & 3.3(-4) & 5.0(-4) \\
CH$_3$SH & 4.8(-4) & 8.5(-3) & 2.9(-4) & 2.9(-4) & 4.4(-4) \\
\enddata
\tablecomments{$A(B)=A \times 10^B$.}
\end{deluxetable}

\section{Comparison with observations}\label{results:obs}

\begin{figure*}
         {\includegraphics[width=0.5\hsize]{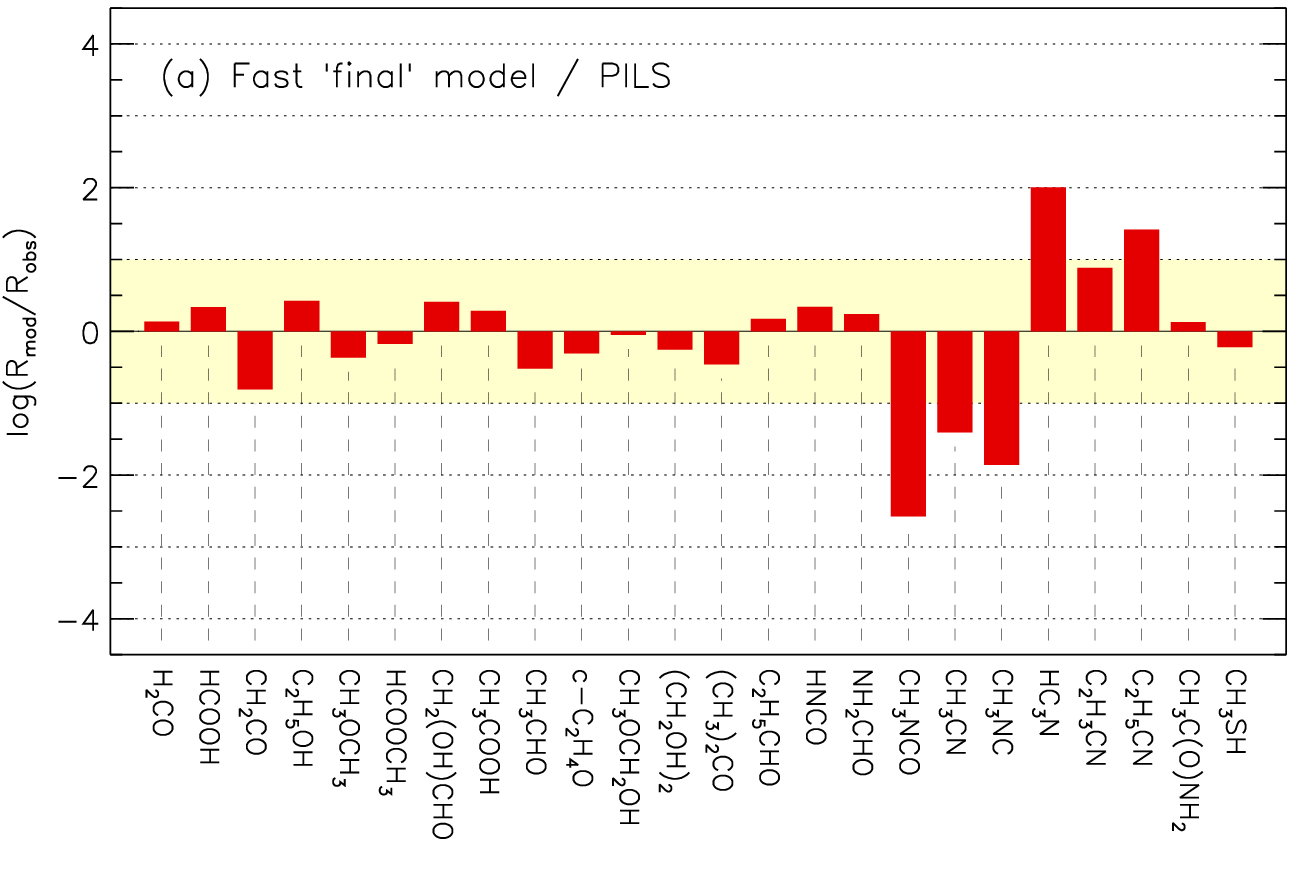}}
         {\includegraphics[width=0.5\hsize]{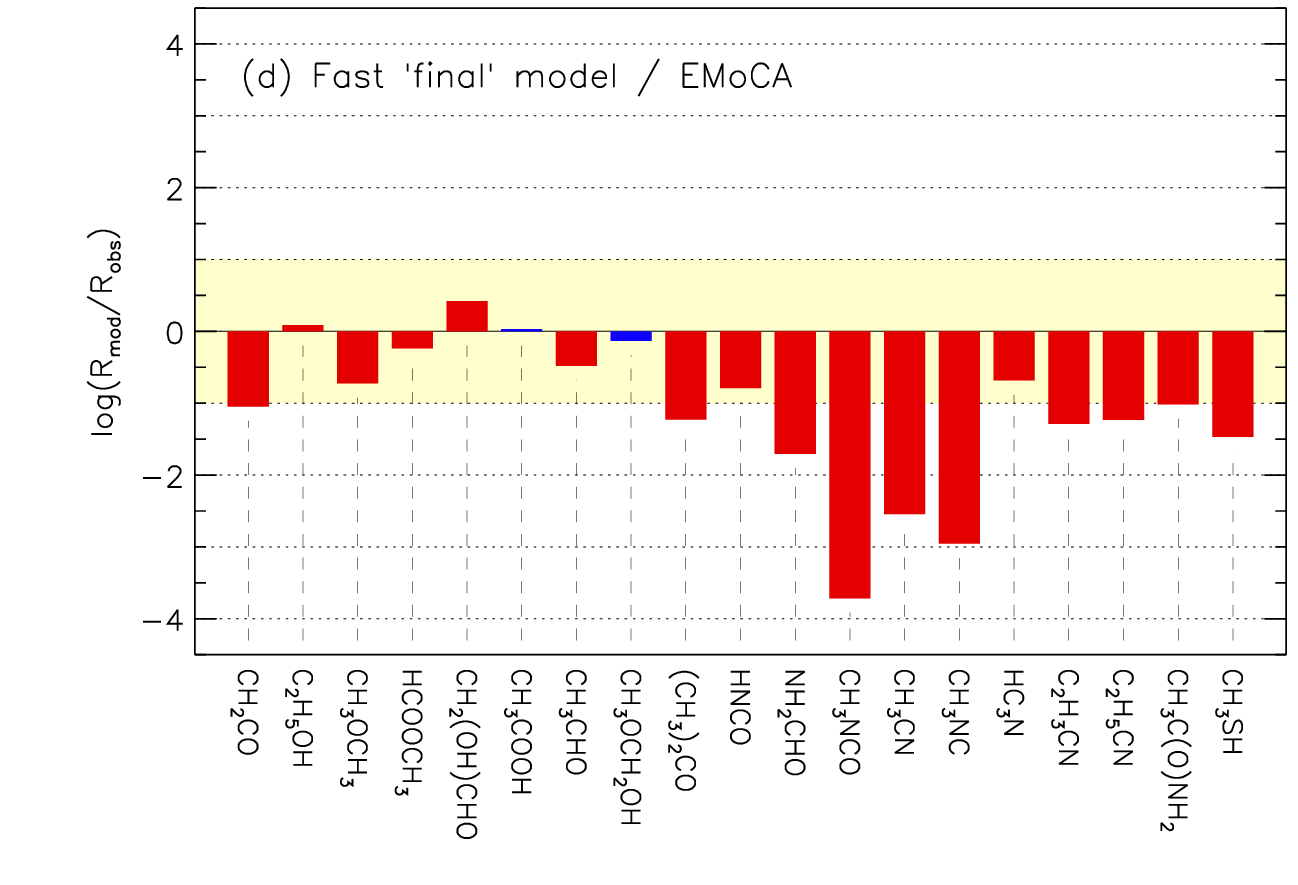}}
         {\includegraphics[width=0.5\hsize]{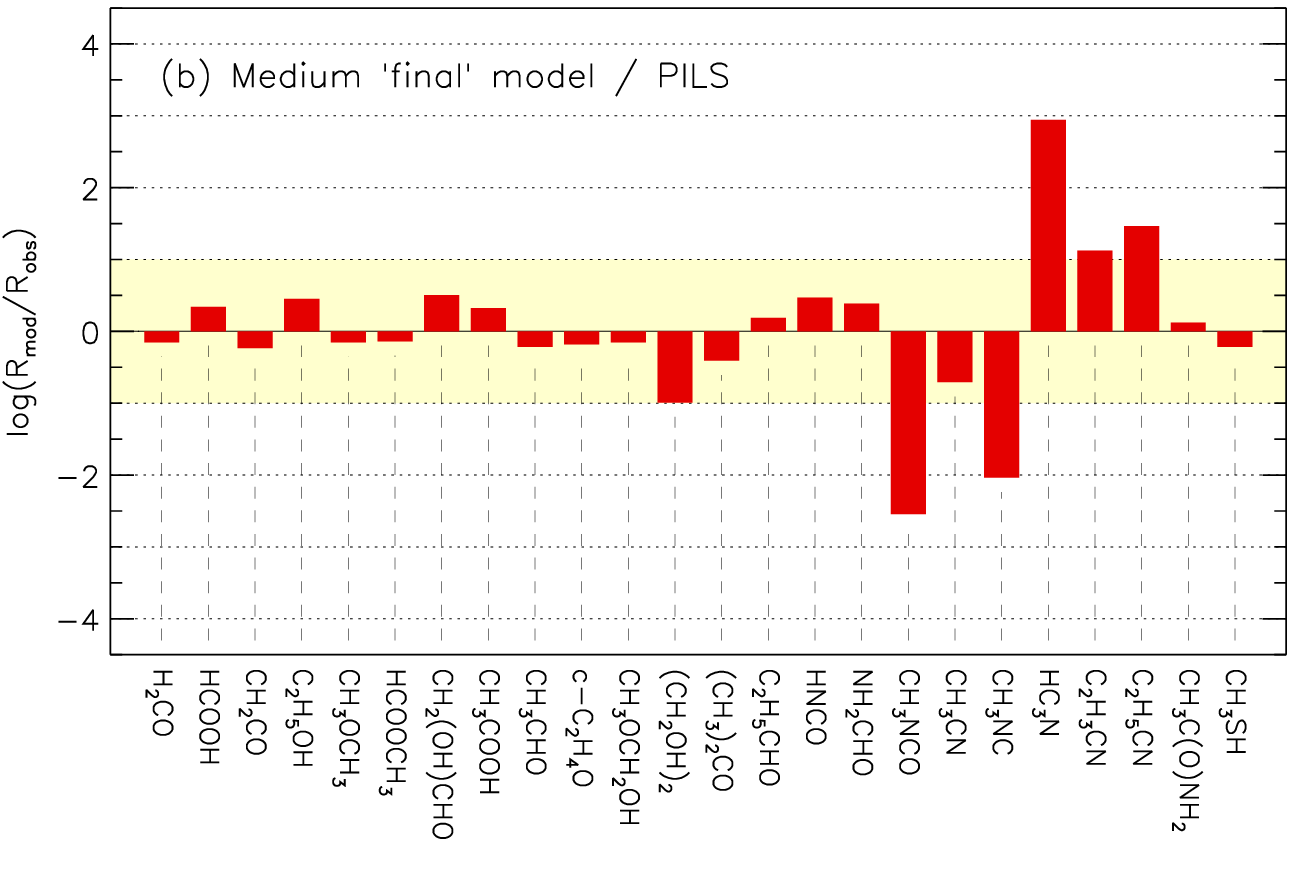}}
         {\includegraphics[width=0.5\hsize]{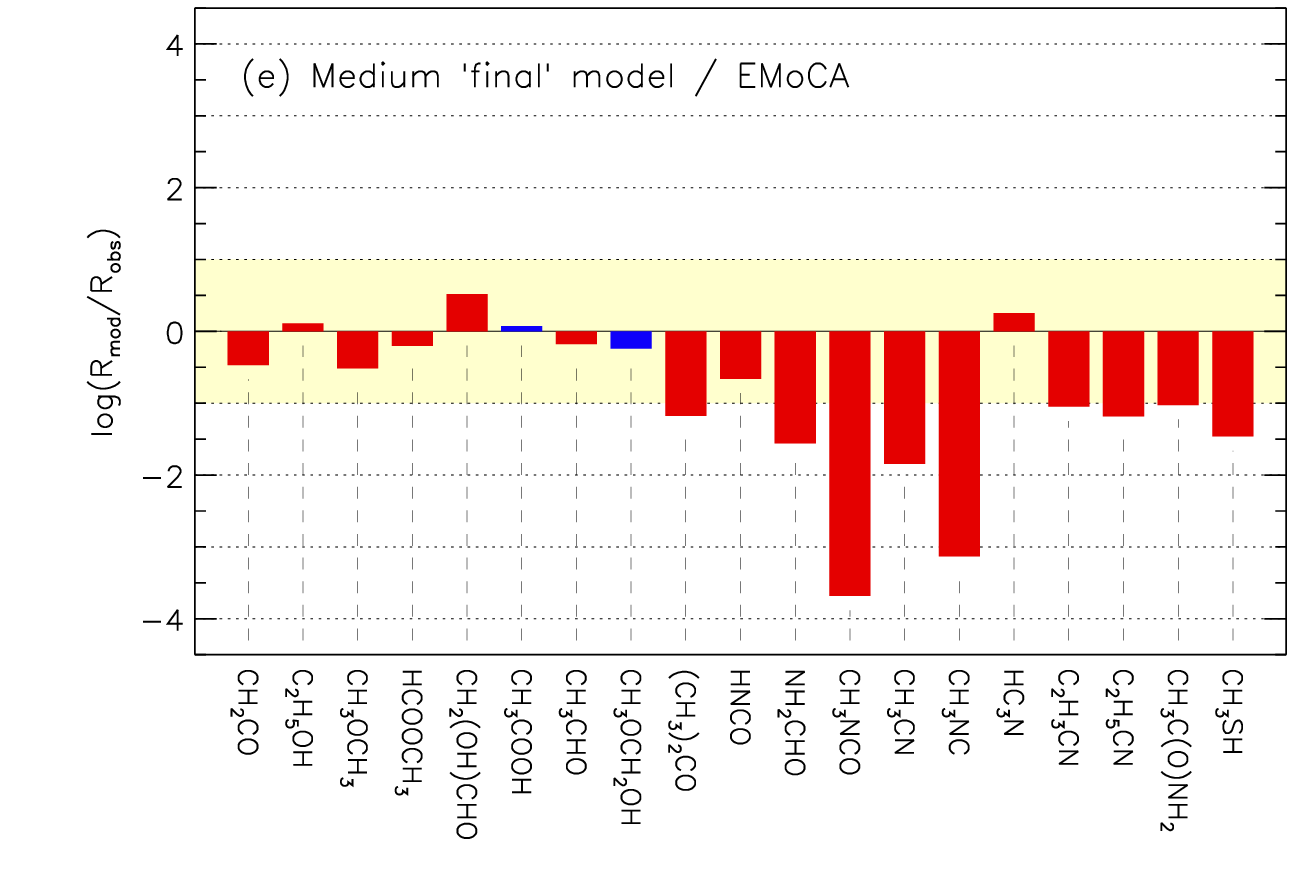}}
         {\includegraphics[width=0.5\hsize]{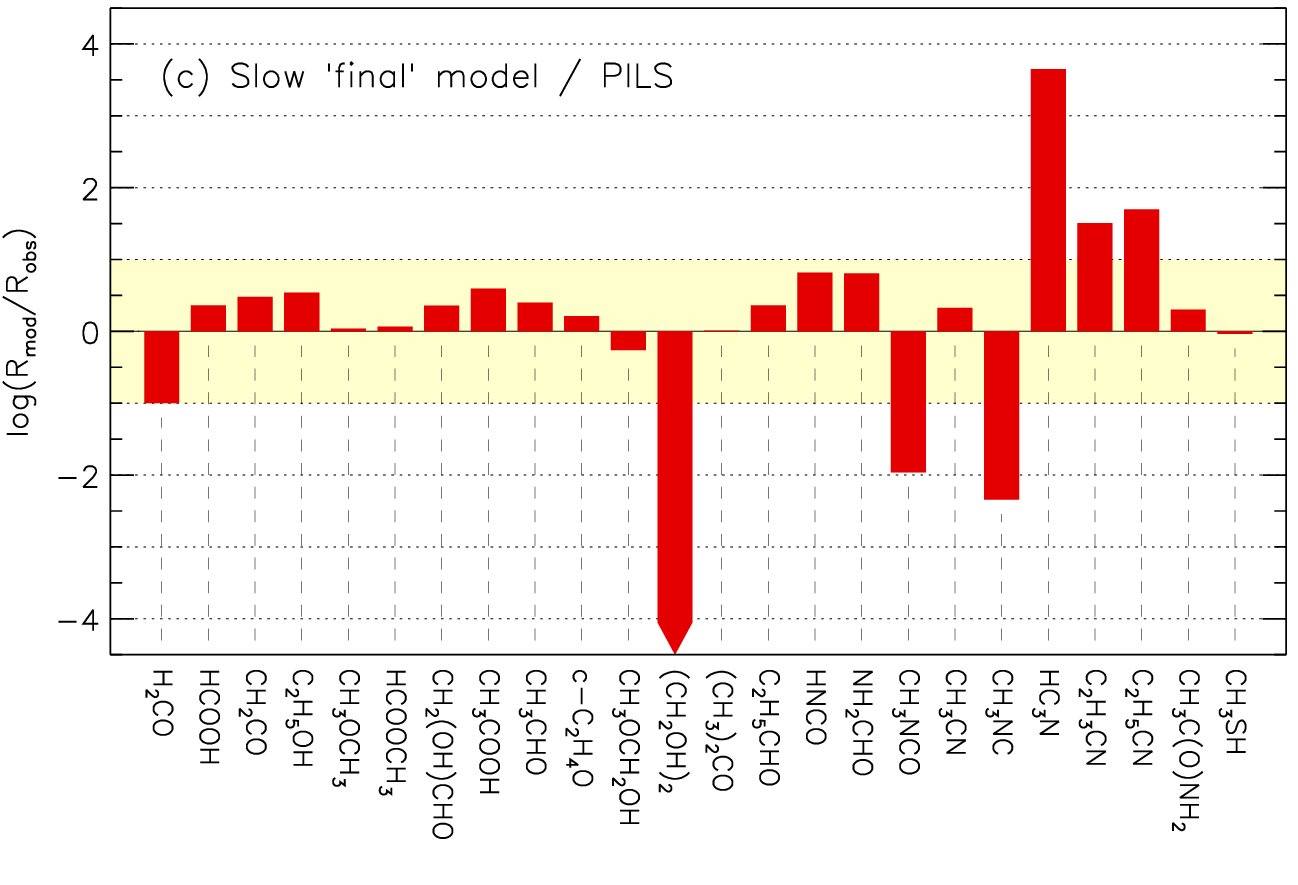}}
         {\includegraphics[width=0.5\hsize]{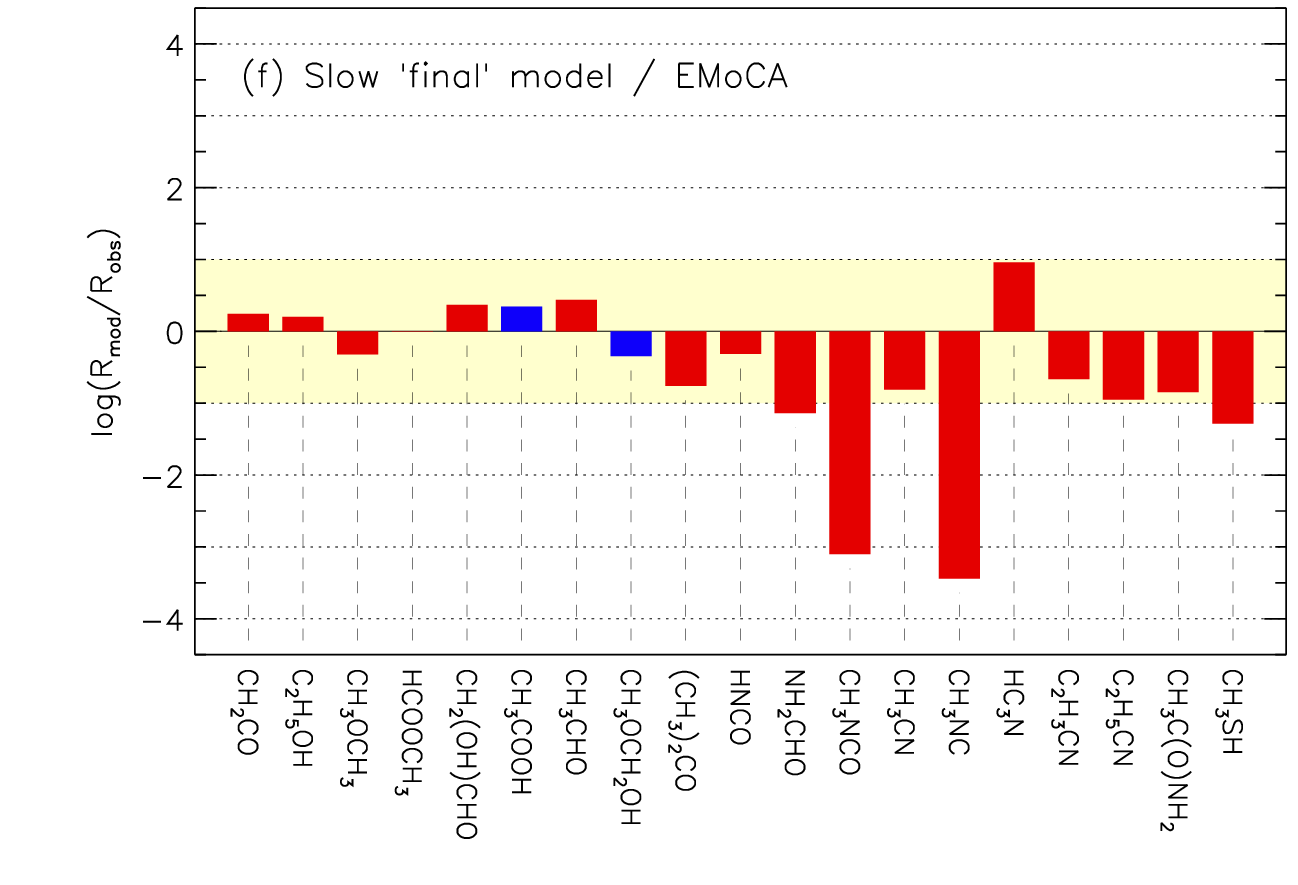}}
\caption{Comparison between the peak gas-phase molecular abundances achieved in the {\tt final} models (normalized to the models' peak methanol abundances), $R_{\mathrm{mod}}$, and observational values of the same quantities (based on column densities), $R_{\mathrm{obs}}$, using data taken from Table \ref{tab:obs}. Bars indicate the logarithm of $R_{\mathrm{mod}} / R_{\mathrm{obs}}$. The shaded area represents values where the models and observations vary by one order of magnitude or less. Bars ending in an arrow indicate values that exceed the bounds of the plots. Blue bars are based on observational upper limits. Panels on the left correspond to the PILS dataset, panels on the right to the EMoCA dataset.}
\label{compare-final}
\end{figure*}

Due to the large number of molecules to consider, our comparison of the {\tt final} model results with observations concentrates primarily on the data collated for (mostly) complex organic molecules by \citet{Jorgensen20} for the low-mass star-forming source IRAS 16293-2422B (also known as IRAS 16293B) and the high-mass Galactic Center hot core Sgr B2(N2), which are based on the PILS \citep{Jorgensen16} and EMoCA \citep{Belloche16} ALMA line surveys, respectively. Table \ref{tab:obs} lists the same species and fractional abundance data with respect to methanol given by those authors in their Table 2. Alongside are listed the peak abundance values from each of the three {\tt final} models, as a ratio with the peak model value for methanol (which does not necessarily occur exactly at the same moment as for the other species). The same modeled and observational data are shown graphically in Fig.~\ref{compare-final} in ratio to each other, as described in more detail below.

The agreement of the models with many of the molecular abundance ratios is impressive. There is some variation between the model values for different warm-up timescales, but the observational values from the two sources generally fall within, or come close to, the ranges produced by the models. Formaldehyde (H$_2$CO) toward IRAS 16293B is particularly well reproduced by the {\em fast} and {\em medium} models, while formic acid (HCOOH) is over-produced by a factor 2 in all cases; the latter molecule is formed predominantly in the early cold stage, so there is little variation in the peak value between models. Ketene (CH$_2$CO) abundance spans a broad range in the models; the observational value for IRAS 16293B falls somewhere between the {\em fast} and {\em medium} model values, while the Sgr B2(N2) value falls between the {\em medium} and {\em slow} model values.

The {\em fast} model value for ethanol (C$_2$H$_5$OH) is too high compared with IRAS 16293B by around a factor 2, and the {\em slow} model value is too high by a similar factor compared with Sgr B2(N2). The modeled dimethyl ether (CH$_3$OCH$_3$) abundance is a little too low by the same factor, in either case. These small divergences notwithstanding, it is apparent that -- based on the molecules considered so far -- the IRAS 16293B data appear to be best reproduced by the {\em fast} warm-up timescale model. Similarly, the Sgr B2(N2) data are closest to the {\em slow} warm-up timescale model values.

Although the variations between abundances observed toward IRAS 16293B and Sgr B2(N2) are smaller for methyl formate, glycolaldehyde and acetic acid, the same sort of correspondence between models and observations of these molecules is apparent: for molecules where higher values are found in IRAS 16293B than in Sgr B2(N2), higher values are achieved in the {\em fast} model, and lower values are achieved in the {\em slow} model; conversely, where lower values are found in IRAS 16293B than in Sgr B2(N2), lower values are achieved in the {\em fast} model, and higher values are achieved in the {\em slow} model.

Other molecules showing the same correspondence are (CH$_3$)$_2$CO, HNCO, NH$_2$CHO, CH$_3$NCO (a very poor match in absolute terms), CH$_3$CN, CH$_3$NC (a poor match), HC$_3$N, C$_2$H$_3$CN, C$_2$H$_5$CN, CH$_3$CONH$_2$ and CH$_3$SH. Indeed, almost all molecules in the list show such behavior, even if the absolute values are not always in good agreement. This is especially noteworthy when contrasting O-bearing versus N-bearing COM species. As discussed by \citet{Jorgensen20}, the former seem to track well with methanol in different sources, while the latter track better with HNCO (which those authors chose as a representative N-bearing molecule); however, the ratios between methanol and HNCO themselves, and thus between O- and N-bearing species more generally, do seem to vary strongly between sources. This nominally agrees with the present model findings, in which N-bearing species are far more strongly affected by the post mantle-desorption timescale than are the O-bearing molecules. The choice of overall warm-up timescale adopted in the models tracks with this post-desorption timescale. Abundance enhancements caused by gas-phase nitrogen-related chemistry at high temperatures may therefore provide an explanation for observational differences between sources in their ratios of O- to N-bearing molecules.


\begin{figure*}
         {\includegraphics[width=0.5\hsize]{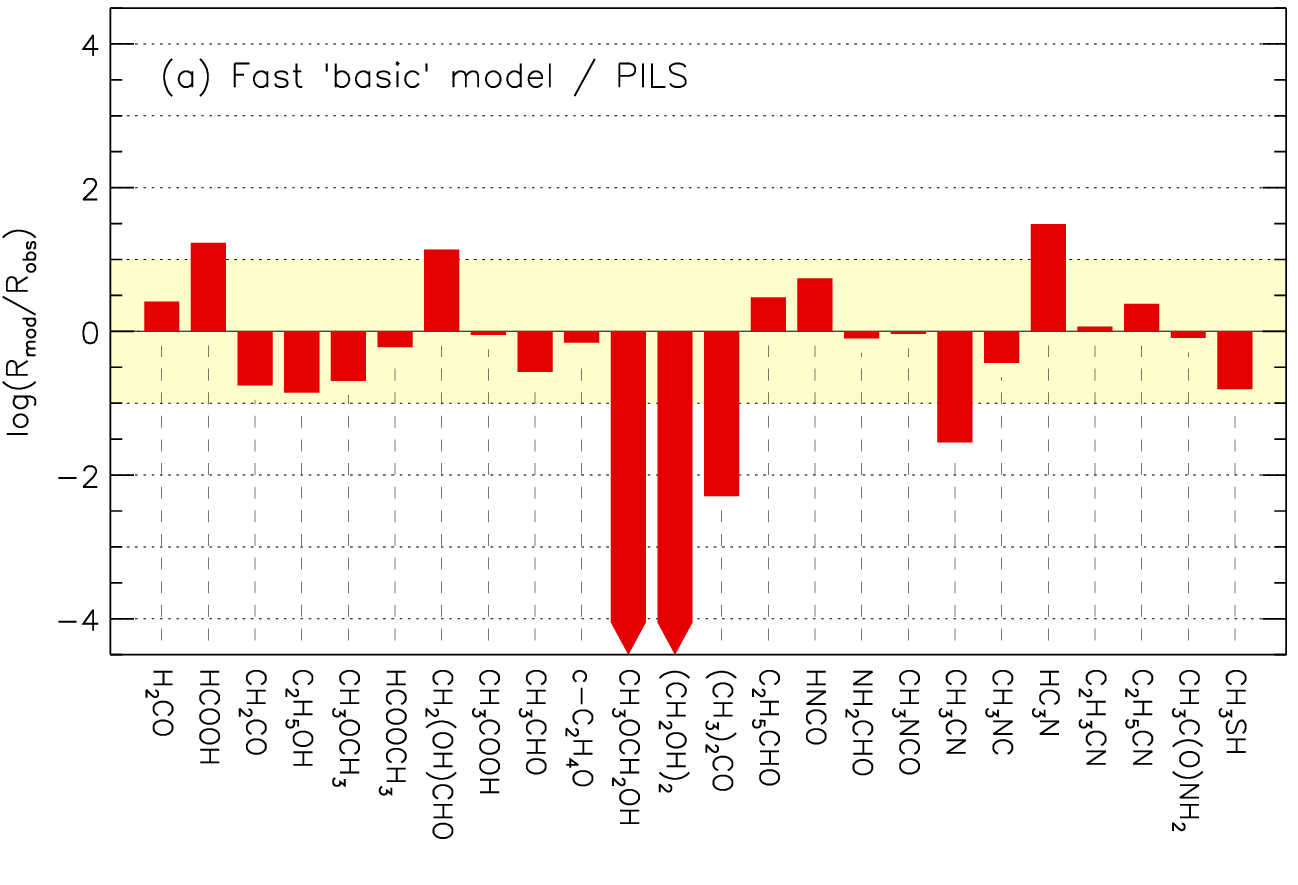}}
         {\includegraphics[width=0.5\hsize]{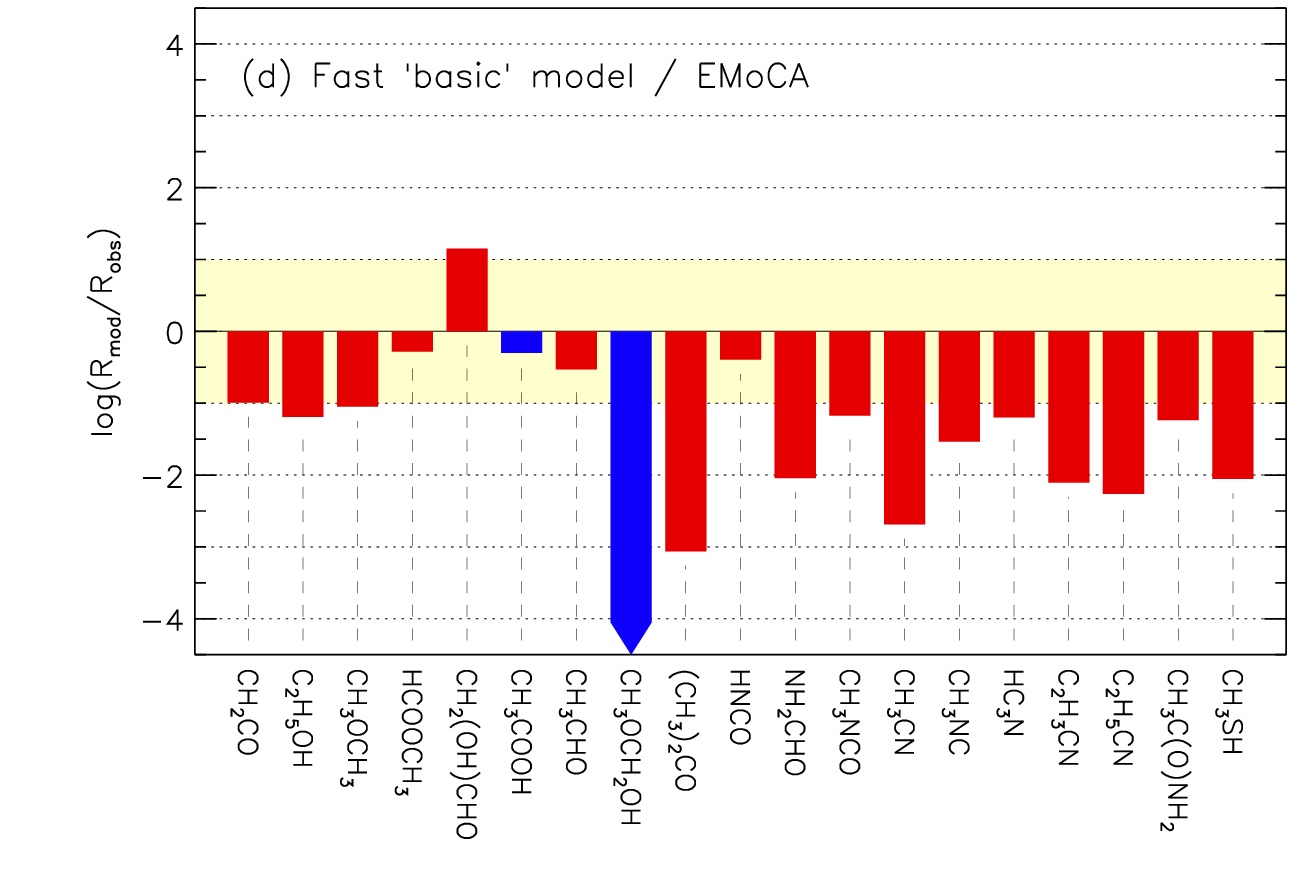}}
         {\includegraphics[width=0.5\hsize]{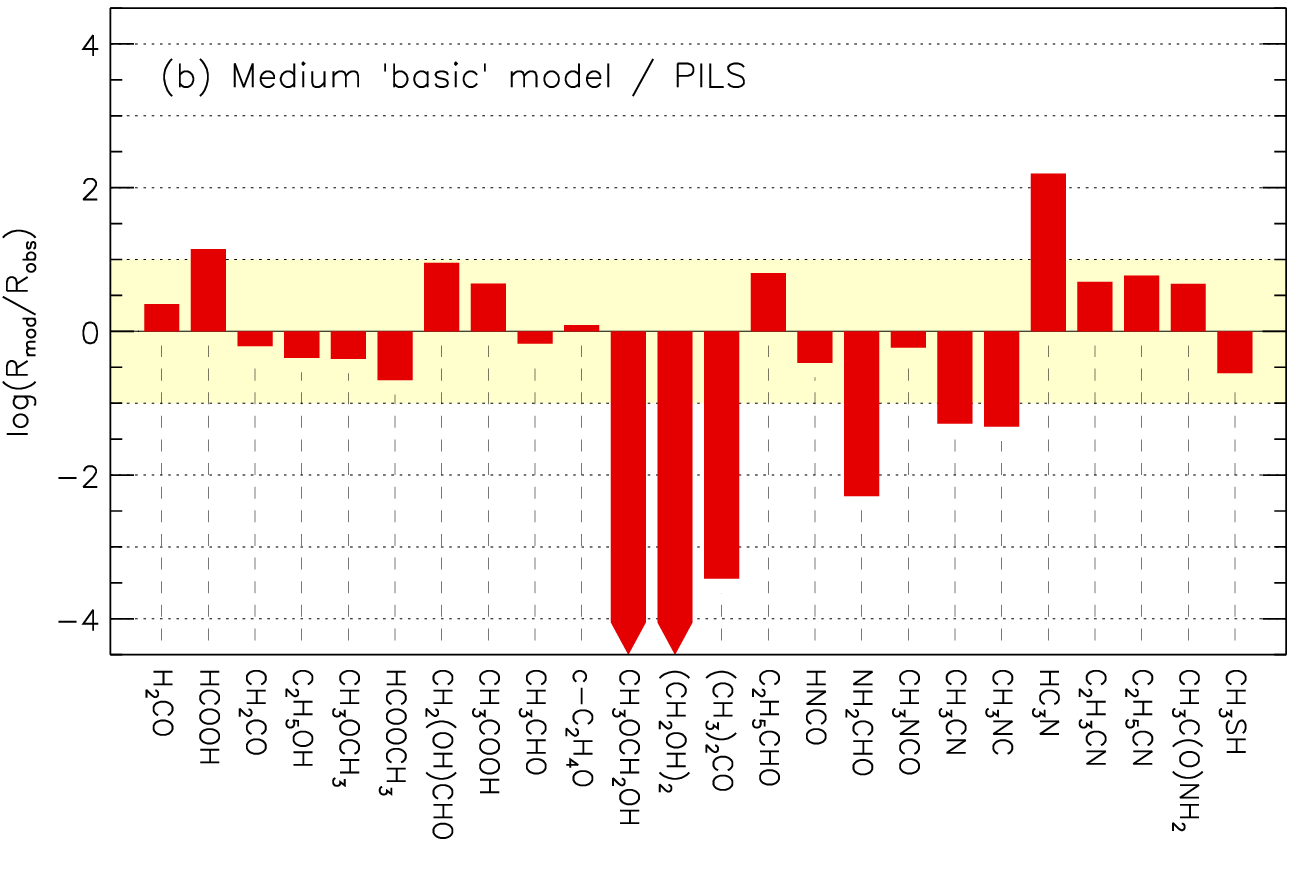}}
         {\includegraphics[width=0.5\hsize]{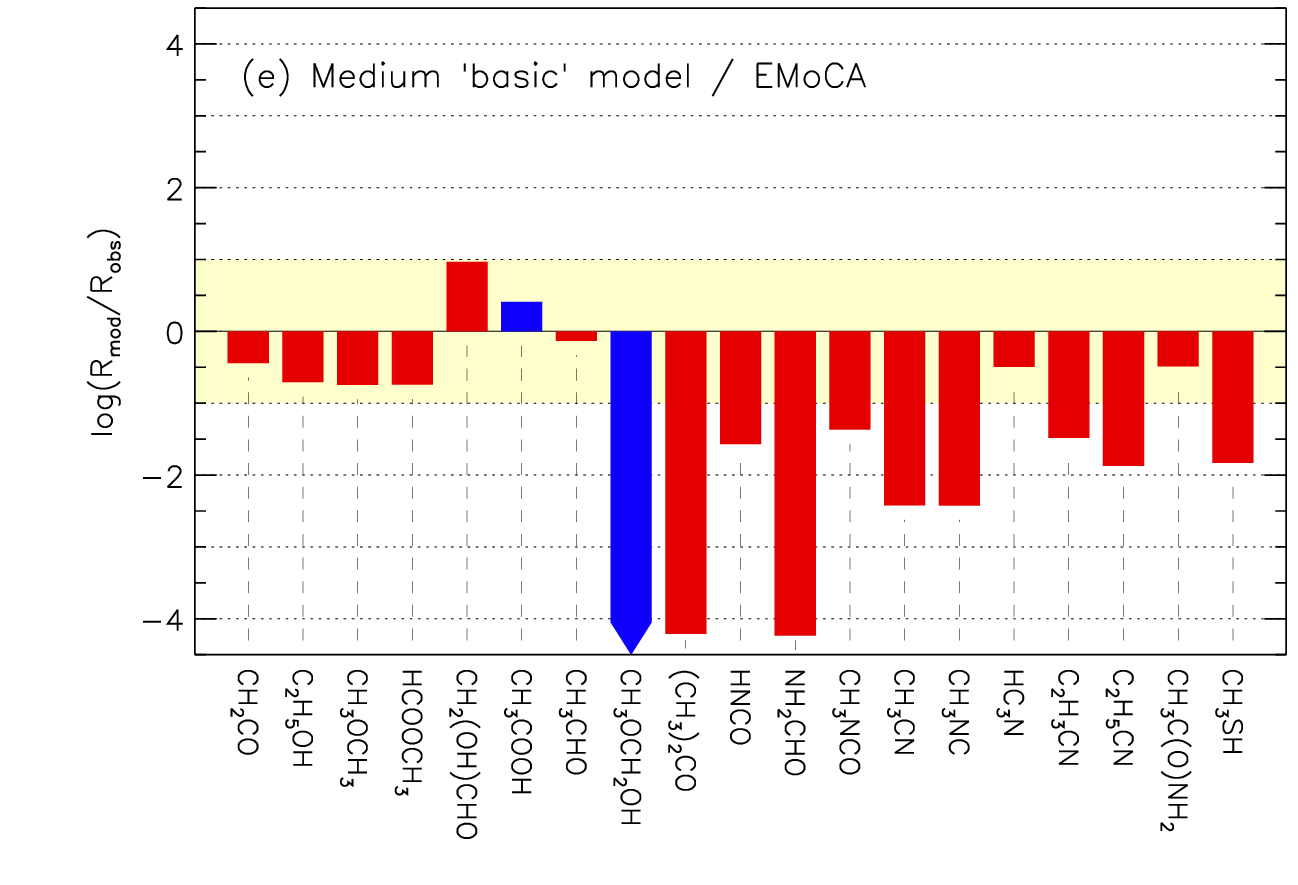}}
         {\includegraphics[width=0.5\hsize]{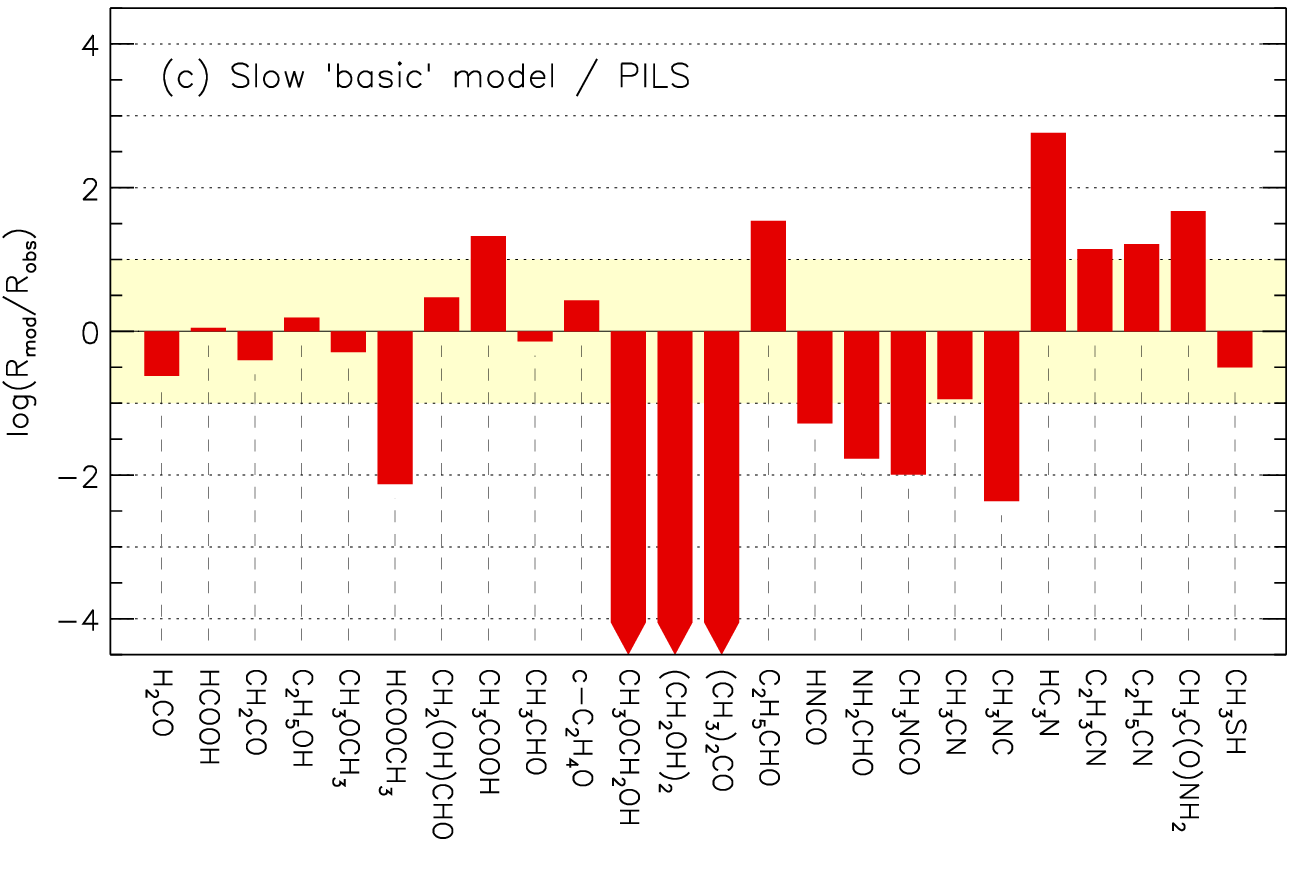}}
         {\includegraphics[width=0.5\hsize]{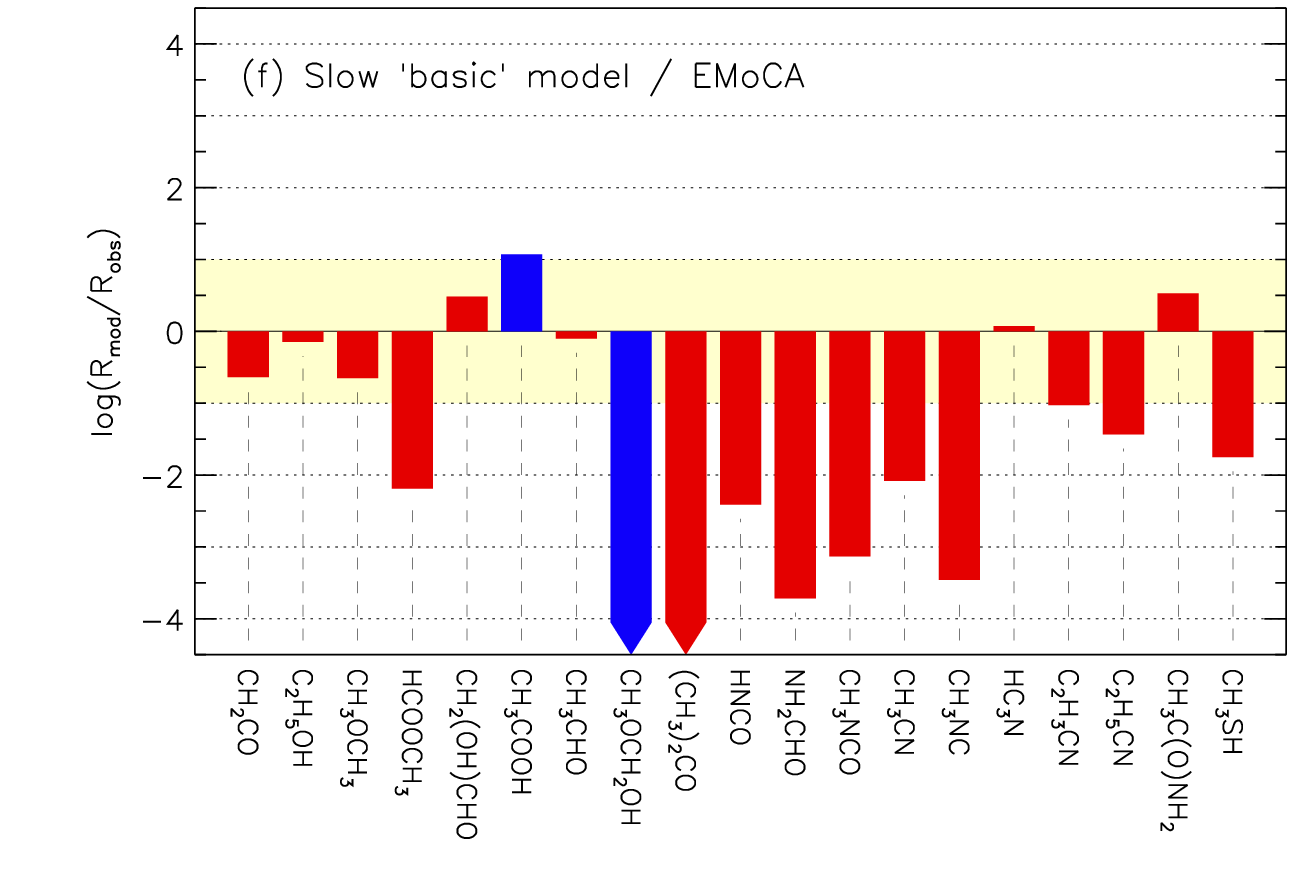}}
\caption{Comparison between the peak gas-phase molecular abundances achieved in the {\tt basic} models (normalized to the models' peak methanol abundances) and observational values of the same quantities; other details as per Fig.~\ref{compare-final}.}
\label{compare-basic}
\end{figure*}


Fig.~\ref{compare-final} plots the ratio of the modeled values to the observational values given in Table~\ref{tab:obs}. Bar charts are shown for each model/source combination. Model values correspond to the peak molecular fractional abundance with respect to peak methanol; observational values are based on corresponding column densities. The shaded background region indicates a model value within one order of magnitude of the observational value. Comparing the models with the IRAS 16293B data (left panels), most of the commonly observed hot-core molecules are reproduced comfortably within a factor 10 by all three models; an exception is methyl cyanide (CH$_3$CN). A few of the other molecules are also substantially over- or under-produced in all the models, e.g.~HC$_3$N.

Ethylene glycol, (CH$_2$OH)$_2$, varies strongly between models; the {\em fast} model provides a very good match with the IRAS 16293B observations, but the {\em slow} model does not match at all well. No observational value is available for Sgr B2(N2). 

Methoxymethanol, CH$_3$OCH$_2$OH, the structural isomer of ethylene glycol, is very well reproduced in all models, for both sources (see also Table \ref{tab:abuns}). In the case of Sgr B2(N2) only an upper limit is available, but the model values all fall somewhat below this limit (by a factor 3 in the {\em slow} model). This marks a substantial success for the new modeling approach; \citet{Mcguire17}, who first detected this species toward the high-mass source NGC 6334I, noted the inability of chemical models to predict its high abundance -- a failure of purely diffusive grain-surface chemistry. In the present work, methoxymethanol is produced almost exclusively in stage 1, during the second temperature regime, i.e.~the cold, late stages of collapse (see Table \ref{tab:rates}); production occurs through 3-B reactions between CH$_3$O and CH$_2$OH. The addition of H to formaldehyde to produce CH$_3$O, and the abstraction of H from methanol by atomic H to produce CH$_2$OH, make approximately equal contributions as the driving processes, while the fractional surface coverages of either radical are very similar during the moments of peak production (around 1\% coverage). Therefore, radiative processing -- either by UV or by possible direct cosmic-ray impacts -- seems not to be required to produce methoxymethanol; rather, it is formed as a by-product of the surface build-up of methanol itself.  \citep[However, see][whose experiments show that photochemistry in methanol ices can produce CH$_2$OH and CH$_3$O radicals, which go on to form methoxymethanol]{Schneider19}. Note that the present chemical network for methoxymethanol, unlike that for ethylene glycol, does not include a mechanism for H abstraction by H atoms; this is left to future work, which will require another substantial expansion of the chemical network.

We note that the modeled abundances of ethylene oxide, c-C$_2$H$_4$O, a structural isomer of acetaldehyde that was recently detected toward IRAS 16293B by \citet{Lykke17}, agree very well with the observational value. This species is formed exclusively on the grains through the addition of an O atom to ethylene (see Sec.~\ref{ethox}), and is produced substantially at low temperatures and early times. \citet{Lykke17} did not detect the other structural isomer in the group, vinyl alcohol (C$_2$H$_3$OH), obtaining an upper limit around three times less than the ethylene oxide abundance; our model is consistent with this value, producing an abundance roughly 4--10 times lower than ethylene oxide. All three structural isomers are therefore quite well reproduced by the model.

The right-hand panels of Fig.~\ref{compare-final} compare the models with the observational data for Sgr B2(N2). Again, most of the commonly-observed hot-core molecules are reproduced to within a factor 10 in all models. The match is much closer for the {\em slow} model results (barring the molecules that show generally poor agreement).

For comparison, Fig.~\ref{compare-basic} shows the observational match produced by the {\tt basic} models in the present work, which are comparable with past hot-core simulations using the {\em MAGICKAL} code with only diffusive grain-surface/ice-mantle mechanisms \citep[e.g.][]{Garrod13b}. Disagreement for the {\tt basic} models is more uniform across all the molecules, especially comparing with the Sgr B2(N2) data. As found by \citet{Bonfand19}, who used a far more detailed physical treatment for several sources within Sgr B2 core N, and by \citet{Barger20b}, who conducted coupled RHD/chemical simulations of generic hot cores, many COMs such as methyl formate are under-produced when only diffusive mechanisms are considered, while glycolaldehyde is uniformly overproduced.

Although methyl cyanide (CH$_3$CN) exhibits the same general trend as shown by various molecules, as discussed above, its absolute abundance is arguably on the low side, by around an order of magnitude, when comparing the {\em fast} model with IRAS 16293B and the {\em slow} model with Sgr B2(N2). Methyl isocyanide is under-produced yet further, while vinyl and ethyl cyanide are also under-produced. It is plausible that the over-production of HC$_3$N is related to the general under-production of the cyanide species; more efficient grain-surface hydrogenation of HC$_3$N at early times could certainly increase the vinyl and ethyl cyanide abundances. However, more detailed physical/dynamical treatments than those employed here would also be instructive in understanding the relationships between these molecules.

Methyl isocyanate (CH$_3$NCO) is also substantially under-produced here. We use the chemical network proposed by \citet{Belloche17}, who also considered associated mechanisms for the production of N-methyl formamide (CH$_3$NHCHO). Methyl isocyanate abundance was found by those authors to be dependent on grain-surface reaction mechanisms with poorly-defined activation-energy barriers; these may require re-assessment in light of the inclusion of nondiffusive chemistry in the models. Gas-phase mechanisms for CH$_3$NCO production that are not included in the present models are also plausible \citep{Halfen15}.

Methyl mercaptan (CH$_3$SH) is moderately well-reproduced in the models, although it is in much better agreement in the case of IRAS 16293B. However, the abundances of sulfur-bearing species are highly dependent on the initial elemental abundance of S, and the chemical form of most of this sulfur in dense regions is still uncertain \citep[see e.g.][]{Vidal18,Shingledecker20}.

\subsection{Methyl formate, glycolaldehyde and acetic acid ratios}

As demonstrated in Fig.~\ref{compare-basic}, which shows data from the {\tt basic} models, past treatments that have considered only diffusive chemistry have tended to produce wildly incorrect ratios between the structural isomers methyl formate (MF), glycolaldehyde (GA) and acetic acid (AA). In particular, the models tended to produce excessive quantities of glycolaldehyde, as the result of the preferential production of the CH$_2$OH radical via methanol photodissociation, in a ratio 5:1 with respect to CH$_3$O \citep{Oberg09}, which ratio is retained in the present models. The introduction of nondiffusive mechanisms for COM production, including three-body surface processes, reduces the influence of this parameter.

Based on the data in Table \ref{tab:obs}, IRAS 16293B has ratios MF:GA:AA = 93:12:1, while Sgr B2(N2) has MF:GA:AA = 60:7:1 (the latter using the upper limit for AA). We may compare these with the ratios produced by the models, based on the individual peak values for each molecule; 33:16:1 ({\em fast}),  32:18:1 ({\em medium}), and 27:7:1 ({\em slow}). Importantly, the modeled abundance ratios are seen to be in the correct sequence. The value for methyl formate may be a little low compared with the other two, although in fact the numbers shown in Table \ref{tab:obs} provide very good agreement with the MF:CH$_3$OH ratio. Rather, both the GA and AA values may be elevated, although not by much more than a factor of 2--3. Indeed, it may be asking too much for such a physically/dynamically simple model to reproduce too closely the exact observational ratios. It is well known that each of the three isomers have different binding energies, with MF by far the lowest \citep[see][whose values are used in these models]{Burke15}. The introduction of spatial temperature and density gradients into a multi-dimensional physical model of hot cores could plausibly alter the ratios of the three species by a factor of 2--3.

\citet{El-Abd19} studied these ratios on lines of sight toward the NGC 6334I complex, also collating other literature values. They found that, while the MF:AA ratio among various sources (low and high mass) was stable at roughly 10:1, there was bimodal behavior in the MF:GA ratio. The latter were classified as either ``small'', i.e.~in keeping with the roughly 10:1 ratio of MF:GA found in IRAS 16293B and Sgr B2(N2), as cited in Table \ref{tab:obs}; or ``large'', at least an order of magnitude higher, based mainly on upper limits to glycolaldehyde detection. The large ratios were exclusively associated with high-mass sources. While the models we present here show a slight bias toward larger MF:GA ratios under the {\em slow} warm-up conditions (which we crudely associate more with the high-mass source Sgr B2(N2) than with the low-mass IRAS 16293B), none of the models demonstrate the extreme ratios observed by \citet{El-Abd19}. However, based on the behavior of the models, we may propose that more extreme values of the MF:GA ratio could be caused by prolongued periods at dust temperatures in the $\sim$115--165~K regime. Under such conditions water ice strongly desorbs, exposing glycolaldehyde to destructive grain-surface reaction pathways prior to its own desorption at somewhat higher temperatures, while methyl formate desorbs rapidly with the water. It is yet unclear whether such a scenario could induce bimodal behavior, nor how strong the effect on the MF:GA ratio might be. Computational investigations of a broad physical parameter space using the new model are ongoing.
\\

\subsection{Amides and related species}\label{amides}

Two observed amides -- formamide (NH$_2$CHO) and acetamide (CH$_3$CONH$_2$) -- are listed in Table~\ref{tab:obs}, along with the related molecule isocyanic acid (HNCO). For the latter, the models seem to reproduce the observations reasonably well, although the IRAS 16293B value is a little lower than the lowest model value, while the Sgr B2(N2) value is a little higher than the highest in the models. Formamide, for which HNCO is sometimes considered a precursor, is also reproduced well for IRAS 16293B, but the Sgr B2(N2) value exceeds the model value by around a factor 15. Similarly, acetamide is well reproduced in the IRAS 16293B case, and less well in the Sgr B2(N2) case, although the disagreement is less than an order of magnitude. It is notable that all three of these species have a significant photochemical production mechanism available, which is especially strong in the {\em slow} warm-up timescale model. In the case of acetamide, production is dominated by CH$_3$ addition to the NH$_2$CO radical, which is produced either from H addition to HNCO or the reaction of NH$_2$ with CO in the bulk ice. Bulk reactions involving the radical CH$_3$CO also make a modest contribution. Each of these processes ultimately is driven by radical production via photodissociation of bulk ices. It is possible that allowing these mechanisms to operate for a longer period, or under a stronger cosmic-ray induced UV field, could result in greater production of all three species.

\citet{Ligterink20} presented ALMA observations of a selection of amides, as well as cyanamide (NH$_2$CN), for a dozen sight lines toward NGC 6334I. Table \ref{tab:amides} shows {\tt final} model results for the same molecules, normalized to NH$_2$CHO as per the observations. Minimum and maximum values toward NGC 6334I sources are also shown, which are not necessarily obtained on the same lines of sight for each molecule; a crude range is shown here to allow a simple comparison with the models. Modeled NH$_2$CN abundances fall comfortably within the range of the observations, although in fact most of the observational values toward NGC 6334I cluster close to the lower value. \citet{Coutens18} made the first detection of NH$_2$CN toward a solar-type star, IRAS 16293B, obtaining an even higher ratio with NH$_2$CHO of 0.20. The {\em fast} model achieves a peak value that falls an order of magnitude short of this. However, the present models are substantially more successful than those presented by Coutens et al., who considered only diffusive radical-addition mechanisms, which become efficient at elevated temperatures; those results were short by 2--3 orders of magnitude. In our models, as outlined in Sec.~\ref{cyanamide}, NH$_2$CN is formed very early on, through bulk-ice photochemistry driven by external UV photons, when the ice mantles are just beginning to build up. If this is indeed the only efficient mechanism for its formation (which is by no means certain), then high observational values could indicate a large degree of ice photoprocessing, at least during the relatively early stages of hot-core evolution. Values obtained toward Orion KL are yet greater \citep{White03} than those of IRAS 16293B.

\begin{deluxetable}{lcccccc}\label{tab:amides}
\tabletypesize{\footnotesize}
\tablecaption{Comparison of model results with observational data for sources in NGC 6334I, corresponding to the minimum and maximum values (or upper or lower limits) measured for any of the observed sources.}
\tablewidth{0pt}
\tablehead{ \colhead{}  &  \multicolumn{3}{c}{Observations} \\
\cline{2-4} 
\colhead{Molecule}  &  \colhead{Ref.} & \colhead{Minimum}  &  \colhead{Maximum}  &  \colhead{{\em Fast} model} &  \colhead{{\em Medium} model} &  \colhead{{\em Slow} model} }
\startdata
NH$_2$CN/NH$_2$CHO               & a & 6.8(-3) & 7.8(-2) & 2.2(-2) & 1.6(-2) & 9.5(-3) \\
HNCO/NH$_2$CHO                      & a &  1.0(0) &  6.1(0) & 4.6(0)  & 4.5(0) & 3.8(0) \\
CH$_3$CONH$_2$/NH$_2$CHO    & a & 5.4(-2) & 5.0(-1) & 1.9(-1) & 1.4(-1) & 7.8(-2) \\
CH$_3$NHCHO/NH$_2$CHO         & a & 8.5(-2) & $<$2.5(-1) & 1.4(-2) & 1.1(-2) & 4.6(-3) \\
NH$_2$CONH$_2$/NH$_2$CHO    & a & $<$3.7(-3) & $<$5.0(-2) & 4.6(-1) & 3.2(-1) & 1.7(-1)    \smallskip \smallskip  \\
CH$_3$NH$_2$/NH$_2$CHO        & b & 8.2(0) & $>$6.0(1) & 2.0(1) & 1.3(1) & 4.8(0) \\
CH$_3$NH$_2$/CH$_3$OH          & b & 5.4(-4)$^{c}$, 4.8(-4)$^{d}$ & 5.9(-3)$^{c}$, 2.5(-3)$^{d}$ & 3.7(-2) & 3.2(-2) & 3.1(-2) \\
CH$_3$NH$_2$/CH$_3$CN           & b & 5.4(-1)$^{e}$, 3.1(-1)$^{f}$ & 1.3(0)$^{e}$, 1.9(0)$^{f}$ & 2.3(2) & 4.1(1) & 3.6(0) \\
\enddata
\tablecomments{$A(B)=A \times 10^B$. $^{a}$\citet{Ligterink20}, $^{b}$\citet{Bogelund19}, $^c$based on $^{13}$CH$_3$OH, $^d$based on CH$_3$$^{18}$OH, $^e$based on $^{13}$CH$_3$CN, $^f$based on CH$_3$C$^{15}$N.}
\end{deluxetable}

Modeled fractional abundances of HNCO, acetamide and N-methyl formamide (CH$_3$NHCHO) with respect to formamide all seem to be in reasonable agreement with the \citet{Ligterink20} observations. In considerable disagreement is the abundance of urea (NH$_2$CONH$_2$), which is at least a factor of a few too high, and is indeed at least 46 times too high compared with the lowest upper limit for the NGC 6334I sight lines. \citet{Belloche19}, who made the first secure interstellar detection of urea, toward a position associated with Sgr B2(N1), also found urea to be overproduced in their models, by around an order of magnitude. Those authors again included only diffusive chemistry in their grain-surface treatment, but suggested that the efficiency of proton transfer from protonated urea to ammonia in the gas phase could affect the destruction rate of urea. The present model indeed includes proton transfer to ammonia for other species, while omitting this process for urea, based on its thermodynamic unfavorability. The explicit introduction of the reverse process, i.e.~proton transfer {\em to} urea {\em from} protonated ammonia might improve the agreement by providing an enhanced rate of destruction through electronic recombination.

Methylamine (CH$_3$NH$_2$), which is not included in Table~\ref{tab:obs}, has only rarely been detected in the ISM; until recently, it had only been securely identified toward Sgr B2 \citep{Kaifu74,Fourikis74}. In the past few years, several detections have been made, including ALMA observations toward NGC6334I by \citet{Bogelund19}. The results of B{\o}gelund et al. are included in Table~\ref{tab:amides}, with methylamine abundance shown in ratio to several other molecules. As noted by those authors, the models of \citet{Garrod13a} appeared to provide acceptable agreement with the observations, and the present models show good agreement with the CH$_3$NH$_2$/NH$_2$CHO ratios. Ratios of methylamine with CH$_3$OH and CH$_3$CN, on the other hand, are somewhat high, perhaps by as much as an order of magnitude or so, dependent on the model and the observational datapoint. \citet{Ligterink18} recently searched unsuccessfully for methylamine toward IRAS 16293B with ALMA. Their upper limit for the CH$_3$NH$_2$/NH$_2$CHO ratio was 0.053, which is at least two orders of magnitude lower than that of our models, and a factor of nearly 400 lower than our {\em fast} model (which we crudely identify as the closest general match for IRAS 16293B abundances). It is possible that methylamine is overproduced on the grains, through early-time photochemistry. However, the fact that methylamine is undetected toward IRAS 16293B while cyanamide -- which is formed in a similar manner in the models -- is especially abundant in that source would apparently rule out that possibility. Differences in gas-phase destruction could serve to differentiate the two molecules. As suggested for urea above, gas-phase proton transfer from protonated ammonia, followed by dissociative electronic recombination, could be an effective methylamine destruction mechanism that is not yet included in the models and which may not be available to cyanamide, based on its proton affinity.

Hydroxylamine (NH$_2$OH) was recently detected for the first time by \citet{Rivilla20}, toward a quiescent molecular cloud. However, this molecule has yet to be detected toward a star-forming source \citep{Pulliam12,Ligterink18}. Experimental work by \citet{Congiu12} suggests that this molecule may form on grain surfaces through the repetitive addition of H to NO \citep[see also the discussion of][]{Rivilla20}. \citet{Garrod13a} found that the inclusion of an effective H + HNO $\rightarrow$ H$_2$NO reaction on the grains ultimately overproduces gas-phase NH$_2$OH. Our {\tt final} model gives a peak fractional abundance that is slightly higher, at around a few 10$^{-6}n_{\mathrm{H}}$, using the same NO-related network. That network, however, may be lacking multiple destruction mechanisms for NO, HNO, and H$_2$NO, which could otherwise reduce the overall production of NO and related species. The addition of such reactions is beyond the scope of this work.

\section{Discussion}\label{disc}

A large number of additions/changes to the {\em MAGICKAL} astrochemical model have been explored here, but the most important is undoubtedly the combination of (i) the deactivation of bulk diffusion for the majority of solid-phase species with (ii) the adoption of nondiffusive chemical mechanisms for grain-surface and bulk chemistry. Comparing the new {\tt final} model with the {\tt basic} (which is very similar to past models that used purely diffusive grain chemistry), the peak gas-phase abundances of many COMs are perhaps not so different. However, the stage at which those molecules are formed (on the grains) has often either been advanced to very early times and low temperatures or pushed back to the stage when the ice mantles are strongly desorbing from the grains -- in particular, when water starts to desorb. These results stand in stark contrast to past models in which the production of typical COMs occurred primarily through diffusive reactions of radicals on the dust grains at intermediate ($\sim$20--50~K) temperatures. The latter mechanisms are relatively diminished in the new models, although nondiffusive photochemistry continues within the bulk ices at all temperatures, and radical-radical addition reactions are still the main pathways to COM formation.

Based on the new models, several distinct stages of COM production in hot cores may be identified, listed more or less in order of occurrence:

\begin{enumerate}
\item Early ice photoprocessing (via external UV)
\item Cold surface chemistry
\item Cosmic ray-induced photoprocessing
\item Hot surface chemistry (water-desorption stage)
\item Hot gas-phase chemistry
\end{enumerate}

(Note that these stages do not necessarily correspond to the temperature regimes identified in Sec.~\ref{results:temps}; some COM production stages occur within the same nominal temperature regime).

The first stage occurs as the ice mantles are just beginning to form, and corresponds to a translucent state (i.e.~low visual extinction) of the cloud/core that external UV photons may penetrate; the {\tt final} model begins at 3 magnitudes of visual extinction and a gas density of 3000 cm$^{-3}$. Several molecules are notably affected by this process, including ethanol, cyanamide, methylamine, and acetic acid. The production of the latter also leads to the formation of glycine, the simplest amino acid; indeed, most glycine production occurs at this early stage in the new models. As the collapse stage proceeds and visual extinction increases, the external UV processing weakens. This first stage of COM formation may be the most variable and/or uncertain within an individual source; without conducting chemical models that explicitly include the early dynamics of collapse as a function of position/extinction, one cannot be certain how widespread the effect of external UV photons on young ice mantles may be. 

The second stage of production is associated with the build-up of the ice mantles at low temperatures. While diffusive chemistry is responsible for the formation of the main, simple ice components, including water, ammonia, methane, and methanol (with CO mostly being accreted directly from the gas phase), nondiffusive reactions are operating in tandem to allow more complex species to form at the same time, albeit in lower abundance. COM formation becomes especially prevalent when the combined effects of falling dust temperature and reduced gas-phase atomic-H abundances lead to longer surface lifetimes for radicals, including those related to methanol formation, i.e.~HCO, CH$_3$O and CH$_2$OH. Longer lifetimes against reaction with diffusive atomic H lead to commensurately greater radical surface coverages. When other radicals are then formed in close proximity to those high-coverage reactants, an immediate three-body (3-B) reaction may occur, producing a complex organic molecule. This nondiffusive chemistry may be identified with that observed experimentally by \citet{Fedoseev15}, \citet{Chuang16} and others. The effectiveness of this broad COM production mechanism in the model is also enhanced by the adoption of a larger representative grain-surface diffusion barrier for atomic H, which increases radical lifetimes and surface coverage.

The third stage of COM production may in some sense be identified with the intermediate-temperature bulk-ice chemistry that was most responsible for COM formation in past models. As in those models, the production of COMs in the bulk ice in the {\tt final} model presented here is driven by photodissociation of stable molecules by the cosmic ray-induced secondary UV field. The nondiffusive photodissociation-induced (PDI) reaction mechanism allows the newly formed radicals or atoms in the bulk ice to find a reaction partner without diffusion, based on which reaction partners may be immediately present when the molecule dissociates. In the old models, bulk diffusion was required to mediate those reactions, which led to a diffusion-related temperature dependence in the bulk-ice chemistry that is somewhat removed with the switch-off of bulk diffusion and the inclusion of nondiffusive processes. The new models still allow H and H$_2$ to diffuse within the bulk ice, allowing atomic H especially to initiate diffusive chemical reactions that can also result in follow-on (3-B) reactions. As these H atoms are ultimately produced by molecular photodissociation, the bulk 3-B reactions are also the result of photochemistry. A full treatment of the bulk-ice photochemistry therefore requires both the PDI and 3-B mechanisms to be included in the model. Although the temperature dependence associated with most bulk diffusion is removed in the new models, the gradual increase in dust/ice temperature over time allows some activation energy barrier-mediated reactions to occur more rapidly (albeit through nondiffusive meetings). The production of certain species in the bulk ice can therefore ramp up or down as dust temperatures increase in the hot core.

The effect of the more advanced PDI2 treatment introduced in the present work is important to consider. By requiring the nondiffusive photoproducts of a single dissociation event to recombine with each other if neither finds an immediate reaction partner, the number of unreacted radicals present in the ice is reduced. This in turn reduces the influence of the photodissociation-induced reaction process as a whole. Photodissociation events that produce a radical and a hydrogen atom, meanwhile, are not subject to this requirement, which allows the H atom to escape (if it does not immediately react with some other species nearby). This H atom can recombine with any radicals it finds in the bulk ice, which also contributes to the reduction in unreacted radicals that could otherwise form COMs.

While the third stage of COM production is technically active all of the time that there are ice mantles present, it becomes important mainly at the intermediate times/temperatures between the formation of the mantles and their desorption. Its contribution is generally less than that of the two stages preceding it. However, some molecules, notably HNCO and NH$_2$CHO, show substantial production at these intermediate temperatures directly as the result of CR-induced UV photochemistry in the bulk ice. Furthermore, the results presented here have assumed a canonical value for the cosmic-ray ionization rate, of order $10^{-17}$ s$^{-1}$. Higher values could lead to much stronger photochemical production of COMs in the ices.

The fourth stage of COM production is largely a new feature of these models, and is related to the removal of bulk diffusion for most species; this leads to their trapping in the bulk ice, due to their inability to diffuse to the ice surface and desorb at their own preferred temperature. Molecules such as formaldehyde, which in past models would desorb at around 40~K, are largely retained within the ices until water -- as the main ice constituent -- begins to desorb. A range of radicals is also trapped until that time, but when the water is lost and new material beneath becomes the new surface, the radicals that are uncovered are free to diffuse on that surface. They can re-combine with each other to produce stable molecules, or they may attack other species -- notably those with low barriers against H-abstraction, such as aldehyde (-CHO) group-bearing species. The products of these reactions may desorb, or they may be of great enough binding energy to be retained on the grain/ice surface to higher temperatures. This process, though rarely dominant, in many cases makes a contribution to COM formation and destruction. It is also somewhat timescale-dependent, and the strength of its effect is not easy to disentangle from the incipient gas-phase chemistry occurring at around the same time, as chemical species released from the ices may also react in the gas; for this reason it should be noted that the COM production occuring during the water-desorption stage, as seen in Figs.~\ref{roc1}-\ref{roc2} and \ref{roc3}--\ref{roc6}, is not representative solely of grain-surface chemistry.

The last of the COM production stages is the traditional hot gas-phase chemistry, which occurs during and after the desorption of the ice mantles. Gas-phase mechanisms retain an influence on, for example, the production of dimethyl ether, as well as several cyanide species (see below). A number of relatively new gas-phase production mechanisms were also introduced to the {\em MAGICKAL} chemical network. The reaction of an oxygen atom with the radical CH$_3$OCH$_2$ to produce methyl formate, as proposed by \citet{Balucani15}, contributes as much as around 30\% of total MF production. Although not dominant in these models, this reaction is nevertheless important in determining the overall abundance ratio of MF to its structural isomers. The CH$_3$OCH$_2$ radical itself, however, is sourced from the dust grains directly, fed by the release of trapped radicals in the bulk ice. 

Gas-phase production of NH$_2$CHO through the reaction of NH$_2$ and formaldehyde was also incorporated into the new models, using the rate calculated by \citet{Skouteris17}. In our models, this mechanism contributes around 9\% of the total formamide production, with the rest coming from grain-surface and bulk-ice production.

\subsection{The comparability of models with observations}

Much of the discussion of COM abundances in the present work has related to peak values. 
The comparison of the present 0-D model results directly with observational data using the peak-value method, although well established in past work, leaves much to be desired. The models predict that different molecules may peak at different times/temperatures, which is often related to their desorption from the dust grains, followed by gas-phase destruction. COMs with grain-surface origins thus tend to peak in the gas phase when their own desorption from the grains is strongest.
However, the inclusion of proton-transfer reactions between gas-phase molecular ions and ammonia, as proposed by \citet{Taquet16}, has a strong influence on the survival lifetimes of COMs, extending them substantially by avoiding the destruction of complex organic ions through electronic recombination.
The avoidance of such a strong decline in gas-phase abundances over time may therefore mitigate the imprecision of comparing peak values, because high gas-phase abundances become less tied to particular times or temperatures. Nevertheless, truly meaningful comparison with observations, either generically or with reference to particular sources, can only be achieved using realistic physical/dynamical treatments that provide not only a final spatial profile for each molecule of interest but a physical history of each parcel of gas, so that the chemistry may be accurately traced over time. This is especially important to judge the influence of gas-phase chemistry on COM abundances, as the physics will determine the timescales and gas densities in which it may proceed. Past modeling efforts, such as those of \cite{Garrod13a}, \cite{Bonfand19} and \cite{Willis20}, have used observationally-tuned physical profiles to calculate molecular column densities and/or emission spectra. While such methods indeed allow direct comparison with observed quantities, their dynamical treatments are essentially empirical. Fully multi-dimensional dynamical treatments of hot cores should ultimately provide the best physical inputs for the chemical models and the most realistic physical profiles from which molecular column densities and spectra may be simulated. \cite{Barger20b} recently published a set of 1-D coupled RHD/chemical models of hot cores, using an older version of {\em MAGICKAL} (i.e.~without any of the new chemical features). Application of the present version of {\em MAGICKAL} to multi-dimensional dynamical simulations is now in progress. Such approaches will allow consideration of physical/dynamical effects such as protostellar radiation, turbulence, and dynamically driven spatial variations in molecular abundances.

The {\tt final} models presented here in fact show remarkably good agreement with observational values for a range of complex organic molecules detected toward Sgr B2(N2) and IRAS 16293B. The agreement is particularly strong between the {\em fast} model and IRAS 16293B, and the {\em slow} model and Sgr B2(N2). The same trend in relative ratios between the models and observations continues even where the absolute values are in relatively poor agreement.
This behavior may appear contrary to the expected shorter dynamical timescales involved in high- versus low-mass star formation, although the identification of longer {\em chemical} timescales with higher mass sources is not new; \citet{GWH08} found a similar trend in the agreement of their models with observations. They argued, following \citet{Aikawa08}, that the important timescale for COM production/survival relates to the time spent by a parcel of gas in a particular temperature regime, rather than the lifetime of the hot core overall. This thermal/chemical timescale should be dependent on the size of the hot-core region (i.e.~the size of the circumstellar disk in which much of the COM material should reside), and inversely dependent on the rate at which the parcel of gas moves through that structure. Each of these features should be influenced by the above-mentioned dynamical parameters that may be tested directly using more sophisticated physical models.

The overall dynamical timescale of the source will nevertheless provide an upper limit to the chemical timescale. In the high-mass case, the high UV flux from a nascent O-type star is expected to begin photoevaporating the circumstellar disk on a timescale of $\sim$10$^5$ yr \citep{ZY07} post formation of a ZAMS star. The {\em slow} warm-up timescale used in our models, which seems to agree best with Sgr B2(N2) data, progresses from 8 to 400~K over a period of 1.43 Myr, which is arguably too long even taking into account that much of this time corresponds to low temperatures. However, the model timescale that determines many of the differences between the {\em fast} and {\em slow} warm-up models is, again, not the overall timescale of the model from start to finish but rather the time spent in the temperature range around 100--200~K, corresponding crudely to the period when the ices are desorbing, and immediately after. This timescale lasts no more than a few 10$^5$ yr in the {\em slow} model, while the timescale required to reach the maximum of 400~K is far less important, as the peak abundances of many observable COMs in the {\em slow} model are achieved below a temperature of 200~K. The temperature behavior used in the models is in any case empirical; the trends in agreement of the different model timescales between the low- and high-mass cases should therefore be viewed in the context of the chemical models alone, rather than determining an accurate timescale for the overall evolution of the source. We note also that the peak density used in all the models is the same ($2 \times 10^8$ cm$^{-3}$); higher gas densities during the 100--200~K period would be expected to accelerate aspects of the gas-phase chemistry. Again, detailed dynamical models will help to alleviate some of these uncertainties.

\subsection{C$_2$H$_4$O$_2$ structural isomers}

A particular success in the comparison of the model results with observational data is the much improved reproduction of observed relative ratios of the C$_2$H$_4$O$_2$ structural isomers methyl formate, glycolaldehyde and acetic acid. A great challenge to past models was the overproduction of glycolaldehyde in particular, with arguably an underproduction of methyl formate. This situation arose largely due to the reliance on photoprocessing of methanol in the ice as the main production mechanism for both species. The models use product branching ratios for methanol ice photodissociation of CH$_2$OH:CH$_3$O:CH$_3$ = 5:1:1, as determined by experiment \citep{Oberg09}. Reaction of CH$_3$O or CH$_2$OH in the ice with an HCO radical then results in either MF or GA formation. The introduction of the nondiffusive, three-body (3-B) surface-reaction mechanism (which allows both species to be formed during, rather than after, the build-up of the ice mantles), combined with the reduced influence of pure photochemistry in the new models, greatly weakens the preference for GA formation over MF.

Production of either molecule via the grain-surface 3-B mechanism relies on sequential diffusive H-addition reactions that convert CO (originating in the gas phase) into grain-surface methanol. The hydrogenation of formaldehyde (H$_2$CO) in particular has a strong preference for the production of the CH$_3$O radical over the alternative CH$_2$OH; further reactions with the HCO radical thus also favor MF production. However, the CH$_2$OH radical can still be produced by H-abstraction from methanol by H atoms. The CH$_2$OH so formed may react with HCO to form GA, as well as reacting with other CH$_2$OH to produce ethylene glycol. 

Further to this, the newly-added reactions surrounding the entire glyoxal -- glycolaldehyde -- ethylene glycol system allow interconversion between each. In our model, repetitive H-abstraction from ethylene glycol (EG) on the grains is often the main net production route for glycolaldehyde; the EG is formed through CH$_2$OH + CH$_2$OH radical addition. This contrasts with the experiments of \citet{Chuang16}, which involved co-deposition of various H:CO:H$_2$CO:CH$_3$OH mixtures. Their results instead suggest GA as a precursor to EG; their comparison of the different setups indicated that hydrogenation of GA to form EG should be more efficient than H abstraction from methanol followed by the recombination of two CH$_2$OH radicals. 

Our chemical network now includes all of the reaction mechanisms suggested by \citet{Chuang16}, as well as the interconversion reactions of \citet{Simons20}. It is possible that the barrier treatments for one or more reactions in our network are inaccurate. However, the difference in laboratory versus interstellar (model) conditions should also be considered. In our models, at the moment when glycolaldehyde production on the grain surfaces reaches its peak ($\sim8 \times 10^5$ yr, see Fig.~\ref{roc1}), the rate at which GA is successfully attacked by mobile atomic H is around 50 times higher than the rate at which it is incorporated into the bulk ice (where it is protected from gas-phase H-atoms). Thus, a surface GA molecule will react with H atoms around 50 times before it is covered over with other material (the laboratory values are likely substantially lower). Although much of this H-related chemistry involves the abstraction and re-addition of atomic H (due to the low barrier to abstraction from the aldehyde group), some of the reactions inevitably lead to conversion to EG or glyoxal. The latter two species also experience a similar degree of processing by atomic H before being incorporated finally into the ice mantle. In absolute terms, this interconversion between all three species is not at all rapid; at this point in the model, the instantaneous timescale for the reaction of an individual GA molecule with a hydrogen atom is around 30 yr, while the timescale for that molecule to be incorporated into the ice mantle is roughly 1,800 yr. Due to the large number of hydrogen reactions experienced by each molecule, it is likely that the interconversion process is at least as important as the mechanism by which any of the three species of interest is initially formed. A direct comparison between models and experiments may therefore be of limited value.

The production of methyl formate at early times, versus glycolaldehyde, is also buoyed by our inclusion of a formalized treatment for the three-body excited formation mechanism (3-BEF) first considered by JG20. In this case, the energy of formation of a newly-formed CH$_3$O radical allows it to overcome the activation energy of reaction with a surface CO molecule, leading to the MF precursor radical CH$_3$OCO. No such opportunity is open to glycolaldehyde production, as CH$_2$OH production directly through H addition to formaldehyde has a high barrier and is thus quite rare. Production of CH$_2$OH through H-abstraction from methanol is not expected to produce sufficient energy to overcome the barrier to the reaction of CH$_2$OH with CO.

In fact, in the {\tt final} version of the model, only around 5-10\% of the production of MF at early times is the result of the 3-BEF mechanism. The model in which the 3-BEF mechanism was introduced ({\tt bas\_stk\_loPD\_H2\_T16\_no-bd\_3B3\_EF}) showed a substantial increase in MF production; however, further changes to the model, especially the increase in the atomic-H surface diffusion barrier, lengthen the lifetimes of radicals on the grains to the point that the HCO + CH$_3$O $\rightarrow$ HCOOCH$_3$ radical-addition reaction becomes highly efficient through the regular three-body mechanism, leaving the 3-BEF mechanism (via CH$_3$OCO) less influential. MF also continues to be produced at a moderate rate in the bulk ices at intermediate temperatures ($\sim$20--50~K), through the CH$_3$O + CO reaction. In this case, the formation and trapping of CH$_3$O in the bulk ice gives it ample time to react with the abundant CO molecule, in spite of a reaction barrier.

Ultimately, the modest divergence in MF and GA abundances produced through grain-surface and solid-phase chemistry is widened during the water-desorption stage, as GA undergoes net destruction via surface radical reactions, while MF desorbs rapidly to avoid this fate. A degree of additional MF production is caused by radical reactions on the warm grain/ice surface, but much of the production at that stage is due to the reaction of gas-phase O with the CH$_3$OCH$_2$ radical.

Acetic acid (AA) is produced very early in the models, driven initially by photochemistry in the bulk ice caused by external UV photons, and then by three-body reactions on the grain surfaces at very low temperatures and high gas densities. However, in both cases, the precursor radical CH$_3$CO is produced as the result of a three-body reaction of either CH$_2$ or CH$_3$ with CO. Both of the latter processes involve an activation energy barrier, but are made more efficient by the availability of the chemical energy released by the preceding reaction to overcome the barrier, or by the two species being locked together in the ice, unable to diffuse away, such that they must inevitably react, even if after a long delay. The very early production of AA differentiates it from the other two structural isomers, which have much more in common with each other. AA is arguably a little higher in peak abundance than it needs to be to agree with observational ratios; a slightly weaker external UV-induced bulk-ice photochemistry at early times might explain this, while the molecule would still be formed through cold surface chemistry at later times.

Thus, a range of processes contribute to the final ratios of MF:GA:AA, and there may be room for further differentiation with the adoption of alternative physical conditions at the early/cold stages (e.g.~choice of initial visual extinction and/or minimum dust temperature), or the use of augmented cosmic-ray ionization rates that would encourage photochemistry over shorter periods.

\subsection{Nitrogen- and oxygen-bearing molecules}

While the gas-phase abundances of O-bearing COMs during the proper, hot-core stage of the models are strongly representative of low-temperature dust-grain chemistry, it is apparent that certain N-bearing molecules, especially the cyanides, are strongly enhanced by gas-phase chemistry at high temperatures (i.e.~$>$100~K). The greatest gas-phase contributions to HCN and CH$_3$CN production come at very high temperatures ($\gtrsim$300~K), but substantial production can also occur in the gas phase at temperatures around 120--160~K, when the ice-mantle material is in the process of being lost to the gas. In the models with longer warm-up timescales, the period of time spent in that temperature range is greater, leading to greater cyanide production. This ceases when the mantle desorption reaches completion, as the arrival of a large amount of new material into the gas opens up alternative chemical pathways that compete with HCN and CH$_3$CN production. One of the beneficiaries of this changeover in the gas-phase chemistry is in fact vinyl cyanide, which grows in abundance as the mantle-desorption period ends. Thus, the models indicate that HCN, CH$_3$CN and C$_2$H$_3$CN production is highly dependent on gas-phase mechanisms, and that longer warm-up timescales -- or, more accurately, longer post-100~K timescales -- enhance the abundances of those species.

This reliance on gas-phase production may help to explain the difference in behavior shown by O- and N-bearing molecules when comparing the low-mass source IRAS 16293B and the high-mass source Sgr B2(N2). The latter shows abundances of N-bearing species with respect to methanol that are at least an order of magnitude greater than those of IRAS 16293B, while many O-bearing species appear very similar between the two sources. In comparing the models with observations, we have identified our {\em slow} warm-up model most closely with Sgr B2(N2) and the {\em fast} model with IRAS 16293B, and the trends for the various nitrogen-bearing molecules seem to agree with this designation. We may venture, then, that differences in nitrogen-bearing species between sources relate to the length of time available for gas-phase chemistry at high temperatures. The different warm-up timescale models all use the same gas density, and this parameter could also affect the degree of cyanide production; however, the rate of desorption of the precursor species should be the most important rate affecting production, and this is a function of dust temperature and not gas density.

Because there is no gas-phase production mechanism for ethyl cyanide in our model, the variation in the abundance of this molecule produced by the different warm-up timescales is limited to around a factor 2. The inclusion of a gas-phase mechanism directly linking ethyl cyanide production to that of the other cyanides could enhance the effect seen in the models. A candidate would be the radiative association reaction CH$_3^+$ + CH$_3$CN $\rightarrow$ C$_2$H$_5$CNH$^+$ + h$\nu$, which according to \citet{McEwan89} may be fast at the $>$100~K temperatures at which the other cyanide chemistry is operative.

The oxygen-related COM chemistry, due to its greater activity during the shared collapse stage of the models, naturally shows less variation between the different warm-up model timescales. The coupling of the chemical models with detailed dynamical treatments that are specifically tailored to low- versus high-mass star formation may be useful in distinguishing differences in the early-time production of COMs in either case.

The behavior of HNCO and NH$_2$CHO in the models is particularly interesting. As noted by other authors \citep[see][]{Jorgensen20}, these two molecules appear to be observationally very closely linked. However, it is unclear whether this relationship is due to a direct chemical link or to a similarity between parallel production mechanisms. 
Experimental evidence disagrees as to whether HNCO may be directly hydrogenated on grain surfaces all the way to NH$_2$CHO \citep{Noble15,Haupa19}. But the models presented here indicate that parallel production is perfectly plausible; the time-dependent production rates for the two molecules look similar in all models, with a cold formation mechanism on grain surfaces during the collapse stage, and substantial photochemical production in the bulk ices at intermediate temperatures. Our chemical network does not include a mechanism for repetitive hydrogenation of HNCO all the way to NH$_2$CHO (although reactions involving other radicals allow the donation of an H-atom to the NH$_2$CO radical to produce formamide; see Sec.~\ref{app2}). Rather, the reactions of NH and CO (for HNCO), NH$_2$ and HCO, NH and HCO and NH$_2$ and H$_2$CO (for NH$_2$CHO) share reactants and precursors with the same origins (gas-phase CO, N and H) and their reactions occur in similar ways at similar times.

\subsection{Methylene and methylidyne chemistry}

Grain-surface reactions involving ground-state (triplet) methylene (CH$_2$) were incorporated into the chemical network as a test for the production mechanism for dimethyl ether (DME) proposed by JG20; reaction of CH$_2$ with methanol (CH$_3$OH) was suggested as a means by which DME could be formed on grain surfaces during the build-up of the ice mantles, allowing some fraction of it to be desorbed into the gas through a non-thermal mechanism. Using literature activation energies for the methylene--methanol reactions (Table \ref{CH2-rxns}), this process appears to be very effective, not just in producing DME but ethanol also. These reactions contribute the most during ice mantle build-up, while gas-phase atomic C is still abundant and the grains are warm enough for CH$_2$ to be somewhat mobile. In the case of ethanol, the effect is drowned out by the early-time photochemistry in the bulk ice involving OH and hydrocarbon radicals. Other species, notably acetaldehyde (CH$_3$CHO), are also affected by the inclusion of methylene reactions. The efficiency of these processes in forming COMs is nevertheless uncertain, as they are a two-step process of H-abstraction followed by radical recombination that may have multiple product branches. The inclusion of excited (singlet) methylene in the network could also be important, as this form of CH$_2$ may react directly and without a barrier in many cases.

Methylidyne (CH) may also react with stable species, often without a barrier. A number of such reactions were included in the network (see below, in the context of glycine). However, the inclusion of the barrierless reaction of atomic C with H$_2$ to a large extent allows CH to be passed over as an intermediary to methane formation. The influence of those reactions is diminished as a result. 

The surface association reaction between C and H$_2$ to form CH$_2$ on amorphous solid water (ASW), the standard ice used in studies of interstellar ice chemistry, has not been studied directly.  Quantum chemistry calculations on the gas-phase problem \citep{Harding93} 
and a number of experiments, including recent studies using liquid He droplets \citep{Kras16, Henning19}, have not reached a definitive result concerning the barrier or lack of one on ASW in particular.  We assume here that the reaction on the grain/ice surface is barrierless, since much of the evidence is in favor of this view; see also \citet{Sim20}, who carried out an astrochemical modeling study focused on the C + H$_2$ association reaction.

A greater focus on reactions involving all hydrogenation states of carbon, both in the chemical networks and in the laboratory, may be important to our understanding of COM production. Similarly, the networks include only a relatively sparse chemistry for the CN radical. This species is highly reactive and could therefore initiate more grain-surface/ice chemistry than the present networks might indicate.

\subsection{Glycine formation}

The models presented here use essentially the same chemical network for glycine -- the simplest amino acid -- as that of \citet{Garrod13a}, aside from the addition of the CH + NH$_3$ reaction \citep{Blitz12}. The recombination of several different radicals provides three main pathways to glycine formation on the dust grains. The \citet{Garrod13a} models suggested peak glycine fractional abundances with respect to total H of around $8 \times10^{-11}$ -- $8 \times10^{-9}$, dependent on model timescale; here, our {\tt final} model produces values of a few $10^{-10}$ for all timescales. This is due to the shift in production to much earlier times in the models, when the ices are beginning to form and external UV is capable of driving photochemistry in the ice. Thus, while this suggests a greater uniformity of glycine production, the strength of the early photochemistry that produces it is intrinsically dependent on the initial physical conditions assumed in the collapse model, which are shared by all the warm-up models.

\citet{Ioppolo20} conducted laboratory experiments to explore the possibility of cold glycine formation on dust grains, without the need for radiative processing; they found that glycine could be readily produced by reactions between CH$_2$NH$_2$ and COOH radicals, which is one of the pathways used in the \citet{Garrod13a} network. The associated models presented by \citet{Ioppolo20} concentrated also on non-radiative mechanisms, and thus did not include photochemistry of any kind in the bulk ice mantles. Those models indeed seemed to agree with the associated experimental picture of glycine production during the early stages of ice formation, fed by the reaction of CH with ammonia to produce the radical CH$_2$NH$_2$. Although that route continues in our present models, the removal of the barrier to the reaction C + H$_2$ $\rightarrow$ CH$_2$ on the grains allows CH production to be skipped over, as molecular hydrogen is far more abundant than atomic H at all times in the models. The various models of \citet{Ioppolo20} either did not include the C + H$_2$ reaction or assumed a high estimated activation energy barrier of 2500~K. 

Production of glycine through the CH$_2$NH$_2$ + COOH mechanism is also reduced in the {\tt final} model because of the lower efficiency assumed for COOH production from the surface reaction of CO and OH, which is necessary to ensure that formic acid abundance agrees with observations. The models still lack any independently determined value, based either on measurement or calculation, to describe the efficiency of stable COOH production from excited COOH$^*$ on dust grains, versus CO$_2$ + H.

The early, external UV-driven photochemical route to glycine depends on the dissociation of solid-phase acetic acid and the subsequent addition of NH$_2$ to the CH$_2$COOH radical. If this mechanism were taken away, for example as the result of the ices forming under higher visual extinction conditions, then the total production of glycine would be reduced by a factor around 3--10, all else being unchanged. The dominant production mechanism would in that case shift to the high-temperature release of the CH$_2$NH$_2$ and COOH radicals trapped in the ice, which would rapidly recombine on the hot dust/ice surface shortly prior to desorption.

Overall, the production of glycine in these models is perhaps surprisingly weak in comparison with other COMs, especially when considered in the context of the detection of abundant glycine in the comet 67P \citep{Altwegg16}, although the production of glycine in the earliest, deepest layers of the ice in our model seems in qualitative agreement with the strong association between glycine and dust particles in 67P \citep{Hadraoui19}. Our low overall fractional abundance of glycine is nevertheless in keeping with the current lack of an interstellar detection of this molecule. As noted by \citet{Garrod13a}, the high desorption temperature of glycine would make its emission region relatively compact compared with other COMs. The temperature at which glycine reaches its peak abundance is also a little higher in the new models, due to the somewhat later release of all material from the grains, which is caused by the removal of the bulk diffusion mechanism.

\subsection{Effects of the removal of bulk diffusion from the model}

A major change to our model setup is the removal of the diffusion process for all species within the bulk-ice mantle other than H and H$_2$, which in turn eliminates bulk-ice chemistry that would be driven by the diffusion of those species. Now that nondiffusive reactions are included in the model, reactions within the bulk may nevertheless occur without the need for diffusion. However, the removal of bulk diffusion also prevents molecules in the ice mantle from diffusing toward the surface of the ice, whence they would otherwise be able to desorb into the gas phase. In past models (as well as, e.g., the {\tt basic} model presented here), the warm-up of the icy dust grains resulted in an increasingly rapid bulk-diffusion rate toward the surface, such that by the time the sublimation temperature of a particular species (based on its binding energy) was reached, it could freely diffuse to the surface and desorb, regardless of its initial position. This allowed all molecules in the ice mantles to desorb precisely at their own sublimation temperatures, without any trapping. For example, all CO molecules in the ice could desorb at $\sim$25~K (Fig.~\ref{basic_medium}) with none retained on the grains.

The removal of bulk diffusion in the new models leads instead to almost precisely the opposite behavior; volatiles such as CO mostly remain trapped in the ice, beyond their nominal sublimation temperatures, because they only reach the surface layer once other material above them has itself desorbed. Thus, only around 1\% of CO, CH$_4$, and other small molecules are released into the gas phase at those low temperatures (e.g.~Fig.~\ref{final_medium}), corresponding mainly to the loss of molecules present in the upper ice layer. Some fraction of all species is retained on the grains until water itself desorbs strongly, while a small amount of water may be trapped in the remaining ice by molecules with yet higher binding energies. The removal of bulk diffusion, combined with the presence of a broad mixture of solid-phase species of various binding energies, thus broadens the range of temperatures over which desorption of any one species may occur, shifting the moment of {\em complete} desorption to somewhat later times than in past models.

When considered purely from the point of view of desorption, the two bulk-diffusion approaches thus lead either to minimal trapping of volatiles, or to extreme trapping. It could be argued that these outcomes may be compared with experiment to determine the best modeling approach. In fact, temperature-programmed desorption (TPD) experiments using mixtures of volatile molecules with water commonly demonstrate behavior somewhere between the two extremes; TPD experiments by \citet{Collings04} show that co-deposited CO:H$_2$O mixtures in the ratio 1:20 (105 ML total) result in substantial release of CO at low temperatures, with some remaining trapped in the bulk ice, to be finally released via the ``volcano'' desorption process (i.e.~water-ice phase change-induced desorption) and by co-desorption with water itself at a slightly higher temperature. The follow-up modeling paper by \citet{Viti04} parameterized the degree of loss/trapping at each stage, suggesting that 19.5\% of CO should be retained up to the volcano stage. Those authors further categorized various chemical species according to their measured or expected desorption kinetics, for use in astrochemical models. While such categorizations are useful for models focused on gas-phase chemistry, imposing empirical desorption behavior on more detailed gas-grain chemical models such as {\em MAGICKAL} is challenging.

Furthermore, while experimental evidence is invaluable to our understanding of interstellar ice composition and structure, specific outcomes from the laboratory cannot necessarily be extrapolated directly to astrophysical timescales or conditions. A major area of uncertainty in both the experimental and astrophysical regimes is the influence of ice porosity. As shown by e.g.~\citet{Dohnalek03}, non-perpendicular deposition angles lead to substantial porosity in water ice. Although the effect is most extreme for very oblique angles, those authors show that a 45 degree angle produces porosity as high as 18\% (their Fig. 7, based on thickness measurements) in $\sim$50 ML ice films. Background deposition of water (i.e.~ambient gas-phase deposition) produces around 33\% porosity (based on refractive index). It should not be a surprise, therefore, that co-deposited binary mixtures could also show substantial amounts of porosity that would affect the desorption kinetics in TPD experiments. Unfortunately, the degree of porosity in experimental mixed ices is rarely determined, and the details of the deposition that might indirectly indicate the degree of porosity are not frequently described. We note that the fraction of CO lost versus trapped in the experiments by \citet{Collings04}, using background deposition, is quite similar whether the CO is co-deposited with water or whether it is deposited onto a pre-formed water ice. This indicates that much of the low-temperature loss of CO corresponds to the desorption of molecules temporarily held in pores produced when the water-ice structure itself is formed via direct deposition.

This large degree of porosity in experimental ices may differ greatly from that of interstellar ices, the majority of which are expected to be formed through chemical activity on the grain/ice surfaces, rather than direct deposition of stable molecules from the gas phase (although the influence of direct deposition in protoplanetary disks, for example, may be rather more important than in the purely interstellar build-up of ices that we consider here). The formation of water ices in particular is understood to occur through the adsorption of gas-phase atomic oxygen, followed by hydrogenation reactions that can involve both atomic and molecular hydrogen \citep[e.g.][]{Cuppen17}, although lesser reaction pathways involving O$_2$ and related species can also produce water. Chemical models of interstellar ice formation by \citet{Garrod13b} show that the ice mantles formed under conditions appropriate to cold, dark clouds should be largely non-porous, due to the ability of oxygen atoms to diffuse on the surface to find the strongest binding sites, prior to their conversion into water. There have been no observational detections of porous interstellar water ice to date \citep[e.g.][]{Keane01}.

In another laboratory study, \citet{Fayolle11} explicitly deposited their ices perpendicular to the surface, ensuring that the porosity should be quite low. Although they only investigated CO:H$_2$O ices with thicknesses up to 25 ML, by varying the ice thickness they found that low-temperature desorption of CO and CO$_2$ from mixed ices of modest porosity is limited to a fixed number of monolayers close to the surface -- whether that be caused by direct loss of CO or CO$_2$ from pores, or whether due to weaker binding leading to easier rearrangement that could allow material that would otherwise be trapped near the surface to desorb. They found that 24\% of CO was trapped in a 20 ML-thick ice of 5:1 H$_2$O:CO mixture (comparable to interstellar ratios). For CO$_2$, 53\% was trapped in an 18 ML-thick ice, and 64\% in 32 ML of identical CO$_2$:H$_2$O ratio. In a different setup, they adopted the same 1:20 CO$_2$:H$_2$O mixture as \citet{Collings04} to achieve greater than 95\% trapping, in agreement with that study. Indeed, as per \citet{Collings04} and \citet{Viti04}, only a limited selection of small, volatile molecules (i.e.~CO, N$_2$, O$_2$, CS, NO, CH$_4$) should be able to achieve substantial low-temperature desorption from water-ice mixtures of thicknesses on the order of 100 ML, indicating either that the pores are small in size, or that they become closed off or collapse before the larger, more strongly-bound species can desorb \citep{Collings03}. The very low-porosity ices expected to be formed in the ISM via surface chemistry could plausibly allow even less volatile desorption at temperatures below the sublimation temperature of water.

Aside from those volatile species listed, most other molecules (including COMs) should remain on the grains until high temperatures are achieved; our {\tt final} model produces just such a result, which is not the case if bulk diffusion is included, although our model does not distinguish volcano desorption specifically. Even formaldehyde (H$_2$CO), which in our past models was released easily at low temperatures from the grains, now shows behavior in agreement with the analysis of \citet{Collings04} and \citet{Viti04}, who classify this species as water-like, i.e.~showing only co-desorption with water.

The retention of CO ice up to high temperatures in these models might seem at odds with, for example, the existence of snowlines in protoplanetary disks. We note, as mentioned above, that the possibility of desorption and gradual, temperature-dependent re-deposition of dust-grain ice mantles in such environments would allow for a much greater degree of stratification and probably water-ice porosity than may be expected for ice mantles formed under more typical interstellar conditions. Our models of hot core chemistry are not intended for protoplanetary applications, although they might be made more applicable by the consideration of, for example, multiple layers of bulk ice material versus just one (see Sec.~\ref{alternative}), allowing more CO to occupy explicitly the upper ice layers.

The chemical models presented here, and indeed most such gas-grain simulations, are clearly not capable in their present form of dealing appropriately with behavior associated with porosity (including any effects that phase changes might have on ice structure and/or trapping), to whatever degree it affects interstellar ices. But the (past) inclusion of bulk diffusion in the models should not be viewed as a potential solution, and indeed may lead to less realistic desorption behavior for all molecules other than the small set of volatile species that can desorb at low temperatures. As noted by \citet{Theule20}, any genuine bulk diffusion should be much slower than that implied by the experiments that purport to measure it, indicating that another mechanism is at work, i.e.~surface diffusion or ``percolation'' through pores. Bulk diffusion is in some sense functionally similar to the kind of pore-surface diffusion that would lead to the exodus of volatile species from pores, but its implementation in the models implicitly allows material at any location in the bulk ice to be affected, producing potentially misleading behavior. Until the degree and effect of porosity in laboratory ices is better understood at a microscopic, mechanistic level, attempts to include porosity in astrochemical models may be of limited value. 

Turning to the question of chemical activity in the bulk ices, it should also be noted that the past inclusion of bulk diffusion in the model by necessity implies long-distance mobility within the bulk ices -- it cannot be separated from short hops or slight rearrangements on a local level. Because these types of rate-based treatment cannot distinguish the positions of reactive species, the formulation used for diffusive reactions assumes that every reactant can be visited by another on some finite timescale, and that no chemical species located within the bulk is out of bounds. Furthermore, as temperatures increase toward the sublimation temperature of any particular species, the ability of some bulk-ice molecule to diffuse, and thus react, within the bulk ice becomes so great that in fact it would (in this crude formulation) meet with every single bulk-ice reactant over a short period of time. We do not believe that this is a realistic picture of bulk-ice reactivity.

Small degrees of bulk diffusion, perhaps involving excited photoproducts \citep[e.g.][]{Butscher16} or minor rearrangements of local ice structures, may nevertheless play a role in bulk chemistry. However, the diffusive description of reactivity in models such as {\em MAGICKAL} will not admit that constraint; limited, short-range diffusion is difficult (arguably impossible) to incorporate into models such as ours, which implicitly assume uniform diffusion rates regardless of the siting of individual atoms and molecules, as described above. Microscopic Monte Carlo kinetics methods might allow a more meaningful investigation of such effects, by allowing the positions of all atoms and molecules to be tracked at all times.

The influence of thermal reactions in the bulk ices (i.e.~activation energy barrier-mediated reactions typically between stable molecules) could be significant in specific cases. For example, the formation of ammonium carbamate (NH$_4^+$NH$_2$COO$^-$) and carbamic acid (NH$_2$COOH) in NH$_3$:CO$_2$ mixed ices has been shown experimentally to proceed through heating, without necessarily requiring energetic processing \citep{Bossa08}. Both reactants are common interstellar ice constituents, and therefore should have a high probability of becoming located contiguously as the ice mantles build up, with no bulk diffusion required. However, the accurate incorporation of these spontaneous, but often slow, reactions into typical astrochemical models is non-trivial; the degree of mixing between reactive species (i.e.~the number of contiguous reactive pairs) is critical to the total production rate. This mixing parameter is not a quantity that most models naturally treat in a meaningful way -- rather, they consider only the fractional population of each species in the ice. This mixing may be crudely approximated using the product of the two fractional populations, but in cases where the reactants are not the main ice constituent (e.g.~a water-dominated ice), the approximation could strongly diverge from the true value over time, as the reactions reduce the number of pairs to leave a substantial fraction of each reactant spatially separated from a reaction partner. A similar technical problem presents itself in the interesting case of reactions between radicals in the bulk ice that require some reorientation barrier to be overcome before reaction may proceed \citep{Martin-Domenech20}. 

In short, bulk-ice reactions that occur spontaneously, with no other process initiating the arrival of the reactants (such as diffusion, photodissociation, cosmic-ray impact, etc.), will not be well reproduced by typical population-level rate-based astrochemical models, whether or not they include the nondiffusive treatments described here and by JG20. Modeling these kinds of reactions demands a treatment that can explicitly account for the siting of the reactants, or that at minimum treats the number of contiguous pairs as a time-dependent quantity to be calculated along with the regular fractional abundances. Microscopic Monte Carlo kinetics models are well suited to this task; however, the much greater computational demands involved mean that their application to hot-core chemistry is not a likely prospect in the near-term.

\subsection{Alternative COM production mechanisms}
\label{alternative}

Although we have attempted to produce as comprehensive as possible a model of gas-grain chemistry in hot cores, some plausible mechanisms for COM production are necessarily omitted. These include the chemical effects of direct cosmic-ray impacts into the grains or their ices, which can result in heating, local and generalized molecular desorption, and/or radiolysis \citep[i.e.~ionization-driven chemistry;][]{Kalvans15,Shingledecker17,Dartois18,Shingledecker18a,Shingledecker18b,Arumainayagam19}. 
However, we note that cosmic-ray collisions with microscopic dust-grains are highly stochastic events. For this reason, it is unclear how accurately their effects may be incorporated into population-level chemical kinetics models such as {\em MAGICKAL}, as opposed to microscopically-exact treatments that do not require rates to be averaged over time or over the entire grain population.  

Furthermore, the cosmic ray-induced radiolysis models of \citet{Shingledecker18b} rely to some extent on the prior production of radicals in the ices by UV-driven dissociation mechanisms, in spite of cosmic-ray penetration/ionization providing the driving rate for COM production in those models. This means that the ices are already primed for chemistry through UV production of radicals, in advance of the impingement of a cosmic ray. The adoption in those models of the nondiffusive mechanisms that we propose here, especially the PDI2 treatment, would presumably have the effect of (i) reducing the prevalence of pre-existing radicals (just as it does in our models) and thus reducing radiolysis-induced COM production, while (ii) allowing the production of COMs directly as a result of photon irradiation. Thus, it is unclear how strong the relative influence of radiolysis would be when modeled in tandem with all nondiffusive ice processes.

As noted by \citet{Schneider19}, while photolysis may induce radical recombination chemistry, radiolysis induced by ionizing photons (versus cosmic rays) may also play a role in forming COMs. In the present models we do not consider any ionization process in the ice, either explicitly or implicitly, although recent modeling studies of laboratory experiments have begun to do so \citep{Mullikin21}. We propose that such models would be improved by the adoption of the nondiffusive reaction rates presented here.

As in most other models, the dust-grain treatment used here adopts only a single, representative grain size of 0.1~$\mu$m. The influence of a distribution of grain sizes on interstellar ice chemistry has been tested in the past \citep{Acharyya11,Pauly16,Pauly18}, but the effects related to size alone are quite modest. Rather, the most important factor is the temperature distribution of the grains as a function of size \citep{Pauly16}, which is most varied under low-extinction, low-density conditions, which are applicable only during the earliest times in the build-up of the ice mantles. During the hot stage of the core, gas densities should be high enough to ensure a uniform temperature across all but the smallest grains, through coupling with the gas, although the inclusion of smaller or larger grains in the model could plausibly affect the degree of trapping of volatile molecules to temperatures above their nominal desorption temperature, due to the different ice thicknesses produced. However, the computational cost involved in including a size distribution is considerable, as it requires replication of the entire set of surface/mantle chemical species for every additional grain size, while the time required to populate and invert the Jacobian matrix goes approximately with the square of the number of species in the model.

The use of a single bulk-ice layer in the present model ignores differences in the local chemical composition of the ice at different depths; for example, one might expect that the greater CO and methanol content in the upper ice layers would lead to a stronger local production of oxygen-bearing COMs, which is not reflected in the generalized bulk-ice composition considered here. Similarly, one might expect the deeper layers of ice to have a greater content of amine group-bearing species, due to the early production of ammonia on the grains. Such behavior could be resolved by stratifying the bulk ice into multiple, distinct layers. However, this would introduce computational challenges similar to those encountered with the adoption of multiple grain sizes, as each additional layer would add a new set of bulk-ice species that would continue to be involved in chemical reactions or other processes. \citet{Taquet12} produced a model purporting to distinguish mantle ice layers, but this aspect of their treatment was in principle no different to the kind of three-phase approach used here, which is also capable of capturing the stratification of the ice mantles as they are deposited \citep[see Fig. \ref{final_layers}, and similar figures by][]{GP11}, while they did not include a chemically active bulk-ice mantle. \citet{Garrod19} constructed a model of comet ice chemistry involving a total of 25 chemically-coupled layers of fixed thickness. A similar approach could be adopted for limited application to interstellar environments, permitting, for example, a more nuanced treatment of volatile trapping than the present model allows.

A number of recently proposed gas-phase mechanisms for COM production also have yet to be tested in our models \citep[e.g.][]{Skouteris18}.

\subsection{The selective importance of new mechanisms and parameter values}

Much of the work presented here concerns the effects on the model of various new mechanisms and updated parameter values, including many that are not strictly ``new'' but which had not yet been implemented in {\em MAGICKAL}. While all of the adjustments have been applied and tested individually or sequentially to determine their relative importance, we consider each one to be necessary in its own right: either to bring the model up to date with new data (e.g.~changes to binding energies and diffusion barriers); to include missing aspects of the kinetics (e.g.~new nondiffusive mechanisms); or to fill out the chemical network for particularly important species (e.g.~glycolaldehyde). The {\tt final} model setup is intended to serve as a new baseline platform for simulating the chemistry of hot cores and related environments. While the updates to physical and chemical parameters are of essentially no computational cost, most of the more costly alterations to the mechanics of the model, such as the nondiffusive mechanisms, appear to be essential to the results. We therefore would not advocate for the selective application of any of the new mechanisms or values, and our investigation of their effects was not intended to distinguish between those to be retained or rejected. Rather, we have sought to build as complete a model as possible, noting the effects of each new feature.

Table \ref{models} indicates the order in which the new model features were tested. While this ordering may appear somewhat arbitrary, certain features were intentionally grouped together: The first six models (i.e.~{\tt basic} up to {\tt bas\_stk\_loPD\_H2\_T16}) involved mostly simple changes, while retaining the overall mechanics and behavior of the old, purely diffusive models; the last of these, {\tt bas\_stk\_loPD\_H2\_T16}, could be considered the most advanced version of the purely diffusive models.

The following eight models (all those marked ``{\tt no-bd}'', i.e.~``no bulk diffusion'') involve fundamental changes to the mechanics of the grain-surface/mantle chemical treatment that were tested individually, and then collectively in model {\tt all}. While different in behavior from the purely diffusive setups, each of these models is directly comparable with those older treatments.

The last six models (from {\tt all\_Edif} to the {\tt final} model) may be considered to be refinements to the framework represented by model {\tt all}. Each of the adjustments made in those models was judged to be most fruitfully applied {\em after} the incorporation of nondiffusive chemical treatments. For example, the adopted rate of diffusion for hydrogen influences the fractional coverage of grain-surface radicals at low temperatures, affecting COM production. This effect would be of little importance without the inclusion of nondiffusive mechanisms.

Although in general we recommend the application of all the changes made in the {\tt final} model, it is indeed true that not all of them show an equally strong influence when the model is applied particularly to the set of conditions used here to approximate the evolution of a hot core. Of the nondiffusive processes, the Eley-Rideal mechanism is generally of little influence, although as may be seen in Fig.~\ref{roc2}, in some cases it provides a moderate supplement to the Langmuir-Hinshelwood (diffusive) production of certain grain-surface species. The cost of implementation of the E-R mechanism is very small compared with the other nondiffusive mechanisms, so it is arguable whether its omission would be of any advantage. The influence of the 3-BEF mechanism is also limited mainly to certain key reaction chains in which the product of a highly exothermic initiating reaction is allowed to react with an abundant stable molecule such as CO (see JG20). For those considering implementing the 3-BEF mechanism in another gas-grain chemical code, the pre-selection of important initiating/follow-on reactions could reduce the computational cost. A small decrease in computational cost could also be obtained by reducing the number of 3-B reaction cycles from 3 down to 2, as the difference in results is minor for most species. However, nondiffusive production of highly complex species that rely on multiple reaction steps could be adversely affected; although low in abundance, molecules such as dihydroxyacetone, (CH$_2$OH)$_2$CO, can be suppressed by as much as $\sim$50\% without the third cycle.

The new treatment to account for the influence of H$_2$ surface coverage on binding energies has no major effect on the abundances of COMs in the present physical setups. The relative effect on H$_2$ abundances within the ice mantles is very strong, however, reducing peak H$_2$ abundance to $\sim$1 molecule per grain, a reduction by a factor of 20,000. The inclusion of the new treatment is likely to have a yet stronger effect on models that reach even lower dust temperatures (less than 8~K), in which the H$_2$ fraction may otherwise build up to extreme values \citep[see][]{GP11}.

The increase in the generic ratio of diffusion barrier to binding energy for atoms is most influential through the altered mobility of atomic H; slower H diffusion on the grain/ice surface leads to longer lifetimes for reactive radicals, which under very cold conditions produces stronger 3-B production of COMs. For the other atoms in the model, the effect is marginal, as those species are already relatively immobile at the low dust temperatures that pertain when the ices are building up. The availability of nondiffusive mechanisms also tends to make slow diffusive reactions less important, as the nondiffusive alternative will take over below some minimum rate. The change to the barriers is trivial to implement, however, and has no noticeable computational cost.

\section{Conclusions}\label{concs}

We have presented a range of new features and treatments that have been incorporated into the astrochemical model {\em MAGICKAL}, as applied to the chemistry of hot, star-forming cores. Changes to the model have been tested individually and collectively to test their influence on observable outcomes. The key feature of the new models is the inclusion of nondiffusive reaction processes on dust-grain surfaces and in the ice mantles, removing the need for purely diffusive grain-surface or bulk-ice reaction processes in the production of COMs. Much of the COM production remains dominated by the reactions of radicals, either with other radicals or with relatively stable species in the ice, but the temperature dependence is much altered. The main conclusions of this study are listed below:

\begin{enumerate}
\item The inclusion of nondiffusive grain-surface and bulk-ice reactions advances the production of many COMs to earlier times and lower temperatures than in previous model implementations.

\item Many COMs are formed as by-products during the chemical formation of simple molecular ice mantles, under cold, high gas-density conditions that favor relatively high surface coverages of reactive radicals. This result agrees with the findings of recent ice deposition experiments \citep[e.g.][]{Fedoseev15}.

\item Some COMs are formed yet earlier, through photoprocessing of the nascent ice mantles by the external UV field under translucent conditions.

\item The overall influence of cosmic ray-induced photodissociation in the bulk ice is somewhat reduced versus past models (assuming a canonical cosmic-ray ionization rate).

\item Modest production (and destruction) of certain COM species occurs during the period when water and other strongly-bound species desorb from the icy grains at high temperatures ($\sim$120 -- 160~K), releasing trapped, reactive radicals onto the hot surface.

\item Some new gas-phase mechanisms appear to contribute moderately to COM production, although they do not dominate.

\item The finalized models show excellent agreement with a broad range of observed COM abundances in multiple sources.

\item Abundance ratios of the structural isomers methyl formate (MF), glycolaldehyde (GA) and acetic acid agree qualitatively with typical observational values, and show a good quantitative match with some sources. However, the chemical mechanisms that act to differentiate the final gas-phase MF and GA abundances are numerous and subtle.

\item The models show good agreement with observed abundances for several recently-detected or unusual molecular species, such as methoxymethanol, ethylene oxide, and cyanamide.

\item The fast warm-up models agree best for the representative low-mass source IRAS 16293B, while the slow models are better for the high-mass source Sgr B2(N2); this would indicate that a representative parcel of gas in the high-mass case would remain longer in the hot region than in the low-mass case.

\item Differences between O- and N-bearing species' observational abundances may be related to strong production of nitriles in the gas phase, during and after ice-mantle desorption, while O-bearing COMs are formed at early times, on the cold grains. The time-period available for post-desorption gas-phase chemistry would play a major role in determining the N/O abundance ratios.

\item The models demonstrate that isocyanic acid (HNCO) and formamide (NH$_2$CHO) naturally share similar pathways to their formation on dust grains, while not requiring a direct means of interconversion in either direction (although this is not ruled out). Both species retain a substantial degree of production via cosmic ray-induced bulk-ice photochemistry.

\item In these models, the simplest amino acid, glycine, is formed most strongly either at the very earliest stage of COM formation, through ice processing by external UV photons, or much later, via hot grain-surface radical chemistry as the ice mantles desorb.

\item The extended grain-surface network used in the new models indicates the importance of a number of new reactions of methylene (CH$_2$) and methylidyne (CH) with stable surface species. These reactions, as well as others potentially involving singlet (excited) methylene, should be investigated further.

\item Consideration of the time-dependent production rates of COMs, summed over all physical phases, provides a convenient means to assess the origins of those molecules through the varying physical conditions employed in the models.

\end{enumerate}


\acknowledgements

We thank the two anonymous referees for their thorough reviews and helpful comments on the manuscript. We thank A. Belloche for providing additional comments on the paper. This work was funded by the Astronomy \& Astrophysics program of the National Science Foundation (grant No.~AST 19-06489).


\appendix

\section{Updates to chemical network of basic model}

\subsection{Ethylene oxide and vinyl alcohol}\label{ethox}

Ethylene oxide, c-C$_2$H$_4$O, was detected toward the high-mass star-forming source Sgr B2(N) by \citet{Dickens97}, and was more recently found toward the solar-type protostar IRAS 16293-2422 by \citet{Lykke17}. With this more recent detection in mind, the chemical network was updated to include a simple treatment for ethylene oxide and vinyl alcohol, C$_2$H$_3$OH, both of which are structural isomers of acetaldehyde, CH$_3$CHO. Gas-phase destruction reactions of both species with reaction partners C$^+$, He$^+$, H$_3^+$, H$_3$O$^+$ and HCO$^+$ were included, as per the methods described by \citet{GWH08}. Here, the production of these two new species is assumed to occur through dust-grain surface/ice-mantle reactions, which are shown in Table \ref{C2H4O-rxns} along with the surface/mantle destruction mechanisms assumed in this model. The first reaction also appears in Table \ref{CH-rxns}, as described in Sec.~\ref{sec:CH-rxns}. The usual grain-surface adsorption and desorption mechanisms were also added to the network for the new molecules.

The fourth reaction in the table, which is the sole production mechanism for ethylene oxide in the model, has been studied on laboratory grain-surface analogs by \citet{Ward11} and by \citet{Bergner19}. Ward \& Price deposited O atoms onto pre-deposited ethene (C$_2$H$_4$), finding that ethylene oxide was the only product of the reaction. Modeling the kinetics of the reaction, they determined a modest barrier to reaction of $190\pm45$~K, although it is unclear how much of this is due to a diffusion barrier rather than an activation energy. Bergner et al. considered the photodissociation of O-bearing molecules in mixed ices containing C$_2$H$_4$; in that case the oxygen was produced in the excited O($^1$D) singlet state. They found that ethylene oxide and vinyl alcohol were produced in approximately equal amounts, independent of temperature and with no apparent activation energy barrier. Here, we assume that all oxygen on the grains is in the O($^3$P) ground state, that the reaction in question may proceed without a barrier, and that it produces only ethylene oxide. The inclusion of electronically-excited O atoms in astrochemical models will require a more detailed treatment \citep[][{\em in press}]{Carder}.

\subsection{Conversion of HNCO to NH$_2$CHO}\label{app2}

In past models, the repetitive hydrogenation of HNCO has been one of multiple routes leading to the formation of formamide, NH$_2$CHO. However, laboratory experiments by \citet{Noble15} indicate that, while the initial hydrogenation step, H + HNCO $\rightarrow$ NH$_2$CO, may occur, subsequent reactions of hydrogen atoms with the NH$_2$CO radical would not produce formamide. More recently, in experiments by \citet{Haupa19}, in which formamide and its deuterated forms are exposed to H-atom bombardment, those authors suggest that hydrogenation all the way to formamide is indeed possible. In the present work, we choose to switch off the H + NH$_2$CO $\rightarrow$ NH$_2$CHO reaction. The initial reaction between H and HNCO remains in the network, with a barrier of 1390~K \citep{Nguyen96}, allowing NH$_2$CO to be available for reactions with other surface/mantle radicals.

\begin{deluxetable}{lllr} \label{C2H4O-rxns}
\tabletypesize{\footnotesize}
\tablecaption{\label{tab-gas} Grain-surface reactions involving ethylene oxide and vinyl alcohol.}
\tablewidth{0pt}
\tablehead{
\colhead{Reaction} & \colhead{} & \colhead{} & \colhead{$E_A$ (K)}
}
\startdata
CH  +   CH$_2$OH          & $\rightarrow$ &   C$_2$H$_3$OH   &  0  \\
OH  +  C$_2$H$_3$        & $\rightarrow$ &   C$_2$H$_3$OH   &  0  \\
H    +  C$_2$H$_3$OH    & $\rightarrow$ &   C$_2$H$_4$OH   &  1840$^a$ \\
O    +  C$_2$H$_4$        & $\rightarrow$ &   c-C$_2$H$_4$O   &  0
\enddata
\tablecomments{References: 
$^a$\citet{Rao11}.
}
\end{deluxetable}

\section{Technical adjustments to basic model}

\subsection{Tunneling through activation-energy barriers}

A long-standing simplification in the gas-grain models, reaching back to the treatment of \citet{Hasegawa92}, involves the switchover between thermal activation and tunneling for surface/bulk reactions involving light species, which becomes important at low temperatures. In this simple approach, the faster of either means of activation is adopted when calculating reaction rates. This leads to a hard switchover between a highly temperature-dependent rate (thermal) and a temperature-independent rate (tunneling). Detailed calculations of reactions at low temperature find a smoother transition. \citet{Taquet13} used a more detailed approach to calculating the barrier penetration in astrochemical models, based on tunneling through an Eckart potential. Here we adopt the method of \citet{Bell}, commonly known as Bell's modification of the Arrhenius equation, which permits the use of the same expressions for pure tunneling or thermal activation as we have used in the past, but which provides a smoother and broader transition between the two. This method gives an activation efficiency factor of the form:

\begin{equation}
\kappa_{\mathrm{act}} = \frac{ a_{\mathrm{tunn}} \exp(-E_{A}/T) - (E_{A}/T) \exp(-a_{\mathrm{tunn}}) }{ a_{\mathrm{tunn}} - E_{A}/T }
\end{equation}

\noindent where the activation energy $E_{A}$ is in units of K, and $a_{\mathrm{tunn}}$ is the argument of the exponential function that describes tunneling through a one-dimensional barrier. Although this method provides a neater solution, it was found in the present models to have only a marginal impact on outcomes.

\subsection{Switching function for modified reaction rates}

\begin{figure}
         {\includegraphics[width=0.5\hsize]{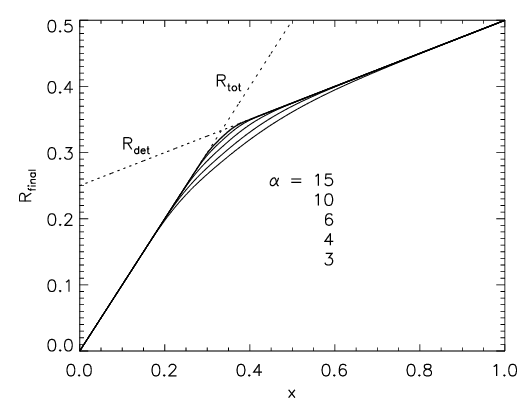}}
\caption{Switching function for smooth transition between two rates (each of which is a function of arbitrary variable $x$), as used in the modified-rate treatment to ensure that the finalized rate, $R_{\mathrm{final}}$, never exceeds the lower of the two rates. Higher values of $\alpha$ produce a tighter switchover between the two values.}
\label{switching}
\end{figure}

The modified-rate method for adjusting diffusive surface-reaction rates proposed by \citet{Garrod08} involves the calculation of an unadjusted {\em deterministic} rate, and a modified rate that uses the small-grain approximation to simulate a {\em stochastic} rate; the modified rate is adopted by default, but may not exceed the deterministic rate, since the latter represents the theoretical upper limit to the reaction rate. In most situations this switchover does not cause any problem; however, under certain conditions, the numerical solver may encounter a discontinuity in the derivative of the final rate in the vicinity of the switchover, which makes the differential equations very stiff and thus slow to integrate. To resolve this technical problem, a switching function is introduced for all modified-rate calculations when determining the finalized rate. Using a standard deterministic rate of some surface reaction $R_{\mathrm{det}}$, and a modified rate $R_{\mathrm{tot}}$ \citep[in the nomenclature used by][]{Garrod08}, the finalized rate is now given by:
\begin{eqnarray}
&& R_{\mathrm{hi}} = \max(R_{\mathrm{det}}, R_{\mathrm{tot}}) \\
&& R_{\mathrm{lo}} = \min(R_{\mathrm{det}}, R_{\mathrm{tot}}) \\
&& R_{\mathrm{final}} = R_{\mathrm{lo}} - \frac{R_{\mathrm{lo}}}{1+2{\alpha}}  \exp\left(1 - \left[ \frac{R_{\mathrm{hi}}}{R_{\mathrm{lo}}}\right]^{\alpha} \right) \\
\end{eqnarray}
\noindent where index $\alpha$ determines the sharpness of the function. Both rates are assumed to be positive. Note that $R_{\mathrm{final}}$ never exceeds the lower of the two rates, which is a key condition of the modified-rate treatment, and only diverges from the lower rate when the two rates are of similar magnitude. With sufficiently high $\alpha$, this adjustment to the modified-rate method has no significant effect on the results, while avoiding the solver hanging up at an arbitrary point in the calculations. Figure \ref{switching} shows rates $R_{\mathrm{det}}(x)$ and $R_{\mathrm{tot}}(x)$, both functions of an arbitrary variable, $x$, that can be considered proportional to the average population of some grain-surface species. Plotted with these rates is $R_{\mathrm{final}}$ as calculated using a selection of $\alpha$ values. In the present models, $\alpha = 6$ is adopted throughout. This value results in a maximum deviation from the lower of the two rates of $1/(1 + 2 \alpha) \simeq 7.7$\%.

\section{Additional figures and tables}\label{app-figs}

\begin{figure*}
\includegraphics[width=0.703\hsize]{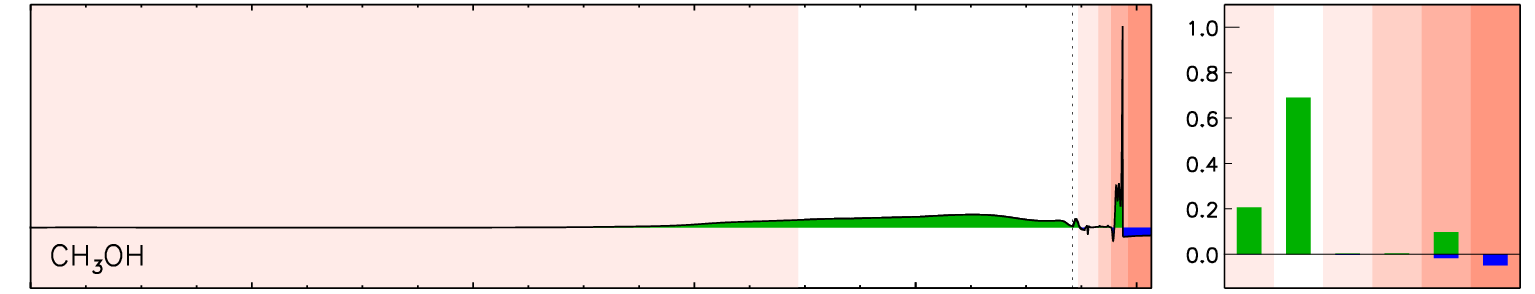}

\includegraphics[width=0.703\hsize]{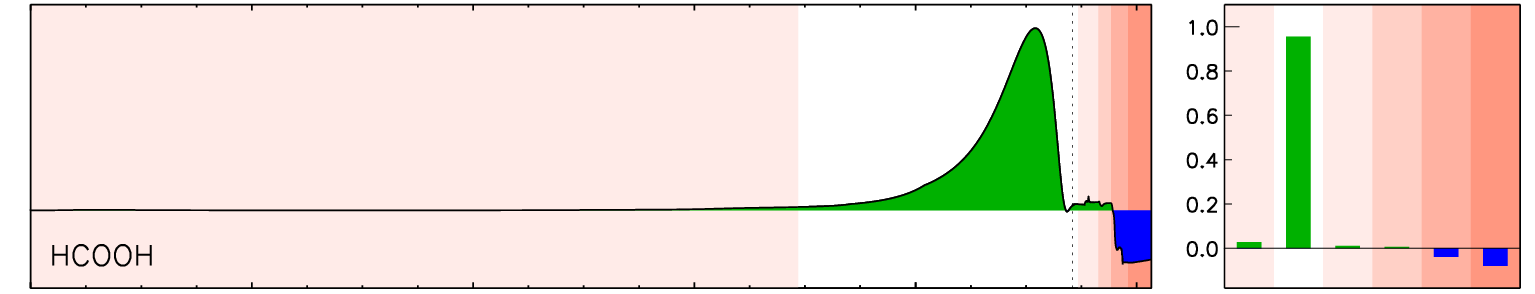}

\includegraphics[width=0.703\hsize]{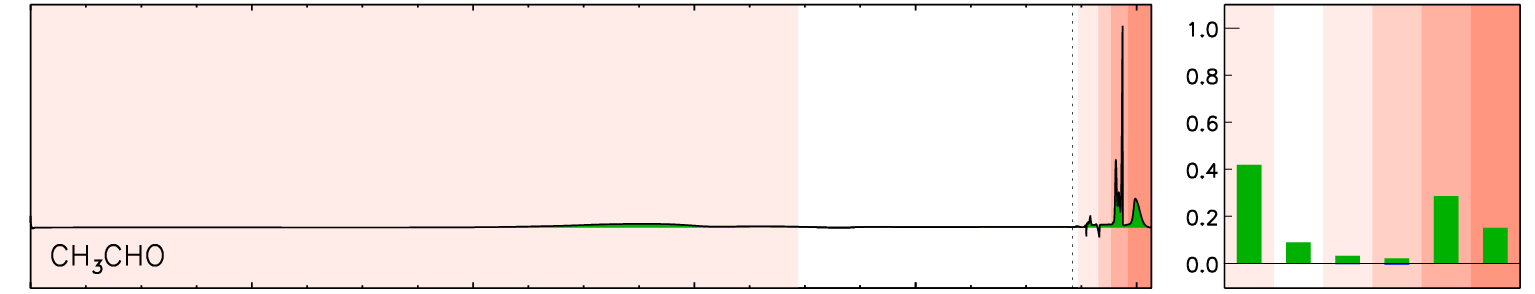}

\includegraphics[width=0.703\hsize]{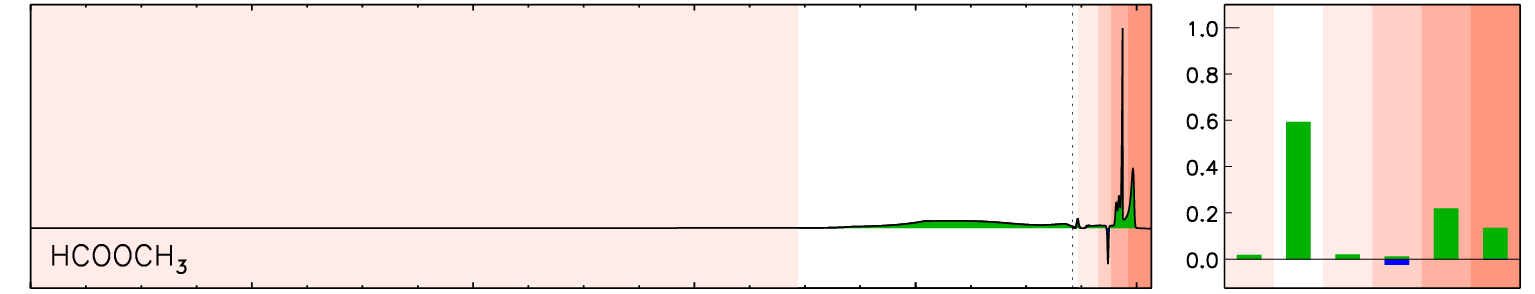}

\includegraphics[width=0.703\hsize]{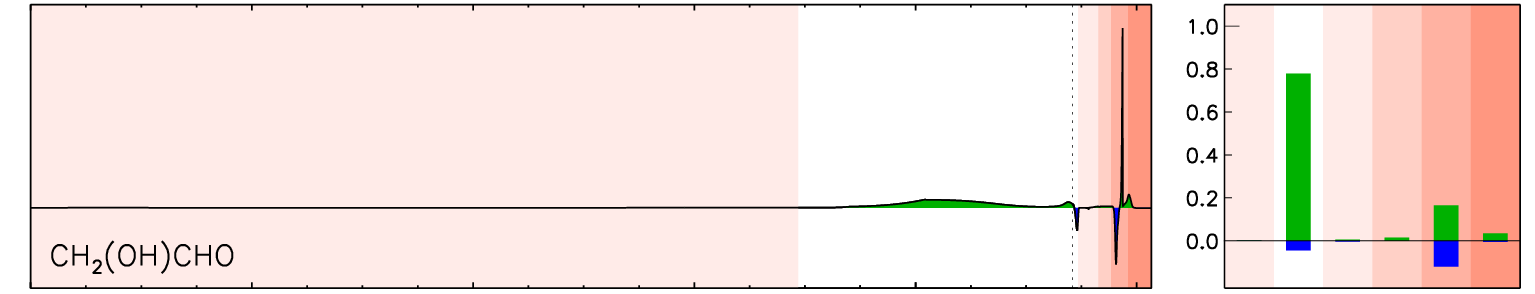}

\includegraphics[width=0.703\hsize]{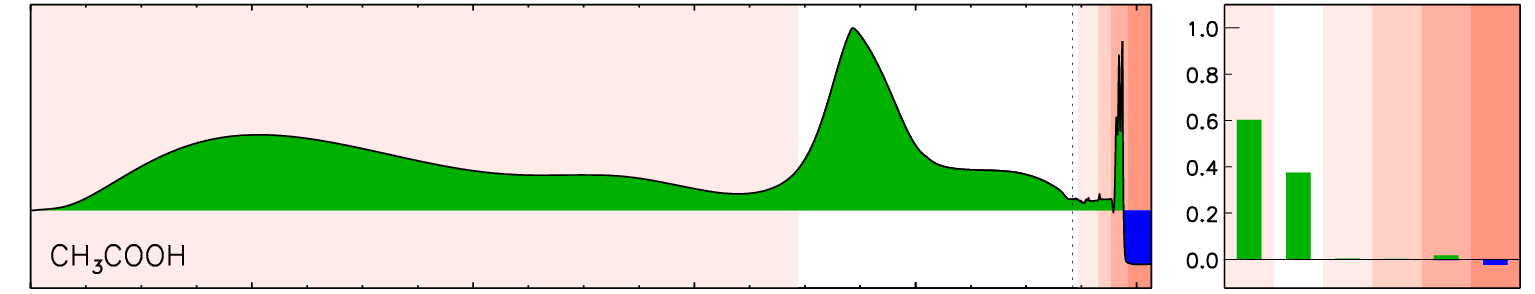}

\includegraphics[width=0.703\hsize]{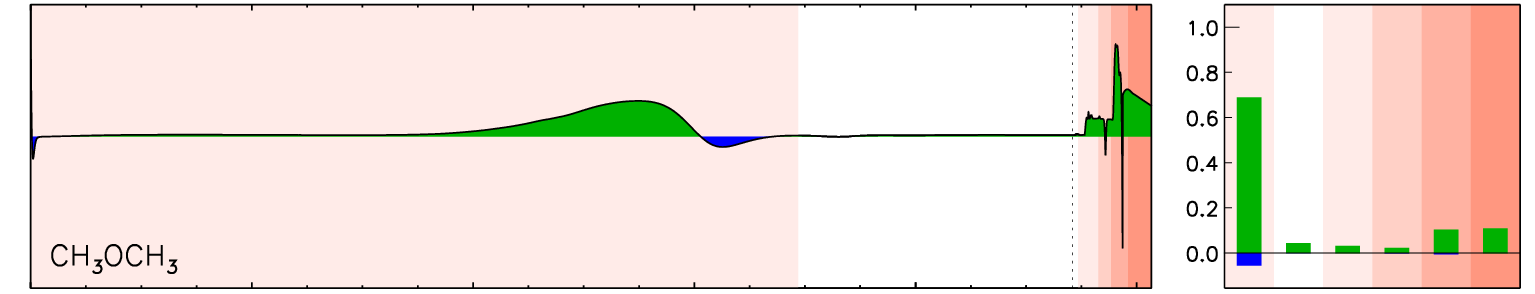}

\includegraphics[width=0.703\hsize]{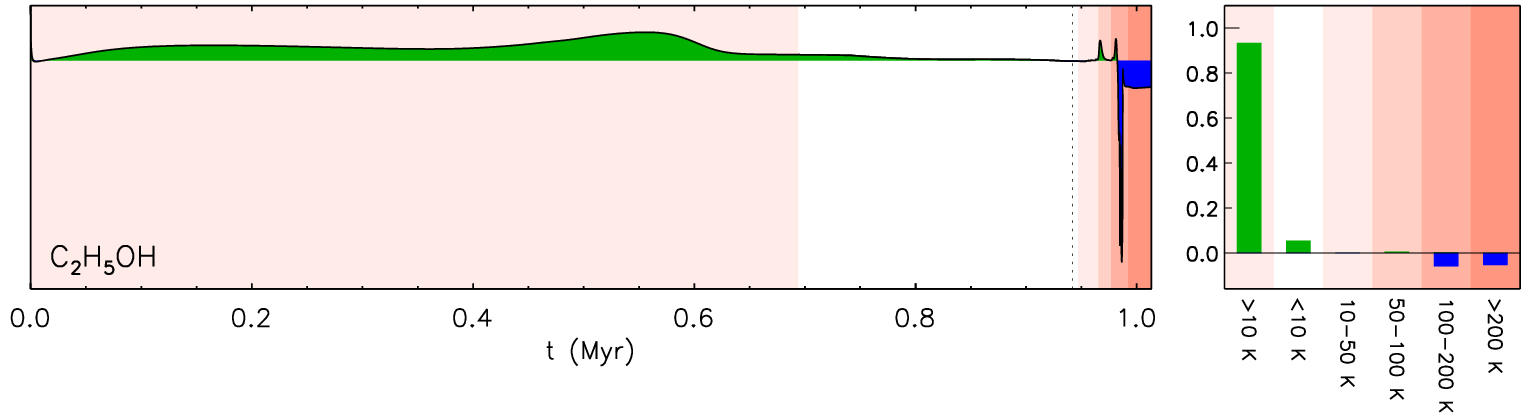}
\caption{{\bf {\em Left:}} Net rate of change (arbitrary units) in the abundances of selected oxygen-bearing molecules, summed over all chemical phases, during stages 1 and 2. Data correspond to the {\tt final} model, using the {\em fast} warm-up timescale in stage 2; the vertical dotted line indicates the start of the warm-up phase. Net gain is shown in green, net loss in blue. Background colors indicate the dust temperature regime; from left to right these are: $>$10~K, $<$10~K, 10--50~K, 50--100~K, 100--200~K, 200--400~K. The initial dust temperature is $\sim$14.7~K. {\bf {\em Right:}} Net rates of change integrated over each temperature range. Positive (formation) and negative (destruction) rates are integrated independently; both are normalized to the total integrated formation rate.}
\label{roc3}
\end{figure*}

\begin{figure*}
\includegraphics[width=0.703\hsize]{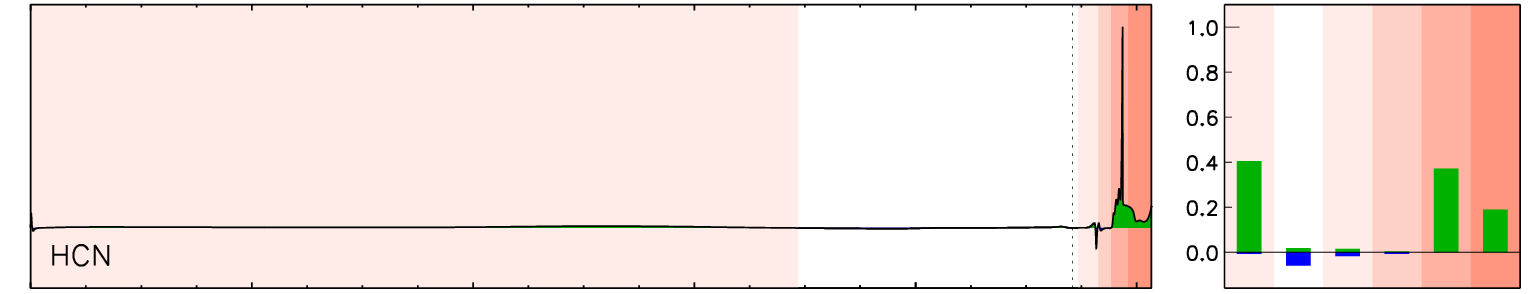}

\includegraphics[width=0.703\hsize]{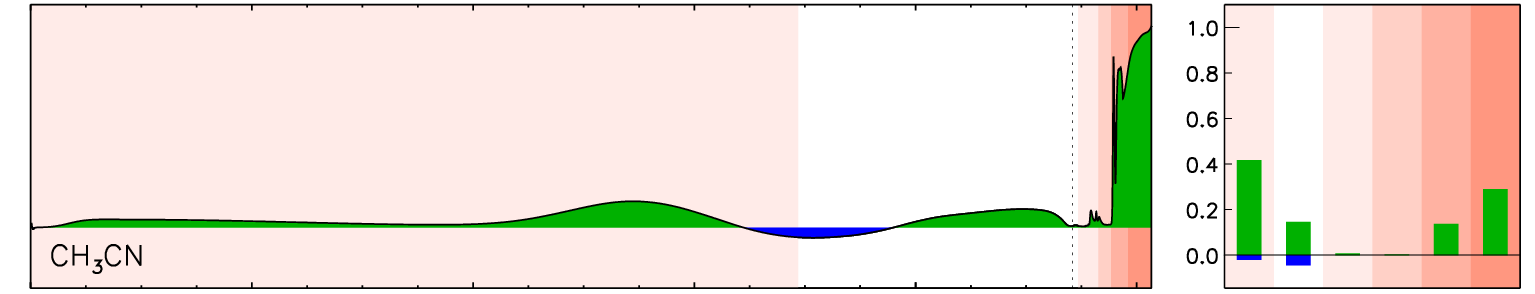}

\includegraphics[width=0.703\hsize]{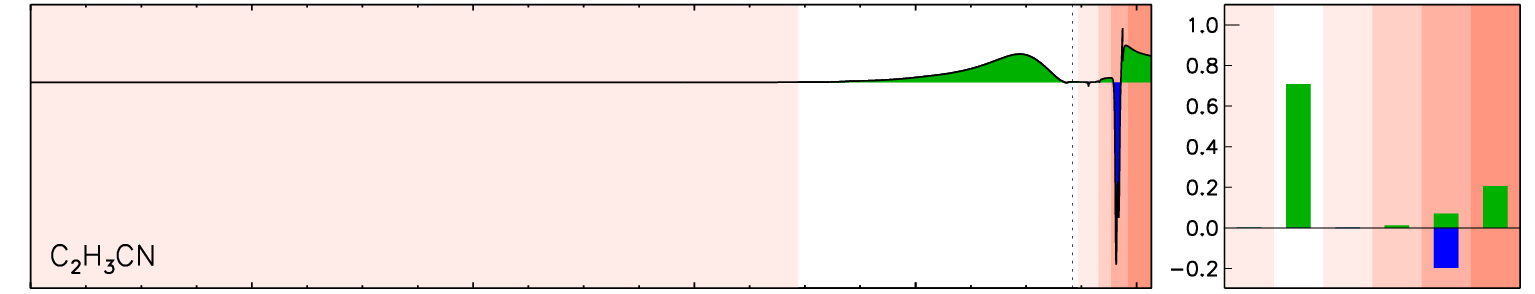}

\includegraphics[width=0.703\hsize]{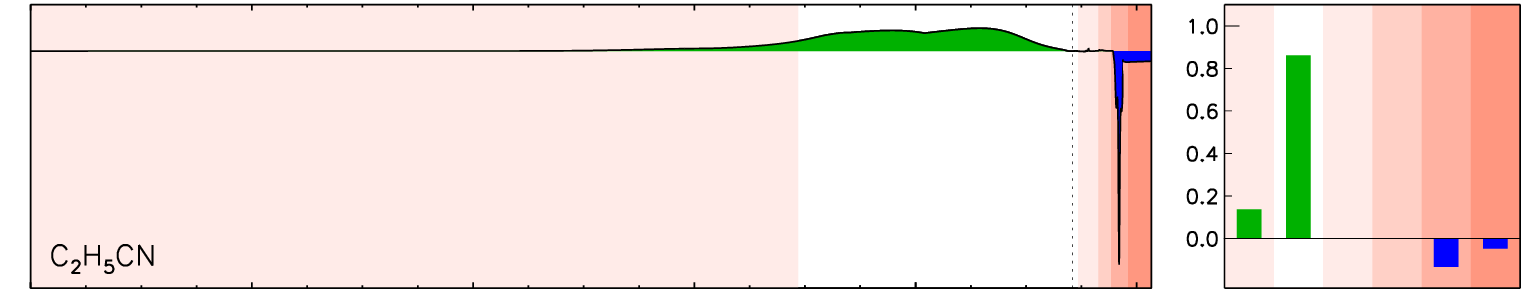}

\includegraphics[width=0.703\hsize]{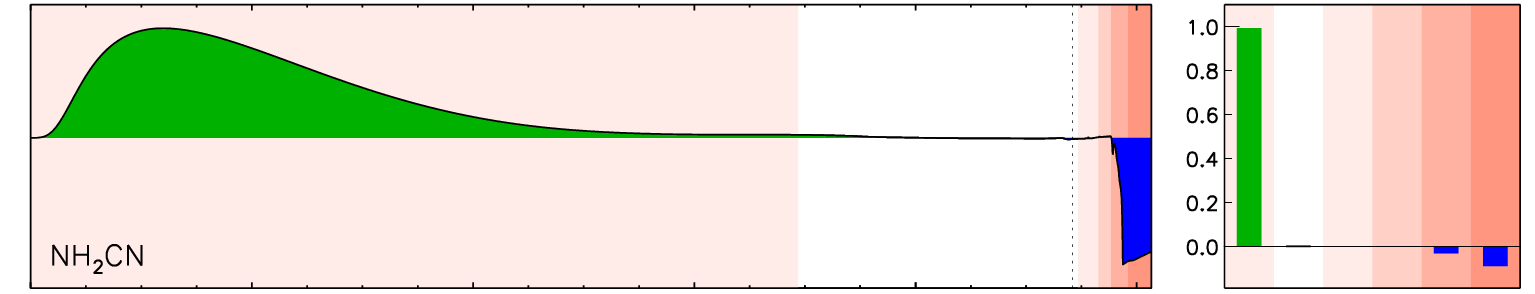}

\includegraphics[width=0.703\hsize]{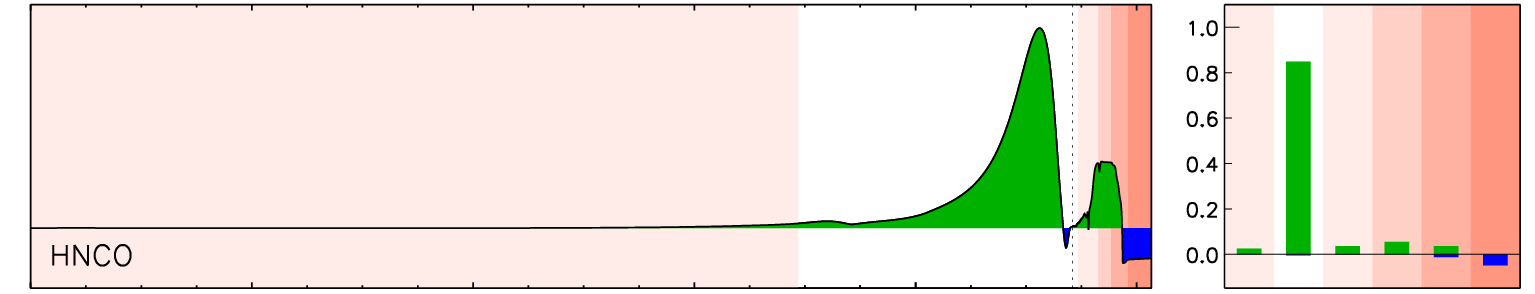}

\includegraphics[width=0.703\hsize]{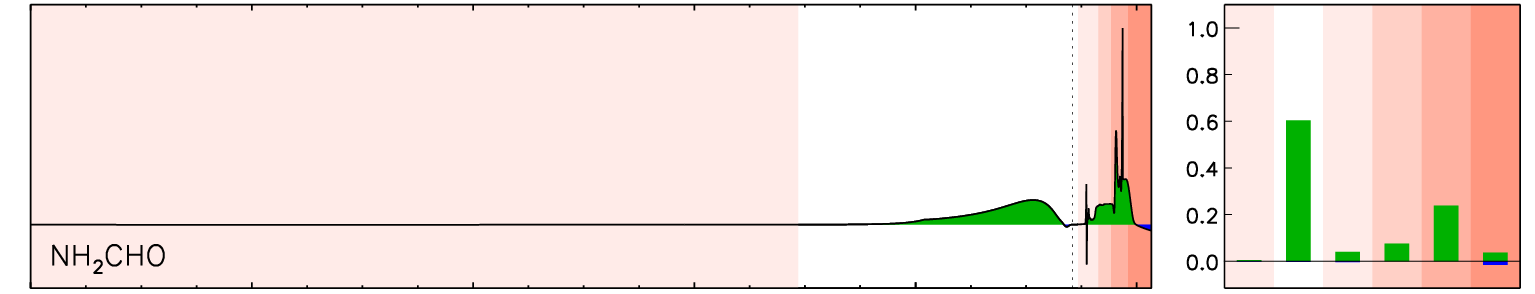}

\includegraphics[width=0.703\hsize]{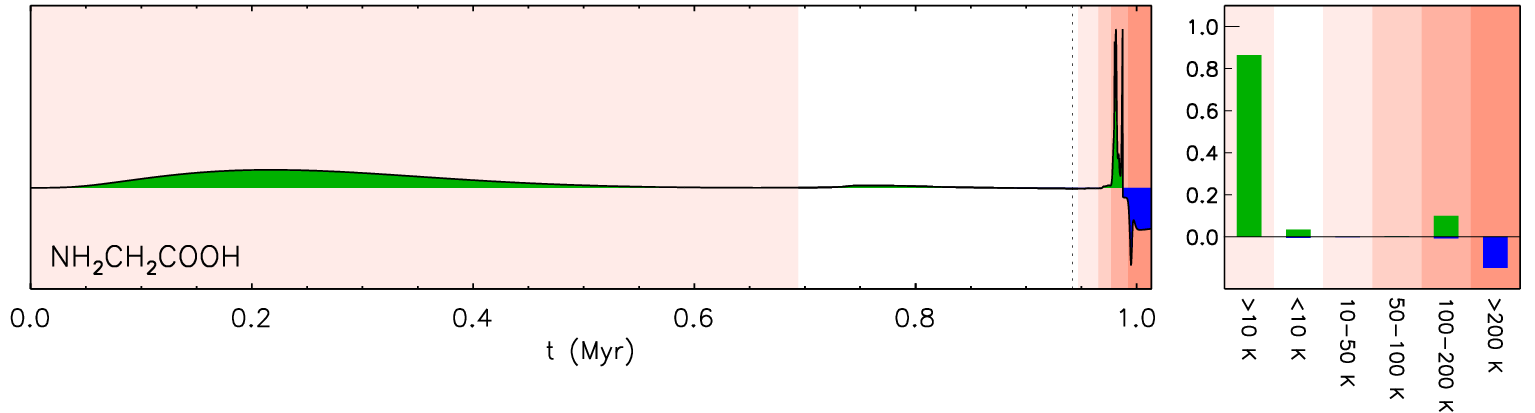}
\caption{Net production rates as a function of time; description as per Fig.~\ref{roc3}. Data correspond to selected nitrogen-bearing molecules in the {\tt final} model, using the {\em fast} warm-up timescale in stage 2.}
\label{roc4}
\end{figure*}

\begin{figure*}
\includegraphics[width=0.703\hsize]{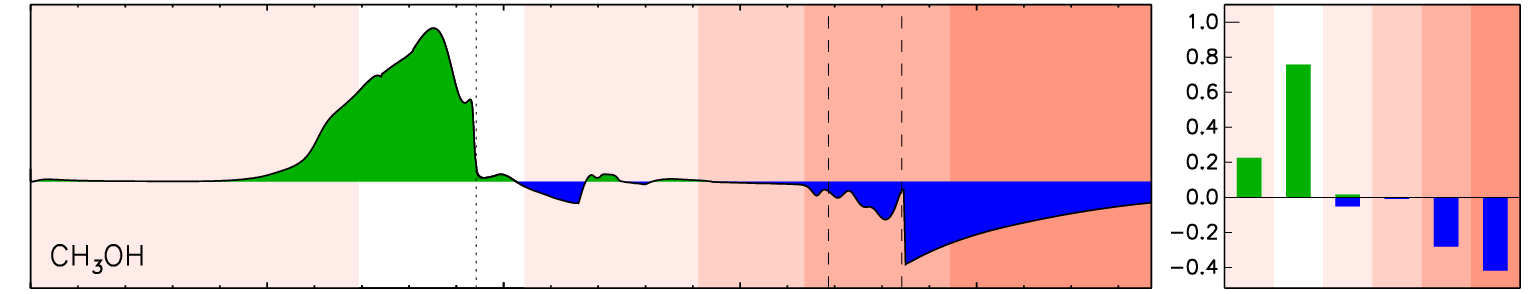}

\includegraphics[width=0.703\hsize]{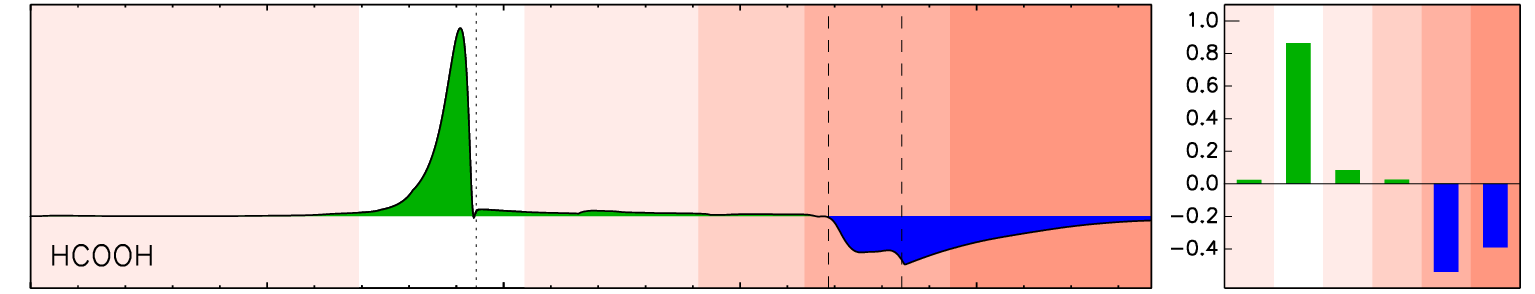}

\includegraphics[width=0.703\hsize]{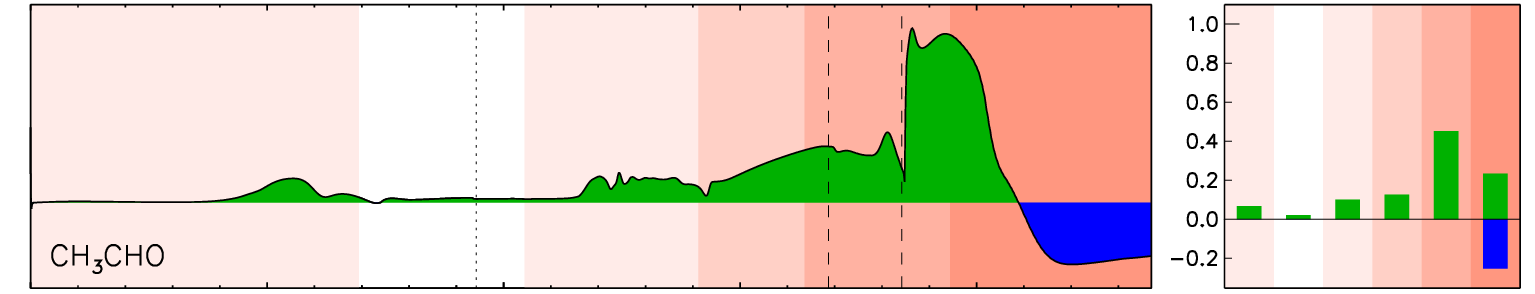}

\includegraphics[width=0.703\hsize]{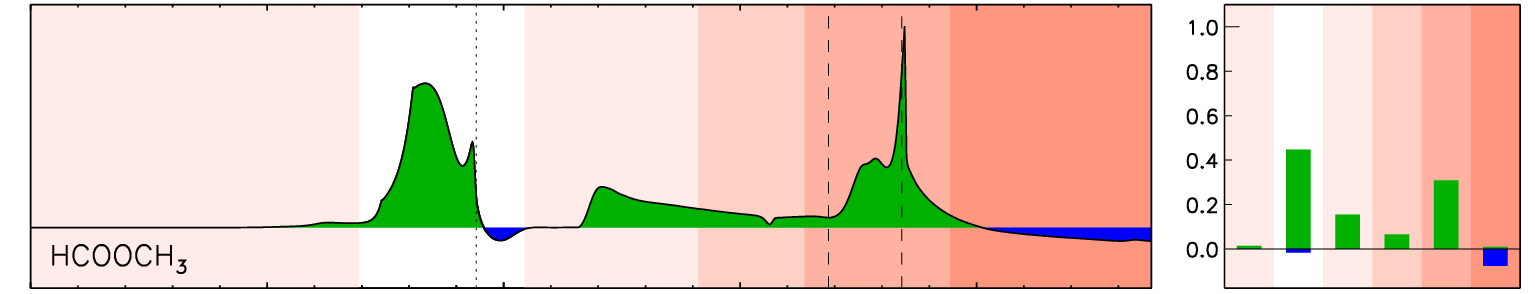}

\includegraphics[width=0.703\hsize]{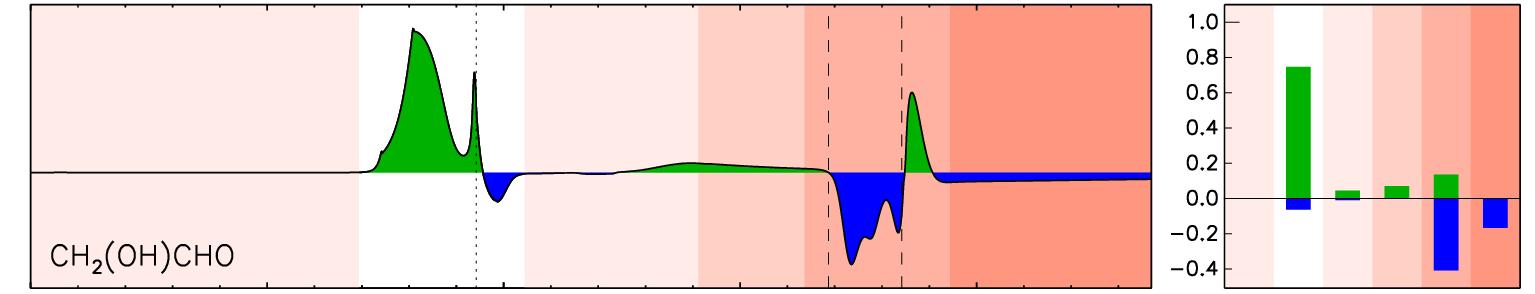}

\includegraphics[width=0.703\hsize]{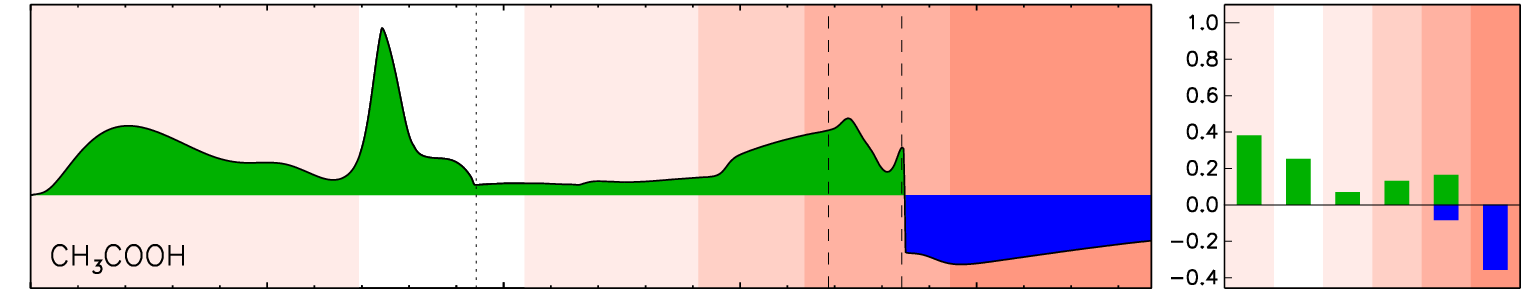}

\includegraphics[width=0.703\hsize]{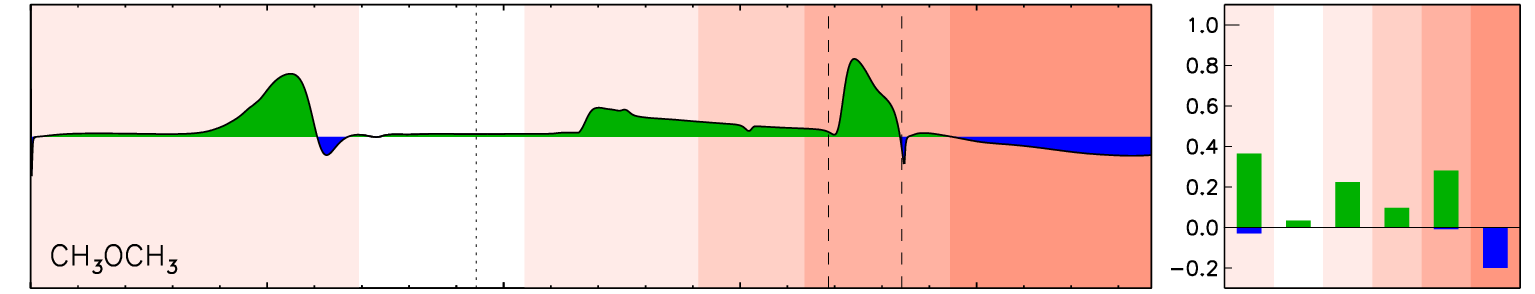}

\includegraphics[width=0.703\hsize]{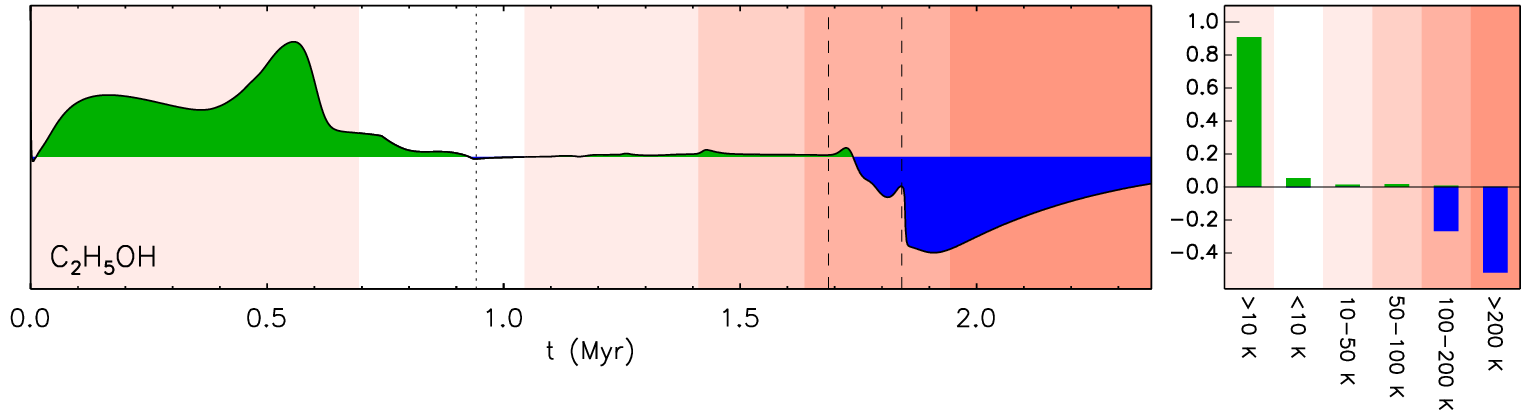}
\caption{Net production rates as a function of time; description as per Fig.~\ref{roc1}. Data correspond to selected oxygen-bearing molecules in the {\tt final} model, using the {\em slow} warm-up timescale in stage 2.}
\label{roc5}
\end{figure*}

\begin{figure*}
\includegraphics[width=0.703\hsize]{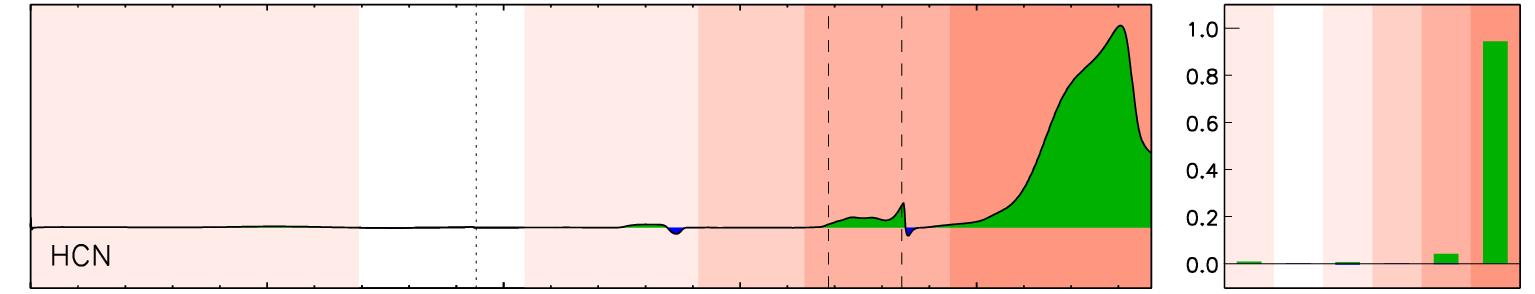}

\includegraphics[width=0.703\hsize]{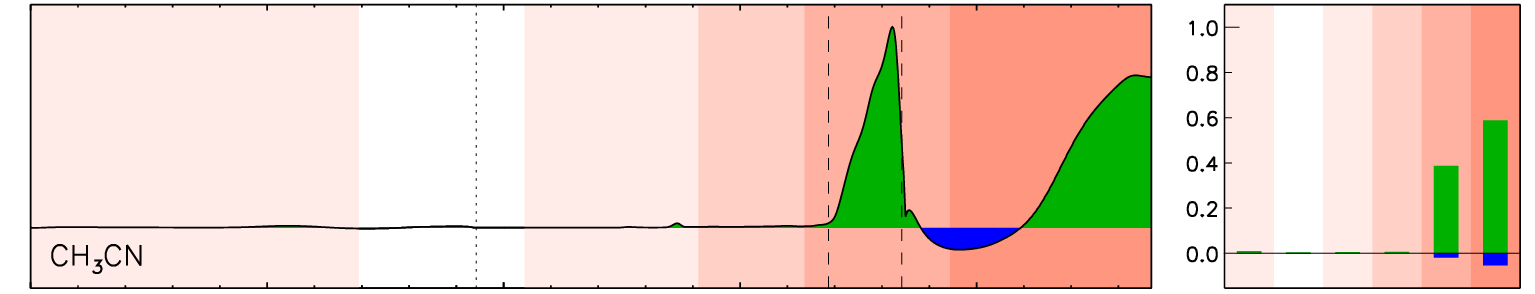}

\includegraphics[width=0.703\hsize]{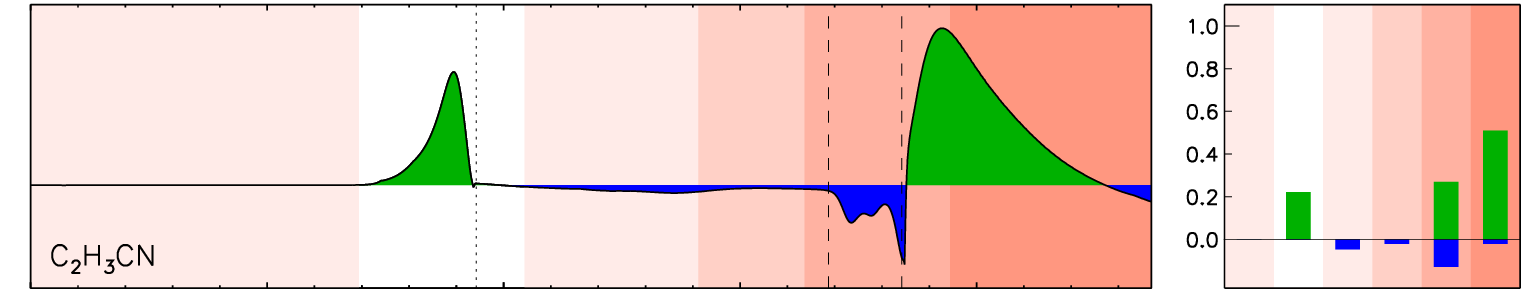}

\includegraphics[width=0.703\hsize]{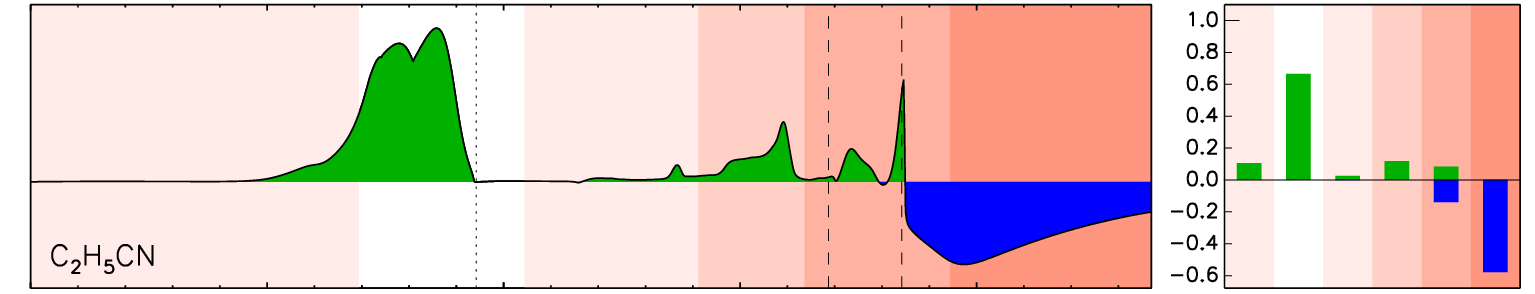}

\includegraphics[width=0.703\hsize]{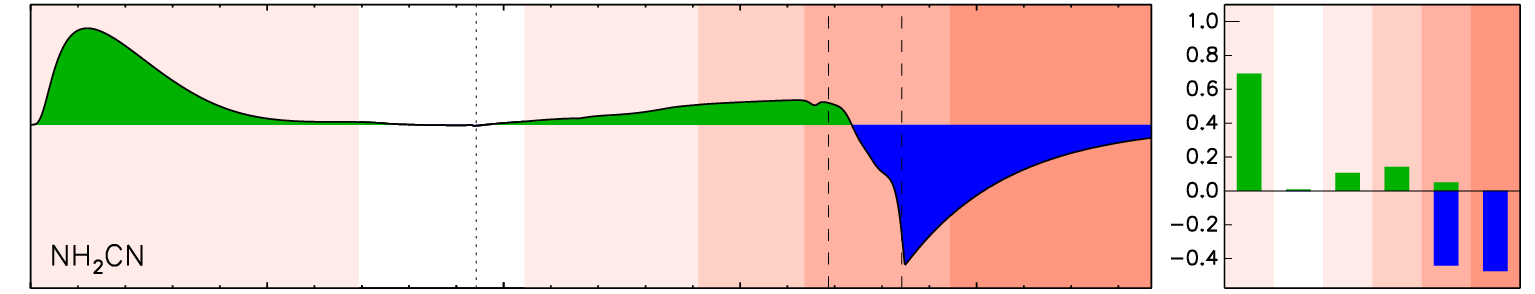}

\includegraphics[width=0.703\hsize]{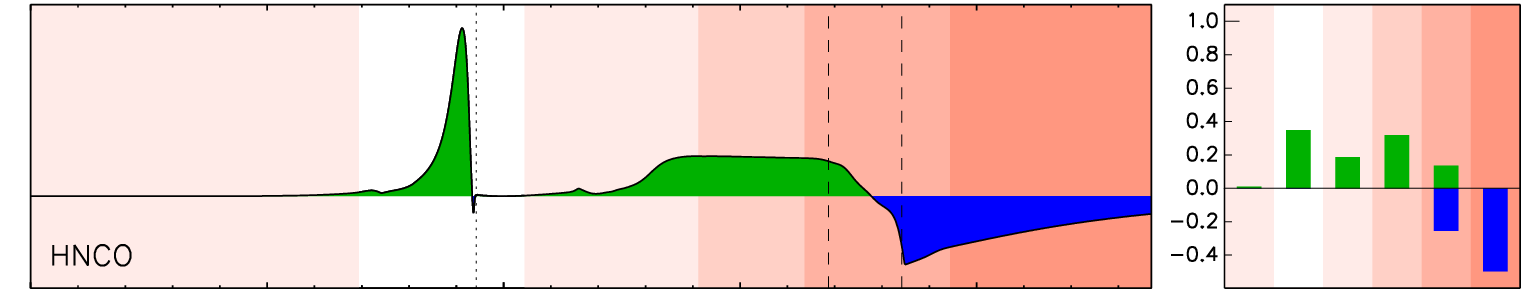}

\includegraphics[width=0.703\hsize]{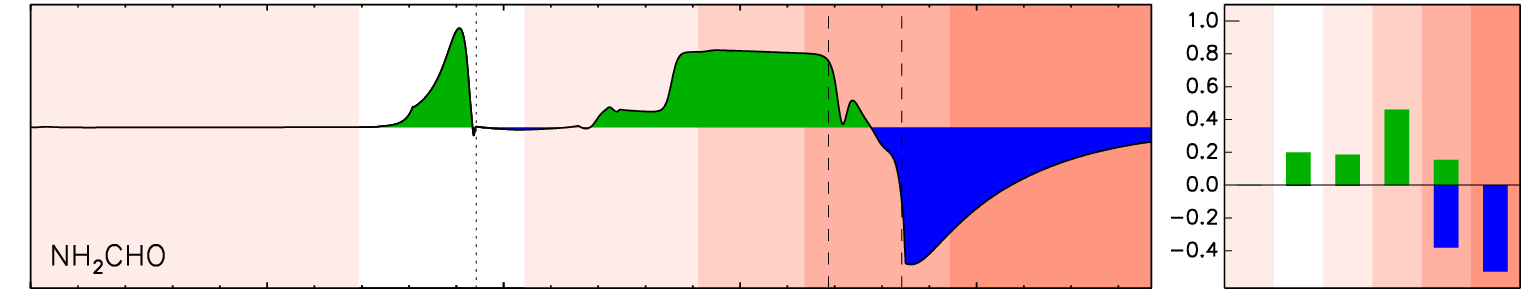}

\includegraphics[width=0.703\hsize]{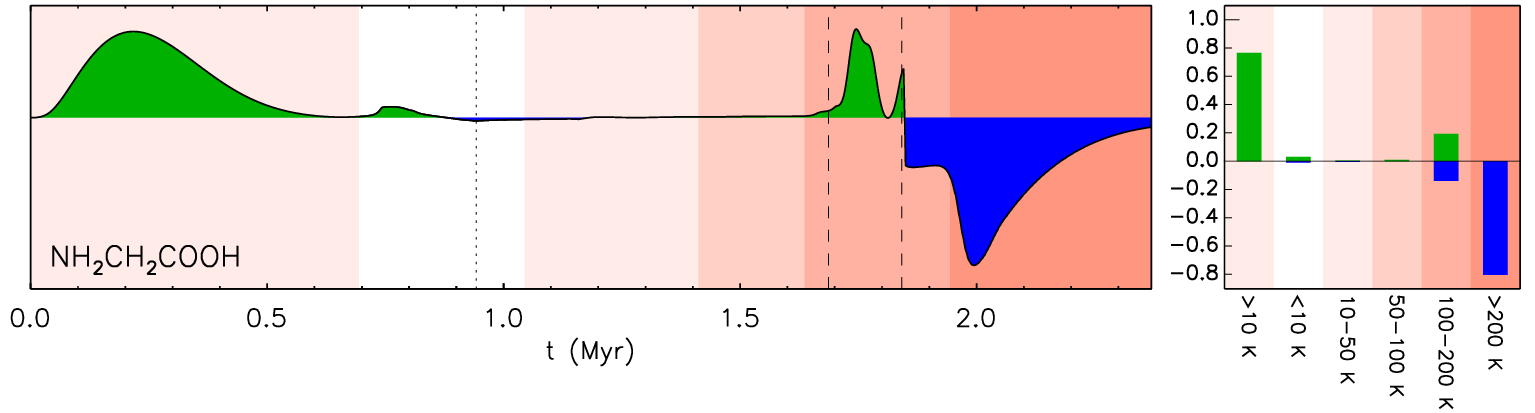}
\caption{Net production rates as a function of time; description as per Fig.~\ref{roc1}. Data correspond to selected nitrogen-bearing molecules in the {\tt final} model, using the {\em slow} warm-up timescale in stage 2.}
\label{roc6}
\end{figure*}

\clearpage

\startlongtable
\begin{deluxetable}{lcrrrrrrrrrrrrrrrrrrrr}\label{tab:rates}
\tabletypesize{\scriptsize}
\tablecaption{\label{tab-roc} Net formation percentages summed over all phases (gas, surface and ice mantle) for a selection of molecules, integrated over the six temperature regimes and rounded to the nearest whole number. Data corresponds to the collapse stage combined with each indicated warm-up timescale, using the {\tt final} model setup. Values less than 0.5\% are denoted `-'. Selected data are also plotted as bar charts in Figs. \ref{roc1} -- \ref{roc2} and \ref{roc3} -- \ref{roc6} (the destruction percentages plotted in those figures are not shown here). The six temperature regimes correspond to those identified in those figures: regime 1: $>$$10$~K (cold); regime 2:  $<$$10$~K; regime 3:  $10 - 50$~K; regime 4:  $50 - 100$~K; regime 5:  $100 - 200$~K; regime 6:  $>$$200$~K.}
\tablewidth{0pt}
\tablehead{&\\
 && \multicolumn{6}{c}{stage 1 + stage 2 ({\em fast})} && \multicolumn{6}{c}{stage 1 + stage 2 ({\em medium})} && \multicolumn{6}{c}{stage 1 + stage 2 ({\em slow})} \\
\cline{3-8} \cline{10-15} \cline{17-22}
Molecule && \multicolumn{6}{c}{Temperature regime} && \multicolumn{6}{c}{Temperature regime} && \multicolumn{6}{c}{Temperature regime} \\
&& 1 & 2 & 3 & 4 & 5 & 6 && 1 & 2 & 3 & 4 & 5 & 6 && 1 & 2 & 3 & 4 & 5 & 6 
}
\startdata
H$_2$O && 51 & 44 & - & - & 2 & 1 && 50 & 43 & - & - & 3 & 2 && 46 & 40 & - & - & 6 & 6 \\
CO && 90 & - & - & - & 5 & 2 && 86 & - & - & - & 7 & 4 && 72 & - & - & - & 12 & 15 \\
CO$_2$ && 32 & 66 & - & - & - & - && 32 & 66 & 1 & - & - & - && 30 & 63 & 4 & 1 & - & - \\
CH$_4$ && 92 & 7 & - & - & - & - && 92 & 7 & - & - & - & - && 92 & 7 & - & - & - & - \\
HCO$^+$ && 91 & - & 6 & 2 & - & - && 90 & - & 6 & 2 & - & - && 90 & - & 7 & 2 & - & - \\
H$_2$CO && 8 & 87 & 1 & 2 & - & - && 7 & 79 & 5 & 5 & 1 & - && 6 & 64 & 19 & 8 & - & - \\
CH$_3$OH && 20 & 69 & - & - & 9 & - && 20 & 70 & - & - & 8 & - && 22 & 75 & 1 & - & - & - \\
NH$_3$ && 64 & 34 & - & - & - & - && 64 & 33 & - & - & 1 & - && 63 & 33 & - & - & 2 & - \\
H$_2$O$_2$ && 5 & 86 & 2 & 4 & - & - && 4 & 70 & 8 & 13 & 2 & - && 2 & 42 & 22 & 27 & 5 & - \\
OH && 31 & 13 & 24 & 24 & 6 & - && 14 & 10 & 37 & 31 & 6 & - && 6 & 13 & 54 & 22 & 2 & - \\
O$_2$ && 6 & 89 & 1 & 2 & - & - && 6 & 86 & 2 & 3 & - & - && 6 & 88 & 1 & 3 & - & - \\
O$_3$ && - & 99 & - & - & - & - && - & 99 & - & - & - & - && - & 99 & - & - & - & - \\
NO && 1 & 51 & - & - & 46 & - && 1 & 53 & - & - & 44 & - && 1 & 59 & - & - & 37 & - \\
HNO && 4 & 88 & 1 & 4 & - & - && 3 & 76 & 5 & 11 & 2 & - && 2 & 57 & 13 & 22 & 3 & - \\
HNCO && 2 & 84 & 3 & 5 & 3 & - && 1 & 64 & 9 & 14 & 8 & - && 1 & 34 & 18 & 31 & 13 & - \\
OCN && 6 & 92 & - & - & - & - && 5 & 91 & - & - & - & 2 && 4 & 75 & - & 7 & 4 & 7 \\
SO && - & 35 & - & - & 48 & 15 && - & 46 & - & - & 50 & 2 && - & 97 & - & - & 1 & - \\
SO$_2$ && - & 54 & - & - & 39 & 4 && - & 49 & - & - & 49 & - && - & 68 & 1 & - & 28 & - \\
HCS && - & 23 & - & - & 50 & 25 && - & 23 & - & - & 49 & 27 && - & 23 & - & - & 53 & 22 \\
H$_2$CS && 3 & - & - & - & 88 & 8 && 1 & - & - & - & 70 & 27 && - & - & - & - & 77 & 21 \\
CH$_3$SH && 35 & 29 & 35 & - & - & - && 34 & 28 & 35 & - & - & - && 30 & 26 & 37 & 6 & - & - \\
C$_2$H$_5$SH && 24 & 63 & 4 & 5 & 1 & - && 21 & 60 & 8 & 7 & 1 & - && 16 & 60 & 18 & 3 & 1 & - \\
C$_2$H && 98 & - & - & - & - & - && 93 & - & 2 & 3 & - & - && 77 & - & 8 & 11 & 2 & - \\
C$_2$H$_2$ && 11 & 42 & - & - & 12 & 31 && 10 & 39 & - & 2 & 35 & 10 && 6 & 23 & 4 & 8 & 26 & 31 \\
C$_2$H$_4$ && 23 & 19 & 3 & 2 & 12 & 38 && 21 & 20 & 9 & 7 & 20 & 19 && 12 & 17 & 17 & 17 & 11 & 22 \\
C$_2$H$_6$ && 95 & - & 2 & 1 & - & - && 85 & - & 9 & 3 & 1 & - && 63 & - & 28 & 7 & - & - \\
C$_3$H$_6$ && 8 & 17 & 1 & 1 & 42 & 28 && 3 & 8 & 2 & 2 & 31 & 51 && 1 & 5 & 5 & 3 & 18 & 66 \\
C$_3$H$_8$ && 71 & 26 & - & - & 1 & - && 69 & 26 & - & - & 3 & - && 66 & 24 & - & - & 8 & - \\
CN && 85 & 1 & 6 & 4 & 1 & - && 65 & 3 & 16 & 12 & 2 & - && 36 & 5 & 34 & 21 & 2 & - \\
HCN && 39 & 1 & 1 & - & 37 & 19 && 7 & - & - & - & 21 & 69 && - & - & - & - & 4 & 94 \\
HNC && 41 & 34 & 3 & 3 & 17 & - && 33 & 27 & 4 & 4 & 29 & - && 34 & 28 & 6 & 3 & 27 & - \\
H$_2$CN && 32 & 55 & 8 & 3 & - & - && 25 & 48 & 18 & 7 & 1 & - && 17 & 44 & 26 & 10 & 1 & - \\
CH$_3$CN && 41 & 14 & - & - & 13 & 28 && 9 & 3 & - & - & 30 & 55 && - & - & - & - & 38 & 58 \\
CH$_3$NC && 94 & 1 & - & - & 3 & - && 95 & 1 & - & - & 2 & - && 96 & 1 & - & - & - & - \\
HC$_3$N && 2 & 24 & - & - & 13 & 60 && - & 2 & - & - & 14 & 81 && - & - & - & - & 35 & 64 \\
C$_2$H$_3$CN && - & 70 & - & 1 & 6 & 20 && - & 42 & - & 1 & 12 & 42 && - & 22 & - & - & 26 & 50 \\
C$_2$H$_5$CN && 13 & 86 & - & - & - & - && 13 & 84 & - & - & - & - && 10 & 66 & 2 & 11 & 8 & - \\
n-C$_3$H$_7$CN && 29 & 1 & - & - & 68 & - && 9 & - & - & - & 89 & - && 3 & - & - & 1 & 82 & 12 \\
i-C$_3$H$_7$CN && 39 & 18 & - & 2 & 39 & - && 21 & 9 & - & 4 & 63 & - && 15 & 7 & 7 & 10 & 58 & - \\
n-C$_4$H$_9$CN && 1 & - & - & - & 97 & - && 1 & - & - & - & 97 & - && - & - & - & - & 95 & 3 \\
i-C$_4$H$_9$CN && 13 & - & - & - & 85 & - && 1 & - & - & - & 95 & 2 && - & - & - & - & 96 & 2 \\
s-C$_4$H$_9$CN && - & - & - & - & 99 & - && - & - & - & - & 98 & - && 1 & 1 & - & 8 & 87 & - \\
t-C$_4$H$_9$CN && 94 & 3 & - & - & 1 & - && 91 & 3 & 2 & 1 & 1 & - && 67 & 4 & 11 & 10 & 6 & - \\
CH$_2$CO && 27 & - & - & 33 & 38 & - && 9 & - & - & 42 & 47 & - && 2 & - & - & 50 & 46 & - \\
CH$_3$CHO && 41 & 9 & 3 & 2 & 28 & 15 && 21 & 5 & 5 & 5 & 27 & 35 && 6 & 2 & 10 & 12 & 45 & 23 \\
c-C$_2$H$_4$O && 33 & 46 & 5 & 14 & - & - && 19 & 27 & 7 & 45 & - & - && 12 & 18 & 6 & 62 & - & - \\
C$_2$H$_3$OH && 42 & 5 & - & - & 36 & 14 && 22 & 2 & 1 & 1 & 52 & 19 && 9 & 1 & 3 & 3 & 66 & 14 \\
C$_2$H$_5$CHO && 32 & 64 & 1 & - & 2 & - && 31 & 64 & 2 & - & 1 & - && 27 & 62 & 8 & 1 & - & - \\
CH$_2$NH && 1 & 1 & - & - & 32 & 63 && 1 & 1 & 1 & - & 70 & 24 && - & 2 & 4 & 3 & 73 & 15 \\
CH$_3$NH$_2$ && 99 & - & - & - & - & - && 98 & - & - & - & - & - && 93 & - & 3 & 2 & - & - \\
NH$_2$CHO && - & 60 & 3 & 7 & 23 & 3 && - & 44 & 10 & 22 & 22 & - && - & 19 & 18 & 45 & 15 & - \\
NH$_2$CN && 99 & - & - & - & - & - && 97 & - & - & 1 & - & - && 69 & - & 10 & 14 & 5 & - \\
NH$_2$OH && 39 & 58 & - & - & 1 & - && 38 & 58 & - & - & 1 & - && 38 & 58 & 3 & - & - & - \\
CH$_3$NCO && - & 12 & 4 & 83 & - & - && 1 & 40 & 22 & 30 & 4 & - && - & 18 & 29 & 38 & 12 & - \\
CH$_3$NHCHO && 34 & 18 & 2 & 39 & 4 & - && 30 & 17 & 4 & 43 & 4 & - && 35 & 24 & 16 & 6 & 16 & - \\
C$_2$H$_5$OH && 93 & 5 & - & - & - & - && 93 & 5 & - & - & - & - && 90 & 5 & 1 & 1 & - & - \\
CH$_3$OCH$_3$ && 68 & 4 & 3 & 2 & 10 & 11 && 44 & 3 & 7 & 4 & 21 & 18 && 36 & 3 & 22 & 9 & 28 & - \\
HCOOCH$_3$ && 1 & 59 & 2 & 1 & 21 & 13 && 1 & 56 & 5 & 4 & 31 & - && 1 & 44 & 15 & 6 & 30 & - \\
HCOOC$_2$H$_5$ && 18 & 75 & 3 & 1 & - & - && 16 & 66 & 10 & 4 & 2 & - && 9 & 49 & 30 & 8 & 1 & - \\
HCOOH && 2 & 95 & 1 & - & - & - && 2 & 92 & 3 & 1 & - & - && 2 & 86 & 8 & 2 & - & - \\
CH$_3$COOH && 60 & 37 & - & - & 1 & - && 56 & 35 & 1 & 1 & 4 & - && 38 & 25 & 7 & 13 & 16 & - \\
C$_2$H$_5$COOH && 63 & 23 & - & 1 & 11 & - && 51 & 19 & - & - & 27 & - && 58 & 23 & 5 & 2 & 9 & - \\
CH$_3$COCHO && 3 & 58 & 35 & - & - & - && 3 & 48 & 43 & 3 & 2 & - && 2 & 38 & 45 & 10 & 4 & - \\
(CH$_3$)$_2$CO && 90 & - & 5 & 3 & - & - && 77 & - & 21 & - & - & - && 37 & - & 61 & - & - & - \\
CH$_3$CONH$_2$ && 64 & 31 & 2 & - & - & - && 61 & 30 & 3 & 3 & 1 & - && 36 & 20 & 12 & 20 & 10 & - \\
CH$_3$OCOCH$_3$ && 12 & 86 & - & - & - & - && 11 & 82 & 2 & 2 & 1 & - && 8 & 68 & 13 & 6 & 3 & - \\
CH$_2$(OH)COCH$_3$ && 10 & 70 & 3 & 10 & 6 & - && 8 & 61 & 5 & 6 & 17 & - && 6 & 49 & 15 & 14 & 14 & - \\
HOCOOH && 3 & 95 & - & - & - & - && 3 & 94 & 1 & - & - & - && 3 & 89 & 4 & 2 & 1 & - \\
HCOCOCHO && - & 96 & 3 & - & - & - && - & 99 & - & - & - & - && - & 99 & - & - & - & - \\
OHCCHO && - & 94 & - & - & 4 & 1 && - & 89 & - & - & 10 & - && - & 74 & - & - & 25 & - \\
OHCCOOH && - & 98 & - & - & - & - && - & 97 & 1 & - & - & - && - & 92 & 4 & 2 & - & - \\
NH$_2$COCHO && - & 95 & - & - & 2 & - && - & 95 & 1 & - & 2 & - && - & 91 & 4 & 2 & 1 & - \\
NH$_2$COOH && 2 & 91 & - & - & 5 & - && 2 & 89 & 2 & 1 & 4 & - && 1 & 73 & 7 & 9 & 7 & - \\
NH$_2$CONH$_2$ && 3 & 78 & - & - & 15 & - && 3 & 76 & 3 & 3 & 12 & - && 2 & 51 & 11 & 16 & 17 & - \\
NH$_2$CH$_2$CHO && 81 & 10 & 1 & - & 6 & - && 76 & 10 & 3 & 1 & 8 & - && 62 & 11 & 13 & 7 & 4 & - \\
NH$_2$CH$_2$CN && 88 & 5 & - & - & 4 & - && 78 & 4 & - & 2 & 13 & - && 60 & 4 & 9 & 11 & 14 & - \\
NH$_2$CH$_2$COOH && 86 & 3 & - & - & 9 & - && 69 & 2 & - & - & 27 & - && 76 & 3 & - & - & 19 & - \\
NH$_2$C$_2$H$_5$ && 96 & - & - & - & 2 & - && 88 & - & - & - & 10 & - && 98 & - & - & - & 1 & - \\
N$_2$H$_4$ && 91 & 4 & 1 & 2 & - & - && 66 & 3 & 11 & 14 & 4 & - && 16 & 2 & 33 & 37 & 10 & - \\
CH$_3$ONH$_2$ && 11 & 85 & - & - & 2 & - && 11 & 82 & 1 & - & 4 & - && 10 & 79 & 5 & 1 & 2 & - \\
CH$_2$(OH)NH$_2$ && 16 & 78 & - & - & 4 & - && 16 & 78 & - & - & 4 & - && 15 & 75 & 5 & 2 & 1 & - \\
CH$_3$OCOCHO && - & 99 & - & - & - & - && - & 99 & - & - & - & - && - & 99 & - & - & - & - \\
CH$_3$OCOOH && - & 98 & - & - & - & - && - & 97 & 1 & - & - & - && - & 94 & 3 & - & - & - \\
CH$_3$OCONH$_2$ && - & 94 & - & - & 3 & - && - & 96 & - & - & 1 & - && - & 93 & 3 & 1 & - & - \\
(CH$_3$O)$_2$CO && - & 99 & - & - & - & - && - & 99 & - & - & - & - && - & 99 & - & - & - & - \\
CH$_3$OCOCH$_2$OH && - & 99 & - & - & - & - && - & 99 & - & - & - & - && - & 99 & - & - & - & - \\
CH$_3$OOH && 5 & 93 & - & - & - & - && 5 & 92 & 1 & - & - & - && 5 & 88 & 6 & - & - & - \\
(CH$_3$O)$_2$ && - & 99 & - & - & - & - && - & 99 & - & - & - & - && - & 99 & - & - & - & - \\
CH$_3$OCH$_2$OH && - & 99 & - & - & - & - && - & 99 & - & - & - & - && - & 98 & - & - & - & - \\
(CH$_2$OH)$_2$ && - & 53 & 1 & 1 & 42 & - && - & 55 & 5 & 5 & 32 & - && - & 53 & 20 & 10 & 14 & - \\
CH$_2$(OH)CHO && - & 77 & - & 1 & 16 & 3 && - & 61 & 1 & 3 & 25 & 8 && - & 74 & 4 & 7 & 13 & - \\
CH$_2$(OH)COCHO && - & 87 & - & - & 7 & 4 && - & 93 & - & - & 6 & - && - & 98 & - & - & - & - \\
CH$_2$(OH)COOH && - & 99 & - & - & - & - && - & 98 & - & - & - & - && - & 91 & 4 & 3 & - & - \\
CH$_2$(OH)CONH$_2$ && - & 74 & - & - & 24 & - && - & 74 & - & 1 & 23 & - && - & 80 & 4 & 5 & 8 & - \\
(CH$_2$OH)$_2$CO && - & 54 & - & - & 45 & - && - & 73 & - & - & 26 & - && - & 99 & - & - & - & - \\
CH$_2$(OH)$_2$ && 8 & 90 & - & - & - & - && 8 & 88 & 2 & - & - & - && 7 & 78 & 9 & 3 & - & - \\
\enddata
\end{deluxetable}

\end{document}